\documentclass[11pt,a4paper,notitlepage]{report}

\usepackage[utf8]{inputenc}
\usepackage{amsmath,amssymb,amsfonts}
\newcommand{\ket}[1]{\left|#1\right\rangle}
\newcommand{\bra}[1]{\left\langle#1\right|}
\DeclareMathOperator{\Tr}{Tr}
\usepackage{bm}
\usepackage{geometry}
\usepackage{setspace}
\usepackage{hyperref}
\usepackage{tikz}
\usepackage{caption}
\usepackage{enumitem}
\usepackage{mathtools}

\usepackage{tabularx}
\usepackage{array}

\usepackage{xcolor}

\usetikzlibrary{arrows.meta,calc,positioning,decorations.pathmorphing,decorations.pathreplacing,decorations.markings}
\newcommand{\Od}{\Omega_{d-1}}

\newcommand{\Seff}{S_{\mathrm{eff}}}

\newcommand{\braket}[2]{\left\langle #1\,\middle|\,#2\right\rangle}

\def\be{\begin{equation}}
\def\ee{\end{equation}}

\newcommand{\cO}{\mathcal{O}}
\newcommand{\detp}{\det\nolimits{}^{\prime}}
\def\sn{{\rm sn}}
\def\cn{{\rm cn}}
\def\dn{{\rm dn}}
\def\ELE{{\mathbb{E}}}
\def\KK{{\mathbb{K}}}
\newcommand{\ba}{\begin{eqnarray}}
\newcommand{\ea}{\end{eqnarray}}

\title{\textbf{\huge Lectures on Semiclassical Methods}\\[10pt]
{\LARGE\textit{for}}\\[10pt]
\textbf{\huge Composite Operators}}
\author{Francesco Sannino\thanks{\texttt{sannino@qtc.sdu.dk}}\\[6pt]
    \normalsize\textit{Quantum Theory Center ($\hbar$QTC) \& D-IAS, University of Southern Denmark, Odense, Denmark}\\[2pt]
   \normalsize\textit{Dipartimento di Fisica ``E.~Pancini'', Universit\`a di Napoli Federico~II, Napoli, Italy}}
\date{}

\begin{document}
\maketitle

\begin{abstract}
\noindent
These  lecture notes are intended as a coherent introduction to
conformal field theory in general, and composite operators in particular,
through a semiclassical framework for computing scaling dimensions, with  
emphasis on operators of the form $\phi^n$. In doing so, they aim to fill a gap in the  
literature and to help decode some of the relevant concepts. The
physical idea is that at large $n$ an (heavy) operator creates a highly occupied state. Through the
state--operator correspondence, this state lives on the cylinder
$\mathbb{R}\times S^{d-1}$, and its scaling dimension is the corresponding energy of the theory on the cylinder.  The notes are organized as a self-contained route from conformal symmetry to
semiclassical dynamics. 
 Part~I reviews the conformal group, primary operators,
radial quantization, the state--operator correspondence, and operator mixing.
Part~II builds the semiclassical framework, first in the free scalar theory, where
the dimension of $\phi^n$ is recovered in three independent ways, and then through
the double-scaling limit, the action variable, and Bohr--Sommerfeld quantization.
Part~III develops the general machinery of periodic saddles, Floquet theory,
fluctuation determinants, the Gel'fand--Yaglom method, and the Gutzwiller trace
formula. Part~IV applies the framework to the $O(N)$ $\phi^4$ theory in
$d=4-\epsilon$ at the Wilson--Fisher fixed point, deriving the classical elliptic
solution, the Lam\'e fluctuation spectrum, the zero modes, and the one-loop
contribution to the large-$n$ scaling dimensions. 
Beyond the explicit computation, the notes emphasize the role of composite
operators as probes of collective sectors of quantum field theory, with extensions
to gauge theories, conformal windows, and asymptotically safe field theories.
\end{abstract}
\tableofcontents
\newpage

\chapter{ Quantum Fields, Criticality, and Phases of Matter}
\label{chap:why_these_notes_matter}

Quantum field theory is the language of elementary
particles: the Standard Model, our best description of
the fundamental interactions, is a quantum field theory. However it is also the language of collective phenomena: it
surfaces whenever a system with many microscopic degrees of freedom organises
itself into a simple long-distance pattern. It is, in this sense, not only a
theory of the smallest constituents of matter but also a theory of emergence.

The aim of these notes is to explain a set of ideas that sit at the
intersection of conformal symmetry, semiclassical physics, and the theory of
critical phenomena.  At a technical level, we shall learn how to compute
scaling dimensions of composite operators, especially in regimes where the
operators are heavy and ordinary perturbation theory is not the most natural
language.  But the broader purpose is more conceptual.  Scaling dimensions,
operator spectra, and correlation functions are not merely formal quantities.
They encode the response of a quantum many-body system to disturbances, the
possible phases of matter, the way order is established or lost, and the
universal properties of phase transitions.

Consider the Ising model \cite{Lenz:1920,Ising:1925,Peierls:1936,Kramers:1941a,Kramers:1941b,Onsager:1944zz}
: its microscopic formulation relies on discrete lattice spin variables subject to nearest-neighbor interactions. At low temperature the spins align into an
ordered magnetic phase; at high temperature thermal noise wins and the system
is disordered. In between sits a critical point, and as one approaches it most
of the microscopic detail simply stops mattering: the particular lattice,
the short-distance couplings, the chemistry of the material. As the system approaches its critical temperature, an emerging divergent correlation length eradicates intrinsic scales, leading to a universal continuum theory.  At  unitary relativistic critical points scale invariance is enhanced to conformal invariance, and the infrared theory is
described by a CFT,  wherein the specific microscopic details become irrelevant to the long-distance physics.
 That is the lesson of universality: systems that
look nothing alike up close can share the same critical exponents, the same
scaling laws, and the same long-distance correlation
functions~ \cite{DiFrancesco:1997nk,YellowBook:1988,Ginsparg:1988ui,Blumenhagen:2009zz}, see also more recent reviews~\cite{arXiv:1511.04074,arXiv:2207.09474,arXiv:1907.05147}.

Conformal field theory is the sharpest language we have for that universal
regime. At a second-order transition the correlation length diverges and the
system loses any intrinsic scale; scale invariance, usually enlarged to full
conformal invariance, then fixes much of the structure. The local
operators of the CFT are the ways one can disturb the critical system: raise
the temperature, switch on a magnetic field, probe the energy density, or excite
some more elaborate composite {operator}. Their scaling dimensions say how each disturbance
grows or fades with distance, so the operator spectrum is, in effect, a
fingerprint of the universality class~\cite{arXiv:1602.07982}.

Computing scaling dimensions is therefore physically relevant. The dimensions of
the low-lying operators fix the familiar critical exponents. In the Ising
class, for instance, the spin and energy operators govern the magnetic and
thermal response. Heavier, more intricate operators encode higher moments,
composite fluctuations, anisotropies, and deformations away from the critical
point. Knowing the spectrum is the same as knowing every way the system can be
pushed off criticality: which perturbations are relevant, which irrelevant,
which marginal. That classification is the backbone of the
renormalisation-group picture of phases of matter.

The same story plays out well beyond Ising. The $O(N)$ universality classes
describe order parameters with a continuous symmetry: $N=2$ covers
superfluids and planar magnets, $N=3$ the Heisenberg magnet, and large $N$
gives a clean laboratory for collective behaviour. A scalar with a quartic
interaction is trivial to write down yet captures a great deal of physics: how
an ordered phase sets in, how Goldstone modes appear when a continuous symmetry
breaks, how fluctuations spoil the naive mean-field estimate. Here the CFT lives
at the critical point between phases, and the renormalisation group explains why
so many different microscopic systems flow to it. Quantum magnets and
antiferromagnets tell a parallel tale: the microscopic actors are quantum spins,
but the long-distance physics is carried by collective order-parameter fields,
and depending on the dimension and symmetry one finds ordered, disordered, or
topological phases and the critical points between them. The questions are
always the same: what is the operator spectrum, what are the correlators,
which perturbations destabilise the fixed
point~\cite{arXiv:0711.3015,arXiv:1008.3477,Pelissetto:2000ek}?

These methods come into their own when an operator is heavy, built from a large
number of fields, so that the state it creates is highly occupied. A large
occupation number lets the dominant configuration behave classically even though
the theory underneath is fully quantum: to leading order the state is a classical
field configuration, and the quantum corrections come from the fluctuations
around it. This is the same semiclassical logic that underlies the WKB
approximation and the treatment of solitons.

Our application is to heavy composite operators. An operator like $\phi^n$ is
more than a product of fields: through the state--operator correspondence it
defines a state on the cylinder, and when $n$ is large that state is heavily
occupied. Computing its scaling dimension turns into computing the energy of a
many-particle state on the cylinder, and in favourable limits that energy
follows from a classical periodic solution together with the fluctuations around
it. An abstract piece of conformal data becomes a concrete dynamical problem.

That last step rests on the state--operator correspondence, which works as a theoretical bridge: a local operator inserted at the
origin of flat space becomes a state on the cylinder, a scaling dimension
becomes an energy, a correlator becomes a transition amplitude, and the
renormalisation-group notion of scaling becomes ordinary time evolution along
the cylinder. With that dictionary in hand we can bring quantum-mechanical,
semiclassical, and spectral intuition to bear on the structure of a CFT.

The payoff is sharpest for heavy operators. Ordinary perturbation theory
expands around a handful of quanta; a heavy operator creates a state with many,
and its physics is closer to a collective configuration than to a few-particle
excitation. The natural reorganisation is a double-scaling limit, with the
number of fields large, the microscopic coupling small, and their product held
fixed. What
comes out is neither the usual weak-coupling series nor a full non-perturbative
solution, but a controlled semiclassical expansion tailored to a densely
populated sector of the theory.

{A related  programme --- the \emph{large-charge
expansion}~\cite{arXiv:1505.01537,arXiv:1611.02912,arXiv:2008.03308} ---
addresses operators carrying a large conserved global charge $Q$ in CFTs
with a $U(1)$ (or higher) global symmetry.  There the classical saddle on
the cylinder is a time-independent, charge-stabilised superfluid
configuration $\phi \sim e^{i\mu\tau}$, and the $1/Q$ expansion follows
from the Goldstone effective field theory of the spontaneously broken
symmetry.  The present notes treat the complementary case of \emph{neutral}
composite operators $\mathcal{O}_n\sim(\phi_a\phi_a)^{n/2}$: without a
conserved charge to stabilise a static saddle, the classical solution is
the time-dependent periodic orbit of Parts~III--IV, and the role of $Q$ is
played by the Bohr--Sommerfeld integer $n = I(E)/(2\pi)$.  The two
programmes share the cylinder geometry and the leading power
$\Delta\sim n^{d/(d-1)}$, but require complementary semiclassical machinery.}

This is also why the subleading terms are worth the effort. The leading
classical piece picks out the dominant collective configuration; the one-loop
determinant measures whether that configuration is quantum-mechanically stable.
The fluctuations carry the spectrum of small disturbances around the saddle,
and they decide how robust the state is, how degeneracies split, and how
universal quantum effects correct the classical answer. In the vocabulary of quantum
matter, this is the step from a mean-field description of an ordered state to
the fluctuation-corrected description that holds the true universal physics.

The reach of these methods is therefore wider than the few coefficients of an
anomalous dimension. They give a way to organise the densely populated sectors
of a quantum field theory, tying conformal data to classical dynamics, periodic
orbits, fluctuation operators, and spectral determinants. They also extend
naturally to settings we will not treat in detail: gauge theories, where
composite operators are the natural observables (bound states, currents, the
stress tensor, order parameters), and other families of heavy operators. Heavy composite operators are not exotic; they are simply probes of
the multi-particle, collective, semiclassical regimes that standard perturbation
theory reaches only with difficulty.

It is best, then, not to read the coming chapters as a string of separate
calculations. Deriving the conformal generators, proving the state--operator
correspondence, counting Wick contractions, building semiclassical saddles,
evaluating fluctuation determinants: these are facets of one picture, showing
how local operators encode the physics of quantum matter and how a heavy
operator turns into a classical configuration whose spectrum carries universal
information about phases of matter. The shift in viewpoint is the real content
of the notes: from asking how to correct a single operator order by order, to
asking how the whole space of operators organises the behaviour of a quantum
system, with conformal field theory as the common language and semiclassics as
a systematic tool for its heavily populated sectors.

\section{How to Navigate These Lectures}
\label{sec:how_to_navigate_lectures}

These notes can be read at more than one level. Someone meeting conformal field
theory for the first time can take them as a gradual introduction to the
operator language of critical phenomena. A reader who already knows CFT can skip
ahead to the semiclassical construction of heavy operators. Someone coming from
quantum matter can treat them as a map between phases, critical points, and the
operator data that labels them. They can also be read three ways at once: as a
CFT course for critical phenomena, where the cast is primary operators,
correlators, scaling dimensions, and relevant deformations; as a course on
semiclassical methods, where it is heavy operators, large occupation numbers,
classical saddles, periodic orbits, and fluctuation determinants; or as a bridge
between the particle-physics and condensed-matter viewpoints, since the same
data a particle physicist reads as composite operators and anomalous dimensions
a condensed-matter physicist reads as universal responses, phase stability, and
collective excitations. The rest of this section explains how the five parts fit
together.

\textbf{Part~I} builds the conformal-field-theory dictionary. We start from
conformal transformations and the conformal algebra, work through primary
operators, descendants, and the constraints conformal symmetry imposes on
correlators, and then establish the state--operator correspondence and the
treatment of composite operators and their mixing. This algebraic structure is strictly necessary to demonstrate the restrictive nature of conformal invariance at criticality. The conformal algebra establishes the foundational constraints on correlation functions and provides the geometric justification for radial quantization. Specifically, a local operator inserted at the origin of flat Euclidean space maps directly to a state on the cylinder 
\begin{equation}
  \mathcal{O}(0)
  \quad \longleftrightarrow \quad
  \ket{\mathcal{O}} ,
\end{equation}
and its scaling dimension becomes the energy of that state,
\begin{equation}
  \Delta_{\mathcal{O}} = R\,E_{\mathcal{O}} ,
\end{equation}
with $R$ the radius of the spatial sphere. (We usually set $R=1$, so that
$\Delta=E$.) This turns a question about local probes of a critical system into
a question about the spectrum of a quantum theory on a sphere, the bridge
between conformal kinematics and semiclassical dynamics.

\textbf{Part~II} sets up the semiclassical \emph{canovaccio}.This part  
establishes the rigorous semiclassical framework. We first validate the methodology on the free scalar theory, demonstrating that the scaling dimension of $\phi^n$ can be derived consistently via three independent methods: direct Wick contractions, cylinder Hamiltonian diagonalization, and flat-space saddle point evaluation. The convergence of these methods justifies the subsequent extension to the non-linear interacting theory. We then lay out the interacting blueprint that carries
us through the rest of the notes: the double-scaling limit, the action variable,
and the Legendre transform. In the free theory the underlying
classical motion is harmonic; switching on the interaction makes it non-linear.

\textbf{Part~III} is the semiclassical core, where the focus moves from light to
heavy operators. A heavy $\phi^n$ creates a highly occupied state on the
cylinder, and that large occupation number is the semiclassical parameter. The
leading scaling dimension comes from a classical periodic solution and the first
correction from the fluctuations around it,
\begin{equation}
  \Delta_n
  =
  \Delta_n^{\rm cl}
  +
  \Delta_n^{\rm 1-loop}
  +
  \cdots ,
\end{equation}
the coefficients depending on the theory and on the scaling limit. Quantising
those fluctuations is the technical heart of the part: fluctuation operators,
stability, spectral determinants, and, because the background is periodic,
Floquet theory, the Gel'fand--Yaglom construction, and the Gutzwiller trace
formula. The job is to measure how the
quantum theory responds to small disturbances of the collective configuration,
that is, the corrections beyond mean field.

\textbf{Part~IV} puts the construction to work on the critical $O(N)$ $\phi^4$
theory in $d=4-\epsilon$. These models are the standard examples of
continuous-symmetry universality classes and also a controlled setting for heavy
operators. The non-linear periodic motion, the Jacobi elliptic functions, and
the Lam\'e fluctuation spectrum that appear are not incidental complications.
They are the natural language of collective excitations in an interacting field
theory, and they deliver the coefficients $C_0$ and $C_1$ explicitly.

The Conclusions chapter closes Part~IV: it retraces the whole argument and draws
it together in a single pipeline diagram, from the state--operator map to the
scaling dimension. \textbf{Part~V} then gathers the supporting material in four
appendices: the identity components of the conformal group, the conformal scalar
and its effective mass on the cylinder, the flat metric in spherical
coordinates, and the classical action of the $\cn$ saddle underlying the
$\phi^4$ computation.

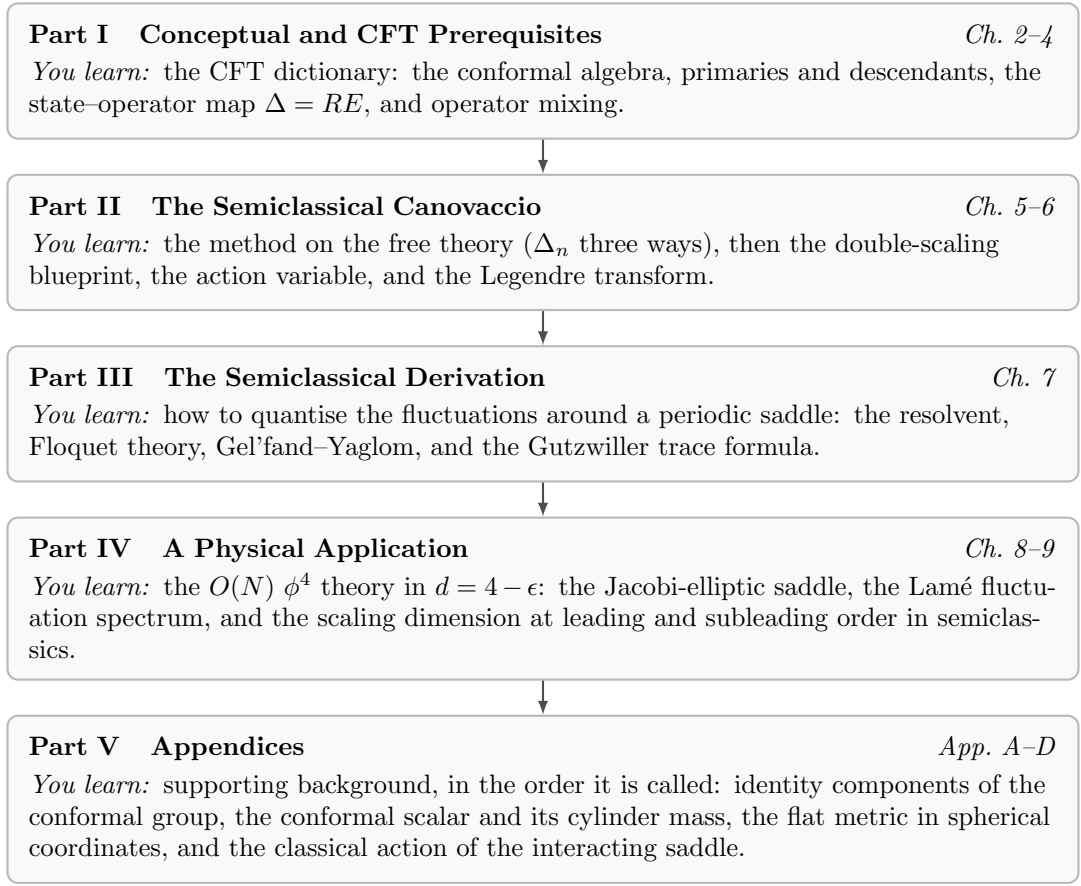
\begin{figure}[htbp]
\centering
\begin{tikzpicture}[
  partband/.style={draw=gray!55, fill=gray!5, thick, rounded corners,
    text width=13.6cm, align=left, inner sep=8pt, font=\small},
  arr/.style={->, thick, >=latex, gray!60!black},
]
\node[partband] (p1) {%
  \textbf{Part~I\quad Conceptual and CFT Prerequisites}\hfill\textit{Ch.\ 2--4}\\[2pt]
  \emph{You learn:} the CFT dictionary: the conformal algebra, primaries and
  descendants, the state--operator map $\Delta=RE$, and operator mixing.};
\node[partband] (p2) [below=0.45cm of p1] {%
  \textbf{Part~II\quad The Semiclassical Canovaccio}\hfill\textit{Ch.\ 5--6}\\[2pt]
  \emph{You learn:} the method on the free theory ($\Delta_n$ three ways), then
  the double-scaling blueprint, the action variable, and the Legendre transform.};
\node[partband] (p3) [below=0.45cm of p2] {%
  \textbf{Part~III\quad The Semiclassical Derivation}\hfill\textit{Ch.\ 7}\\[2pt]
  \emph{You learn:} how to quantise the fluctuations around a periodic saddle:
  the resolvent, Floquet theory, Gel'fand--Yaglom, and the Gutzwiller trace
  formula.};
\node[partband] (p4) [below=0.45cm of p3] {%
  \textbf{Part~IV\quad A Physical Application}\hfill\textit{Ch.\ 8--9}\\[2pt]
  \emph{You learn:} the $O(N)$ $\phi^4$ theory in $d=4-\epsilon$: the
  Jacobi-elliptic saddle, the Lam\'e fluctuation spectrum, and the scaling
  dimension at leading and subleading order in semiclassics.};
\node[partband] (p5) [below=0.45cm of p4] {%
  \textbf{Part~V\quad Appendices}\hfill\textit{App.\ A--D}\\[2pt]
  \emph{You learn:} supporting background, in the order it is called: identity
  components of the conformal group, the conformal scalar and its cylinder mass,
  the flat metric in spherical coordinates, and the classical action of the interacting  saddle.};
\draw[arr] (p1) -- (p2);
\draw[arr] (p2) -- (p3);
\draw[arr] (p3) -- (p4);
\draw[arr] (p4) -- (p5);
\end{tikzpicture}
\caption{The five parts of these notes and what each delivers. Parts~I--II build
the language and the method, Part~III carries out the semiclassical derivation,
Part~IV applies it to a concrete theory and closes with the Conclusions, and
Part~V provides the supporting appendices.}
\label{fig:ch1_parts}
\end{figure}

Figure~\ref{fig:ch1_parts} gathers the five parts in one place, with the
chapters each one spans and what it delivers.

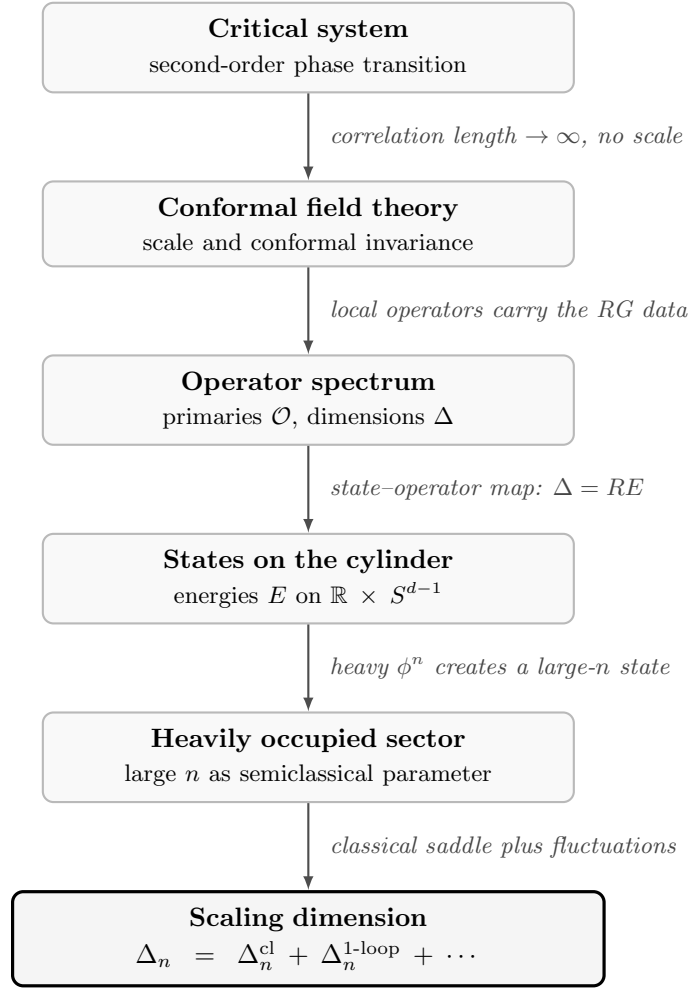
\begin{figure}[htbp]
\centering
\begin{tikzpicture}[
  stage/.style={draw=gray!55, fill=gray!5, thick, rounded corners,
                text width=6.6cm, align=center, inner sep=6pt, font=\small},
  result/.style={draw=black, fill=gray!8, very thick, rounded corners,
                 text width=7.4cm, align=center, inner sep=6pt, font=\small},
  arr/.style={->, thick, >=latex, gray!60!black},
  lab/.style={font=\footnotesize\itshape, text=gray!55!black, anchor=west},
]
\node[stage]  (s1) {\textbf{Critical system}\\[1pt]\footnotesize second-order phase transition};
\node[stage]  (s2) [below=1.15cm of s1] {\textbf{Conformal field theory}\\[1pt]\footnotesize scale and conformal invariance};
\node[stage]  (s3) [below=1.15cm of s2] {\textbf{Operator spectrum}\\[1pt]\footnotesize primaries $\mathcal{O}$, dimensions $\Delta$};
\node[stage]  (s4) [below=1.15cm of s3] {\textbf{States on the cylinder}\\[1pt]\footnotesize energies $E$ on $\mathbb{R}\times S^{d-1}$};
\node[stage]  (s5) [below=1.15cm of s4] {\textbf{Heavily occupied sector}\\[1pt]\footnotesize large $n$ as semiclassical parameter};
\node[result] (s6) [below=1.15cm of s5]
  {\textbf{Scaling dimension}\\[2pt]$\Delta_n=\Delta_n^{\rm cl}+\Delta_n^{\rm 1\text{-}loop}+\cdots$};

\draw[arr]  (s1) -- (s2) node[lab,midway,xshift=4pt] {correlation length $\to\infty$, no scale};
\draw[arr]  (s2) -- (s3) node[lab,midway,xshift=4pt] {local operators carry the RG data};
\draw[arr]  (s3) -- (s4) node[lab,midway,xshift=4pt] {state--operator map: $\Delta=RE$};
\draw[arr]  (s4) -- (s5) node[lab,midway,xshift=4pt] {heavy $\phi^n$ creates a large-$n$ state};
\draw[arr]  (s5) -- (s6) node[lab,midway,xshift=4pt] {classical saddle plus fluctuations};
\end{tikzpicture}
\caption{The logical spine of these notes: from a critical system to the
scaling dimension of a heavy composite operator. Each step is the subject of a
later chapter.}
\label{fig:ch1_pipeline}
\end{figure}

A useful way to keep the logic in view is the diagram in
Figure~\ref{fig:ch1_pipeline}, which traces the descent from a critical system
to the scaling dimension of a heavy operator. Each arrow names the mechanism
that carries one stage into the next.

This pipeline is the backbone of the lectures, and each technical chapter
develops one link in it. So if the early material on conformal transformations,
renormalisation-group ideas, harmonic oscillators on the cylinder,
Bohr--Sommerfeld quantisation, elliptic functions, and fluctuation determinants
seems at first to belong to different subjects, it does not: every piece serves
one end, understanding how universal field-theoretic data emerge from collective
dynamics. The variety of methods is a measure of the problem's richness, not a
change of topic.

\part{Conceptual and CFT Prerequisites}

\chapter{Conformal Symmetry and Primary Operators}\label{chap:prereq}

\label{app:conformal}

This chapter provides a self-contained treatment of conformal symmetry in $d$
Euclidean dimensions and establishes the CFT dictionary that underpins the rest of these
notes.  We proceed in three stages.

We begin by deriving the conformal Killing equations and integrating them to obtain the
full conformal group: translations, rotations, dilations, and special conformal
transformations.  The corresponding generators close into the conformal algebra
$\mathfrak{so}(d+1,1)$, and we work out the commutation relations explicitly.

We then study the representations of this algebra relevant to quantum field theory.
A \emph{primary operator} $\mathcal{O}$ of spin $\ell$ and scaling dimension $\Delta$
is, by definition, annihilated by the special conformal generators $K_\mu$; all other
states in the multiplet---\emph{descendants}---are reached by acting with the momentum
generators $P_\mu$.  We derive the transformation law of primary operators under finite
conformal transformations and extract the strong constraints it places on two- and
three-point functions.

The chapter closes by establishing the \emph{state--operator correspondence}:
on the cylinder $\mathbb{R}_\tau\times S^{d-1}_R$ obtained by the Weyl rescaling
$\delta_{\mu\nu}\to  {(R/r)^2}\,\delta_{\mu\nu}$, every primary operator $\mathcal{O}$ of
dimension $\Delta$ maps to a state of energy $E = \Delta/R$.  This identification
$\Delta = RE$ is the bridge between the operator spectrum of the CFT and the
Hamiltonian mechanics problem studied in subsequent chapters~\cite{Cardy:1984epx}.

\section{Definition and infinitesimal generators}
\label{app:conf:def}

\subsection{Conformal maps}

A \emph{conformal transformation} of flat Euclidean $\mathbb{R}^d$ (with metric $\delta_{\mu\nu}$)
is a smooth map $x\mapsto x'(x)$ that preserves angles, i.e.\ that leaves the metric
\emph{invariant up to a local positive rescaling}:
\begin{equation}
  \frac{\partial x'^\mu}{\partial x^\rho}\,
  \frac{\partial x'^\nu}{\partial x^\sigma}\,
  \delta_{\mu\nu}
  = \Omega(x)^2\,\delta_{\rho\sigma}\,,
  \qquad \Omega(x)>0\,.
  \label{eq:conf_def}
\end{equation}
The function $\Omega(x)$ is called the \emph{conformal (or Weyl) factor}.

\subsection{Infinitesimal form and the conformal Killing equation}

We now derive the infinitesimal version of the finite conformal condition
\begin{equation}
  \frac{\partial x'^\mu}{\partial x^\rho}\,
  \frac{\partial x'^\nu}{\partial x^\sigma}\,
  \delta_{\mu\nu}
  =
  \Omega(x)^2\,\delta_{\rho\sigma},
  \qquad \Omega(x)>0 .
\end{equation}
Write the conformal transformation as a small deformation of the identity,
\begin{equation}
  x'^\mu = x^\mu + \epsilon^\mu(x),
  \qquad |\epsilon^\mu|\ll 1 .
\end{equation}
Then its Jacobian is
\begin{equation}
  \frac{\partial x'^\mu}{\partial x^\rho}
  =
  \frac{\partial}{\partial x^\rho}
  \left(x^\mu+\epsilon^\mu(x)\right)
  =
  \delta^\mu{}_\rho+\partial_\rho\epsilon^\mu .
\end{equation}
Substituting this into \eqref{eq:conf_def} gives
\begin{equation}
  \left(
    \delta^\mu{}_\rho+\partial_\rho\epsilon^\mu
  \right)
  \left(
    \delta^\nu{}_\sigma+\partial_\sigma\epsilon^\nu
  \right)
  \delta_{\mu\nu}
  =
  \Omega(x)^2\,\delta_{\rho\sigma}.
\end{equation}
Expanding the left-hand side to first order in \(\epsilon^\mu\), we find
\begin{align}
  &\left(
    \delta^\mu{}_\rho+\partial_\rho\epsilon^\mu
  \right)
  \left(
    \delta^\nu{}_\sigma+\partial_\sigma\epsilon^\nu
  \right)
  \delta_{\mu\nu}
  \nonumber\\
  &\qquad =
  \delta^\mu{}_\rho\delta^\nu{}_\sigma\delta_{\mu\nu}
  +
  \delta^\mu{}_\rho
  \partial_\sigma\epsilon^\nu
  \delta_{\mu\nu}
  +
  \partial_\rho\epsilon^\mu
  \delta^\nu{}_\sigma
  \delta_{\mu\nu}
  +
  \mathcal{O}(\epsilon^2).
\end{align}
The first term is simply
\begin{equation}
  \delta^\mu{}_\rho\delta^\nu{}_\sigma\delta_{\mu\nu}
  =
  \delta_{\rho\sigma}.
\end{equation}
For the two linear terms we lower the index on \(\epsilon^\mu\) with the flat
metric,
\begin{equation}
  \epsilon_\rho \equiv \delta_{\rho\nu}\epsilon^\nu .
\end{equation}
Thus
\begin{align}
  \delta^\mu{}_\rho
  \partial_\sigma\epsilon^\nu
  \delta_{\mu\nu}
  &=
  \delta_{\rho\nu}\partial_\sigma\epsilon^\nu
  =
  \partial_\sigma\epsilon_\rho,
  \\
  \partial_\rho\epsilon^\mu
  \delta^\nu{}_\sigma
  \delta_{\mu\nu}
  &=
  \delta_{\mu\sigma}\partial_\rho\epsilon^\mu
  =
  \partial_\rho\epsilon_\sigma .
\end{align}
Therefore the conformal condition becomes
\begin{equation}
  \delta_{\rho\sigma}
  +
  \partial_\rho\epsilon_\sigma
  +
  \partial_\sigma\epsilon_\rho
  =
  \Omega(x)^2\,\delta_{\rho\sigma}
  +
  \mathcal{O}(\epsilon^2).
\end{equation}

Since the transformation is infinitesimal, the Weyl factor is also close to
one.  We therefore write
\begin{equation}
  \Omega(x)^2 = 1+f(x),
\end{equation}
where \(f(x)\) is of first order in the infinitesimal transformation.  Hence
\begin{equation}
  \Omega(x)^2\delta_{\rho\sigma}
  =
  \left(1+f(x)\right)\delta_{\rho\sigma}
  =
  \delta_{\rho\sigma}+f(x)\delta_{\rho\sigma}.
\end{equation}
Cancelling the common zeroth-order term \(\delta_{\rho\sigma}\), and dropping
terms of order \(\mathcal{O}(\epsilon^2)\), gives
\begin{equation}
  \partial_\rho\epsilon_\sigma
  +
  \partial_\sigma\epsilon_\rho
  =
  f(x)\delta_{\rho\sigma}.
\end{equation}
Renaming the dummy indices \(\rho,\sigma\to\mu,\nu\), we obtain
\begin{equation}
  \partial_\mu\epsilon_\nu
  +
  \partial_\nu\epsilon_\mu
  =
  f(x)\delta_{\mu\nu}.
  \label{eq:CKE}
\end{equation}
This is the infinitesimal conformal Killing equation.

We now determine \(f(x)\) by taking the trace of \eqref{eq:CKE}.  Contract both
sides with \(\delta^{\mu\nu}\):
\begin{equation}
  \delta^{\mu\nu}
  \left(
    \partial_\mu\epsilon_\nu
    +
    \partial_\nu\epsilon_\mu
  \right)
  =
  \delta^{\mu\nu}f(x)\delta_{\mu\nu}.
\end{equation}
The right-hand side is
\begin{equation}
  \delta^{\mu\nu}f(x)\delta_{\mu\nu}
  =
  f(x)\delta^\mu{}_\mu
  =
  d\,f(x),
\end{equation}
because \(\delta^\mu{}_\mu=d\) in \(d\) dimensions.  The left-hand side is
\begin{align}
  \delta^{\mu\nu}\partial_\mu\epsilon_\nu
  +
  \delta^{\mu\nu}\partial_\nu\epsilon_\mu
  &=
  \partial^\nu\epsilon_\nu
  +
  \partial^\mu\epsilon_\mu
  \nonumber\\
  &=
  \partial_\mu\epsilon^\mu
  +
  \partial_\mu\epsilon^\mu
  \nonumber\\
  &=
  2\,\partial_\mu\epsilon^\mu .
\end{align}
Therefore
\begin{equation}
  2\,\partial_\mu\epsilon^\mu
  =
  d\,f(x),
\end{equation}
and hence
\begin{equation}
  f(x)
  =
  \frac{2}{d}\,\partial_\mu\epsilon^\mu .
  \label{eq:CKE_trace}
\end{equation}
Substituting this back into \eqref{eq:CKE}, the conformal Killing equation (CKE) takes
the standard form
\begin{equation}
  \partial_\mu\epsilon_\nu
  +
  \partial_\nu\epsilon_\mu
  =
  \frac{2}{d}
  \left(\partial_\rho\epsilon^\rho\right)\delta_{\mu\nu}.
\end{equation}
Equivalently, subtracting the trace part,
\begin{equation}
  \partial_\mu\epsilon_\nu
  +
  \partial_\nu\epsilon_\mu
  -
  \frac{2}{d}
  \left(\partial_\rho\epsilon^\rho\right)\delta_{\mu\nu}
  =
  0,
\end{equation}
whose left-hand side is explicitly traceless.

Differentiating \eqref{eq:CKE} with respect to $x^\rho$:
\begin{align}
  \partial_\rho\partial_\mu\epsilon_\nu + \partial_\rho\partial_\nu\epsilon_\mu
  &= \partial_\rho f\,\delta_{\mu\nu}\,.\label{eq:CKE_diff1}
\end{align}
Writing the same equation with $(\mu\nu\rho)\to(\nu\rho\mu)$ and $(\nu\rho\mu)\to(\rho\mu\nu)$
and combining (antisymmetrize two, add the third):
\begin{equation}
  2\,\partial_\mu\partial_\nu\epsilon_\rho
  = \partial_\mu f\,\delta_{\nu\rho} + \partial_\nu f\,\delta_{\mu\rho}
    - \partial_\rho f\,\delta_{\mu\nu}\,.
  \label{eq:CKE_2nd}
\end{equation}
 
Contracting \eqref{eq:CKE_2nd} on $\mu,\nu$:
\begin{equation}
  2\,\partial^\mu\partial_\mu\epsilon_\rho
  = (2-d)\,\partial_\rho f\,,
\end{equation}
i.e.\ $\Box\epsilon_\rho = \frac{2-d}{2}\partial_\rho f$.  Differentiating
$f = \frac{2}{d}\partial\cdot\epsilon$ once more:
$\Box f = \frac{2}{d}\,\partial\cdot\Box\epsilon = \frac{(2-d)}{d}\,\Box  f$,
which for $d>2$ gives
\begin{equation}
  (d-1)\,\Box f = 0\,,
  \qquad d>2\,.
\end{equation}
Since $d>1$, $f$ is harmonic: $\Box f = 0$.
 
Taking the divergence $\partial^\rho$ of \eqref{eq:CKE_2nd} gives
\[
2\,\partial_\mu\partial_\nu(\partial\cdot\epsilon)
=
2\,\partial_\mu\partial_\nu f-\delta_{\mu\nu}\Box f .
\]
Using $\partial\cdot\epsilon=\frac d2 f$ and $\Box f=0$, this becomes
\[
d\,\partial_\mu\partial_\nu f
=
2\,\partial_\mu\partial_\nu f .
\]
Hence, for $d>2$,
\[
(d-2)\partial_\mu\partial_\nu f=0,
\qquad\Rightarrow\qquad
\partial_\mu\partial_\nu f=0.
\]
Thus $f$ is at most linear in $x$:
\begin{equation}
  f(x) = A + B_\mu x^\mu\,,
  \label{eq:f_linear}
\end{equation}
for constants $A$ and $B_\mu$.  

We now show explicitly that this implies that
$\epsilon^\mu(x)$ is at most quadratic in $x$. Recall \eqref{eq:CKE_2nd} and using \eqref{eq:f_linear}, we have
\begin{equation}
  \partial_\mu f = B_\mu ,
\end{equation}
so \eqref{eq:CKE_2nd}  becomes
\begin{equation}
2\,\partial_\mu\partial_\nu\epsilon_\rho
  =
  B_\mu\delta_{\nu\rho}
  +
  B_\nu\delta_{\mu\rho}
  -
  B_\rho\delta_{\mu\nu}.
  \label{eq:second_derivative_constant}
\end{equation}
The right-hand side is independent of $x$. Hence all second derivatives of
$\epsilon_\rho$ are constants. Differentiating once more gives
\begin{equation}
  \partial_\lambda\partial_\mu\partial_\nu\epsilon_\rho=0.
\end{equation}
Therefore $\epsilon_\rho(x)$ cannot contain cubic or higher powers of $x$.
It is at most quadratic:
\begin{equation}
  \epsilon_\rho(x)
  =
  a_\rho
  +
  C_{\rho\mu}x^\mu
  +
  Q_{\rho\mu\nu}x^\mu x^\nu ,
  \label{eq:epsilon_general_quadratic}
\end{equation}
where $a_\rho$, $C_{\rho\mu}$ and $Q_{\rho\mu\nu}$ are constant tensors. Since
$x^\mu x^\nu$ is symmetric under $\mu\leftrightarrow\nu$, only the symmetric
part of $Q_{\rho\mu\nu}$ in its last two indices contributes, so we may take
\begin{equation}
  Q_{\rho\mu\nu}=Q_{\rho\nu\mu}.
\end{equation}

We now impose the original conformal Killing equation,
\begin{equation}
  \partial_\mu\epsilon_\nu+\partial_\nu\epsilon_\mu
  =
  f(x)\,\delta_{\mu\nu}.
  \label{eq:CKE_again}
\end{equation}

From \eqref{eq:epsilon_general_quadratic},
\begin{equation}
  \partial_\mu\epsilon_\nu
  =
  C_{\nu\mu}
  +
  2 Q_{\nu\mu\lambda}x^\lambda .
\end{equation}
Therefore
\begin{equation}
  \partial_\mu\epsilon_\nu+\partial_\nu\epsilon_\mu
  =
  C_{\nu\mu}+C_{\mu\nu}
  +
  2\left(
  Q_{\nu\mu\lambda}+Q_{\mu\nu\lambda}
  \right)x^\lambda .
  \label{eq:CKE_expanded}
\end{equation}
Comparing with
\begin{equation}
  f(x)\delta_{\mu\nu}
  =
  \left(A+B_\lambda x^\lambda\right)\delta_{\mu\nu},
\end{equation}
we obtain separately the constant and linear constraints
\begin{align}
  C_{\mu\nu}+C_{\nu\mu}
  &=
  A\,\delta_{\mu\nu},
  \label{eq:C_constraint}
  \\
  2\left(
  Q_{\nu\mu\lambda}+Q_{\mu\nu\lambda}
  \right)
  &=
  B_\lambda\,\delta_{\mu\nu}.
  \label{eq:Q_constraint}
\end{align}

The first condition fixes the symmetric part of $C_{\mu\nu}$. We decompose
\begin{equation}
  C_{\mu\nu}
  =
  \omega_{\mu\nu}
  +
  \lambda\,\delta_{\mu\nu},
  \qquad
  \omega_{\mu\nu}=-\omega_{\nu\mu}.
\end{equation}
Then
\begin{equation}
  C_{\mu\nu}+C_{\nu\mu}=2\lambda\,\delta_{\mu\nu},
\end{equation}
so comparison with \eqref{eq:C_constraint} gives
\begin{equation}
  A=2\lambda.
\end{equation}
Thus the constant and linear pieces of $\epsilon_\mu$ are
\begin{equation}
  a_\mu+\omega_{\mu\nu}x^\nu+\lambda x_\mu .
\end{equation}

It remains to determine the quadratic part. Equation
\eqref{eq:second_derivative_constant} already gives the most efficient route.
Since
\begin{equation}
  \partial_\mu\partial_\nu\epsilon_\rho
  =
  2 Q_{\rho\mu\nu},
\end{equation}
we have from \eqref{eq:second_derivative_constant}
\begin{equation}
  4 Q_{\rho\mu\nu}
  =
  B_\mu\delta_{\nu\rho}
  +
  B_\nu\delta_{\mu\rho}
  -
  B_\rho\delta_{\mu\nu}.
\end{equation}
Hence
\begin{equation}
  Q_{\rho\mu\nu}
  =
  \frac14
  \left(
  B_\mu\delta_{\nu\rho}
  +
  B_\nu\delta_{\mu\rho}
  -
  B_\rho\delta_{\mu\nu}
  \right).
  \label{eq:Q_solution}
\end{equation}
The quadratic contribution to $\epsilon_\rho$ is therefore
\begin{align}
  Q_{\rho\mu\nu}x^\mu x^\nu
  &=
  \frac14
  \left(
  B_\mu\delta_{\nu\rho}
  +
  B_\nu\delta_{\mu\rho}
  -
  B_\rho\delta_{\mu\nu}
  \right)x^\mu x^\nu
  \nonumber\\
  &=
  \frac14
  \left(
  B_\mu x^\mu x_\rho
  +
  B_\nu x^\nu x_\rho
  -
  B_\rho x^2
  \right)
  \nonumber\\
  &=
  \frac12 (B\cdot x)x_\rho
  -
  \frac14 B_\rho x^2 .
  \label{eq:quadratic_piece_B}
\end{align}

It is conventional to introduce the special conformal parameter $b_\rho$ by
\begin{equation}
  B_\rho=-4 b_\rho .
\end{equation}
Then the quadratic piece becomes
\begin{equation}
  Q_{\rho\mu\nu}x^\mu x^\nu
  =
  b_\rho x^2
  -
  2(b\cdot x)x_\rho .
\end{equation}
Putting all terms together, the most general infinitesimal conformal Killing
vector in $d>2$ is
\begin{equation}
 {
  \epsilon_\rho(x)
  =
  a_\rho
  +
  \omega_{\rho\nu}x^\nu
  +
  \lambda x_\rho
  +
  b_\rho x^2
  -
  2(b\cdot x)x_\rho
  }
  \label{eq:general_CKV_lower}
\end{equation}
with
\begin{equation}
  \omega_{\rho\nu}=-\omega_{\nu\rho}.
\end{equation}
Equivalently, with an upper index,
\begin{equation}
  \boxed{
  \epsilon^\mu(x)
  =
  a^\mu
  +
  \omega^\mu{}_{\nu}x^\nu
  +
  \lambda x^\mu
  +
  b^\mu x^2
  -
  2(b\cdot x)x^\mu
  } .
  \label{eq:general_CKV_upper}
\end{equation}

The four terms correspond respectively to translations, rotations, dilatations,
and special conformal transformations (SCT, plural SCTs).

 \subsection{Classification of conformal Killing vectors}
 
The constants $a^\mu$, $\omega_{\mu\nu}$, $\lambda$, and $b^\mu$ generate,
respectively, translations, rotations, dilatations, and special conformal
transformations.

Let us verify directly how each term appears in the conformal Killing equation
\begin{equation}
  \partial_\mu\epsilon_\nu+\partial_\nu\epsilon_\mu
  =
  f(x)\,\delta_{\mu\nu}.
  \label{eq:CKE_classification}
\end{equation}

\begin{enumerate}
\item \textbf{Translations.}
For
\begin{equation}
  \epsilon^\mu=a^\mu,
\end{equation}
with constant $a^\mu$, one has
\begin{equation}
  \partial_\mu\epsilon_\nu+\partial_\nu\epsilon_\mu=0.
\end{equation}
Thus
\begin{equation}
  f=0.
\end{equation}
Translations therefore contribute $d$ independent parameters.

\item \textbf{Rotations.}
For
\begin{equation}
  \epsilon^\mu=\omega^\mu{}_\nu x^\nu,
  \qquad
  \omega_{\mu\nu}=-\omega_{\nu\mu},
\end{equation}
one finds
\begin{equation}
  \partial_\mu\epsilon_\nu=\omega_{\nu\mu},
  \qquad
  \partial_\nu\epsilon_\mu=\omega_{\mu\nu}.
\end{equation}
Hence
\begin{equation}
  \partial_\mu\epsilon_\nu+\partial_\nu\epsilon_\mu
  =
  \omega_{\nu\mu}+\omega_{\mu\nu}
  =
  0.
\end{equation}
Thus again
\begin{equation}
  f=0.
\end{equation}
Rotations contribute
\begin{equation}
  \frac{d(d-1)}{2}
\end{equation}
independent parameters.

\item \textbf{Dilatations.}
For
\begin{equation}
  \epsilon^\mu=\lambda x^\mu,
\end{equation}
one has
\begin{equation}
  \partial_\mu\epsilon_\nu=\lambda\delta_{\mu\nu},
\end{equation}
and therefore
\begin{equation}
  \partial_\mu\epsilon_\nu+\partial_\nu\epsilon_\mu
  =
  2\lambda\delta_{\mu\nu}.
\end{equation}
Thus
\begin{equation}
  f=2\lambda.
\end{equation}
Dilatations contribute one independent parameter.

\item \textbf{Special conformal transformations.}
For
\begin{equation}
  \epsilon^\mu=b^\mu x^2-2(b\cdot x)x^\mu,
\end{equation}
we lower the index,
\begin{equation}
  \epsilon_\nu=b_\nu x^2-2(b\cdot x)x_\nu.
\end{equation}
Then
\begin{align}
  \partial_\mu\epsilon_\nu
  &=
  2b_\nu x_\mu
  -2b_\mu x_\nu
  -2(b\cdot x)\delta_{\mu\nu},
  \\
  \partial_\nu\epsilon_\mu
  &=
  2b_\mu x_\nu
  -2b_\nu x_\mu
  -2(b\cdot x)\delta_{\mu\nu}.
\end{align}
Adding the two expressions gives
\begin{equation}
  \partial_\mu\epsilon_\nu+\partial_\nu\epsilon_\mu
  =
  -4(b\cdot x)\delta_{\mu\nu}.
\end{equation}
Therefore
\begin{equation}
  f(x)=-4(b\cdot x).
\end{equation}
Equivalently, in the notation $f(x)=A+B_\mu x^\mu$, the SCT corresponds to
\begin{equation}
  A=0,
  \qquad
  B_\mu=-4b_\mu.
\end{equation}
Special conformal transformations contribute $d$ independent parameters.
\end{enumerate} 

The total number of independent parameters is therefore
\begin{equation}
  d+\frac{d(d-1)}{2}+1+d
  =
  \frac{(d+1)(d+2)}{2},
\end{equation}
which is the dimension of $\mathrm{SO}(d+1,1)$, the Euclidean conformal group in
$d$ dimensions. At this stage we have only matched the parameter count; that
the infinitesimal transformations \eqref{eq:general_CKV_upper} actually
\emph{generate} the Lie algebra $\mathfrak{so}(d+1,1)$ will be established in
Section~\ref{app:conf:so_iso}.

\section{The conformal algebra}
\label{app:conf:algebra}

Having identified the four families of infinitesimal conformal Killing
vectors---translations, rotations, dilatations, and special conformal
transformations---we now turn to the \emph{algebraic} structure they generate.
The parameter count $(d+1)(d+2)/2$ derived above only tells us how many
independent infinitesimal transformations there are; it does not yet specify
how successive transformations compose, nor how the four families interact.
That information is encoded in the Lie algebra of commutators, and it is what
controls every subsequent result of this prerequisite material on conformal
symmetry: the classification of representations into primaries and
descendants (defined via eigenvalues of $D$ and annihilation by $K_\mu$), the
constraints on correlation functions, the Ward identities, and the
state--operator correspondence on the cylinder.

In the remainder of this section we proceed in three steps. First, we
represent the generators as first-order differential operators acting on
fields, simply by reading off the coefficients of $a^\mu$, $\omega_{\mu\nu}$,
$\lambda$, and $b^\mu$ in the general conformal Killing vector
\eqref{eq:general_CKV_upper}. Second, we compute all commutators of these
operators directly. Third, we exhibit the explicit isomorphism with
$\mathfrak{so}(d+1,1)$ promised at the end of the previous subsection, by
embedding
$P_\mu,K_\mu,M_{\mu\nu},D$ into a single antisymmetric tensor $J_{AB}$ in
$d{+}2$ dimensions with signature $(-,+,\ldots,+,+)$. The first two steps
furnish the algebra; the third closes the link between the local geometric
analysis of \S\,1.1 and the global Lie group $\mathrm{SO}(d+1,1)$.

\subsection{Differential operators (generators acting on functions)}

The generators acting on scalar functions $\phi(x)$ are the differential operators
$G = -i\epsilon^\mu\partial_\mu$ for each type:
\begin{align}
  P_\mu &= -i\partial_\mu & &\text{(momentum / translations)}\label{eq:gen_P}\\
  M_{\mu\nu} &= i(x_\mu\partial_\nu - x_\nu\partial_\mu) & &\text{(angular momentum / rotations)}\label{eq:gen_M}\\
  D &= -ix^\mu\partial_\mu & &\text{(dilatation)}\label{eq:gen_D}\\
  K_\mu &= i(2x_\mu\,x^\nu\partial_\nu - x^2\partial_\mu) & &\text{(special conformal)}\label{eq:gen_K}
\end{align}

\subsection{Derivation of the algebra}

We compute all commutators directly.  We write $[A,B]\phi = A(B\phi)-B(A\phi)$.

\paragraph{$[P_\mu, P_\nu]$:}
$[{-i\partial_\mu},{-i\partial_\nu}]\phi = -\partial_\mu\partial_\nu\phi + \partial_\nu\partial_\mu\phi = 0$,
so
\begin{equation}
  [P_\mu, P_\nu] = 0\,.
\end{equation}

\paragraph{$[D, P_\mu]$:}
$[D,P_\mu]\phi = (-ix^\nu\partial_\nu)(-i\partial_\mu\phi) - (-i\partial_\mu)(-ix^\nu\partial_\nu\phi)
= -x^\nu\partial_\nu\partial_\mu\phi + \partial_\mu(x^\nu\partial_\nu\phi)$.
Using $\partial_\mu(x^\nu\partial_\nu\phi) = \delta_\mu{}^\nu\partial_\nu\phi + x^\nu\partial_\mu\partial_\nu\phi = \partial_\mu\phi + x^\nu\partial_\nu\partial_\mu\phi$, we get
\begin{equation}
  [D,P_\mu] = i P_\mu\,.
\end{equation}

\paragraph{$[D, K_\mu]$:}
Unlike $P_\mu = -i\partial_\mu$, the special conformal generator
$K_\mu = i(2x_\mu x^\nu\partial_\nu - x^2\partial_\mu)$ contains explicit factors
of $x^\nu$ in addition to a derivative. We compute the two orderings separately,
distributing the $x$-derivatives carefully:
\begin{align}
  DK_\mu\phi
  &= (-ix^\nu\partial_\nu)
     \bigl[i(2x_\mu x^\rho\partial_\rho - x^2\partial_\mu)\phi\bigr]
   = x^\nu\partial_\nu
     \bigl[(2x_\mu x^\rho\partial_\rho - x^2\partial_\mu)\phi\bigr]
     \nonumber\\
  &= 4\,x_\mu x^\rho\partial_\rho\phi
     + 2\,x_\mu x^\nu x^\rho\partial_\nu\partial_\rho\phi
     - 2\,x^2\partial_\mu\phi
     - x^2 x^\nu\partial_\nu\partial_\mu\phi\,, \\[4pt]
  K_\mu D\phi
  &= i(2x_\mu x^\rho\partial_\rho - x^2\partial_\mu)(-ix^\nu\partial_\nu\phi)
   = (2x_\mu x^\rho\partial_\rho - x^2\partial_\mu)(x^\nu\partial_\nu\phi)
     \nonumber\\
  &= 2\,x_\mu x^\rho\partial_\rho\phi
     + 2\,x_\mu x^\nu x^\rho\partial_\nu\partial_\rho\phi
     - x^2\partial_\mu\phi
     - x^2 x^\nu\partial_\mu\partial_\nu\phi\,.
\end{align}
The two-derivative terms coincide (using $\partial_\mu\partial_\nu =
\partial_\nu\partial_\mu$) and cancel in the difference. Subtracting:
\begin{equation}
  [D,K_\mu]\phi
  = (2\,x_\mu x^\rho\partial_\rho - x^2\partial_\mu)\phi
  = -i\,K_\mu\phi\,,
\end{equation}
where in the last step we recognised
$2x_\mu x^\rho\partial_\rho - x^2\partial_\mu = -iK_\mu$. Hence
\begin{equation}
  [D, K_\mu] = -i K_\mu\,.
\end{equation}
Together with $[D,P_\mu] = +iP_\mu$, the two commutators show that $D$ acts
diagonally on $P_\mu$ and $K_\mu$ with eigenvalues $\pm i$. The physical
meaning of the opposite signs---that, on a primary operator of dimension
$\Delta$, $P_\mu$ raises $\Delta$ by $1$ and $K_\mu$ lowers it by $1$---will
emerge once the notion of scaling dimension itself is introduced in
Section~\ref{app:conf:primaries}.

\paragraph{$[K_\mu, P_\nu]$:}
This is the crucial commutator.  We compute:
\begin{align}
  K_\mu P_\nu\phi &= i(2x_\mu x^\rho\partial_\rho - x^2\partial_\mu)(-i\partial_\nu\phi)
  = (2x_\mu x^\rho\partial_\rho\partial_\nu - x^2\partial_\mu\partial_\nu)\phi\,,\nonumber\\
  P_\nu K_\mu\phi &= -i\partial_\nu\bigl[i(2x_\mu x^\rho\partial_\rho - x^2\partial_\mu)\phi\bigr]
  = \partial_\nu\bigl(2x_\mu x^\rho\partial_\rho\phi - x^2\partial_\mu\phi\bigr)\nonumber\\
  &= 2\delta_{\mu\nu}\,x^\rho\partial_\rho\phi
    + 2x_\mu\partial_\nu\phi
    + 2x_\mu x^\rho\partial_\nu\partial_\rho\phi
    - 2x_\nu\partial_\mu\phi - x^2\partial_\nu\partial_\mu\phi\,.
\end{align}
The two-derivative terms ($2x_\mu x^\rho\partial_\rho\partial_\nu$ vs.\
$2x_\mu x^\rho\partial_\nu\partial_\rho$, and $x^2\partial_\mu\partial_\nu$
vs.\ $x^2\partial_\nu\partial_\mu$) coincide and cancel in the difference.
Subtracting:
\begin{equation}
  [K_\mu,P_\nu]\phi
  = -2\delta_{\mu\nu}\,x^\rho\partial_\rho\phi
    - 2x_\mu\partial_\nu\phi
    + 2x_\nu\partial_\mu\phi\,.
\end{equation}
We now express the right-hand side in generator notation. From
$D=-ix^\rho\partial_\rho$ we read off $x^\rho\partial_\rho = iD$, and from
$M_{\mu\nu}=i(x_\mu\partial_\nu - x_\nu\partial_\mu)$ we read off
$x_\nu\partial_\mu - x_\mu\partial_\nu = iM_{\mu\nu}$. Hence
 \begin{equation}
  \boxed{[K_\mu, P_\nu] = -2i\bigl(\delta_{\mu\nu}D - M_{\mu\nu}\bigr)}\,.
  \label{eq:KP_comm}
\end{equation}

\paragraph{$[M_{\mu\nu}, P_\rho]$, $[M_{\mu\nu}, K_\rho]$, $[M_{\mu\nu}, M_{\rho\sigma}]$:}
These follow from the Leibniz rule. The rotation $M_{\mu\nu}$ acts as an infinitesimal $\mathrm{SO}(d)$ rotation on any covariant vector index. Hence for any vector $V_\rho$:
$[M_{\mu\nu}, V_\rho] = i(\delta_{\mu\rho}V_\nu - \delta_{\nu\rho}V_\mu)$.
This gives:
\begin{align}
  [M_{\mu\nu}, P_\rho] &= i(\delta_{\mu\rho}P_\nu - \delta_{\nu\rho}P_\mu)\,,\label{eq:MP}\\
  [M_{\mu\nu}, K_\rho] &= i(\delta_{\mu\rho}K_\nu - \delta_{\nu\rho}K_\mu)\,,\label{eq:MK}\\
  [M_{\mu\nu}, M_{\rho\sigma}] &= i(\delta_{\mu\rho}M_{\nu\sigma}
    - \delta_{\mu\sigma}M_{\nu\rho}
    - \delta_{\nu\rho}M_{\mu\sigma}
    + \delta_{\nu\sigma}M_{\mu\rho})\,,\label{eq:MM}\\
  [M_{\mu\nu}, D] &= 0\,,\quad
  [K_\mu, K_\nu] = 0\,.\label{eq:MD_KK}
\end{align}

\subsection{The group $\mathrm{SO}(d+1,1)$, its Lie algebra, and the explicit isomorphism}
\label{app:conf:so_iso}

The parameter count of \S\,1.1 already established
$\dim\,\mathrm{Conf}(\mathbb{R}^d) = (d+1)(d+2)/2 = \dim\,\mathrm{SO}(d+1,1)$,
so the two are at least the same in size. Matching dimensions, however, is
not the same as matching algebras: there exist Lie groups of equal dimension
with totally different brackets. The aim of this subsection is to upgrade
the dimension match into a genuine isomorphism by exhibiting an explicit map
between the conformal generators $\{P_\mu, K_\mu, M_{\mu\nu}, D\}$ and the
generators of $\mathfrak{so}(d+1,1)$, and verifying that the conformal
commutators \eqref{eq:KP_comm} and
\eqref{eq:MP}--\eqref{eq:MD_KK}, together with  the commutators
$[D,P_\mu]=+iP_\mu$ and $[D,K_\mu]=-iK_\mu$, all follow from the
$\mathfrak{so}(d+1,1)$ relations.

\paragraph{The group.}
The indefinite-orthogonal group $\mathrm{SO}(p,q)$ is the connected component
containing the identity inside the group of linear transformations of
$\mathbb{R}^{p+q}$ that preserve a non-degenerate symmetric bilinear form of
signature $(p,q)$. Concretely: equip $\mathbb{R}^{p+q}$ with coordinates
$X^A$ ($A=1,\ldots,p+q$) and a metric $\eta_{AB}$ with $p$ positive and $q$
negative eigenvalues, and define
\begin{equation}
  \mathrm{SO}(p,q) \;=\;
  \bigl\{\Lambda\in\mathrm{GL}(p+q,\mathbb{R})  {\;\big|\;}
  \Lambda^{T}\eta\,\Lambda = \eta,\ \det\Lambda=+1\bigr\}_{0}\,,
\end{equation}
where the subscript $0$ denotes the \emph{identity component}: the
connected component of the determinant-one orthogonal set that contains
the identity element. A self-contained discussion is collected in
Appendix~\ref{app:identity_component}; for the present subsection it is
enough to know that, for positive-definite signature ($q=0$), the
subscript is redundant and one recovers the familiar rotation group
$\mathrm{SO}(p)$, whereas for indefinite signature ($p,q\ge 1$) it does
real work and the resulting Lie group is non-compact.

For our purposes we take $p=d+1$, $q=1$. Pick coordinates $X^A$ with
$A\in\{-1,0,1,\ldots,d\}$ ($d+2$ values in total) and metric
\begin{equation}
  \eta_{AB} \;=\; \mathrm{diag}\bigl(\underbrace{-1}_{A=-1},
   \underbrace{+1}_{A=0}, \underbrace{+1,\ldots,+1}_{A=1,\ldots,d}\bigr)\,,
\end{equation}
so that the unique negative direction is the one labelled $A=-1$. The
dimension of $\mathrm{SO}(d+1,1)$ as a real manifold is the number of
independent antisymmetric $(d+2)\times(d+2)$ real matrices, namely
$(d+1)(d+2)/2$, matching the parameter count of \S\,1.1.

\paragraph{The Lie algebra.}
The Lie algebra $\mathfrak{so}(d+1,1)$ consists of the antisymmetric
matrices $\Lambda_{AB}=-\Lambda_{BA}$ (indices lowered with $\eta$). A
convenient basis is given by the $(d+1)(d+2)/2$ generators
$J_{AB}=-J_{BA}$  {(in this realisation the compact rotation generators
are Hermitian, while the non-compact boosts mixing the timelike index are
anti-Hermitian)} with commutation relations
\begin{equation}
  [J_{AB},\,J_{CD}]
  \;=\;
  -i\bigl(
     \eta_{AC}J_{BD}
   - \eta_{AD}J_{BC}
   - \eta_{BC}J_{AD}
   + \eta_{BD}J_{AC}
  \bigr)\,.
  \label{eq:so_structure}
\end{equation}
The overall sign of the right-hand side is a convention: with Hermitian
generators the bracket is conventionally $\pm i\,(\text{structure constants})$
and either choice defines the same real Lie algebra (the two are related by
$J\to -J$ on a chosen subset of generators). The sign in
\eqref{eq:so_structure} is the one consistent with the operator realisation
$J_{AB}=-i(X_A\partial_B - X_B\partial_A)$, with $X_A=\eta_{AB}X^B$ and
$[\partial_A,X_C]=\eta_{AC}$, and with the conformal generators
\eqref{eq:gen_P}--\eqref{eq:gen_K}.

\paragraph{The embedding.}
We now identify the conformal generators with components of $J_{AB}$:
\begin{equation}
\boxed{\
  J_{\mu\nu} = M_{\mu\nu}\,,\quad
  J_{-1,\mu} = \tfrac{1}{2}(P_\mu + K_\mu)\,,\quad
  J_{0,\mu} = \tfrac{1}{2}(P_\mu - K_\mu)\,,\quad
  J_{-1,0} = D\,,
\ }
  \label{eq:so_emb}
\end{equation}
with inverses
\begin{equation}
  P_\mu = J_{-1,\mu} + J_{0,\mu}\,,\qquad
  K_\mu = J_{-1,\mu} - J_{0,\mu}\,,\qquad
  D = J_{-1,0}\,,\qquad
  M_{\mu\nu} = J_{\mu\nu}\,.
  \label{eq:so_emb_inverse}
\end{equation}
Two observations make this guess natural. First, the rotational subalgebra
$\mathfrak{so}(d)\subset\mathfrak{so}(d+1,1)$ generated by
$\{J_{\mu\nu}\}_{\mu,\nu=1,\ldots,d}$ has the same dimension and the same
$[M,M]$ relation as $\mathrm{span}\{M_{\mu\nu}\}$, so $J_{\mu\nu}=M_{\mu\nu}$
is forced. Second, the remaining $2d+1$ generators
$\{J_{-1,\mu}, J_{0,\mu}, J_{-1,0}\}$ involve the two extra directions
$A\in\{-1,0\}$ and must therefore correspond to the $2d+1$ remaining
conformal generators $\{P_\mu, K_\mu, D\}$. The light-cone-like combinations
$P_\mu = J_{-1,\mu}+J_{0,\mu}$ and $K_\mu = J_{-1,\mu}-J_{0,\mu}$ then
diagonalise the \emph{adjoint action} of $D$. Recall that, for any element
$X$ of a Lie algebra, the \emph{adjoint map} $\mathrm{ad}_X$ is the linear
operator on the algebra defined by
\begin{equation}
  \mathrm{ad}_X(Y) \;\equiv\; [X,\,Y]\,,
\end{equation}
and an element $Y$ is called an eigenvector of $\mathrm{ad}_X$ with
eigenvalue $\lambda$ when $[X,Y] = \lambda Y$. The verification below will
show that $[D,P_\mu] = +iP_\mu$ and $[D,K_\mu] = -iK_\mu$, so $P_\mu$ and
$K_\mu$ are precisely the eigenvectors of
$\mathrm{ad}_D = \mathrm{ad}_{J_{-1,0}}$ with eigenvalues $+i$ and $-i$
respectively. This is the precise sense in which they diagonalise
$\mathrm{ad}_D$.

\paragraph{Verification.}
We now check that the abstract relations \eqref{eq:so_structure} reproduce
the conformal commutators. The five building blocks needed are
\begin{align}
  [J_{-1,0},\,J_{-1,\mu}]
  &= -i\,\eta_{-1,-1}\,J_{0,\mu} = +i\,J_{0,\mu}\,,
  \label{eq:bb_1}\\[2pt]
  [J_{-1,0},\,J_{0,\mu}]
  &= +i\,\eta_{0,0}\,J_{-1,\mu} = +i\,J_{-1,\mu}\,,
  \label{eq:bb_2}\\[2pt]
  [J_{-1,\mu},\,J_{0,\nu}]
  &= -i\,\delta_{\mu\nu}\,J_{-1,0} = -i\,\delta_{\mu\nu}\,D\,,
  \label{eq:bb_3}\\[2pt]
  [J_{-1,\mu},\,J_{-1,\nu}]
  &= -i\,\eta_{-1,-1}\,J_{\mu\nu} = +i\,M_{\mu\nu}\,,
  \label{eq:bb_4}\\[2pt]
  [J_{0,\mu},\,J_{0,\nu}]
  &= -i\,\eta_{0,0}\,J_{\mu\nu} = -i\,M_{\mu\nu}\,.
  \label{eq:bb_5}
\end{align}
In each line only one term of \eqref{eq:so_structure} survives because the
remaining $\eta$-components vanish (the metric is block-diagonal between the
$\{-1,0\}$ pair and the indices $\{1,\ldots,d\}$).

\smallskip
\textbf{(i) $[D,P_\mu]=+iP_\mu$.}
Using \eqref{eq:so_emb_inverse}, \eqref{eq:bb_1} and \eqref{eq:bb_2},
\begin{equation}
  [D,P_\mu]
  = [J_{-1,0},\,J_{-1,\mu}+J_{0,\mu}]
  = iJ_{0,\mu}+iJ_{-1,\mu}
  = i(J_{-1,\mu}+J_{0,\mu})
  = iP_\mu\,.
\end{equation}

\textbf{(ii) $[D,K_\mu]=-iK_\mu$.}
\begin{equation}
  [D,K_\mu]
  = [J_{-1,0},\,J_{-1,\mu}-J_{0,\mu}]
  = iJ_{0,\mu}-iJ_{-1,\mu}
  = -i(J_{-1,\mu}-J_{0,\mu})
  = -iK_\mu\,.
\end{equation}

\textbf{(iii) $[K_\mu,P_\nu]=-2i(\delta_{\mu\nu}D-M_{\mu\nu})$.}
Distributing and using \eqref{eq:bb_3}--\eqref{eq:bb_5} (and
$[J_{0,\mu},J_{-1,\nu}]=-[J_{-1,\nu},J_{0,\mu}]=+i\delta_{\mu\nu}D$),
\begin{align}
  [K_\mu,P_\nu]
  &= [J_{-1,\mu}-J_{0,\mu},\,J_{-1,\nu}+J_{0,\nu}] \nonumber\\
  &= [J_{-1,\mu},J_{-1,\nu}] + [J_{-1,\mu},J_{0,\nu}]
     - [J_{0,\mu},J_{-1,\nu}] - [J_{0,\mu},J_{0,\nu}] \nonumber\\
  &= (+iM_{\mu\nu}) + (-i\delta_{\mu\nu}D) - (+i\delta_{\mu\nu}D)
     - (-iM_{\mu\nu}) \nonumber\\
  &= 2iM_{\mu\nu} - 2i\delta_{\mu\nu}D
   = -2i\bigl(\delta_{\mu\nu}D - M_{\mu\nu}\bigr)\,,
\end{align}
matching the corrected boxed result \eqref{eq:KP_comm}.

\textbf{(iv) $[M_{\mu\nu},P_\rho]=i(\delta_{\mu\rho}P_\nu-\delta_{\nu\rho}P_\mu)$.}
Apply \eqref{eq:so_structure} with $A=\mu, B=\nu, C=-1, D=\rho$:
\begin{equation}
  [J_{\mu\nu},J_{-1,\rho}]
  = -i\bigl(\eta_{\mu,-1}J_{\nu\rho} - \eta_{\mu\rho}J_{\nu,-1}
     - \eta_{\nu,-1}J_{\mu\rho} + \eta_{\nu\rho}J_{\mu,-1}\bigr)\,,
\end{equation}
which, using $\eta_{\mu,-1}=\eta_{\nu,-1}=0$ and $J_{\nu,-1}=-J_{-1,\nu}$,
collapses to $i(\delta_{\mu\rho}J_{-1,\nu}-\delta_{\nu\rho}J_{-1,\mu})$. The
same expression with $-1\to 0$ gives
$i(\delta_{\mu\rho}J_{0,\nu}-\delta_{\nu\rho}J_{0,\mu})$. Adding the two,
\begin{equation}
  [M_{\mu\nu},P_\rho] = i\bigl(\delta_{\mu\rho}P_\nu-\delta_{\nu\rho}P_\mu\bigr)\,,
\end{equation}
which is exactly \eqref{eq:MP}. The verification for $[M_{\mu\nu},K_\rho]$
is identical (subtract instead of add) and reproduces \eqref{eq:MK}.

\smallskip
The remaining commutators $[M,M]$ \eqref{eq:MM},
$[M,D]=[K,K]=[P,P]=0$ \eqref{eq:MD_KK}, and $[D,M]=0$ all follow from
\eqref{eq:so_structure} by analogous  calculations: each reduces to
a single non-zero $\eta$-component and the corresponding $J$-component on
the right-hand side.

 \bigskip
 Therefore, every conformal commutator follows from \eqref{eq:so_structure} via the
embedding \eqref{eq:so_emb}, and conversely every $\mathfrak{so}(d+1,1)$
relation between the chosen $J_{AB}$ basis is one of the conformal
commutators (or zero). The map \eqref{eq:so_emb} is therefore a Lie-algebra
isomorphism, and at the group level
\begin{equation}
  \boxed{\mathrm{Conf}(\mathbb{R}^d)_{0} \;\cong\; \mathrm{SO}(d+1,1)\,,}
\end{equation}
where the subscript $0$ again denotes the identity component
(Appendix~\ref{app:identity_component}). The full Euclidean conformal
group is the larger, disconnected $\mathrm{O}(d+1,1)$, which also contains
the inversion $x^\mu\mapsto x^\mu/|x|^2$ (Figure~\ref{fig:cft_sct});
the inversion is not reachable from the identity by continuous
deformation and therefore sits outside the identity component.

\subsection{Geometry of the SCT}
\label{app:conf:sct_geometry}

The figure on the next page summarises the special conformal
transformation (SCT) as an inversion--translation--inversion composition.
We close \S\,1.2 by deriving each of its three panels and by explaining
why straight lines in the Cartesian grid are mapped into circles.

\begin{figure}[h]
\centering
\includegraphics[width=\textwidth]{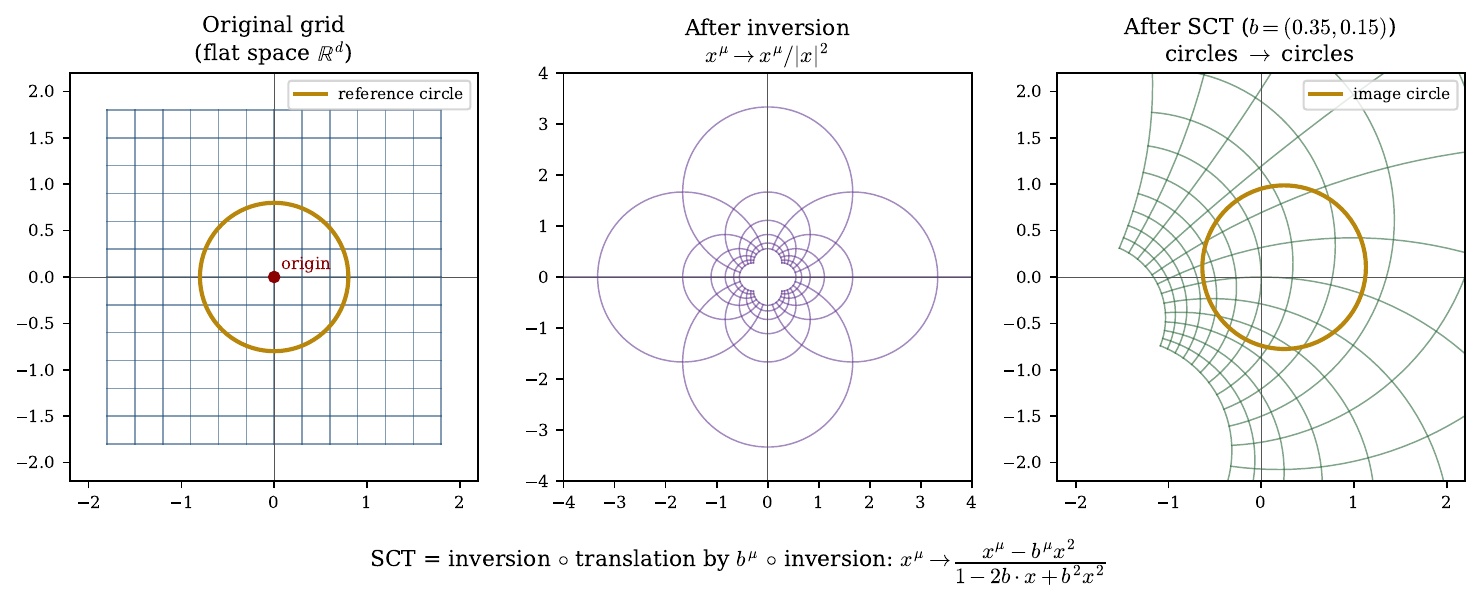}
\caption{%
\textbf{Special conformal transformation as inversion--translation--inversion.}
\textit{Left:} A regular Cartesian grid in flat space.
\textit{Centre:} After the inversion $x^\mu\mapsto x^\mu/|x|^2$; straight lines
not passing through the origin are mapped to circles passing through the origin.
\textit{Right:} The full special conformal transformation with parameter
$b=(0.35,0.15)$. The reference circle is mapped to another circle, illustrating
that conformal transformations preserve angles and map generalized circles
(lines or circles) to generalized circles, while not preserving lengths or
centres. Infinitesimally, this inversion--translation--inversion composition
generates the vector field
$\epsilon^\mu=b^\mu x^2-2(b\cdot x)x^\mu$, associated with $K_\mu$.}
\label{fig:cft_sct}
\end{figure}

\paragraph{The inversion map.}
Define the conformal inversion
\begin{equation}
  I:\ \mathbb{R}^d\setminus\{0\}\to\mathbb{R}^d\setminus\{0\}\,,
  \qquad
  I(x)^\mu = \frac{x^\mu}{|x|^2}\,.
\end{equation}
The map is an involution,
\begin{equation}
  I\circ I = \mathrm{id}\,,
\end{equation}
because
\begin{equation}
  I(I(x))^\mu
  =
  \frac{x^\mu/|x|^2}{|x|^{-2}}
  =
  x^\mu .
\end{equation}
Its Jacobian is
\begin{equation}
  \frac{\partial I^\nu}{\partial x^\mu}
  =
  \frac{1}{|x|^2}
  \left(
    \delta^\nu{}_\mu
    -
    2\,\hat{x}^\nu\hat{x}_\mu
  \right),
  \qquad
  \hat{x}^\mu = \frac{x^\mu}{|x|}\,.
\end{equation}
Thus the differential of inversion is the product of a
position-dependent dilation, with scale factor $|x|^{-2}$, and a
Householder reflection
\begin{equation}
  R^\nu{}_\mu
  =
  \delta^\nu{}_\mu
  -
  2\,\hat{x}^\nu\hat{x}_\mu .
\end{equation}
Since $R$ is orthogonal, $R^T R=\mathbf{1}$, inversion is conformal:
\begin{equation}
  \delta_{\rho\sigma}
  \frac{\partial I^\rho}{\partial x^\mu}
  \frac{\partial I^\sigma}{\partial x^\nu}
  =
  \frac{1}{|x|^4}\,\delta_{\mu\nu}.
\end{equation}
Equivalently, the Weyl factor of inversion is
\begin{equation}
  \Omega_I(x)=\frac{1}{|x|^2}.
\end{equation}
Points on the unit sphere $|x|=1$ are fixed pointwise, while the interior
and exterior of the unit sphere are exchanged:
\begin{equation}
  |x|<1 \longleftrightarrow |I(x)|>1,
  \qquad
  |x|>1 \longleftrightarrow |I(x)|<1.
\end{equation}
The leftmost two panels of Figure~\ref{fig:cft_sct} display precisely
this action: the Cartesian grid in flat space and its image under
inversion.

\paragraph{Why lines become circles through the origin.}
The interesting feature of the centre panel is that straight grid lines
which do not pass through the origin are mapped into circles passing
through the origin. Lines through the origin are exceptional: they are
mapped into themselves.

The simplest way to see this is to consider a vertical line in two
dimensions,
\begin{equation}
  x_1=a,
  \qquad
  a\neq 0.
\end{equation}
Let
\begin{equation}
  x'^\mu = I(x)^\mu = \frac{x^\mu}{|x|^2}.
\end{equation}
Since inversion is an involution, we may equivalently write
\begin{equation}
  x^\mu = \frac{x'^\mu}{|x'|^2}.
\end{equation}
Therefore the equation $x_1=a$ becomes
\begin{equation}
  \frac{x'_1}{|x'|^2}=a.
\end{equation}
Multiplying by $|x'|^2=(x'_1)^2+(x'_2)^2$ gives
\begin{equation}
  x'_1
  =
  a\left((x'_1)^2+(x'_2)^2\right),
\end{equation}
or
\begin{equation}
  (x'_1)^2+(x'_2)^2-\frac{1}{a}x'_1=0.
\end{equation}
Completing the square,
\begin{equation}
  \left(x'_1-\frac{1}{2a}\right)^2+(x'_2)^2
  =
  \frac{1}{4a^2}.
\end{equation}
Thus the image is a circle with centre
\begin{equation}
  \left(\frac{1}{2a},0\right)
\end{equation}
and radius
\begin{equation}
  R=\frac{1}{2|a|}.
\end{equation}
The distance of the centre from the origin is also $1/(2|a|)$, and hence
the circle passes through the origin. This is why the non-central
vertical grid lines in the centre panel become circles tangent to, and
passing through, the origin.

The same argument applies to horizontal lines. For example, the line
$x_2=a$, with $a\neq0$, is mapped to
\begin{equation}
  (x'_1)^2+(x'_2)^2-\frac{1}{a}x'_2=0,
\end{equation}
or
\begin{equation}
  (x'_1)^2+
  \left(x'_2-\frac{1}{2a}\right)^2
  =
  \frac{1}{4a^2},
\end{equation}
again a circle passing through the origin.

More generally, consider an arbitrary affine hyperplane in
$\mathbb{R}^d$,
\begin{equation}
  n\cdot x = c,
  \qquad
  n\in\mathbb{R}^d,\qquad c\neq 0.
\end{equation}
In two dimensions this is an arbitrary affine line not passing through
the origin. Again writing $x=x'/|x'|^2$, its image under inversion is
defined by
\begin{equation}
  n\cdot \frac{x'}{|x'|^2}=c.
\end{equation}
Multiplying by $|x'|^2$ gives
\begin{equation}
  c\,|x'|^2-n\cdot x'=0.
\end{equation}
Since $c\neq0$, we may divide by $c$:
\begin{equation}
  |x'|^2-\frac{n}{c}\cdot x'=0.
\end{equation}
Completing the square,
\begin{equation}
  \left|
    x'-\frac{n}{2c}
  \right|^2
  =
  \frac{|n|^2}{4c^2}.
\end{equation}
Thus the image is a sphere of centre
\begin{equation}
  x'_0=\frac{n}{2c}
\end{equation}
and radius
\begin{equation}
  R=\frac{|n|}{2|c|}.
\end{equation}
Since
\begin{equation}
  |x'_0|=\frac{|n|}{2|c|}=R,
\end{equation}
the sphere passes through the origin. In $d=2$ this sphere is precisely
a circle through the origin. Hence every affine line in the plane that
does not pass through the origin is mapped by inversion into a circle
passing through the origin.

By contrast, a line through the origin has equation
\begin{equation}
  n\cdot x=0.
\end{equation}
After inversion this becomes
\begin{equation}
  n\cdot \frac{x'}{|x'|^2}=0,
\end{equation}
and therefore
\begin{equation}
  n\cdot x'=0.
\end{equation}
Thus lines through the origin are mapped into themselves. This explains
why the horizontal and vertical axes in the centre panel remain straight.

\paragraph{Inversion preserves generalised circles.}
The previous examples are special cases of a general fact. A
\emph{generalised circle} in $\mathbb{R}^d$---more precisely, a
generalised sphere for $d>2$---is the locus
\begin{equation}
  A\,|x|^2 + B\cdot x + C = 0,
  \qquad
  A,C\in\mathbb{R},
  \qquad
  B\in\mathbb{R}^d .
  \label{eq:gen_circle}
\end{equation}
When $A=0$, this is a hyperplane. When $A\neq0$, it is a sphere. Indeed,
for $A\neq0$ one may complete the square:
\begin{equation}
  A\left|
    x+\frac{B}{2A}
  \right|^2
  +
  C
  -
  \frac{|B|^2}{4A}
  =
  0.
\end{equation}
Thus the centre is
\begin{equation}
  x_0=-\frac{B}{2A},
\end{equation}
and the squared radius is
\begin{equation}
  R^2
  =
  \frac{|B|^2}{4A^2}
  -
  \frac{C}{A}.
\end{equation}

Now apply inversion. Since $x=x'/|x'|^2$, substituting into
\eqref{eq:gen_circle} gives
\begin{equation}
  A\,\frac{1}{|x'|^2}
  +
  B\cdot\frac{x'}{|x'|^2}
  +
  C
  =
  0.
\end{equation}
Multiplying by $|x'|^2$, one obtains
\begin{equation}
  C\,|x'|^2+B\cdot x' + A =0.
\end{equation}
Hence inversion maps a generalised circle into another generalised
circle, with
\begin{equation}
  (A,B,C)\longmapsto (C,B,A).
\end{equation}
This immediately explains the cases seen in the figure:
\begin{itemize}[leftmargin=2em]
  \item A line not passing through the origin has $A=0$ and $C\neq0$.
  Under inversion it becomes a sphere with $A'=C\neq0$ and $C'=0$.
  Since $C'=0$, the resulting sphere passes through the origin.

  \item A hyperplane passing through the origin has $A=0$ and $C=0$.
  Under inversion it remains the same hyperplane.

  \item A sphere passing through the origin has $A\neq0$ and $C=0$.
  Under inversion it becomes a hyperplane not passing through the origin.

  \item A generic sphere, with $A\neq0$ and $C\neq0$, is mapped into
  another generic sphere.
\end{itemize}

Thus the centre panel of Figure~\ref{fig:cft_sct} should be read as
follows: every grid line not passing through the origin becomes a circle
passing through the origin, while the grid lines that do pass through
the origin remain straight.

\paragraph{The SCT as inversion--translation--inversion.}
Let
\begin{equation}
  T_{-b}(x)=x-b
\end{equation}
be translation by the constant vector $-b\in\mathbb{R}^d$, and consider
the composition
\begin{equation}
  S_b
  \equiv
  I\circ T_{-b}\circ I .
  \label{eq:SCT_composition}
\end{equation}
 Direct computation gives
\begin{equation}
  T_{-b}(I(x))
  =
  \frac{x}{|x|^2}-b,
\end{equation}
and therefore
\begin{equation}
  \bigl|T_{-b}(I(x))\bigr|^2
  =
  \left|
    \frac{x}{|x|^2}-b
  \right|^2
  =
  \frac{1-2\,b\cdot x+|b|^2|x|^2}{|x|^2}.
\end{equation}
Applying the final inversion gives
\begin{equation}
  S_b(x)
  =
  \frac{T_{-b}(I(x))}
       {\bigl|T_{-b}(I(x))\bigr|^2}.
\end{equation}
Hence
\begin{equation}
  \boxed{
  S_b(x)
  =
  \frac{x-b\,|x|^2}
       {1-2\,b\cdot x+|b|^2|x|^2}
  }.
  \label{eq:SCT_finite}
\end{equation}
In components,
\begin{equation}
  S_b(x)^\mu
  =
  \frac{x^\mu-b^\mu x^2}
       {1-2\,b\cdot x+b^2 x^2}.
\end{equation}
This is the finite special conformal transformation in the sign
convention used in the caption of Figure~\ref{fig:cft_sct}.

Expanding to first order in $b$, one finds
\begin{align}
  S_b(x)^\mu
  &=
  \left(x^\mu-b^\mu x^2\right)
  \left(1+2\,b\cdot x+\mathcal{O}(b^2)\right)
  \nonumber\\
  &=
  x^\mu
  +
  \left(
    2(b\cdot x)x^\mu
    -
    b^\mu x^2
  \right)
  +
  \mathcal{O}(b^2).
\end{align}
Thus the infinitesimal SCT vector field is
\begin{equation}
  \delta_b x^\mu
  =
  2(b\cdot x)x^\mu-b^\mu x^2.
\end{equation}
This is the standard conformal Killing vector associated with special
conformal transformations, up to the overall sign convention used for
the generator $K_\mu$. Equivalently, using $T_{+b}$ instead of $T_{-b}$
would replace $b$ by $-b$ throughout:
\begin{equation}
  I\circ T_{+b}\circ I:
  \qquad
  x^\mu
  \mapsto
  \frac{x^\mu+b^\mu x^2}
       {1+2\,b\cdot x+b^2x^2}.
\end{equation}
The two formulae describe the same family of transformations with
opposite parametrisations of the SCT parameter.

\paragraph{The right panel.}
Each of the three operations appearing in $S_b=I\circ T_{-b}\circ I$
maps generalised circles into generalised circles. The two inversions do
so by the argument above, and the translation does so trivially.
Therefore their composition also maps generalised circles into
generalised circles.

Consequently, the reference circle in the right panel of
Figure~\ref{fig:cft_sct} is sent to another circle. The underlying
Cartesian grid is bent by the same conformal map. Angles are preserved
locally, but lengths and the positions of circle centres are not
preserved. The Weyl factor of the SCT follows from
\eqref{eq:SCT_finite}:
\begin{equation}
  \Omega_{S_b}(x)
  =
  \frac{1}
       {1-2\,b\cdot x+b^2x^2}.
\end{equation}
Since this factor is position dependent, the SCT is not a rigid motion.
This is the visual difference between the left and right panels: the
right panel still preserves angles and maps circles into circles, but it
does so with a non-uniform local rescaling of lengths.

\section{Primary operators and their transformation rules}
\label{app:conf:primaries}

In the previous sections we constructed the conformal algebra. We now turn
to its representations on local operators, and in particular to the special
operators that generate irreducible conformal multiplets: the
\emph{primary operators}. A conformal multiplet is the set of operators obtained
from one primary by repeated action of the translation generators $P_\mu$,
equivalently by taking derivatives. In a unitary CFT the dilatation spectrum is
bounded from below, so each multiplet contains a lowest-weight operator at the
origin, annihilated by the special conformal generators $K_\mu$ and with definite
scaling dimension $\Delta$. This operator is the primary
$\mathcal{O}_{\Delta,s}$, where $s$ denotes its Lorentz spin, while the remaining
operators in the same multiplet are its descendants.

The importance of primaries is that they contain the independent dynamical data
of the theory. Descendants are fixed by conformal symmetry once the primary is
known, whereas the list of primaries, their scaling dimensions $\Delta$, their
spins $s$, and their OPE coefficients specify the CFT data. The transformation
rules derived below make this statement concrete: scalar primaries transform
with the conformal Jacobian factor $\Omega(x)^{-\Delta}$, while tensor primaries
also acquire the corresponding local $\mathrm{SO}(d)$ rotation. These rules are
the starting point for the constraints on two- and three-point functions in
Section~\ref{app:conf:corr}, and, via the state--operator correspondence of
Section~\ref{app:soc}, for the identification of $\Delta$ with the energy on
the cylinder. This is the bridge to the semiclassical computation of scaling
dimensions developed later in the lectures.

\begin{figure}[h]
\centering
\includegraphics[width=0.72\textwidth]{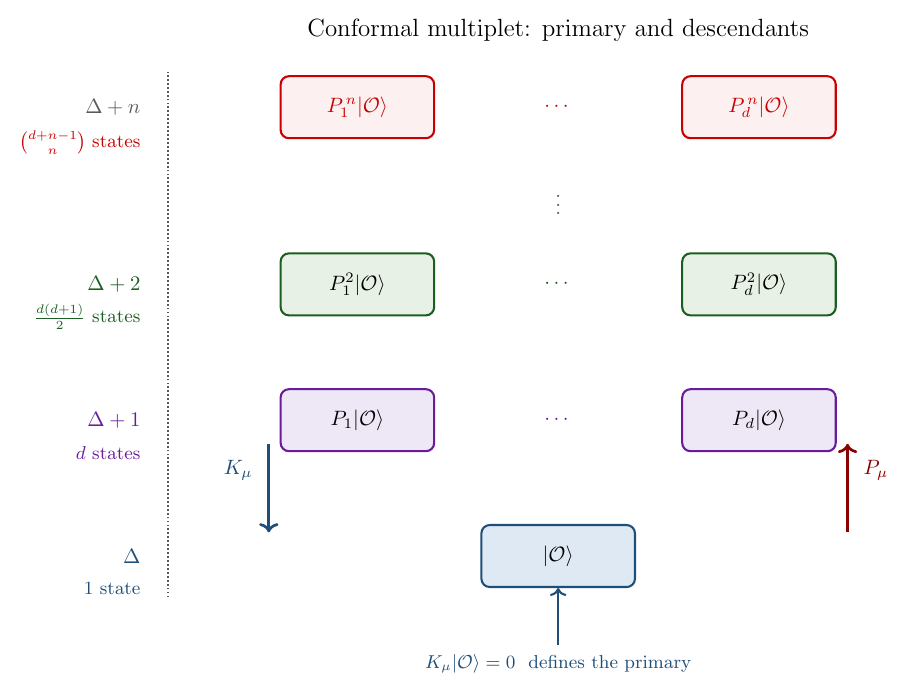}
\caption{%
\textbf{The conformal multiplet.}
Starting from a primary state $|\mathcal{O}\rangle$ (bottom, blue) satisfying
$K_\mu|\mathcal{O}\rangle = 0$, repeated action of the momentum operator $P_\mu$
(red arrows, upward) generates descendants at dimensions $\Delta,\Delta+1,\Delta+2,\ldots$
 {The number of independent states at level $n$ equals the number of symmetric
tensors of rank $n$, namely $\binom{d+n-1}{n}$.}  The special-conformal generator $K_\mu$ (blue arrow, downward)
lowers the dimension; its vanishing on $|\mathcal{O}\rangle$ defines the primary as the
lowest-weight state of the representation.  In the state--operator correspondence each node
is a local operator; the levels are the scaling dimensions measured on the cylinder.}
\label{fig:cft_multiplet}
\end{figure}


\subsection{Representations of the conformal algebra}

A \emph{conformal multiplet} is a representation of the conformal algebra on
the space of local operators.  A \emph{primary operator}
$\mathcal{O}_{\Delta,s}(x)$ of scaling dimension $\Delta$ and spin $s$ is
characterized at the origin by
\begin{itemize}
  \item being an eigenoperator of dilatations:
  \begin{equation}
    [D,\mathcal{O}_{\Delta,s}(0)]
    =
    -i\Delta\,\mathcal{O}_{\Delta,s}(0),
  \end{equation}
  \item being annihilated by the special conformal generators:
  \begin{equation}
    [K_\mu,\mathcal{O}_{\Delta,s}(0)] = 0 .
  \end{equation}
\end{itemize}
The second condition says that the primary is the lowest-weight operator in
the conformal multiplet. Acting repeatedly with the translation generators
$P_\mu=-i\partial_\mu$ produces the \emph{descendants},
\begin{equation}
  P_{\mu_1}\cdots P_{\mu_k}\mathcal{O}_{\Delta,s}(0)
  \quad\Longleftrightarrow\quad
  \partial_{\mu_1}\cdots\partial_{\mu_k}
  \mathcal{O}_{\Delta,s}(0).
\end{equation}
Thus a conformal multiplet consists of one primary together with all of its
descendants.

\subsection{Finite transformation of a scalar primary}

Let us first derive the finite transformation law of a scalar primary under
dilatations.  By definition, at the origin a scalar primary satisfies
\begin{equation}
  [D,\mathcal{O}(0)] = -i\Delta\,\mathcal{O}(0).
  \label{eq:D_primary_origin}
\end{equation}
Away from the origin, the dilatation generator also moves the insertion point.
For a scalar operator of scaling dimension $\Delta$, the infinitesimal
dilatation $x^\mu\to x'^\mu=x^\mu+\alpha x^\mu$ acts as
\begin{equation}
  \delta_\alpha \mathcal{O}(x)
  =
  -\alpha\left(x^\mu\partial_\mu+\Delta\right)\mathcal{O}(x).
  \label{eq:inf_dilation_primary}
\end{equation}
The first term is the change due to moving the point $x$, while the second term
is the intrinsic scaling of the operator.

To integrate this infinitesimal equation, write a finite dilatation as
\begin{equation}
  x^\mu \longmapsto x'^\mu=\lambda x^\mu,
  \qquad
  \lambda=e^\alpha .
\end{equation}
Define
\begin{equation}
  F(\lambda,x)\equiv \mathcal{O}'(\lambda x).
\end{equation}
Equation \eqref{eq:inf_dilation_primary} implies
\begin{equation}
  \lambda\frac{d}{d\lambda}F(\lambda,x)
  =
  -\Delta\,F(\lambda,x).
\end{equation}
Solving this first-order equation gives
\begin{equation}
  F(\lambda,x)=\lambda^{-\Delta}F(1,x).
\end{equation}
Therefore
\begin{equation}
  \boxed{
  \mathcal{O}'(\lambda x)=\lambda^{-\Delta}\mathcal{O}(x).
  }
  \label{eq:dilation_primary}
\end{equation}
This is the finite scaling law of a scalar primary: under a rescaling of
lengths by $\lambda$, the operator acquires the factor $\lambda^{-\Delta}$.

We now generalize this result to an arbitrary finite conformal transformation
$x\mapsto x'(x)$.  By definition, a conformal transformation rescales the flat
metric locally:
\begin{equation}
  \frac{\partial x'^\mu}{\partial x^\rho}
  \frac{\partial x'^\nu}{\partial x^\sigma}
  \delta_{\mu\nu}
  =
  \Omega(x)^2\,\delta_{\rho\sigma}.
  \label{eq:local_conformal_factor}
\end{equation}
Thus, near any point $x$, the transformation is locally a scale
transformation by the factor $\Omega(x)$, followed by an orthogonal rotation.
A scalar operator is insensitive to the rotation, so only the local scale
factor matters. Since a scalar primary of dimension $\Delta$ picks up
$\lambda^{-\Delta}$ under a scale transformation, it follows locally that
\begin{equation}
  \boxed{
  \mathcal{O}'(x')=\Omega(x)^{-\Delta}\mathcal{O}(x).
  }
  \label{eq:primary_transf_omega}
\end{equation}

Equivalently, taking the determinant of
\eqref{eq:local_conformal_factor} gives
\begin{equation}
  \left|\frac{\partial x'}{\partial x}\right|^2
  =
  \Omega(x)^{2d},
  \qquad\Rightarrow\qquad
  \left|\frac{\partial x'}{\partial x}\right|
  =
  \Omega(x)^d.
\end{equation}
Hence the same transformation law can be written as
\begin{equation}
  \boxed{
  \mathcal{O}'(x')
  =
  \left|\frac{\partial x'}{\partial x}\right|^{-\Delta/d}
  \mathcal{O}(x)
  =
  \Omega(x)^{-\Delta}\mathcal{O}(x).
  }
  \label{eq:primary_transf}
\end{equation}
This is the finite transformation rule for a scalar primary.

\paragraph{Explicit conformal factors.}

\noindent\textbf{Translations.}
For
\begin{equation}
  x'^\mu=x^\mu+a^\mu,
\end{equation}
one has
\begin{equation}
  \frac{\partial x'^\mu}{\partial x^\nu}=\delta^\mu{}_\nu,
  \qquad
  \left|\frac{\partial x'}{\partial x}\right|=1,
  \qquad
  \Omega(x)=1.
\end{equation}
Therefore
\begin{equation}
  \mathcal{O}'(x+a)=\mathcal{O}(x).
\end{equation}

\noindent\textbf{Dilatations.}
For
\begin{equation}
  x'^\mu=\lambda x^\mu,
\end{equation}
one finds
\begin{equation}
  \frac{\partial x'^\mu}{\partial x^\nu}
  =
  \lambda\delta^\mu{}_\nu,
  \qquad
  \left|\frac{\partial x'}{\partial x}\right|
  =
  \lambda^d,
  \qquad
  \Omega(x)=\lambda.
\end{equation}
Equation \eqref{eq:primary_transf} gives
\begin{equation}
  \mathcal{O}'(\lambda x)
  =
  \lambda^{-\Delta}\mathcal{O}(x),
\end{equation}
in agreement with the direct derivation above.

\noindent\textbf{Inversion.}
For the inversion
\begin{equation}
  x'^\mu=\frac{x^\mu}{x^2},
  \qquad x^2\equiv x^\mu x_\mu,
\end{equation}
the Jacobian matrix is
\begin{equation}
  \frac{\partial x'^\mu}{\partial x^\nu}
  =
  \frac{1}{x^2}
  \left(
    \delta^\mu{}_\nu
    -
    2\frac{x^\mu x_\nu}{x^2}
  \right).
  \label{eq:inversion_jac}
\end{equation}
Introducing the unit vector $\hat x^\mu=x^\mu/|x|$, this becomes
\begin{equation}
  \frac{\partial x'^\mu}{\partial x^\nu}
  =
  \frac{1}{x^2}
  \left(
    \delta^\mu{}_\nu
    -
    2\hat x^\mu\hat x_\nu
  \right).
\end{equation}
The matrix in parentheses is a reflection: it has eigenvalue $-1$ along the
direction $\hat x^\mu$ and eigenvalue $+1$ in the $d-1$ directions orthogonal
to $\hat x^\mu$. Therefore its determinant is $-1$, and hence
\begin{equation}
  \left|
  \frac{\partial x'}{\partial x}
  \right|
  =
  \frac{1}{(x^2)^d}.
\end{equation}
Thus
\begin{equation}
  \Omega(x)^d=\frac{1}{(x^2)^d},
  \qquad
  \Omega(x)=\frac{1}{x^2}.
\end{equation}
The scalar-primary transformation law gives
\begin{equation}
  \boxed{
  \mathcal{O}'\!\left(\frac{x}{x^2}\right)
  =
  (x^2)^\Delta\,\mathcal{O}(x).
  }
\end{equation}

\noindent\textbf{Special conformal transformations.}
A special conformal transformation can be written as an inversion, followed by
a translation, followed by another inversion. With the convention
\begin{equation}
  x'^\mu
  =
  \frac{x^\mu-b^\mu x^2}{1-2b\cdot x+b^2x^2},
  \label{eq:SCT_finite_primary}
\end{equation}
the conformal factor is
\begin{equation}
  \Omega_{\rm SCT}(x)
  =
  \frac{1}{1-2b\cdot x+b^2x^2}.
  \label{eq:SCT_Omega}
\end{equation}
This follows because conformal factors multiply under composition. Therefore
\begin{equation}
  \boxed{
  \mathcal{O}'(x')
  =
  \left(1-2b\cdot x+b^2x^2\right)^\Delta
  \mathcal{O}(x).
  }
  \label{eq:SCT_scalar_primary}
\end{equation}

\subsection{Tensor primaries}

For a primary with Lorentz spin $s$, the local conformal transformation is not
only a local rescaling but also a local rotation. The Jacobian can be decomposed
as
\begin{equation}
  \frac{\partial x'^\mu}{\partial x^\nu}
  =
  \Omega(x)\,R^\mu{}_\nu(x),
  \qquad
  R^\mu{}_\rho(x)R^\nu{}_\sigma(x)\delta_{\mu\nu}
  =
  \delta_{\rho\sigma}.
\end{equation}
Thus $R^\mu{}_\nu(x)$ is an orthogonal matrix. A tensor primary transforms with
the same scaling factor as a scalar primary, together with the appropriate
rotation acting on its indices. For a symmetric traceless rank-$s$ tensor,
\begin{equation}
  \boxed{
  \mathcal{O}'_{\mu_1\cdots\mu_s}(x')
  =
  \Omega(x)^{-\Delta}
  R_{\mu_1}{}^{\nu_1}(x)\cdots
  R_{\mu_s}{}^{\nu_s}(x)
  \mathcal{O}_{\nu_1\cdots\nu_s}(x).
  }
  \label{eq:tensor_primary_transf}
\end{equation}
For $s=0$ this reduces to the scalar-primary rule
\eqref{eq:primary_transf}.

\section{Constraints on correlation functions}
\label{app:conf:corr}

\subsection{Invariance condition}

Let
\begin{equation}
  G_n(x_1,\ldots,x_n)
  =
  \left\langle
  \mathcal{O}_1(x_1)\cdots\mathcal{O}_n(x_n)
  \right\rangle
\end{equation}
be an $n$-point function of scalar primary operators with dimensions
$\Delta_i$.  Under a conformal transformation $x_i\mapsto x'_i$, each scalar
primary transforms as
\begin{equation}
  \mathcal{O}'_i(x'_i)
  =
  \Omega(x_i)^{-\Delta_i}\mathcal{O}_i(x_i).
  \label{eq:primary_each_insertion}
\end{equation}
Substituting \eqref{eq:primary_each_insertion} inside the correlator gives
\begin{align}
  \left\langle
  \mathcal{O}'_1(x'_1)\cdots\mathcal{O}'_n(x'_n)
  \right\rangle
  &=
  \left\langle
  \prod_{i=1}^n
  \Omega(x_i)^{-\Delta_i}\mathcal{O}_i(x_i)
  \right\rangle
  \nonumber\\
  &=
  \left[
  \prod_{i=1}^n \Omega(x_i)^{-\Delta_i}
  \right]
  \left\langle
  \mathcal{O}_1(x_1)\cdots\mathcal{O}_n(x_n)
  \right\rangle
  \nonumber\\
  &=
  \left[
  \prod_{i=1}^n \Omega(x_i)^{-\Delta_i}
  \right]
  G_n(x_1,\ldots,x_n).
  \label{eq:transformed_corr_step}
\end{align}
The factors $\Omega(x_i)^{-\Delta_i}$ are ordinary functions of the insertion
points and therefore can be pulled outside the expectation value.
Conformal invariance of the vacuum means that the transformed correlator is the
same physical correlator evaluated at the transformed points,
\begin{equation}
  \left\langle
  \mathcal{O}'_1(x'_1)\cdots\mathcal{O}'_n(x'_n)
  \right\rangle
  =
  G_n(x'_1,\ldots,x'_n).
\end{equation}
Combining this with \eqref{eq:transformed_corr_step}, we obtain the covariance
condition
\begin{equation}
  \boxed{
  G_n(x'_1,\ldots,x'_n)
  =
  \prod_{i=1}^n \Omega(x_i)^{-\Delta_i}\,
  G_n(x_1,\ldots,x_n)
  }.
  \label{eq:corr_invariance}
\end{equation}
We will use this formula to fix the two- and three-point functions.

\subsection{Two-point function}

Consider two scalar primaries $\mathcal{O}_1$ and $\mathcal{O}_2$ with
dimensions $\Delta_1$ and $\Delta_2$:
\begin{equation}
  G_2(x_1,x_2)
  =
  \big\langle
  \mathcal{O}_1(x_1)\mathcal{O}_2(x_2)
  \big\rangle .
\end{equation}

\paragraph{Translations and rotations.}
Translation invariance implies that the correlator depends only on
\begin{equation}
  x_{12}\equiv x_1-x_2.
\end{equation}
Rotational invariance then implies that it depends only on the distance
$r_{12}=|x_{12}|$:
\begin{equation}
  G_2(x_1,x_2)=h(r_{12}).
\end{equation}

\paragraph{Dilatations.}
Under $x^\mu\mapsto x'^\mu=\lambda x^\mu$, we have
$\Omega=\lambda$. Equation \eqref{eq:corr_invariance} gives
\begin{equation}
  h(\lambda r_{12})
  =
  \lambda^{-(\Delta_1+\Delta_2)}h(r_{12}).
\end{equation}
The solution is a power law:
\begin{equation}
  h(r_{12})
  =
  \frac{C_{12}}{r_{12}^{\Delta_1+\Delta_2}}.
  \label{eq:2pt_after_dil}
\end{equation}

\paragraph{Special conformal transformations.}
Dilatations alone allow the power
$\Delta_1+\Delta_2$. Special conformal invariance imposes the stronger
condition that the two operators have the same scaling dimension.

To see this, use inversion, since an SCT is generated by inversion,
translation, and inversion. Under inversion,
\begin{equation}
  x^\mu\mapsto x'^\mu=\frac{x^\mu}{x^2},
  \qquad
  \Omega(x)=\frac{1}{x^2}.
\end{equation}
The distance transforms\footnote{Under inversion,
\begin{equation}
  x^\mu \mapsto x'^\mu=\frac{x^\mu}{x^2},
  \qquad x^2\equiv x^\mu x_\mu .
\end{equation}
Therefore
\begin{equation}
  x'_{12}{}^\mu
  =
  x_1'{}^\mu-x_2'{}^\mu
  =
  \frac{x_1^\mu}{x_1^2}
  -
  \frac{x_2^\mu}{x_2^2}.
\end{equation}
Squaring this expression gives
\begin{align}
  |x'_{12}|^2
  &=
  \left(
  \frac{x_1}{x_1^2}
  -
  \frac{x_2}{x_2^2}
  \right)^2
  =
  \frac{x_2^2+x_1^2-2x_1\cdot x_2}
       {x_1^2x_2^2}
  =
  \frac{(x_1-x_2)^2}{x_1^2x_2^2}
  =
  \frac{|x_{12}|^2}{x_1^2x_2^2}.
\end{align}} as
\begin{equation}
  |x'_{12}|^2
  =
  \frac{|x_{12}|^2}{x_1^2x_2^2}.
  \label{eq:distance_inversion}
\end{equation}

 Using the power-law form obtained from translations, rotations, and
dilatations \eqref{eq:2pt_after_dil}, 
\begin{equation}
  G_2(x_1,x_2)
  =
  \frac{C_{12}}{|x_{12}|^{\Delta_1+\Delta_2}},
\end{equation}
 the inversion leads to
\begin{equation}
  |x'_{12}|
  =
  \frac{|x_{12}|}{\sqrt{x_1^2x_2^2}} .
\end{equation}
Therefore the left-hand side of the covariance condition is
\begin{align}
  G_2(x'_1,x'_2)
  &=
  \frac{C_{12}}{|x'_{12}|^{\Delta_1+\Delta_2}}
  \nonumber\\
  &=
  C_{12}
  \left(
  \frac{\sqrt{x_1^2x_2^2}}{|x_{12}|}
  \right)^{\Delta_1+\Delta_2}
  \nonumber\\
  &=
  \frac{C_{12}}{|x_{12}|^{\Delta_1+\Delta_2}}
  (x_1^2)^{(\Delta_1+\Delta_2)/2}
  (x_2^2)^{(\Delta_1+\Delta_2)/2}.
  \label{eq:2pt_inversion_lhs}
\end{align}

On the other hand, the right-hand side of the covariance condition is
\begin{align}
  \Omega(x_1)^{-\Delta_1}
  \Omega(x_2)^{-\Delta_2}
  G_2(x_1,x_2)
  &=
  \left(\frac{1}{x_1^2}\right)^{-\Delta_1}
  \left(\frac{1}{x_2^2}\right)^{-\Delta_2}
  \frac{C_{12}}{|x_{12}|^{\Delta_1+\Delta_2}}
  \nonumber\\
  &=
  (x_1^2)^{\Delta_1}
  (x_2^2)^{\Delta_2}
  \frac{C_{12}}{|x_{12}|^{\Delta_1+\Delta_2}} .
  \label{eq:2pt_inversion_rhs}
\end{align}

Equating \eqref{eq:2pt_inversion_lhs} and
\eqref{eq:2pt_inversion_rhs}, and cancelling the common factor
$C_{12}/|x_{12}|^{\Delta_1+\Delta_2}$, gives
\begin{equation}
  (x_1^2)^{(\Delta_1+\Delta_2)/2}
  (x_2^2)^{(\Delta_1+\Delta_2)/2}
  =
  (x_1^2)^{\Delta_1}
  (x_2^2)^{\Delta_2}.
  \label{eq:2pt_inversion_compare}
\end{equation}
Since $x_1$ and $x_2$ are independent, the powers of $x_1^2$ and $x_2^2$
must agree separately. Hence
\begin{equation}
  \frac{\Delta_1+\Delta_2}{2}=\Delta_1,
  \qquad
  \frac{\Delta_1+\Delta_2}{2}=\Delta_2.
\end{equation}
Both equations imply
\begin{equation}
  \Delta_1=\Delta_2.
\end{equation}
Thus the two-point function of scalar primaries is
\begin{equation}
  \boxed{
  \big\langle
  \mathcal{O}_1(x_1)\mathcal{O}_2(x_2)
  \big\rangle
  =
  \frac{C_{12}\,\delta_{\Delta_1,\Delta_2}}
       {|x_{12}|^{2\Delta}}.
  }
  \label{eq:2pt}
\end{equation}
For a single normalized operator one usually chooses $C_{12}=1$.

\begin{figure}[h]
\centering
\includegraphics[width=0.90\textwidth]{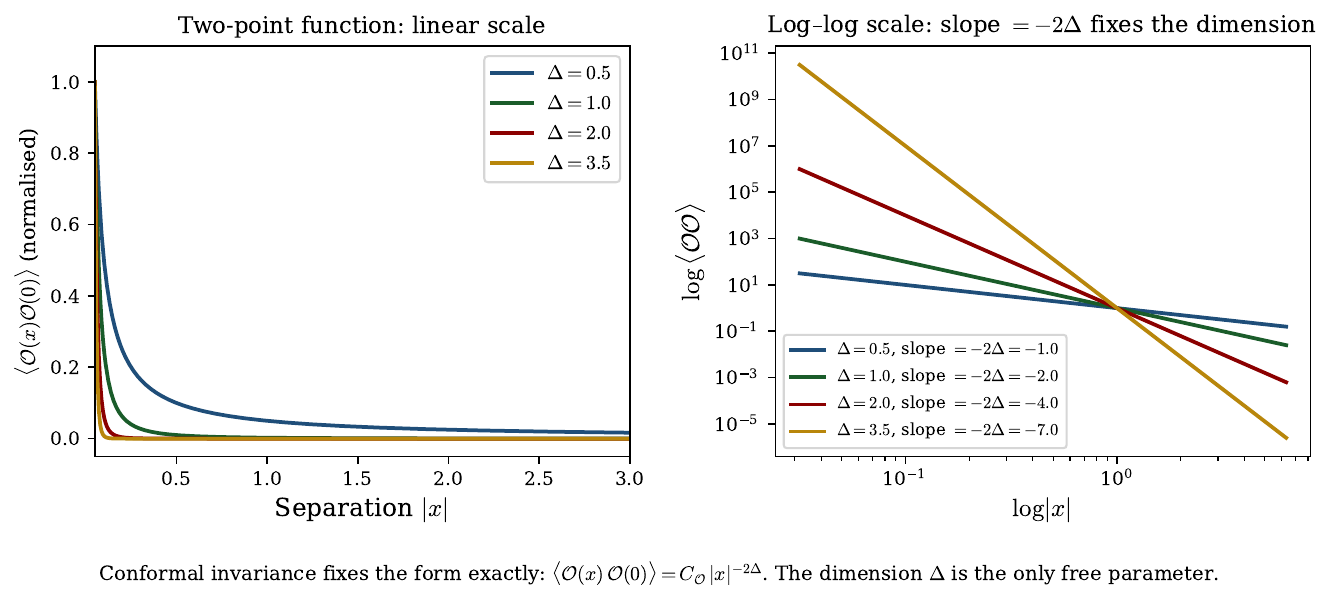}
\caption{%
\textbf{The two-point function fixes the scaling dimension.}
Conformal invariance uniquely determines
$\langle\mathcal{O}(x)\,\mathcal{O}(0)\rangle
= C_{\mathcal{O}}\,|x|^{-2\Delta}$;
the only free parameter is the dimension $\Delta$.
\textit{Left:} Linear scale for $\Delta=0.5,1,2,3.5$; higher dimensions decay
faster.
\textit{Right:} Log--log scale, where the relation becomes a straight line
with slope $-2\Delta$.  Measuring the slope of the two-point function in a
log--log plot is therefore the direct numerical route to extracting scaling
dimensions.}
\label{fig:cft_2pt}
\end{figure}

\subsection{Three-point function}

Now consider three scalar primaries of dimensions
$\Delta_1,\Delta_2,\Delta_3$:
\begin{equation}
  G_3(x_1,x_2,x_3)
  =
  \big\langle
  \mathcal{O}_1(x_1)
  \mathcal{O}_2(x_2)
  \mathcal{O}_3(x_3)
  \big\rangle .
\end{equation}
Translation and rotation invariance imply that the correlator can depend only
on the three distances
\begin{equation}
  r_{ij}=|x_i-x_j|.
\end{equation}
We therefore make the power-law ansatz
\begin{equation}
  G_3
  =
  C_{123}\,
  r_{12}^{a_{12}}
  r_{13}^{a_{13}}
  r_{23}^{a_{23}}.
  \label{eq:3pt_ansatz}
\end{equation}
Dilatation invariance gives
\begin{equation}
  a_{12}+a_{13}+a_{23}
  =
  -(\Delta_1+\Delta_2+\Delta_3).
  \label{eq:3pt_dil_constraint}
\end{equation}

To determine the individual exponents, use inversion. Distances transform as
\begin{equation}
  r'_{ij}
  =
  \frac{r_{ij}}{\sqrt{x_i^2x_j^2}}.
\end{equation}
Therefore the ansatz transforms as
\begin{align}
  G_3(x'_1,x'_2,x'_3)
  &=
  C_{123}
  \left(\frac{r_{12}}{\sqrt{x_1^2x_2^2}}\right)^{a_{12}}
  \left(\frac{r_{13}}{\sqrt{x_1^2x_3^2}}\right)^{a_{13}}
  \left(\frac{r_{23}}{\sqrt{x_2^2x_3^2}}\right)^{a_{23}}
  \nonumber\\
  &=
  G_3(x_1,x_2,x_3)\,
  (x_1^2)^{-(a_{12}+a_{13})/2}
  (x_2^2)^{-(a_{12}+a_{23})/2}
  (x_3^2)^{-(a_{13}+a_{23})/2}.
\end{align}
On the other hand, covariance under inversion gives
\begin{equation}
  G_3(x'_1,x'_2,x'_3)
  =
  (x_1^2)^{\Delta_1}
  (x_2^2)^{\Delta_2}
  (x_3^2)^{\Delta_3}
  G_3(x_1,x_2,x_3).
\end{equation}
Equating the powers of the independent variables $x_i^2$ gives
\begin{align}
  -\frac{a_{12}+a_{13}}{2} &= \Delta_1, \\
  -\frac{a_{12}+a_{23}}{2} &= \Delta_2, \\
  -\frac{a_{13}+a_{23}}{2} &= \Delta_3.
\end{align}
Equivalently,
\begin{align}
  a_{12}+a_{13} &= -2\Delta_1, \\
  a_{12}+a_{23} &= -2\Delta_2, \\
  a_{13}+a_{23} &= -2\Delta_3.
  \label{eq:3pt_sys}
\end{align}
Solving this linear system gives
\begin{align}
  a_{12} &= -\Delta_1-\Delta_2+\Delta_3,\\
  a_{13} &= -\Delta_1-\Delta_3+\Delta_2,\\
  a_{23} &= -\Delta_2-\Delta_3+\Delta_1.
\end{align}
Therefore
\begin{equation}
  \boxed{
  G_3
  =
  \frac{C_{123}}
  {|x_{12}|^{\Delta_1+\Delta_2-\Delta_3}
   |x_{13}|^{\Delta_1+\Delta_3-\Delta_2}
   |x_{23}|^{\Delta_2+\Delta_3-\Delta_1}}.
  }
  \label{eq:3pt_pre}
\end{equation}
The constant $C_{123}$ is not fixed by conformal symmetry. It is an independent
CFT datum: the OPE coefficient.

\subsection{Four-point function and conformal cross-ratios}

For four or more points, conformal symmetry does not fix the correlator
completely. The remaining freedom is encoded in conformal cross-ratios,
namely combinations of coordinates invariant under the full conformal group.
For four points in $d\geq2$, two independent cross-ratios are
\begin{equation}
  u
  =
  \frac{x_{12}^2x_{34}^2}{x_{13}^2x_{24}^2},
  \qquad
  v
  =
  \frac{x_{14}^2x_{23}^2}{x_{13}^2x_{24}^2}.
  \label{eq:cross_ratios}
\end{equation}
Thus a four-point function is fixed only up to an arbitrary function of
$(u,v)$. For example, for identical scalar primaries of dimension $\Delta$,
one may write
\begin{equation}
  \big\langle
  \mathcal{O}(x_1)\mathcal{O}(x_2)
  \mathcal{O}(x_3)\mathcal{O}(x_4)
  \big\rangle
  =
  \frac{1}{x_{12}^{2\Delta}x_{34}^{2\Delta}}\,
  g(u,v),
\end{equation}
where $g(u,v)$ is constrained by crossing symmetry and the operator product
expansion, but not fixed by conformal symmetry alone~\cite{arXiv:1204.3894,arXiv:1601.02883,arXiv:1706.07813,arXiv:1612.08987}.

\section{Conformal Ward identities}
\label{app:conf:ward}

\subsection{Derivation from the invariance condition}

The finite covariance condition \eqref{eq:corr_invariance} implies the
infinitesimal conformal Ward identities. Let
\begin{equation}
  x_i^\mu\to x_i'{}^\mu=x_i^\mu+\epsilon^\mu(x_i),
\end{equation}
where $\epsilon^\mu$ is a conformal Killing vector. To first order,
\begin{equation}
  \Omega(x_i)
  =
  1+\frac{1}{d}\partial_\mu\epsilon^\mu(x_i).
\end{equation}
Expanding \eqref{eq:corr_invariance} to first order gives
\begin{equation}
  G_n(x'_1,\ldots,x'_n)
  =
  G_n(x_1,\ldots,x_n)
  +
  \sum_{i=1}^n
  \epsilon^\mu(x_i)\partial_{x_i^\mu}G_n,
\end{equation}
while
\begin{equation}
  \prod_{i=1}^n
  \Omega(x_i)^{-\Delta_i}
  =
  1
  -
  \sum_{i=1}^n
  \frac{\Delta_i}{d}
  \partial_\mu\epsilon^\mu(x_i).
\end{equation}
Equating both sides and keeping terms linear in $\epsilon$ yields
\begin{equation}
  \boxed{
  \sum_{i=1}^n
  \left[
    \epsilon^\mu(x_i)\partial_{x_i^\mu}
    +
    \frac{\Delta_i}{d}
    \partial_\mu\epsilon^\mu(x_i)
  \right]
  G_n
  =
  0.
  }
  \label{eq:Ward_gen}
\end{equation}
This is the conformal Ward identity for scalar primary correlators.

\paragraph{{Spinning primaries.}}
{For an operator carrying spin, the variation picks up a local
rotation acting through the spin matrix $S^{\mu\nu}$,
\begin{equation*}
  \delta_\epsilon\mathcal{O}(x)
  = -\Bigl[\epsilon^\mu\partial_\mu
  + \tfrac{\Delta}{d}\,(\partial\!\cdot\!\epsilon)
  - \tfrac{i}{2}\,\partial_{[\mu}\epsilon_{\nu]}\,S^{\mu\nu}\Bigr]\mathcal{O}(x),
\end{equation*}
so that \eqref{eq:Ward_gen} is the trace (spin-zero, $S^{\mu\nu}=0$) part of the
general identity; the antisymmetrised derivative $\partial_{[\mu}\epsilon_{\nu]}$ is
the local Lorentz rotation generated by the conformal Killing vector, and signs
follow the conventions of Table~\ref{tab:conformal}.}

\paragraph{Translation Ward identity.}
For $\epsilon^\mu=a^\mu$, one has
$\partial_\mu\epsilon^\mu=0$, and therefore
\begin{equation}
  \sum_{i=1}^n
  \partial_{x_i^\mu}G_n=0.
\end{equation}
This expresses translation invariance.

\paragraph{Dilatation Ward identity.}
For $\epsilon^\mu=\lambda x^\mu$, one has
$\partial_\mu\epsilon^\mu=d\lambda$, so
\begin{equation}
  \boxed{
  \left(
  \sum_{i=1}^n x_i^\mu\partial_{x_i^\mu}
  +
  \sum_{i=1}^n \Delta_i
  \right)G_n
  =
  0.
  }
  \label{eq:dilation_ward}
\end{equation}
Thus the correlator is homogeneous of degree $-\sum_i\Delta_i$.

\paragraph{Special conformal Ward identity.}
For
\begin{equation}
  \epsilon^\mu(x)
  =
  b^\mu x^2-2(b\cdot x)x^\mu,
\end{equation}
one finds
\begin{equation}
  \partial_\mu\epsilon^\mu(x)
  =
  -2d\,b\cdot x.
\end{equation}
Substituting in \eqref{eq:Ward_gen} and factoring out the arbitrary parameter
$b^\mu$ gives
\begin{equation}
  \boxed{
  \sum_{i=1}^n
  \left[
    x_i^2\partial_{x_i^\mu}
    -
    2x_{i\mu}(x_i\cdot\partial_{x_i})
    -
    2\Delta_i x_{i\mu}
  \right]
  G_n
  =
  0.
  }
  \label{eq:SCT_ward}
\end{equation}
Together with translation, rotation, and dilatation invariance, this identity
fixes the scalar two-point function completely and the scalar three-point
function up to the constant $C_{123}$.

\section{Summary of conformal group data}
\label{app:conf:summary}

For quick reference, Table~\ref{tab:conformal} collects the generators, their geometric meaning,
their action on a scalar primary at $x=0$, and the dimension of the parameter space.

\begin{table}[ht]
\centering
\renewcommand{\arraystretch}{1.4}
\begin{tabular}{lcccc}
\hline
Generator & Symbol & Parameters & Action on $\mathcal{O}(0)$ & $\Omega(x)$ \\
\hline
Translations & $P_\mu$ & $d$ & $[P_\mu,\mathcal{O}(0)] = -i\partial_\mu\mathcal{O}(0)$ & $1$ \\
Rotations    & $M_{\mu\nu}$ & $d(d{-}1)/2$ & angular momentum matrices & $1$ \\
Dilatation   & $D$ & $1$ & $[D,\mathcal{O}(0)]=-i\Delta\,\mathcal{O}(0)$ & $\lambda$ \\
SCT          & $K_\mu$ & $d$ & $[K_\mu,\mathcal{O}(0)]=0$ & $(1{-}2b{\cdot}x{+}b^2x^2)^{-1}$ \\
\hline
\textbf{Total} & & $\mathbf{(d+1)(d+2)/2}$ & & \\
\hline
\end{tabular}
\caption{Generators of the conformal group $\mathrm{Conf}(\mathbb{R}^d)\cong\mathrm{SO}(d+1,1)$ in $d$ Euclidean dimensions.
The last column gives the Weyl factor $\Omega(x)$ for finite transformations.}
\label{tab:conformal}
\end{table}

\newpage


\chapter{State--Operator Correspondence}\label{chap:soc}
\label{app:soc}
 
The \emph{state--operator correspondence} is one of the structural pillars of
Euclidean conformal field theory. It states that local operators inserted at the
origin of flat Euclidean space are in one-to-one correspondence with states
obtained by quantizing the theory on a spatial sphere:
\begin{equation}
  \mathcal{O}(0)
  \quad \longleftrightarrow \quad
  \ket{\mathcal{O}} \in \mathcal{H}_{S^{d-1}} .
\end{equation}
Moreover, if $\mathcal{O}$ is a scaling operator of dimension $\Delta_{\mathcal{O}}$,
then the corresponding state on the cylinder $\mathbb{R}\times S^{d-1}_{R}$ has energy
\begin{equation}
  \boxed{
  \Delta_{\mathcal{O}} = R\,E_{\mathcal{O}} .
  }
  \label{eq:SOC_Delta_RE}
\end{equation}
Here $R$ is the radius of the spatial sphere.

The proof has three ingredients:
\begin{enumerate}
  \item flat space written in radial coordinates is Weyl-equivalent to the cylinder;
  \item radial evolution is generated by dilatations;
  \item a path integral with an operator insertion at the origin defines a state on the
  sphere surrounding that insertion.
\end{enumerate}
We first derive the Weyl map to the cylinder; the related geometric fact that a
conformally coupled scalar acquires an effective mass on the cylinder is
established separately in Appendix~\ref{app:conformal_coupling}~\cite{arXiv:1505.00963}.

\section{Flat space, radial coordinates, and the cylinder}
\label{app:soc:weyl_map}

Let $r$ denote the radial coordinate in $\mathbb{R}^d$:
\begin{equation}
  x^\mu = r\,\hat n^\mu,
  \qquad
  r = |x|,
  \qquad
  \hat n \in S^{d-1}.
\end{equation}
The flat Euclidean metric in these radial coordinates is
\begin{equation}
  ds_{\mathbb{R}^d}^2
  =
  dr^2+r^2 d\Omega_{d-1}^2 ,
  \label{eq:flat_radial_metric}
\end{equation}
where $d\Omega_{d-1}^2$ is the round metric on the unit sphere $S^{d-1}$.  The
derivation --- from the Cartesian line element, through the cross-term
cancellation enforced by $\hat n_\mu\,d\hat n^\mu=0$, to the explicit angular
metric in generalized spherical coordinates --- is collected in
Appendix~\ref{app:metric_spherical}.

To obtain a cylinder whose spatial sphere has radius $R$, define the logarithmic
radial coordinate
\begin{equation}
  r = R\,e^{\tau/R}.
  \label{eq:r_tau_def}
\end{equation}
Then
\begin{equation}
  dr = e^{\tau/R}\,d\tau,
  \qquad
  dr^2 = e^{2\tau/R}\,d\tau^2,
\end{equation}
and
\begin{equation}
  r^2 d\Omega_{d-1}^2
  =
  R^2 e^{2\tau/R} d\Omega_{d-1}^2 .
\end{equation}
Substitution into \eqref{eq:flat_radial_metric} gives
\begin{equation}
  ds_{\mathbb{R}^d}^2
  =
  e^{2\tau/R}
  \left(
    d\tau^2+R^2 d\Omega_{d-1}^2
  \right).
  \label{eq:flat_to_cylinder_metric}
\end{equation}
Therefore
\begin{equation}
  ds_{\mathbb{R}^d}^2
  =
  \Omega(\tau)^2 ds_{\rm cyl}^2,
  \qquad
  ds_{\rm cyl}^2
  =
  d\tau^2+R^2 d\Omega_{d-1}^2,
\end{equation}
with Weyl factor
\begin{equation}
  \boxed{
  \Omega(\tau)=e^{\tau/R}=\frac{r}{R}.
  }
  \label{eq:weyl_factor_cylinder}
\end{equation}
Thus
\begin{equation}
  \mathbb{R}^d\setminus\{0\}
  \simeq
  \mathbb{R}_{\tau}\times S^{d-1}_{R}
\end{equation}
up to the Weyl factor \eqref{eq:weyl_factor_cylinder}.

The origin and infinity of flat space become the two asymptotic ends of the
cylinder:
\begin{equation}
  r\to 0
  \quad \Longleftrightarrow \quad
  \tau\to -\infty,
  \qquad
  r\to \infty
  \quad \Longleftrightarrow \quad
  \tau\to +\infty .
\end{equation}

\section{Weyl transformation of primary operators}
\label{app:soc:weyl_primary}

Under a Weyl transformation
\begin{equation}
  g_{\mu\nu}\longrightarrow g'_{\mu\nu}
  =
  e^{2\sigma(x)}g_{\mu\nu},
\end{equation}
a scalar primary operator of scaling dimension $\Delta$ transforms as
\begin{equation}
  \mathcal{O}'(x)
  =
  e^{-\Delta\sigma(x)}\mathcal{O}(x).
  \label{eq:primary_weyl_general}
\end{equation}

In the present case,
\begin{equation}
  g_{\mathbb{R}^d}
  =
  \Omega^2 g_{\rm cyl},
  \qquad
  g_{\rm cyl}
  =
  \Omega^{-2}g_{\mathbb{R}^d}.
\end{equation}
Hence the cylinder metric is obtained from the flat metric by a Weyl rescaling
with factor $\Omega^{-1}$. Therefore a scalar primary becomes
\begin{equation}
  \boxed{
  \mathcal{O}_{\rm cyl}(\tau,\hat n)
  =
  \Omega(\tau)^{\Delta}\,
  \mathcal{O}_{\mathbb{R}^d}(r\hat n)
  =
  \left(\frac{r}{R}\right)^{\Delta}
  \mathcal{O}_{\mathbb{R}^d}(r\hat n).
  }
  \label{eq:O_cyl_from_flat}
\end{equation}
The positive power is important: it converts flat-space power laws into
exponential cylinder propagation factors.

\section{The conformal scalar on the cylinder}
\label{app:soc:conformal_scalar}

When the flat-space conformal scalar is placed on the curved cylinder, Weyl
invariance forces the minimally coupled action to be supplemented by the
conformal curvature coupling $\xi_c R\phi^2$ with $\xi_c=(d-2)/[4(d-1)]$.  On
$\mathbb{R}\times S^{d-1}_R$ the sphere curvature then turns this coupling into
an effective ``conformal mass''
\begin{equation*}
  \mu = \frac{d-2}{2R},
\end{equation*}
a purely geometric consequence of Weyl invariance rather than an explicit
breaking of conformal symmetry.  The full derivation --- the failure of Weyl
invariance of the minimal action, the fixing of $\xi_c$, and the evaluation of
$\xi_c R_{\rm cyl}$ --- is given in Appendix~\ref{app:conformal_coupling}.

\section{Radial quantization}
\label{app:soc:radial_quantization}

We now turn to the Hilbert-space construction. In radial quantization, the role of
Euclidean time is played by the logarithmic radial coordinate
\begin{equation}
  \tau = R\log\frac{r}{R}.
\end{equation}
The constant-$\tau$ slices are spheres
\begin{equation}
  S^{d-1}_{R}.
\end{equation}
Therefore the Hilbert space is obtained by quantizing the theory on
$S^{d-1}_{R}$:
\begin{equation}
  \mathcal{H}_{S^{d-1}_{R}}.
\end{equation}

A state on the sphere at Euclidean time $\tau_0$ may be prepared by a path integral
over the interior region $\tau<\tau_0$. Fixing the boundary value of the field to
be $\varphi(\hat n)$ at $\tau=\tau_0$, one defines the wavefunctional
\begin{equation}
  \Psi[\varphi]
  =
  \int_{\phi(\tau_0,\hat n)=\varphi(\hat n)}
  \mathcal{D}\phi\,
  e^{-S_E[\phi]}.
  \label{eq:radial_wavefunctional}
\end{equation}
If local operators are inserted inside the ball, the same path integral prepares
a state depending on those insertions~\cite{arXiv:1811.00528,arXiv:2010.09730,arXiv:1703.00278}.

\section{Dilatations become cylinder time translations}
\label{app:soc:D_as_H}

A scale transformation in flat space acts as
\begin{equation}
  x^\mu\longrightarrow \lambda x^\mu,
  \qquad
  r\longrightarrow \lambda r.
\end{equation}
Using
\begin{equation}
  r=R e^{\tau/R},
\end{equation}
we find
\begin{equation}
  \lambda r
  =
  R e^{\tau'/R}
  =
  \lambda R e^{\tau/R}.
\end{equation}
Thus
\begin{equation}
  e^{\tau'/R}
  =
  \lambda e^{\tau/R},
\end{equation}
and therefore
\begin{equation}
  \boxed{
  \tau'
  =
  \tau+R\log\lambda.
  }
  \label{eq:dilatation_tau_translation}
\end{equation}
Dilatations in flat space are translations along the cylinder.

Let $D$ denote the dimensionless generator of dilatations and $H_{\rm cyl}$ the
Hamiltonian generating translations in the dimensionful Euclidean cylinder time
$\tau$. A finite cylinder translation by $\Delta\tau$ corresponds to
\begin{equation}
  \lambda=e^{\Delta\tau/R}.
\end{equation}
Therefore
\begin{equation}
  \boxed{
  D=R H_{\rm cyl},
  \qquad
  H_{\rm cyl}=\frac{D}{R}.
  }
  \label{eq:D_RH}
\end{equation}

\section{From a local operator to a state}
\label{app:soc:operator_to_state}

Let $\mathcal{O}(x)$ be a renormalized local operator. Insert it at the origin of
flat space. Surround the origin by a small sphere and perform the path integral
over the ball. In cylinder coordinates, the origin is at $\tau=-\infty$.
Therefore the associated state is
\begin{equation}
  \boxed{
  \ket{\mathcal{O}}
  =
  \lim_{\tau\to-\infty}
  {e^{-\Delta_{\mathcal{O}}\tau/R}\,}\mathcal{O}_{\rm cyl}(\tau,\hat n)\ket{0}.
  }
  \label{eq:operator_to_state_cyl}
\end{equation}
For a scalar primary of dimension $\Delta_{\mathcal{O}}$, this is equivalently
\begin{equation}
  \boxed{
  \ket{\mathcal{O}}
  =
  \lim_{r\to 0}
  \mathcal{O}_{\mathbb{R}^d}(r\hat n)\ket{0}
  \;=\;
  \mathcal{O}(0)\ket{0}.
  }
  \label{eq:operator_to_state_flat}
\end{equation}
{
  The compensating factor ~\eqref{eq:operator_to_state_cyl}  on the cylinder  $e^{-\Delta_{\mathcal{O}}\tau/R}
=(R/r)^{\Delta_{\mathcal{O}}}$,  offsets $\mathcal{O}_{\rm cyl}\to0$ as
$\tau\to-\infty$.}
For a primary, the limit is independent of $\hat n$. Descendants are obtained by
acting with derivatives, equivalently with the translation generators $P_\mu$.

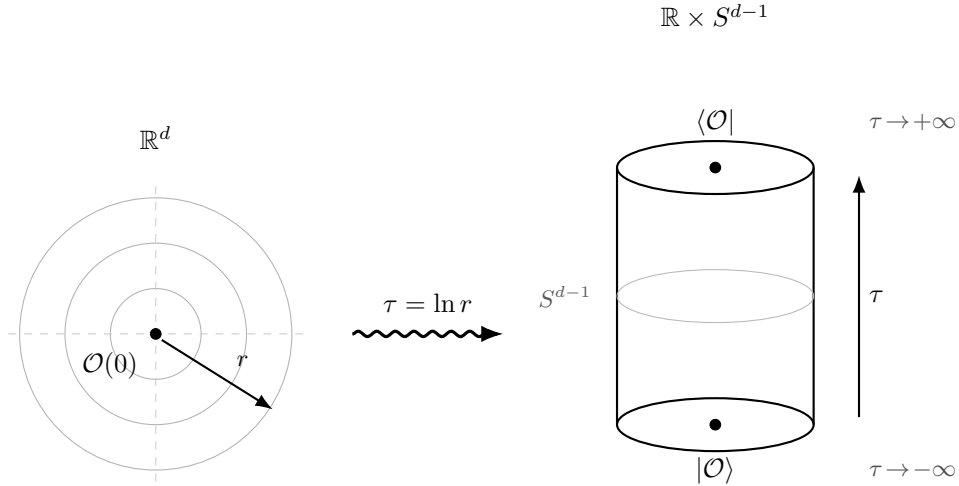
\begin{figure}[h]
\centering
\begin{tikzpicture}[>=Latex]

\begin{scope}[shift={(-3.8,1.6)}]

  \node[font=\bfseries\small] at (0,2.6) {$\mathbb{R}^d$};

  \draw[gray!60, thin] (0,0) circle (0.6);
  \draw[gray!60, thin] (0,0) circle (1.2);
  \draw[gray!60, thin] (0,0) circle (1.8);

  \draw[gray!40, thin, dashed] (0,-1.95) -- (0,1.95);
  \draw[gray!40, thin, dashed] (-1.95,0) -- (1.95,0);

  \filldraw[black] (0,0) circle (2pt);
  \node[font=\small, anchor=north east] at (-0.1,-0.1) {$\mathcal{O}(0)$};

  \draw[->, thick] (0.08,-0.08) -- (1.55,-1.0);
  \node[font=\footnotesize] at (1.15,-0.35) {$r$};

\end{scope}

\draw[->, very thick, decorate,
  decoration={snake, amplitude=1pt, segment length=7pt, post length=3pt}]
  (-1.2,1.6) -- (0.8,1.6);
\node[font=\small, above] at (-0.2,1.75) {$\tau = \ln r$};

\begin{scope}[shift={(3.6,0)}]

  \node[font=\bfseries\small] at (0,5.8) {$\mathbb{R}\times S^{d-1}$};

  \draw[thick] (-1.3,0.4) -- (-1.3,3.8);
  \draw[thick] ( 1.3,0.4) -- ( 1.3,3.8);

  \draw[thick] (0,0.4) ellipse (1.3 and 0.35);

  \draw[thick] (0,3.8) ellipse (1.3 and 0.35);

  \draw[thin, gray!55] (0,2.1) ellipse (1.3 and 0.35);
  \node[font=\footnotesize, gray!70!black, left] at (-1.5,2.1) {$S^{d-1}$};

  \draw[->, thick] (1.9,0.5) -- (1.9,3.7);
  \node[font=\small, right] at (1.9,2.1) {$\tau$};

  \filldraw[black] (0,0.4) circle (2pt);
  \node[font=\small] at (0,-0.2) {$|\mathcal{O}\rangle$};
  \node[font=\footnotesize, gray!60!black, right] at (1.9,-0.2) {$\tau\!\to\!-\infty$};

  \filldraw[black] (0,3.8) circle (2pt);
  \node[font=\small] at (0,4.4) {$\langle\mathcal{O}|$};
  \node[font=\footnotesize, gray!60!black, right] at (1.9,4.4) {$\tau\!\to\!+\infty$};

\end{scope}

\end{tikzpicture}
\caption{Radial quantization: the conformal map $\tau=\ln|x|$ sends $\mathbb{R}^d$
(left, shown with constant-$r$ circles) to the cylinder $\mathbb{R}\times S^{d-1}$ (right).
The operator insertion $\mathcal{O}(0)$ at the origin maps to the state $|\mathcal{O}\rangle$
at $\tau\to-\infty$; spatial infinity maps to $\tau\to+\infty$.
Dilatations in $\mathbb{R}^d$ become $\tau$-translations on the cylinder.}
\label{fig:cylinder}
\end{figure}

\section{Energy of the state created by a primary}
\label{app:soc:energy_primary}

Let $\mathcal{O}$ be a scalar primary of scaling dimension $\Delta_{\mathcal{O}}$.
Its flat-space two-point function is
\begin{equation}
  \langle
  \mathcal{O}_{\mathbb{R}^d}(x_2)
  \mathcal{O}_{\mathbb{R}^d}(x_1)
  \rangle
  =
  \frac{C_{\mathcal{O}}}{|x_2-x_1|^{2\Delta_{\mathcal{O}}}}.
  \label{eq:flat_2pt_soc}
\end{equation}

\begin{figure}[h]
\centering
\begin{tikzpicture}[>=Latex]

\begin{scope}[shift={(-4,1.6)}]

  \node[font=\bfseries\small] at (0,2.6) {$\mathbb{R}^d$};

  \draw[gray!50, thin] (0,0) circle (0.5);
  \draw[gray!50, thin] (0,0) circle (1.0);
  \draw[gray!50, thin] (0,0) circle (1.5);
  \draw[gray!50, thin] (0,0) circle (2.0);

  \draw[gray!30, thin, dashed] (0,-2.15) -- (0,2.15);
  \draw[gray!30, thin, dashed] (-2.15,0) -- (2.15,0);

  \filldraw[blue!70!black] (0.35,0.35) circle (2.5pt);
  \node[font=\small, blue!70!black, above right] at (0.4,0.4)
    {$\mathcal{O}_1(x_1)$};

  \filldraw[red!70!black] (-0.95,1.15) circle (2.5pt);
  \node[font=\small, red!70!black, above left] at (-1.0,1.2)
    {$\mathcal{O}_2(x_2)$};

  \draw[blue!70!black, thin, ->] (0,0) -- (0.32,0.32);
  \node[font=\scriptsize, blue!70!black, below] at (0.18,0.1) {$r_1$};
  \draw[red!70!black, thin, ->] (0,0) -- (-0.92,1.12);
  \node[font=\scriptsize, red!70!black, right] at (-0.35,0.65) {$r_2$};

  \node[font=\small] at (0,-2.6)
    {$\langle \mathcal{O}_2(x_2)\,\mathcal{O}_1(x_1)\rangle$};

\end{scope}

\draw[->, very thick, decorate,
  decoration={snake, amplitude=1pt, segment length=7pt, post length=3pt}]
  (-1.1,1.6) -- (0.9,1.6);
\node[font=\small, above] at (-0.1,1.75) {$\tau_i = \ln r_i$};

\begin{scope}[shift={(3.8,0)}]

  \node[font=\bfseries\small] at (0,5.2) {$\mathbb{R}\times S^{d-1}$};

  \draw[thick] (-1.3,0) -- (-1.3,4.2);
  \draw[thick] ( 1.3,0) -- ( 1.3,4.2);

  \draw[thick] (0,0) ellipse (1.3 and 0.35);

  \draw[thick] (0,4.2) ellipse (1.3 and 0.35);

  \draw[blue!60!black, thin] (0,1.0) ellipse (1.3 and 0.35);
  \filldraw[blue!70!black] (0,1.0) circle (2.5pt);
  \node[font=\footnotesize, blue!70!black, left] at (-1.5,1.0)
    {$\tau_1\!=\!\ln r_1$};

  \draw[red!60!black, thin] (0,2.8) ellipse (1.3 and 0.35);
  \filldraw[red!70!black] (0,2.8) circle (2.5pt);
  \node[font=\footnotesize, red!70!black, left] at (-1.5,2.8)
    {$\tau_2\!=\!\ln r_2$};

  \draw[->, thick, green!50!black, densely dotted] (0,1.15) -- (0,2.65);
  \node[font=\footnotesize, green!50!black, fill=white, inner sep=1pt] at (0,1.9)
    {$e^{-D\,\Delta\tau}$};

  \draw[->, thick] (1.9,0.1) -- (1.9,4.1);
  \node[font=\small, right] at (1.9,2.1) {$\tau$};

  \node[font=\footnotesize, gray!60!black, right] at (1.9,0.0)
    {$\tau\!\to\!-\infty$};
  \node[font=\footnotesize, gray!60!black, right] at (1.9,4.2)
    {$\tau\!\to\!+\infty$};

  \node[font=\small] at (0,-0.9)
    {$\langle \mathcal{O}_2|\,e^{-D(\tau_2-\tau_1)}\,|\mathcal{O}_1\rangle$};

\end{scope}

\end{tikzpicture}
\caption{Correlators as cylinder amplitudes.
\textbf{Left:} Two operator insertions at $|x_1|=r_1$ and $|x_2|=r_2>r_1$ in flat space.
\textbf{Right:} On the cylinder, these become states on the $S^{d-1}$ slices at
$\tau_1=\ln r_1$ and $\tau_2=\ln r_2$;
the two-point function is the matrix element of $e^{-D\,\Delta\tau}$ where $D$ is the
dilatation operator (cylinder Hamiltonian).
Inserting a complete set of eigenstates $D|\Delta\rangle = \Delta|\Delta\rangle$ yields
the OPE decomposition $\sum_\Delta c_\Delta\, (r_2/r_1)^{-\Delta}$.}
\label{fig:correlator_map}
\end{figure}
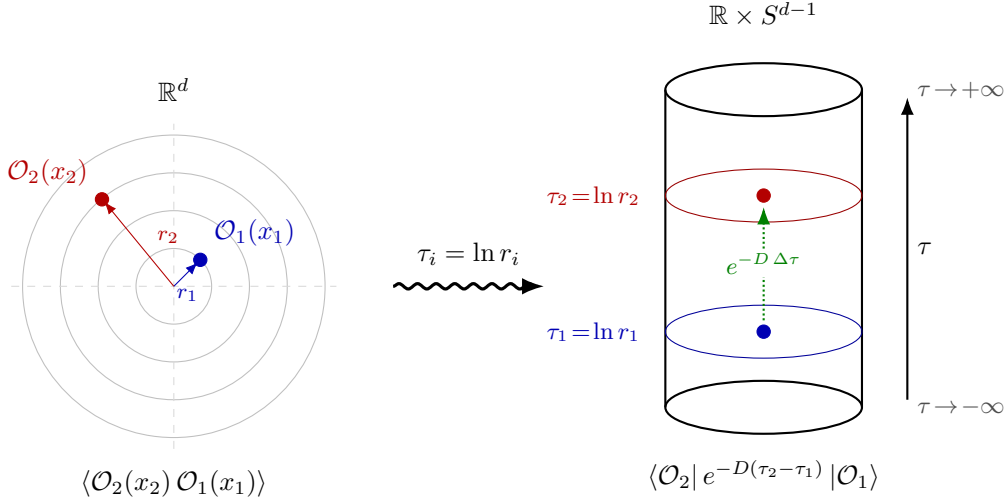

For computational convenience we align both insertions along the same unit
vector $\hat n$ (the general-position picture is shown in
Figure~\ref{fig:correlator_map}; the conclusion is the same by conformal
invariance of the two-point function):
\begin{equation}
  x_i=r_i\hat n,
  \qquad
  r_i=R e^{\tau_i/R},
  \qquad
  \tau_2>\tau_1.
\end{equation}
Then
\begin{equation}
  |x_2-x_1|
  =
  r_2-r_1
  =
  R e^{\tau_2/R}
  \left(
    1-e^{-(\tau_2-\tau_1)/R}
  \right).
\end{equation}
Using \eqref{eq:O_cyl_from_flat}, the cylinder two-point function is
\begin{align}
  \langle
  \mathcal{O}_{\rm cyl}(\tau_2,\hat n)
  \mathcal{O}_{\rm cyl}(\tau_1,\hat n)
  \rangle
  &=
  \left(\frac{r_2}{R}\right)^{\Delta_{\mathcal{O}}}
  \left(\frac{r_1}{R}\right)^{\Delta_{\mathcal{O}}}
  \frac{C_{\mathcal{O}}}{(r_2-r_1)^{2\Delta_{\mathcal{O}}}}
  \nonumber\\
  &=
  C_{\mathcal{O}}\,
  \frac{
    e^{\Delta_{\mathcal{O}}(\tau_1+\tau_2)/R}
  }{
    e^{2\Delta_{\mathcal{O}}\tau_2/R}
    \left(
      1-e^{-(\tau_2-\tau_1)/R}
    \right)^{2\Delta_{\mathcal{O}}}
  }
  \nonumber\\
  &=
  C_{\mathcal{O}}\,
  \frac{
    e^{-\Delta_{\mathcal{O}}(\tau_2-\tau_1)/R}
  }{
    \left(
      1-e^{-(\tau_2-\tau_1)/R}
    \right)^{2\Delta_{\mathcal{O}}}
  }.
  \label{eq:cyl_2pt_intermediate}
\end{align}
For large Euclidean time separation,
\begin{equation}
  \tau_2-\tau_1\gg R,
\end{equation}
this becomes
\begin{equation}
  \langle
  \mathcal{O}_{\rm cyl}(\tau_2,\hat n)
  \mathcal{O}_{\rm cyl}(\tau_1,\hat n)
  \rangle
  \sim
  C_{\mathcal{O}}\,
  e^{-\Delta_{\mathcal{O}}(\tau_2-\tau_1)/R}.
  \label{eq:cyl_2pt_large_tau}
\end{equation}

On the other hand, the spectral decomposition on the cylinder gives
\begin{equation}
  \langle
  \mathcal{O}_{\rm cyl}(\tau_2)
  \mathcal{O}_{\rm cyl}(\tau_1)
  \rangle
  =
  \sum_\alpha
  |\langle \alpha|\mathcal{O}_{\rm cyl}|0\rangle|^2
  e^{-E_\alpha(\tau_2-\tau_1)}.
  \label{eq:spectral_cylinder_soc}
\end{equation}
In the limit $\tau_2-\tau_1\gg R$ the spectral sum
\eqref{eq:spectral_cylinder_soc} is dominated by its slowest-decaying term ---
that is, by the state of \emph{lowest} energy that $\mathcal{O}_{\rm cyl}$
creates from the vacuum.  Acting with $\mathcal{O}_{\rm cyl}$ on $\ket{0}$
produces the primary state $\ket{\mathcal{O}}$ together with its descendants;
the descendants sit at energies $E_{\mathcal{O}}+1/R,\,E_{\mathcal{O}}+2/R,\ldots$
above it (each $P_\mu$ raises the energy by $1/R$) and are exponentially
suppressed relative to the primary.  Hence the leading large-time behaviour of
\eqref{eq:spectral_cylinder_soc} is
\begin{equation}
  \langle
  \mathcal{O}_{\rm cyl}(\tau_2)
  \mathcal{O}_{\rm cyl}(\tau_1)
  \rangle
  \;\sim\;
  |\langle \mathcal{O}|\mathcal{O}_{\rm cyl}|0\rangle|^2\,
  e^{-E_{\mathcal{O}}(\tau_2-\tau_1)},
  \qquad
  \tau_2-\tau_1\gg R,
  \label{eq:spectral_leading_soc}
\end{equation}
controlled by the energy $E_{\mathcal{O}}$ of the primary state itself --- not by
any descendant.  Comparing this with the explicit result
\eqref{eq:cyl_2pt_large_tau} for the same large-time correlator, the two
exponentials agree only if the state created by $\mathcal{O}$ has energy
\begin{equation}
  E_{\mathcal{O}}
  =
  \frac{\Delta_{\mathcal{O}}}{R},
\end{equation}
and matching the prefactors fixes the normalization
$C_{\mathcal{O}}=|\langle\mathcal{O}|\mathcal{O}_{\rm cyl}|0\rangle|^2$.
Thus
\begin{equation}
  \boxed{
  \Delta_{\mathcal{O}}=R\,E_{\mathcal{O}}.
  }
\end{equation}

\section{Algebraic form of the same result}
\label{app:soc:algebraic}

The same conclusion follows from the conformal algebra. At the origin, a primary
satisfies
\begin{equation}
  D\,\mathcal{O}(0)\ket{0}
  =
  \Delta_{\mathcal{O}}\,
  \mathcal{O}(0)\ket{0},
\end{equation}
where $D$ is the dimensionless dilatation generator.  {Here $D$ is taken
in the radial-quantization convention, in which it acts as the (real) cylinder
Hamiltonian $R\,H_{\rm cyl}$ with eigenvalue $+\Delta_{\mathcal{O}}$ and $[D,P_\mu]=P_\mu$;
it differs by a factor of $i$ from the Hermitian generator of
Chapter~\ref{chap:prereq} (where $[D,\mathcal{O}(0)]=-i\Delta\,\mathcal{O}(0)$ and
$[D,P_\mu]=+iP_\mu$), the explicit $i$'s having been absorbed into the definition.}
Since
\begin{equation}
  D=R H_{\rm cyl},
\end{equation}
the state
\begin{equation}
  \ket{\mathcal{O}}=\mathcal{O}(0)\ket{0}
\end{equation}
obeys
\begin{equation}
  H_{\rm cyl}\ket{\mathcal{O}}
  =
  \frac{\Delta_{\mathcal{O}}}{R}\ket{\mathcal{O}}.
\end{equation}
Therefore
\begin{equation}
  E_{\mathcal{O}}=\frac{\Delta_{\mathcal{O}}}{R}.
\end{equation}

Descendants arise by acting with $P_\mu$. Since
\begin{equation}
  [D,P_\mu]=P_\mu,
\end{equation}
each action of $P_\mu$ raises the scaling dimension by one. Thus
\begin{equation}
  P_{\mu_1}\cdots P_{\mu_k}\mathcal{O}(0)
\end{equation}
corresponds to a cylinder state of energy
\begin{equation}
  E_{\rm desc}
  =
  \frac{\Delta_{\mathcal{O}}+k}{R}.
\end{equation}

\section{Bra states and insertions at infinity}
\label{app:soc:bra_states}

The ket state associated with a scalar primary $\mathcal{O}$ is obtained by inserting
$\mathcal{O}$ at the origin of flat space, or equivalently at $\tau=-\infty$ on
the cylinder:
\begin{equation}
  \ket{\mathcal{O}}
  =
  \lim_{r\to 0}
  \mathcal{O}_{\mathbb{R}^d}(r\hat n)\ket{0}.
  \label{eq:ket_from_origin_detailed}
\end{equation}
 The conjugate bra state should therefore be obtained from the opposite end of the
cylinder, namely from $\tau=+\infty$. Since
\begin{equation}
  r=R e^{\tau/R},
\end{equation}
the limit $\tau\to+\infty$ is precisely the flat-space limit $r\to\infty$.

The subtle point is the normalization.  A primary two-point function in flat space
has the form
\begin{equation}
  \langle
  \mathcal{O}_{\mathbb{R}^d}(x)
  \mathcal{O}_{\mathbb{R}^d}(0)
  \rangle
  =
  \frac{C_{\mathcal{O}}}{|x|^{2\Delta_{\mathcal{O}}}}.
  \label{eq:primary_two_point_origin_infinity}
\end{equation}
Taking $x=r\hat n$ gives
\begin{equation}
  \langle
  \mathcal{O}_{\mathbb{R}^d}(r\hat n)
  \mathcal{O}_{\mathbb{R}^d}(0)
  \rangle
  =
  \frac{C_{\mathcal{O}}}{r^{2\Delta_{\mathcal{O}}}}.
  \label{eq:two_point_large_r}
\end{equation}
Therefore the naive limit
\begin{equation}
  \lim_{r\to\infty}
  \bra{0}\mathcal{O}_{\mathbb{R}^d}(r\hat n)
\end{equation}
would vanish when evaluated on $\ket{\mathcal{O}}$. To obtain a finite bra state,
one must multiply the insertion at infinity by $r^{2\Delta_{\mathcal{O}}}$.
Thus, up to the harmless normalization factor $R^{-2\Delta_{\mathcal{O}}}$, one defines
\begin{equation}
  \boxed{
  \bra{\mathcal{O}}
  =
  \lim_{r\to\infty}
  \left(\frac{r}{R}\right)^{2\Delta_{\mathcal{O}}}
  \bra{0}\,
  \mathcal{O}_{\mathbb{R}^d}(r\hat n).
  }
  \label{eq:bra_infinity_soc_detailed}
\end{equation}
Indeed, using \eqref{eq:two_point_large_r},
\begin{align}
  \braket{\mathcal{O}}{\mathcal{O}}
  &=
  \lim_{r\to\infty}
  \left(\frac{r}{R}\right)^{2\Delta_{\mathcal{O}}}
  \langle
  \mathcal{O}_{\mathbb{R}^d}(r\hat n)
  \mathcal{O}_{\mathbb{R}^d}(0)
  \rangle
  \nonumber\\
  &=
  \lim_{r\to\infty}
  \left(\frac{r}{R}\right)^{2\Delta_{\mathcal{O}}}
  \frac{C_{\mathcal{O}}}{r^{2\Delta_{\mathcal{O}}}}
  \nonumber\\
  &=
  \frac{C_{\mathcal{O}}}{R^{2\Delta_{\mathcal{O}}}}.
  \label{eq:bra_ket_norm_with_R}
\end{align}
If desired, one may absorb the factor $R^{-2\Delta_{\mathcal{O}}}$ into the
normalization of the state, or set $R=1$. With the common convention
$C_{\mathcal{O}}=1$ and $R=1$, the state has unit norm:
\begin{equation}
  \braket{\mathcal{O}}{\mathcal{O}}=1.
\end{equation}

\subsection{Derivation using the cylinder and relation to conformal inversion}

The same normalization can be understood directly on the cylinder.  The Weyl map gives
\begin{equation}
  \mathcal{O}_{\rm cyl}(\tau,\hat n)
  =
  \left(\frac{r}{R}\right)^{\Delta_{\mathcal{O}}}
  \mathcal{O}_{\mathbb{R}^d}(r\hat n),
  \qquad
  r=R e^{\tau/R}.
  \label{eq:cyl_operator_for_bra}
\end{equation}
Thus
\begin{equation}
  \mathcal{O}_{\rm cyl}(\tau,\hat n)
  =
  e^{\Delta_{\mathcal{O}}\tau/R}
  \mathcal{O}_{\mathbb{R}^d}(r\hat n).
  \label{eq:cyl_operator_exp_tau}
\end{equation}
On the cylinder, a bra state associated with $\mathcal{O}$ is obtained by sending
the insertion to future Euclidean time:
\begin{equation}
  \bra{\mathcal{O}}
  =
  \lim_{\tau\to+\infty}
  e^{E_{\mathcal{O}}\tau}
  \bra{0}\mathcal{O}_{\rm cyl}(\tau,\hat n).
  \label{eq:bra_cylinder_definition}
\end{equation}
The factor $e^{E_{\mathcal{O}}\tau}$ removes the Euclidean damping
$e^{-E_{\mathcal{O}}\tau}$ associated with propagation to $\tau=+\infty$.
Using
\begin{equation}
  E_{\mathcal{O}}=\frac{\Delta_{\mathcal{O}}}{R},
\end{equation}
together with \eqref{eq:cyl_operator_exp_tau}, we find
\begin{align}
  e^{E_{\mathcal{O}}\tau}
  \bra{0}\mathcal{O}_{\rm cyl}(\tau,\hat n)
  &=
  e^{\Delta_{\mathcal{O}}\tau/R}
  \bra{0}
  \left[
    e^{\Delta_{\mathcal{O}}\tau/R}
    \mathcal{O}_{\mathbb{R}^d}(r\hat n)
  \right]
  \nonumber\\
  &=
  e^{2\Delta_{\mathcal{O}}\tau/R}
  \bra{0}\mathcal{O}_{\mathbb{R}^d}(r\hat n)
  \nonumber\\
  &=
  \left(\frac{r}{R}\right)^{2\Delta_{\mathcal{O}}}
  \bra{0}\mathcal{O}_{\mathbb{R}^d}(r\hat n).
  \label{eq:cylinder_to_flat_bra_factor}
\end{align}
This reproduces precisely \eqref{eq:bra_infinity_soc_detailed}.

There is also a useful intrinsic flat-space way to define the insertion at infinity.
Recall the inversion map
\begin{equation}
  I:\qquad
  x^\mu \longmapsto x'^\mu=\frac{x^\mu}{x^2} \ ,
\end{equation}
with  conformal factor 
\begin{equation}
  \Omega_I(x)=\frac{1}{x^2}.
\end{equation}
A scalar primary transforms as
\begin{equation}
  \mathcal{O}'(x')
  =
  \Omega_I(x)^{-\Delta_{\mathcal{O}}}
  \mathcal{O}(x)
  =
  (x^2)^{\Delta_{\mathcal{O}}}\mathcal{O}(x).
\end{equation}
Since $x^2=r^2$, the factor is
\begin{equation}
  (x^2)^{\Delta_{\mathcal{O}}}=r^{2\Delta_{\mathcal{O}}}.
\end{equation}
This explains why the natural flat-space definition of the operator at infinity is
\begin{equation}
  \boxed{
  \mathcal{O}(\infty)
  \equiv
  \lim_{r\to\infty}
  r^{2\Delta_{\mathcal{O}}}
  \mathcal{O}(r\hat n).
  }
  \label{eq:operator_at_infinity}
\end{equation}
With this notation,
\begin{equation}
  \bra{\mathcal{O}}
  =
  R^{-2\Delta_{\mathcal{O}}}
  \bra{0}\mathcal{O}(\infty).
\end{equation}
The factor of $R^{-2\Delta_{\mathcal{O}}}$ is only due to the convention of
measuring cylinder energies with a sphere of radius $R$.

\subsection{Radial ordering and Hermitian conjugation}

In ordinary canonical quantization, time ordering orders operators according to
their Euclidean time.  In radial quantization, the role of Euclidean time is played
by
\begin{equation}
  \tau = R\log\frac{r}{R}.
\end{equation}
Therefore increasing Euclidean time $\tau$ is the same as increasing the radial
distance $r=|x|$ from the origin.  Radial ordering is simply time ordering with
respect to $\tau$, or equivalently ordering according to the radius.

For two local operators, radial ordering is defined by
\begin{equation}
  \mathcal{R}\!\left\{
  \mathcal{O}_1(x_1)\mathcal{O}_2(x_2)
  \right\}
  =
  \begin{cases}
  \mathcal{O}_1(x_1)\mathcal{O}_2(x_2),
  & |x_1|>|x_2|, \\[4pt]
  \mathcal{O}_2(x_2)\mathcal{O}_1(x_1),
  & |x_2|>|x_1|.
  \end{cases}
  \label{eq:radial_ordering_def}
\end{equation}
Thus the operator with larger radius is placed to the left, just as the operator
at later Euclidean time is placed to the left in ordinary time ordering.

The relation with Hermitian conjugation is also geometric.  A ket state is created
by inserting an operator near the origin:
\begin{equation}
  \ket{\mathcal{O}}
  =
  \mathcal{O}(0)\ket{0}.
\end{equation}
The conjugate bra should be obtained by sending the insertion to the opposite end
of the cylinder, namely to $\tau=+\infty$.  Since
\begin{equation}
  r=R e^{\tau/R},
\end{equation}
the limit $\tau\to+\infty$ is the same as $r\to\infty$.  Hence the bra state is
represented by an insertion at infinity.

This is why Hermitian conjugation in radial quantization is not simply ordinary
complex conjugation at the same point.  It is ordinary Hermitian conjugation
combined with the conformal inversion
\begin{equation}
  x^\mu \longmapsto x'^\mu = \frac{x^\mu}{x^2}.
  \label{eq:inversion_for_radial_conjugation}
\end{equation}
The inversion exchanges the origin and infinity:
\begin{equation}
  x\to 0
  \quad \Longleftrightarrow \quad
  x'=\frac{x}{x^2}\to\infty.
\end{equation}

For a scalar primary of scaling dimension $\Delta_{\mathcal{O}}$, the inversion
has conformal factor
\begin{equation}
  \Omega_I(x)=\frac{1}{x^2}.
\end{equation}
Therefore
\begin{equation}
  \mathcal{O}'(x')
  =
  \Omega_I(x)^{-\Delta_{\mathcal{O}}}\mathcal{O}(x)
  =
  (x^2)^{\Delta_{\mathcal{O}}}\mathcal{O}(x).
\end{equation}
Equivalently, writing $y^\mu=x^\mu/x^2$, so that $|y|=1/|x|$, one obtains
can also be written as
\begin{equation}
  \mathcal{O}'(y)
  =
  |y|^{-2\Delta_{\mathcal{O}}}\mathcal{O}(x).
  \label{eq:primary_under_inversion_y}
\end{equation}
Radial Hermitian conjugation is now  {\it defined} by combining ordinary Hermitian conjugation with inversion. For a real scalar primary, ordinary Hermitian conjugation does not change the operator itself, so the radial conjugate is
\begin{equation}
 {
  \left[
    \mathcal{O}(x)
  \right]^\dagger_{\rm radial}
  =
  |x|^{-2\Delta_{\mathcal{O}}}
  \mathcal{O}\!\left(\frac{x}{x^2}\right).
  }
  \label{eq:radial_conjugation_scalar_correct}
\end{equation}
The factor $|x|^{-2\Delta_{\mathcal{O}}}$ is fixed by the requirement that an operator near the origin is mapped into a finite insertion at infinity. To see this explicitly, recall that taking $x\to 0$ gives $y\to\infty$, and
\begin{align}
  \lim_{x\to 0}
  \left[
    \mathcal{O}(x)
  \right]^\dagger_{\rm radial}
  &=
  \lim_{x\to 0}
  |x|^{-2\Delta_{\mathcal{O}}}
  \mathcal{O}\!\left(\frac{x}{x^2}\right)
  \nonumber\\
  &=
  \lim_{y\to\infty}
  |y|^{2\Delta_{\mathcal{O}}}
  \mathcal{O}(y)
  \nonumber\\
  &\equiv
  \mathcal{O}(\infty).
  \label{eq:radial_conjugation_origin_to_infinity}
\end{align}
Thus radial conjugation maps the ket-creating insertion at the origin into the bra-creating insertion at infinity. Consequently,
\begin{equation}
  \langle \mathcal{O}|\mathcal{O}\rangle
  =
  \langle
  \mathcal{O}(\infty)\mathcal{O}(0)
  \rangle.
\end{equation}
Using
\begin{equation}
  \langle
  \mathcal{O}(x)\mathcal{O}(0)
  \rangle
  =
  \frac{C_{\mathcal{O}}}{|x|^{2\Delta_{\mathcal{O}}}},
\end{equation}
we obtain
\begin{align}
  \langle
  \mathcal{O}(\infty)\mathcal{O}(0)
  \rangle
  &=
  \lim_{r\to\infty}
  r^{2\Delta_{\mathcal{O}}}
  \langle
  \mathcal{O}(r\hat n)\mathcal{O}(0)
  \rangle
  \nonumber\\
  &=
  \lim_{r\to\infty}
  r^{2\Delta_{\mathcal{O}}}
  \frac{C_{\mathcal{O}}}{r^{2\Delta_{\mathcal{O}}}}
  \nonumber\\
  &=
  C_{\mathcal{O}}.
\end{align}
With the normalization $C_{\mathcal{O}}=1$, the corresponding state has unit norm:
\begin{equation}
  \langle \mathcal{O}|\mathcal{O}\rangle=1.
\end{equation}

\section{Role of the Weyl anomaly}
\label{app:soc:weyl_anomaly}

In even spacetime dimension, the quantum theory may have a Weyl anomaly. The
partition function is then not strictly invariant under Weyl transformations of
the background metric. The anomaly is a local functional of the background
geometry and can shift the vacuum, or Casimir, energy on the cylinder{;
it is classified geometrically by Deser and Schwimmer~\cite{Deser:1993yx} into a
``type-A'' part proportional to the Euler density (the $a$-anomaly) and ``type-B''
parts built from Weyl invariants}.

This does not invalidate the local state--operator map. The relation
\begin{equation}
  \Delta_{\mathcal{O}}=R E_{\mathcal{O}}
\end{equation}
is understood after the standard normalization of the cylinder vacuum energy.
Energy differences between operator-created states and the vacuum are unaffected
by this subtlety. {Concretely, on $\mathbb{R}\times S^{d-1}$ the
anomaly contributes only an operator-independent constant to every energy --- in
$d=4$ a scheme-independent Casimir energy fixed by the $a$-anomaly~\cite{Cardy:1984epx}
--- which cancels in the differences $E_{\mathcal{O}}-E_{0}$ that define the scaling
dimensions, so $\Delta_{\mathcal{O}}=R\,E_{\mathcal{O}}$ holds for the normal-ordered
spectrum.}

\section{Dictionary of the correspondence}
\label{app:soc:dictionary}

We now collect the essential identifications in two compact dictionaries.  The
first table summarizes the geometric map from punctured flat space to the
cylinder.  The second table summarizes the corresponding operator and Hilbert
space identifications.  Together they give the operational content of the
state--operator correspondence.

\begin{table}[t]
\centering
\footnotesize
\renewcommand{\arraystretch}{1.25}
\setlength{\tabcolsep}{4pt}
\begin{tabular}{p{0.36\textwidth}p{0.56\textwidth}}
\hline
\textbf{Flat-space geometry} & \textbf{Cylinder geometry} \\
\hline

Radial coordinate:
\[
  r=|x|
\]
&
Cylinder Euclidean time:
\[
  \tau=R\log\frac{r}{R},
  \qquad
  r=R e^{\tau/R}.
\]
Radial evolution is cylinder time evolution.
\\

\hline

Flat metric:
\[
  ds^2_{\mathbb{R}^d}
  =
  dr^2+r^2d\Omega_{d-1}^2
\]
&
Weyl-equivalent cylinder metric:
\[
  ds^2_{\mathbb{R}^d}
  =
  \left(\frac{r}{R}\right)^2 ds^2_{\rm cyl},
  \qquad
  ds^2_{\rm cyl}
  =
  d\tau^2+R^2d\Omega_{d-1}^2 .
\]
\\

\hline

Weyl factor:
\[
  \Omega(\tau)=\frac{r}{R}
\]
&
Metric relation:
\[
  g_{\mathbb{R}^d}=\Omega^2 g_{\rm cyl},
  \qquad
  g_{\rm cyl}=\Omega^{-2}g_{\mathbb{R}^d}.
\]
\\

\hline

Origin and infinity:
\[
  r\to0,
  \qquad
  r\to\infty
\]
&
Cylinder past and future:
\[
  r\to0 \Longleftrightarrow \tau\to-\infty,
  \qquad
  r\to\infty \Longleftrightarrow \tau\to+\infty.
\]
\\

\hline

Scale transformation:
\[
  x^\mu\to\lambda x^\mu,
  \qquad
  r\to\lambda r
\]
&
Cylinder time translation:
\[
  \tau\to\tau+R\log\lambda.
\]
Thus dilatations generate time translations.
\\

\hline

Inversion:
\[
  x^\mu\to\frac{x^\mu}{x^2}
\]
&
Exchanges origin and infinity:
\[
  0 \longleftrightarrow \infty.
\]
This is the geometric origin of radial Hermitian conjugation.
\\

\hline
\end{tabular}
\caption{Geometric dictionary for radial quantization.  Punctured flat space is
Weyl-equivalent to the cylinder $\mathbb{R}\times S^{d-1}_{R}$.}
\label{tab:state_operator_dictionary_geometry}
\end{table}

\begin{table}[t]
\centering
\footnotesize
\renewcommand{\arraystretch}{1.25}
\setlength{\tabcolsep}{4pt}
\begin{tabular}{p{0.36\textwidth}p{0.56\textwidth}}
\hline
\textbf{Flat-space operator data} & \textbf{Cylinder / Hilbert-space data} \\
\hline

Scalar primary in flat space:
\[
  \mathcal{O}_{\mathbb{R}^d}(r\hat n)
\]
&
Scalar primary on the cylinder:
\[
  \mathcal{O}_{\rm cyl}(\tau,\hat n)
  =
  \left(\frac{r}{R}\right)^{\Delta_{\mathcal O}}
  \mathcal{O}_{\mathbb{R}^d}(r\hat n).
\]
\\

\hline

Dilatation generator:
\[
  D
\]
&
Cylinder Hamiltonian:
\[
  D=R H_{\rm cyl},
  \qquad
  H_{\rm cyl}=\frac{D}{R}.
\]
\\

\hline

Primary insertion at the origin:
\[
  \mathcal{O}(0)
\]
&
Primary state:
\[
  \ket{\mathcal O}
  =
  \lim_{r\to0}
  \mathcal{O}_{\mathbb{R}^d}(r\hat n)\ket{0}.
\]
For a primary, the limit is independent of $\hat n$.
\\

\hline

Flat-space two-point function:
\[
  \langle\mathcal{O}(x)\mathcal{O}(0)\rangle
  =
  \frac{C_{\mathcal O}}{|x|^{2\Delta_{\mathcal O}}}
\]
&
Large-time cylinder propagation:
\[
  \langle\mathcal{O}_{\rm cyl}(\tau_2)
  \mathcal{O}_{\rm cyl}(\tau_1)\rangle
  \sim
  C_{\mathcal O}\,
  e^{-\Delta_{\mathcal O}(\tau_2-\tau_1)/R}.
\]
Thus
\[
  E_{\mathcal O}=\frac{\Delta_{\mathcal O}}{R}.
\]
\\

\hline

Scaling dimension:
\[
  \Delta_{\mathcal O}
\]
&
Cylinder energy:
\[
  \boxed{
  \Delta_{\mathcal O}=R E_{\mathcal O}
  }.
\]
This is the core state--operator identity.
\\

\hline

Descendant operator:
\[
  P_{\mu_1}\cdots P_{\mu_k}\mathcal{O}(0)
\]
&
Excited state in the same conformal multiplet:
\[
  E_{\rm desc}
  =
  \frac{\Delta_{\mathcal O}+k}{R}.
\]
Each $P_\mu$ raises the energy by $1/R$.
\\

\hline

Operator at infinity:
\[
  \mathcal{O}(\infty)
  =
  \lim_{r\to\infty}
  r^{2\Delta_{\mathcal O}}
  \mathcal{O}(r\hat n)
\]
&
Bra state:
\[
  \bra{\mathcal O}
  \propto
  \bra{0}\mathcal{O}(\infty).
\]
The factor $r^{2\Delta_{\mathcal O}}$ gives a finite norm.
\\

\hline

Radial ordering:
\[
  \mathcal{R}\{\cdots\}
\]
&
Euclidean time ordering with respect to
\[
  \tau=R\log\frac{r}{R}.
\]
The operator with larger radius is placed to the left.
\\

\hline

Radial conjugation:
\[
  [\mathcal{O}(x)]_{\rm radial}^{\dagger}
  =
  |x|^{-2\Delta_{\mathcal O}}
  \mathcal{O}\!\left(\frac{x}{x^2}\right)
\]
&
Maps the ket insertion near the origin to the bra insertion at infinity:
\[
  \lim_{x\to0}
  [\mathcal{O}(x)]_{\rm radial}^{\dagger}
  =
  \mathcal{O}(\infty).
\]
\\

\hline
\end{tabular}
\caption{Operator dictionary for the state--operator correspondence.  Local
operators in flat space become states on $S^{d-1}_{R}$, and scaling dimensions
become cylinder energies.}
\label{tab:state_operator_dictionary_operators}
\end{table}

The two dictionaries display the essential logic of the proof.  The logarithmic
radial coordinate is cylinder time, dilatations become the cylinder Hamiltonian,
and therefore the scaling dimension of a local operator is the energy of the
corresponding cylinder state measured in units of the radius:
\begin{equation}
  {
  \Delta_{\mathcal O}=R E_{\mathcal O}.
  }
\end{equation}

The table makes the main result manifest.  Radial quantization turns scale
transformations into time translations, so the dimensionless dilatation
generator becomes $D=RH_{\rm cyl}$.  Consequently, a local scaling operator
$\mathcal{O}$ of dimension $\Delta_{\mathcal{O}}$ creates a cylinder state of
energy $E_{\mathcal{O}}=\Delta_{\mathcal{O}}/R$.  This is the precise content of
the state--operator correspondence used in the semiclassical computation of
operator dimensions~\cite{arXiv:1609.00026,arXiv:1403.8052,arXiv:1604.08913,Badel:2019oxl}.

 {\section{Numerical applications of the state--operator correspondence}}
\label{sec:numerical_soc}

{The identification $\Delta = RE$  is a computational
platform that has been implemented directly in numerical studies of critical phenomena.
When one places a quantum critical Hamiltonian on a sphere $S^{d-1}$ of radius $R$ and
diagonalizes it, the energy eigenvalues $E_k$ yield the scaling dimensions
$\Delta_k = RE_k$ without constructing the anomalous-dimension matrix or expanding in any
small parameter. Three families of methods exploit this platform in complementary regimes.}

\subsection{{Fuzzy sphere regularization.}}
{In the fuzzy-sphere approach~\cite{arXiv:2210.13482}, a $(2{+}1)$-dimensional
quantum critical system is placed on $S^2$ with a UV cutoff provided by a finite matrix
algebra (the ``fuzziness'' of the sphere).  At the critical coupling, exact diagonalization
of the Hamiltonian directly yields the tower of cylinder energies $\{E_k\}$; via $\Delta_k
= RE_k$ each level maps to a scaling dimension, and primaries are identified by their
$SO(3)$ quantum numbers.  For the three-dimensional Ising CFT this method has produced
precision determinations of more than a dozen primary operators, including parity-odd
primaries not previously accessible~\cite{arXiv:2210.13482}.  The framework has been
extended to the $O(N)$ Wilson--Fisher universality class, yielding operator spectra for
$N=2,3,4$ in agreement with conformal-bootstrap results~\cite{arXiv:2512.02234}.  The
fuzzy sphere is therefore a direct numerical implementation of the same cylinder
Hamiltonian $H_{\rm cyl}=D/R$ whose spectrum is computed analytically via semiclassics in
Parts~II--IV: both approaches diagonalize $H_{\rm cyl}$, one numerically, the other
through a controlled large-$n$ expansion.}

\subsection{ {Lattice radial quantization.}}
 {A complementary implementation places the $\phi^4$ theory on a discrete
approximation to $\mathbb{R}\times S^{d-1}$, combining finite-element methods on
$S^{d-1}$ with a lattice in the radial direction.  Brower, Cheng and Fleming introduced this framework in ~\cite{arXiv:1212.1757}; subsequent
work developed the finite-element discretization and demonstrated recovery of the conformal
spectrum in the continuum limit~\cite{arXiv:1407.7597,arXiv:2006.15636}.  Recent
implementations have extended the method to extract OPE coefficients~\cite{arXiv:2311.01100},
going beyond scaling dimensions.  Unlike conventional Euclidean lattice field theory, which
works on $\mathbb{R}^d$ and extracts $\Delta_k$ from the large-distance decay of
correlation functions, lattice radial quantization works directly on
$\mathbb{R}\times S^{d-1}$ and reads off $\Delta_k = RE_k$ from the energy
spectrum---the same geometry and the same dictionary as discussed at the beginning of Chapter~\ref{chap:soc}.}

\subsection{ {Traditional Euclidean lattice Monte Carlo.}}
 { Although this lattice methodology does not employ state-operator-correspondence it is useful to review it here since it is the standard approach. It discretizes the theory on a hypercubic Euclidean lattice and
computes correlation functions via Monte Carlo sampling.  Scaling dimensions are extracted
from the power-law decay $\langle\mathcal{O}_n(x)\mathcal{O}_n(0)\rangle\sim
|x|^{-2\Delta_n}$ near the critical point, or equivalently from finite-size scaling on a
torus.  For the three-dimensional Ising universality class this programme has been carried
to high precision~\cite{arXiv:1004.4486}, and the resulting critical exponents are among the
most accurate available for small operator degrees $n$.  The chief limitation for the heavy
operators studied in Parts~II--IV is that constructing $\mathcal{O}_n
\sim(\phi_a\phi_a)^{n/2}$ for large $n$ requires isolating an exponentially rare
multi-particle signal, and the signal-to-noise ratio degrades rapidly with $n$.  The
semiclassical expansion, by contrast, becomes \emph{more} reliable as $n\to\infty$.}
 
\newpage

\chapter{Composite Operators and Mixing}\label{chap:composites}

\section{Composite Operators}
\label{sec:composite_operators}

A \textbf{composite} \textbf{local operator} is a product
of elementary fields and their derivatives, all evaluated at a single spacetime
point, after renormalization to remove ultraviolet subdivergences.  We work in a
$d$-dimensional QFT with real scalar fields $\phi^a(x)$, $a=1,\ldots,N$, governed
by a Lagrangian $\mathcal{L}(\phi,\partial\phi)$.  In a free theory the construction
reduces to normal ordering; in an interacting theory it requires a systematic
counterterm procedure.

\subsection{Definition}
\label{sec:composite_def}

An elementary composite of \textbf{field degree $n$} and \textbf{derivative order $k$}
is any expression of the form
\begin{equation}
  \mathcal{O}_{n,k}(x)
  \;=\;
  \mathcal{N}\!\left[
    c^{a_1\cdots a_n}_{\mu_1\cdots\mu_k}\,
    \bigl(\partial^{\mu_1}\phi^{a_1}\bigr)(x)\cdots
    \bigl(\partial^{\mu_{k}}\phi^{a_n}\bigr)(x)
  \right],
\label{eq:composite_op_def}
\end{equation}
where $c^{a_1\cdots a_n}_{\mu_1\cdots\mu_k}$ is a constant tensor encoding the
Lorentz and internal-symmetry structure, and $\mathcal{N}[\,\cdot\,]$ denotes the
chosen renormalization procedure (normal ordering, minimal subtraction, etc.).  One
may allow several derivatives to act on the same factor, give different field
components different derivative structures, and project onto any irreducible
representation of the Lorentz group $SO(d)$.  The most general composite of degree
$n$ is a linear combination of such elementary composites.

The \textbf{classical (engineering) dimension} of~\eqref{eq:composite_op_def} is
\begin{equation}
  \Delta_{\rm cl}
  \;=\;
  n\,\frac{d-2}{2} + k \,,
\label{eq:composite_classical_dim}
\end{equation}
since each fundamental scalar carries dimension $(d-2)/2$ in $d$ dimensions and each
derivative contributes one unit.  A convenient shorthand labels a spin-$s$ singlet
primary with $2p$ contracted derivatives by the triple $(n,s,p)$, giving
\begin{equation}
  \Delta_{\rm cl}(n,s,p) \;=\; n\,\frac{d-2}{2} + s + 2p\,.
\label{eq:classical_dim_nsp}
\end{equation}

\subsection{Renormalization of composite operators}

In an interacting theory, products of fields at coincident points are ultraviolet
divergent beyond leading order.  Renormalization of a composite operator
$\mathcal{O}_i$ proceeds by adding local counterterms — finite linear combinations
of operators with the same or lower classical dimension — to render the Green
functions involving $\mathcal{O}_i$ finite order by order in perturbation theory.

A central consequence is that operators sharing the same quantum numbers
(classical dimension, Lorentz spin, global-symmetry representation) are not
independently renormalized: {the renormalized operators are finite
linear combinations of the bare ones, and  they mix}.  This \emph{operator mixing} is unavoidable
whenever the set of operators with given quantum numbers has dimension greater than one,
and it is the central complication for heavy composite operators at large~$n$~\cite{arXiv:1601.01310,arXiv:1505.00963,arXiv:1605.08868}.

\section{Composite Primaries}
\label{sec:conformal_primaries}

Section~\ref{app:conf:primaries} develops the theory of conformal primaries in
full, including derivations and geometric interpretation.  The following is a
self-contained summary of the facts needed for the mixing analysis of this chapter.

A local operator $\mathcal{O}_{\Delta,\ell}(x)$ of scaling dimension $\Delta$ and
Lorentz spin $\ell$ is a \textbf{conformal primary} if it sits at the bottom of a
conformal multiplet.  Concretely, two conditions must hold at the origin.  First,
the operator is an eigenstate of dilatations with eigenvalue $-i\Delta$:
\begin{equation}
  [D,\,\mathcal{O}(0)] \;=\; -i\,\Delta\;\mathcal{O}(0)\,.
\label{eq:primary_dilatation}
\end{equation}
Second, it is annihilated by all special conformal generators:
\begin{equation}
  [K_\mu,\,\mathcal{O}(0)] \;=\; 0 \,.
\label{eq:primary_SCT}
\end{equation}
The first condition assigns a definite scaling dimension; the second singles out the
lowest-weight state in the representation.  All other operators in the same multiplet
are \textbf{descendants}: $\partial_{\mu_1}\cdots\partial_{\mu_k}\mathcal{O}$
carries dimension $\Delta+k$ and contains no independent dynamical information —
its correlators are determined by conformal Ward identities once the primary is
specified.

Under a finite conformal transformation $x\mapsto x'$ with local Jacobian
$|\partial x'/\partial x|= \Omega(x)^d$, a scalar primary transforms as
\begin{equation}
  \mathcal{O}'(x') \;=\;
  \left|\frac{\partial x'}{\partial x}\right|^{-\Delta/d}
  \mathcal{O}(x)
  \;=\; \Omega(x)^{-\Delta}\,\mathcal{O}(x)\,.
\label{eq:primary_transform}
\end{equation}
This is the defining covariance property: the operator picks up exactly the
local conformal factor raised to the power $-\Delta$.  Applying this rule twice
— once with a dilatation and once with an inversion — fixes the two-point
function of a scalar primary to be
\begin{equation}
  \langle\mathcal{O}(x)\,\mathcal{O}(0)\rangle
  \;=\; \frac{C_{\mathcal{O}}}{|x|^{2\Delta}}\,,
\label{eq:primary_two_point}
\end{equation}
with $C_{\mathcal{O}}$ a theory-dependent normalization constant.  For a collection
$\{\mathcal{O}_i\}$ of scalar primaries, the same argument shows that two-point
functions between operators of different scaling dimensions must vanish, leaving a
block-diagonal structure:
\begin{equation}
  \langle\mathcal{O}_i(x)\,\mathcal{O}_j(0)\rangle
  \;=\; \frac{C_i\,\delta_{ij}}{|x|^{2\Delta_i}}\,.
\label{eq:primary_orthogonality}
\end{equation}
The orthogonality encoded in $\delta_{ij}$ holds whenever $\Delta_i\neq\Delta_j$;
within a degenerate subspace of operators sharing the same $\Delta$, spin, and
global-symmetry quantum numbers it can always be achieved by a basis rotation.
The two-point function~\eqref{eq:primary_orthogonality} will be the main
diagnostic for identifying physical primaries in the presence of operator mixing.

Among composite operators, a \textbf{composite primary} is one that simultaneously
satisfies the renormalization conditions of a composite operator and the primary
conditions~\eqref{eq:primary_dilatation}--\eqref{eq:primary_SCT}.  In a free CFT,
composite primaries can be constructed explicitly by symmetrizing and
trace-subtracting indices and imposing the equations of motion.  In an interacting
CFT the explicit form is deformed by quantum corrections, but the characterization
via~\eqref{eq:primary_dilatation}--\eqref{eq:primary_SCT} — or equivalently via
diagonalization of the two-point-function matrix — remains the operative definition.

\section{Operator Mixing and the Anomalous Dimension Matrix}
\label{sec:operator_mixing}

\subsection{The mixing matrix}

Let $\bigl\{\mathcal{O}_i^{\rm bare}(x)\bigr\}_{i=1}^{M}$ be a basis of bare
composite operators sharing the same classical quantum numbers: classical dimension
$\Delta_{\rm cl}$, Lorentz spin $\ell$, and global-symmetry representation $\mathbf{r}$.
The renormalized operators at scale $\mu$ are related to the bare operators by the
$M\times M$ \textbf{mixing matrix} $Z(\mu,\lambda)$:
\begin{equation}
  \bigl[\mathcal{O}_i^R\bigr](x;\mu)
  \;=\;
  \sum_{j=1}^{M} \bigl(Z^{-1}\bigr)_{ij}(\mu,\lambda)\;
  \mathcal{O}_j^{\rm bare}(x) \,.
\label{eq:renorm_mixing}
\end{equation}
The matrix $Z_{ij}$ depends on the renormalization scale $\mu$ and on the coupling
constants $\lambda = \{\lambda^\alpha\}$.  Its off-diagonal entries encode the mixing
of bare operators under renormalization.

\subsection{The anomalous dimension matrix}

The \textbf{anomalous dimension matrix} is defined by
\begin{equation}
  \gamma_{ij}(\lambda)
  \;\equiv\;
  \left(\mu\frac{\partial Z}{\partial\mu}\cdot Z^{-1}\right)_{ij}\,,
\label{eq:anomalous_dim_matrix}
\end{equation}
so that the Callan--Symanzik equation for the renormalized composite operators reads
\begin{equation}
  \left[
    \mu\frac{\partial}{\partial\mu}
    + \beta^\alpha(\lambda)\frac{\partial}{\partial\lambda^\alpha}
    + \gamma(\lambda)
  \right]_{ij}
  \bigl\langle\mathcal{O}_i^R(x)\,\mathcal{O}_j^R(0)\bigr\rangle
  \;=\; 0 \,.
\label{eq:CS_composite}
\end{equation}
Here $\beta^\alpha(\lambda) = \mu\,\partial\lambda^\alpha/\partial\mu$ is the
beta-function of coupling $\lambda^\alpha$.  The matrix $\gamma_{ij}$ governs how
the operators mix as the renormalization scale is varied.

\subsection{Diagonalization at a conformal fixed point}

At a conformal fixed point $\lambda = \lambda^*$ where $\beta^\alpha(\lambda^*)=0$,
the Callan--Symanzik equation~\eqref{eq:CS_composite} reduces to a purely
algebraic condition: the two-point function matrix $G_{ij}(x) =
\langle\mathcal{O}_i^R(x)\mathcal{O}_j^R(0)\rangle$ must be simultaneously
diagonalizable with $\gamma(\lambda^*)$.

The \textbf{physical primaries} are the eigenvectors of $\gamma(\lambda^*)$.  Let
$v^{(k)} = (v_i^{(k)})_{i=1}^M$ be the $k$-th normalized eigenvector:
\begin{equation}
  \sum_j \gamma_{ij}(\lambda^*)\,v_j^{(k)}
  \;=\; \gamma_k\;v_i^{(k)}, \qquad
  \mathcal{O}_k^{\rm phys} = \sum_i v_i^{(k)}\,\mathcal{O}_i^R \,.
\label{eq:eigenvector_primaries}
\end{equation}
The \textbf{scaling dimension} of the $k$-th primary is
\begin{equation}
  \boxed{
  \Delta_k \;=\; \Delta_{\rm cl} + \gamma_k
  }
\label{eq:scaling_dim_fixed_point}
\end{equation}
and the two-point function of the physical primaries is diagonal:
\begin{equation}
  \bigl\langle\mathcal{O}_k^{\rm phys}(x)\,\mathcal{O}_l^{\rm phys}(0)\bigr\rangle
  \;=\; \frac{C_k\;\delta_{kl}}{|x|^{2\Delta_k}} \,,
\label{eq:two_point_physical}
\end{equation}
consistent with the orthogonality~\eqref{eq:primary_orthogonality}.
 {The diagonal form is not enforced by $\beta=0$ alone: within a block
of operators sharing the same $\Delta_k$ the two-point matrix must additionally be
diagonalised by a choice of basis (always possible in a unitary theory), and for the non-normal $\gamma(\lambda^*)$ of the remark below it
need not be orthogonalisable at all.} In a non-unitary CFT (e.g.\ at a fixed point with complex couplings, or in a theory
with wrong-sign kinetic terms), the mixing matrix $\gamma(\lambda^*)$ need not be
Hermitian, and its eigenvalues $\gamma_k$ can be complex.  The diagonalization
procedure and~\eqref{eq:scaling_dim_fixed_point} remain formally valid; however,
the physical interpretation of complex scaling dimensions requires care~\cite{arXiv:1610.08472,arXiv:1703.04830,arXiv:1502.01437}.

\section{The Mixing Problem for Heavy Composite Operators}
\label{sec:heavy_mixing}

\paragraph{Single real scalar.}
As a prototype — and as the model that will be analyzed in full detail in §\ref{sec:semiclassical_program} and §\ref{energypath} — consider a single real scalar field $\phi$ with a quartic self-interaction $\lambda\phi^4/4$.  The independent composite primaries of field degree $n$, spin $s$, and $p$ extra pairs of contracted derivatives all share the same classical scaling dimension,
\begin{equation}
  \Delta_{\rm cl}(n,s,p) \;=\; n\,\frac{d-2}{2} + s + 2p \,,
\label{eq:singlet_classical_dim}
\end{equation}
but receive different anomalous dimensions at the fixed point, so they mix under renormalization.  The number of independent operators $M(n,s,p)$ at fixed $(n,s,p)$ grows with $n$ — roughly polynomially for fixed $(s,p)$ — and the direct diagonalization of the $M\times M$ matrix $\gamma_{ij}(\lambda^*)$ becomes impractical.  Two obstacles compound each other:
\begin{enumerate}[leftmargin=2.2em]
  \item the multiplicity $M(n,s,p)$ grows with $n$, so the mixing matrix becomes large;
  \item the matrix elements $\gamma_{ij}(\lambda^*)$ receive contributions at every
        loop order that depend on $n$ in a complicated way, so fixed-order
        perturbation theory does not converge uniformly for large $n$~\cite{Goldberg:1990qk,Badel:2019oxl}.
\end{enumerate}

\paragraph{$O(N)$ generalization.}
Promoting the single scalar to $N$ scalar fields $\phi^i$, $i=1,\ldots,N$, one naturally focuses on
$O(N)$-singlet composite primaries — operators invariant under the full $O(N)$ rotation symmetry.
These are built from even powers of $|\phi|^2 \equiv \phi^i\phi^i$ and its derivatives, and they form
the sector relevant to the Wilson--Fisher fixed point \cite{Wilson:1971dc}, at leading order in $1/N$.  The classical
dimension formula \eqref{eq:singlet_classical_dim} is unchanged (with $n$ now counting the total
field degree), but the singlet constraint further restricts the operator basis relative to the
unconstrained single-scalar case: only $O(N)$-invariant contractions of the $N$ fields contribute
to the mixing block.  The multiplicity $M(n,s,p)$ is accordingly reduced, yet it still grows
with $n$, and both obstacles listed above persist~\cite{Henriksson:2022rnm,arXiv:1904.00032,arXiv:1801.03512}.

\paragraph{Semiclassical resolution.}
The semiclassical framework developed in the remainder of this chapter applies to both the
single-scalar and $O(N)$ theories.  By working directly with periodic classical solutions
of the field equations on the cylinder — whose action is $O(n)$, providing the large parameter
controlling the semiclassical approximation — one obtains the spectrum of physical primaries
$\{\Delta_k\}$ in a controlled expansion in $1/n$.  This approach automatically accounts for
the full mixing: the semiclassical energy eigenstates on the cylinder correspond precisely to
the diagonalized operators $\mathcal{O}_k^{\rm phys}$.  We develop the machinery first for
the single scalar (Sections~\ref{sec:semiclassical_program}--\ref{energypath}), then carry
out the $O(N)$ computation in Chapter~\ref{chap:interacting}.

 {\paragraph{Relation to the large-charge expansion.}
The approach developed here is conceptually related to, but technically
distinct from, the large-charge
expansion~\cite{arXiv:1505.01537,arXiv:1611.02912,arXiv:2008.03308}.  In
the large-charge programme one studies the lowest operator of charge $Q$
under a global $U(1)$ (or other) symmetry; on the cylinder this state
corresponds to a superfluid phase with a rotating, time-independent
classical field $\phi\sim e^{i\mu\tau}$, stabilised by the conserved
Noether charge, and the $1/Q$ expansion follows from the Goldstone EFT of
the spontaneously broken $U(1)$.  The key difference in the present notes
is that $\mathcal{O}_n\sim(\phi_a\phi_a)^{n/2}$ is \emph{neutral}: there
is no conserved charge to hold a static saddle, so the classical solution
is the time-dependent periodic orbit $v(\tau)$, and
the analogue of $Q$ is the Bohr--Sommerfeld integer $n=I(E)/(2\pi)$ --- a
quantisation condition on the action variable, not a Noether charge.  This
is precisely why Floquet theory and the Gel'fand--Yaglom theorem are
essential ingredients here, whereas in the large-charge computation the
fluctuation operators have constant (or slowly varying) coefficients.  Both
approaches use $\Delta=RE$ and yield a $1/n$ (respectively $1/Q$) expansion
sharing the same leading power $\Delta\sim n^{d/(d-1)}$; the technical
machinery differs because the nature of the saddle differs.}


\paragraph{Perturbative counterpart.}
{The perturbative approach to the same problem --- computing scaling dimensions of composite operators in $O(N)$ $\phi^4$ theory --- proceeds by evaluating the anomalous dimension matrix $\gamma_{ij}(\lambda)$ order by order in $\lambda$ and then substituting the Wilson--Fisher fixed-point  \cite{Wilson:1971dc} coupling $\lambda_*(\varepsilon)$.  Recent multiloop results~\cite{Henriksson:2025hwi,Henriksson:2025vyi,Schnetz:2022nsc,Kompaniets:2017yct} push this programme to five loops for all operators up to classical dimension six and Lorentz rank two, providing an independent determination of the low-lying spectrum of the Ising, $O(N)$, and hypercubic CFTs.  The two approaches are complementary: the perturbative $\varepsilon$-expansion is reliable for small $\varepsilon$ at fixed $n$, while the semiclassical $1/n$ expansion is reliable for large $n$ at fixed $\kappa = \lambda n$.  Their overlap region --- moderate $n$ and moderate $\varepsilon$ --- where both descriptions should agree provides a non-trivial consistency check on the semiclassical machinery developed in the rest of this lecture.}

 {\paragraph{Lattice and non-perturbative counterpart.}
A third class of methods operates directly in $d=3$ without expansion in any small
parameter.  Traditional Euclidean lattice Monte Carlo extracts $\Delta_n$ from the
power-law decay of $\langle\mathcal{O}_n(x)\mathcal{O}_n(0)\rangle$ via finite-size
scaling~\cite{arXiv:1004.4486}, while lattice radial
quantization~\cite{arXiv:1212.1757,arXiv:1407.7597,arXiv:2006.15636} and the
fuzzy-sphere method~\cite{arXiv:2210.13482,arXiv:2512.02234} place the theory directly
on $\mathbb{R}\times S^{d-1}$ and diagonalize $H_{\rm cyl}$, reading off
$\Delta_k = RE_k$ without constructing the anomalous-dimension matrix at all.  These
methods sidestep the mixing problem of Section~\ref{sec:operator_mixing} because they
work at the level of physical energy eigenstates rather than at the level of a
perturbative operator basis --- a feature they share with the semiclassical approach of
Parts~II--IV. Their limitation is complementary: they are most powerful for
small-to-moderate $n$ in $d=3$, where signal-to-noise is manageable and exact
diagonalization is feasible; for large $n$ the computational cost grows and the signal
degrades, precisely where the semiclassical $1/n$ expansion becomes the controlled tool.
 }


%

\part{The Semiclassical Canovaccio}

\chapter{Free Theory: Three Roads to $\Delta_n$}\label{chap:freetheory}


The goal of this chapter is to compute, in three independent ways, the scaling
dimension of the composite operator
\[
  \mathcal{O}_n(x) \;\sim\; \phi^n(x)
\]
in a free massless scalar CFT in $d > 2$ Euclidean dimensions.  The answer is
exact in free theory and requires no perturbative expansion:
\[
  \Delta_n \;=\; \frac{d-2}{2}\,n .
\]
We reach this result by three logically distinct routes.  Each illuminates a
different facet of the physics and motivates a different piece of the machinery
that will be needed for the interacting theory.

\paragraph{Road~1 --- Direct Wick contractions.}
The two-point function of $\phi^n$ is evaluated \emph{algebraically} via Wick's
theorem: all propagator pairings are enumerated, giving a factor $n!$, and the
power-law behaviour of the free propagator $G(x)\propto|x|^{-(d-2)}$ then
immediately yields $\Delta_n$ by comparison with the CFT two-point structure.
This road is exact, elementary, and serves as the \emph{benchmark} against
which the other two methods are checked.  It also provides the clearest view of
\emph{why} large $n$ behaves semiclassically: after Stirling's approximation the
factorial becomes an exponential $e^{n(\log n-1)}$, which is the hallmark of a
saddle-point evaluation.

\paragraph{Road~2 --- Flat-space saddle point.}
The operator insertions $\phi^n$ are exponentiated as $e^{n\log\phi}$, and the
path integral is rewritten with the rescaled field $\phi = \sqrt{n}\,\varphi$.
The overall factor of $n$ in the exponent plays the role of $1/\hbar$, so the
path integral is controlled by a saddle point: a classical field configuration
$v(x)$ satisfying a sourced Laplace equation.  The on-shell effective action
evaluates to $1 - \log G(x_f - x_i)$, and the resulting correlator
$[G(x)]^n \propto |x|^{-n(d-2)}$ again yields $\Delta_n = (d-2)n/2$.  This
road generalises: in the interacting theory the same exponentiation and
rescaling lead to a non-trivial saddle, and the effective coupling becomes
$\kappa = \lambda n$ --- the double-scaling variable.

\paragraph{Road~3 --- Cylinder and Bohr--Sommerfeld quantization.}
Via the state--operator correspondence, computing $\Delta_n$ is equivalent to
finding the energy $E_n = \Delta_n / R$ of the state $|\phi^n\rangle$ on the
cylinder $\mathbb{R}_\tau\times S^{d-1}_R$.  In the heavy-operator limit the
dominant configuration is spatially homogeneous on $S^{d-1}_R$ and periodic in
Lorentzian cylinder time.  The curvature of the sphere generates a conformal
mass $\mu = (d-2)/(2R)$ for the conformally coupled scalar, so the reduced
system is a harmonic oscillator.  Bohr--Sommerfeld quantization of the periodic
orbit fixes the amplitude $A$ in terms of $n$, and the resulting energy $E_n =
n\mu$ reproduces $\Delta_n = (d-2)n/2$.  This road is the template for the full
semiclassical programme: in the interacting theory the periodic orbit is a
\emph{cnoidal} (Jacobi-elliptic) wave, but the Bohr--Sommerfeld structure
persists~\cite{arXiv:1607.07439,arXiv:1707.03866,Badel:2019oxl}.

\paragraph{Structure of the chapter.}
Each road is a self-contained section.  The reader primarily interested in
Road~3 (which generalizes most directly to the interacting theory) may read the
preamble of Roads~1 and 2, note their boxed results, and proceed to Road~3 in
full.  A summary at the end of the chapter collects the three derivations side
by side and explains how they connect.

%
\section{Road~1: Direct Wick Contractions}
\label{sec:freetheory:road1}

\subsection{Direct counting: Wick contractions}
\label{sec:freetheory:wick}

The simplest road to $\Delta_n$ is a direct evaluation of the two-point
function using the fundamental rules of free-field theory.  No path integral,
no effective action, no geometry: only propagators and combinatorics.  The
result is exact for all $n\geq 1$ and anchors every subsequent approximation.

\subsubsection{The two-point function of \texorpdfstring{$\phi^n$}{phi\^{}n}}
\label{sec:freetheory:wick:twopoint}

Consider a real free massless scalar $\phi$ in $d>2$ Euclidean dimensions.
The only non-zero building block is the Euclidean propagator
\begin{equation}
\langle \phi(x)\phi(y) \rangle_0 = G(x-y),
\label{eq:EuclideanPropagator}
\end{equation}
where the subscript $0$ denotes the free-theory expectation value.

To evaluate $\langle \phi^n(x)\phi^n(0)\rangle_0$ we apply Wick's theorem: each
$\phi$ at position $x$ must be paired with one $\phi$ at position $0$.  The
first $\phi(x)$ can pair with any of the $n$ fields at the origin; the second
with any of the remaining $n-1$; and so on.  All $n!$ pairings give the same
propagator $G(x)$, so
\begin{equation}
\langle \phi^n(x)\phi^n(0)\rangle_0 = n!\,[G(x)]^n .
\label{eq:WickContractionFormula}
\end{equation}
This is exact: no approximation has been made.

\subsubsection{The free propagator in \texorpdfstring{$d>2$}{d>2}}
\label{sec:freetheory:wick:propagator}

For a massless scalar in $d>2$ flat Euclidean space, the Green function
$-\partial^2 G(x) = \delta^{(d)}(x)$ is solved by
\begin{equation}
G(x) = \frac{1}{(d-2)\,\Omega_{d-1}}\,\frac{1}{|x|^{d-2}},
\qquad
\Omega_{d-1} = \frac{2\pi^{d/2}}{\Gamma(d/2)}.
\label{eq:FreePropagator}
\end{equation}
The key feature is the power law $G(x) \propto |x|^{-(d-2)}$.
Substituting into \eqref{eq:WickContractionFormula},
\begin{equation}
\langle \phi^n(x)\phi^n(0)\rangle_0
= n!\,\left[\frac{1}{(d-2)\Omega_{d-1}}\right]^n \frac{1}{|x|^{n(d-2)}}.
\label{eq:WickTwoPoint}
\end{equation}

\subsubsection{Stirling's approximation and the semiclassical structure}
\label{sec:freetheory:wick:stirling}

For large $n$, Stirling's approximation gives
\begin{equation}
\log n! = n\log n - n + \tfrac{1}{2}\log(2\pi n) + \mathcal{O}(1/n),
\label{eq:StirlingLargeN}
\end{equation}
so to leading order in $n$,
\begin{equation}
n! \;\simeq\; e^{n(\log n - 1)}.
\label{eq:StirlingLeading}
\end{equation}
The correlator \eqref{eq:WickTwoPoint} becomes
\begin{equation}
\langle \phi^n(x)\phi^n(0)\rangle_0
\;\simeq\;
e^{n(\log n-1)}\,\left[\frac{1}{(d-2)\Omega_{d-1}}\right]^n \frac{1}{|x|^{n(d-2)}}.
\label{eq:WickLargeNCorrelator}
\end{equation}

\emph{Remark.}  The appearance of the exponential $e^{n(\log n - 1)}$ is the
hallmark of a \emph{saddle-point} result: if one computes $n!$ by the integral
$\int_0^\infty t^n e^{-t}dt$, steepest descent at $t_* = n$ gives exactly this
leading exponential.  Road~2 elevates this observation into a general principle.

\subsubsection{Reading off the scaling dimension}
\label{sec:freetheory:wick:dimension}

The exact CFT two-point function of a scalar primary of dimension $\Delta_n$ is
\begin{equation}
\langle \mathcal{O}_n(x)\mathcal{O}_n(0)\rangle \propto \frac{1}{|x|^{2\Delta_n}}.
\label{eq:CFTTwoPoint}
\end{equation}
Comparing \eqref{eq:WickTwoPoint} with \eqref{eq:CFTTwoPoint}:
\begin{equation}
2\Delta_n = n(d-2)
\qquad\Longrightarrow\qquad
\boxed{\Delta_n = \frac{d-2}{2}\,n.}
\label{eq:DeltaWick}
\end{equation}
This is exact for all $n\geq 1$ and all $d>2$.  The prefactor $n!$ sets the
\emph{amplitude} of the two-point function but does not affect the
\emph{exponent} of $|x|$, which alone determines the scaling dimension.

\paragraph{What Road~1 teaches.}
The combinatorial derivation yields the exact scaling dimension; however, the geometric origin of the linear $n$ dependence is more transparently resolved via the effective action formalism developed in the subsequent section.
 
\section{Road~2: Flat-Space Saddle-Point Derivation}
\label{sec:freetheory:road2}

\subsection{Free theory: Semiclassical description of large composite operators}
\label{sec:freetheory:saddle}

In this section we provide a detailed derivation of the semiclassical description
of large composite operators in a free conformal scalar field theory, following the discussion in~\cite{Cuomo:2024fuy}.  The goal
is to make explicit how the classical solution on flat space arises, how its
normalization is fixed by a self-consistency condition, and how the associated
scaling dimension follows from the power-law behaviour of the on-shell correlator.

The key insight is that, after exponentiating the operator insertions and
rescaling the field by $\sqrt{n}$, the large-$n$ limit plays the role of
$1/\hbar$: the path integral localises on the saddle of an effective action, and
the saddle can be solved in closed form.

\subsubsection{Setup: free scalar in \texorpdfstring{$d>2$}{d>2}}
\label{sec:freetheory:saddle:setup}

We consider a real, free, massless scalar in Euclidean signature:
\begin{equation}
S_E[\phi] = \frac12 \int d^d x\, (\partial\phi)^2, \qquad d>2,
\label{eq:FreeAction}
\end{equation}
and the two-point correlator of heavy neutral operators:
\begin{equation}
\langle \phi^n(x_f)\phi^n(x_i) \rangle
=
\frac{1}{Z}\int \mathcal{D}\phi\;
\phi^n(x_f)\phi^n(x_i)\, e^{-S_E[\phi]} .
\label{eq:correlator}
\end{equation}

\subsubsection{Exponentiating the insertions}
\label{sec:freetheory:saddle:exponentiate}

A purely algebraic step that becomes powerful at large \(n\) is
\begin{equation}
\phi^n(x) = e^{n\log\phi(x)}.
\label{eq:expinsert}
\end{equation}
Substituting \eqref{eq:expinsert} into \eqref{eq:correlator} yields
\begin{equation}
\langle \phi^n(x_f)\phi^n(x_i)\rangle
=
\frac{1}{Z}\int \mathcal{D}\phi\;
\exp\!\left[
-\frac12\int d^d x\,(\partial\phi)^2
+ n\log\phi(x_f)+n\log\phi(x_i)
\right].
\label{eq:correlator_exp}
\end{equation}

\subsubsection{Field rescaling and the effective action}
\label{sec:freetheory:saddle:rescale}

We now make the key step that reveals the semiclassical structure:
\begin{equation}
\phi(x)=\sqrt{n}\,\varphi(x).
\label{eq:rescale}
\end{equation}
Then
\begin{align}
\frac12\int d^d x\,(\partial\phi)^2 &= n \int d^d x\,\frac12(\partial\varphi)^2,\\
n\log\phi(x) &= n\left(\frac12\log n + \log\varphi(x)\right).
\end{align}
Factoring out the explicit \(n\)-dependence gives
\begin{equation}
\langle \phi^n(x_f)\phi^n(x_i)\rangle
=
n^n \,\frac{1}{Z}\int \mathcal{D}\varphi\;
e^{-n\,\Seff[\varphi]},
\label{eq:saddlerewrite}
\end{equation}
with the effective action
\begin{equation}
\Seff[\varphi]
=
\int d^d x\,\frac12(\partial\varphi)^2
-\log\varphi(x_f)-\log\varphi(x_i).
\label{eq:Seff_def}
\end{equation}
Equation \eqref{eq:saddlerewrite} makes the central point manifest:
\emph{large \(n\) plays the role of \(1/\hbar\)}, so the path integral is
dominated by saddle points of \(\Seff\).

\begin{figure}[h]
\centering
\includegraphics[width=0.85\textwidth]{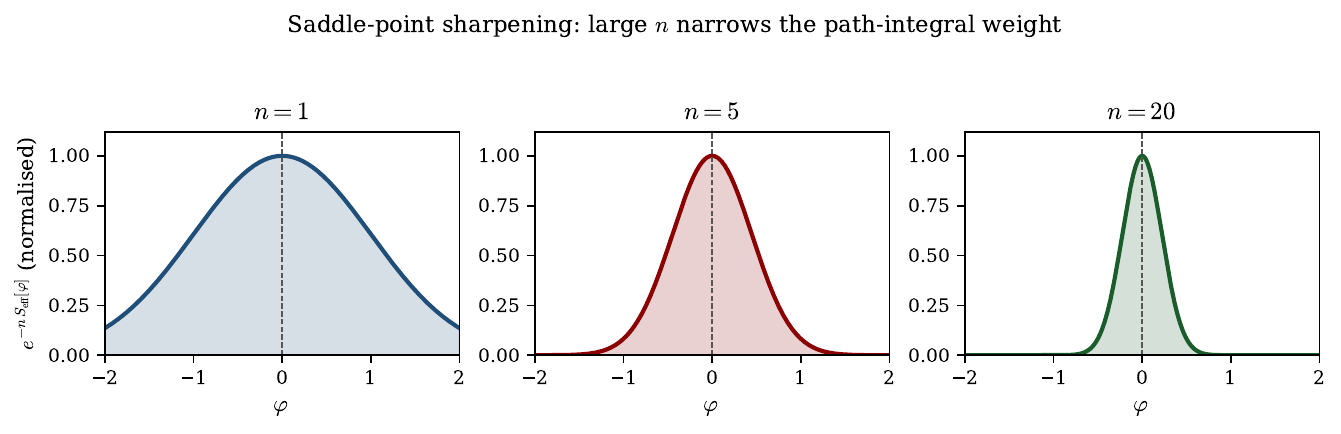}
\caption{%
\textbf{Saddle-point sharpening.}
The path-integral weight $e^{-n\,S_{\mathrm{eff}}[\varphi]}$ (normalised to its maximum)
for a schematic parabolic effective action $S_{\mathrm{eff}} = \tfrac12\varphi^2$.
As $n$ increases from $1$ (left) to $20$ (right), the weight concentrates sharply around
the saddle at $\varphi=0$, making the semiclassical approximation increasingly accurate.
This is why large $n$ plays the role of $1/\hbar$.}
\label{fig:saddle_sharpening}
\end{figure}

\subsection{Saddle-point equation and its solution}
\label{sec:freetheory:saddleeq}

\subsubsection{Derivation of the Euler--Lagrange equation}
\label{sec:freetheory:saddleeq:EL}

Varying \eqref{eq:Seff_def} gives, for a generic variation \(\delta\varphi\),
\begin{equation}
\delta\Seff
=
\int d^d x\, \delta\varphi(x)\,[-\partial^2\varphi(x)]
-\frac{\delta\varphi(x_f)}{\varphi(x_f)}-\frac{\delta\varphi(x_i)}{\varphi(x_i)}.
\end{equation}
Rewriting the last two terms with delta functions,
\begin{equation}
\delta\Seff
=
\int d^d x\, \delta\varphi(x)\left[
-\partial^2\varphi(x)
-\frac{\delta^{(d)}(x-x_f)}{\varphi(x_f)}
-\frac{\delta^{(d)}(x-x_i)}{\varphi(x_i)}
\right].
\end{equation}
Thus the saddle \(v(x)\equiv \varphi_{\rm cl}(x)\) satisfies
\begin{equation}
-\partial^2 v(x)
=
\frac{\delta^{(d)}(x-x_f)}{v(x_f)}
+
\frac{\delta^{(d)}(x-x_i)}{v(x_i)}.
\label{eq:saddleEOM}
\end{equation}

\subsubsection{Green function}
\label{sec:freetheory:saddleeq:green}

Let \(G(x)\) be the Green function of the Laplacian in \(\mathbb{R}^d\):
\begin{equation}
-\partial^2 G(x)=\delta^{(d)}(x).
\label{eq:GreenDef}
\end{equation}
For \(d>2\),
\begin{equation}
G(x)=\frac{1}{(d-2)\Od}\,\frac{1}{|x|^{d-2}},
\qquad
\Od=\frac{2\pi^{d/2}}{\Gamma(d/2)}.
\label{eq:GreenSol}
\end{equation}

\subsubsection{Solution of the saddle equation}
\label{sec:freetheory:saddleeq:solution}

By linearity of \(-\partial^2\), a solution of \eqref{eq:saddleEOM} is
\begin{equation}
v(x)=\frac{G(x-x_f)}{v(x_f)}+\frac{G(x-x_i)}{v(x_i)}.
\label{eq:v_solution}
\end{equation}

\paragraph{Self-consistency and renormalization of \(G(0)\).}
Evaluating \eqref{eq:v_solution} at \(x=x_f\) produces a term \(G(0)\), which is
UV-divergent.  Physically this is the \emph{self-contraction} of \(\phi^n\), and
its removal corresponds to normal ordering / operator renormalization.  After
subtracting \(G(0)\) (equivalently working with renormalized operators), one obtains
the finite condition
\begin{equation}
v(x_f)\,v(x_i)=G(x_f-x_i).
\label{eq:consistency}
\end{equation}
This is the geometric mean statement: the amplitude of the classical field at each
source is the square root of the propagator connecting them.

\subsection{On-shell action and the correlator}
\label{sec:freetheory:onshell}

We now evaluate \(\Seff[v]\) explicitly.

\subsubsection{Reducing \texorpdfstring{$\int(\partial v)^2$}{integral grad v squared} using the equation of motion}
\label{sec:freetheory:onshell:grad}

Start from
\begin{equation}
\int d^d x\,\frac12(\partial v)^2
=
-\frac12\int d^d x\, v\,\partial^2 v,
\end{equation}
where we integrated by parts and dropped boundary terms (justified by the falloff
of \(v\) away from the insertions).  Using \eqref{eq:saddleEOM},
\begin{align}
-\int d^d x\, v\,\partial^2 v
&=
\int d^d x\, v(x)\left[
\frac{\delta^{(d)}(x-x_f)}{v(x_f)}+\frac{\delta^{(d)}(x-x_i)}{v(x_i)}
\right]\nonumber\\
&=
\frac{v(x_f)}{v(x_f)}+\frac{v(x_i)}{v(x_i)}=2.
\end{align}
Hence
\begin{equation}
\int d^d x\,\frac12(\partial v)^2 = 1.
\label{eq:gradterm}
\end{equation}

\subsubsection{Evaluating \texorpdfstring{$\Seff[v]$}{S\_eff[v]}}
\label{sec:freetheory:onshell:eval}

From \eqref{eq:Seff_def} and \eqref{eq:gradterm},
\begin{equation}
\Seff[v]
=
1 - \log v(x_f) - \log v(x_i)
=
1 - \log\!\big(v(x_f)v(x_i)\big).
\end{equation}
Using the consistency condition \eqref{eq:consistency}:
\begin{equation}
\Seff[v] = 1 - \log G(x_f-x_i).
\label{eq:SeffOnShell}
\end{equation}

\subsubsection{Final saddle-point approximation}
\label{sec:freetheory:onshell:final}

Insert \eqref{eq:SeffOnShell} into \eqref{eq:saddlerewrite}:
\begin{align}
\langle \phi^n(x_f)\phi^n(x_i)\rangle
&\simeq
n^n \exp\!\Big(-n\,\Seff[v]\Big)\nonumber\\
&=
n^n \exp\!\Big(-n\big[1-\log G(x_f-x_i)\big]\Big)\nonumber\\
&=
e^{n(\log n-1)}\,\big[G(x_f-x_i)\big]^n.
\label{eq:final_correlator}
\end{align}

\subsection{Extracting the scaling dimension}
\label{sec:freetheory:dimension}

In a CFT, the two-point function of a scalar primary of dimension \(\Delta_n\)
behaves as
\begin{equation}
\langle \mathcal{O}_n(x)\mathcal{O}_n(0)\rangle \propto \frac{1}{|x|^{2\Delta_n}}.
\label{eq:CFT2pt}
\end{equation}
In the free theory, \(G(x)\propto |x|^{-(d-2)}\). Therefore \eqref{eq:final_correlator} implies
\begin{equation}
\langle \phi^n(x)\phi^n(0)\rangle \propto |x|^{-n(d-2)}.
\end{equation}
Comparing with \eqref{eq:CFT2pt} yields the scaling dimension
\begin{equation}
\boxed{
\Delta_n = \frac{d-2}{2}\,n.
}
\label{eq:DeltaFree}
\end{equation}

This concludes the flat-space saddle-point derivation: the heavy neutral operator
\(\phi^n\) is captured by a saddle-point configuration of the effective action
\(\Seff\), and \(\Delta_n\) follows directly from the power-law scaling of the
on-shell correlator.

\paragraph{What Road~2 teaches.}
The exponentiation $\phi^n = e^{n\log\phi}$ and the rescaling $\phi=\sqrt{n}\,\varphi$
are not specific to the free theory.  In the interacting $O(N)$ $\phi^4$ theory at
the Wilson--Fisher fixed point \cite{Wilson:1971dc}, the same two moves produce an effective action whose
saddle is a \emph{non-trivial} field configuration; the expansion in $1/n$ around
that saddle is the systematic semiclassical programme~\cite{arXiv:1808.04380,arXiv:1907.03531}.

\section{Road~3: Cylinder and Bohr--Sommerfeld Quantization}
\label{sec:freetheory:road3}

\subsection{Geometry: state--operator correspondence on the cylinder}
\label{sec:freetheory:geometry}

A central conceptual tool is the state--operator correspondence:
local operators in \(\mathbb{R}^d\) correspond to states on \(S^{d-1}\),
equivalently to quantization on the cylinder \(\mathbb{R}_\tau\times S^{d-1}_R\).
Scaling dimensions are cylinder energies in units of the sphere radius \(R\):
\begin{equation}
\Delta = R\,E.
\label{eq:DeltaEnergy}
\end{equation}

In the geometric picture an operator at the origin maps to a ket state at
$\tau\to-\infty$, and correlators map to cylinder matrix elements as 
illustrated in Figures~\ref{fig:cylinder} and~\ref{fig:correlator_map} in
\S\ref{app:soc}.

In the heavy-operator semiclassics, the dominant configuration on the cylinder
becomes \emph{spatially homogeneous} on \(S^{d-1}\) and \emph{time-dependent}
along \(\mathbb{R}\).  This is the origin of the classical periodic orbit picture
that we now make explicit.

\subsection{Mapping flat space to the cylinder}
\label{sec:freetheory:mapping}

To exploit conformal symmetry, we map flat space $\mathbb{R}^d$ to the Euclidean
cylinder $\mathbb{R}_\tau\times S^{d-1}_R$ with sphere radius $R$.  Using radial
coordinates $r, \hat{n}$ in $\mathbb{R}^d$ (so $x^\mu = r\,\hat{n}^\mu$),
\[
  ds^2_{\mathbb{R}^d} = dr^2 + r^2\,d\Omega_{d-1}^2,
\]
define the cylinder time $\tau$ through $r = R\,e^{\tau/R}$.  Then
\[
  ds^2_{\mathbb{R}^d}
  = e^{2\tau/R}\!\left(d\tau^2 + R^2\,d\Omega_{d-1}^2\right)
  = \Omega(\tau)^2\,ds^2_{\rm cyl},
  \qquad
  \Omega(\tau) = e^{\tau/R} = \frac{r}{R},
\]
so flat space is conformally equivalent to the cylinder metric
$ds^2_{\rm cyl} = d\tau^2 + R^2\,d\Omega_{d-1}^2$.
Under this map, the origin ($r=0$) and spatial infinity ($r\to\infty$) of flat space
correspond to $\tau\to -\infty$ and $\tau\to +\infty$, respectively.  After
analytic continuation to Lorentzian time, $\tau = it$, operator insertions are
mapped to asymptotic past and future on the Lorentzian cylinder.

\subsection{Conformal coupling and effective mass}
\label{sec:conformal_coupling}

A conformally coupled scalar field has action
\begin{equation}
S = \frac12 \int d^d x\,\sqrt{g}\,
\left(
  g^{\mu\nu}\partial_\mu\phi\,\partial_\nu\phi
  + \xi\,\mathcal{R}\,\phi^2
\right),
\qquad
\xi = \frac{d-2}{4(d-1)}.
\end{equation}
On $\mathbb{R}_\tau\times S^{d-1}_R$ with sphere radius $R$, the scalar curvature is
$\mathcal{R} = (d-1)(d-2)/R^2$, which induces an effective mass
\begin{equation}
\mu^2 = \xi\,\mathcal{R} = \frac{(d-2)^2}{4R^2},
\qquad
\mu = \frac{d-2}{2R}.
\label{eq:ConformalMass}
\end{equation}
The conformal mass $\mu$ sets the only frequency scale on the cylinder; it is
the mass the scalar \emph{must} have to propagate conformally on the curved
background.

\subsection{Homogeneous classical solution}
\label{sec:freetheory:classicalsol}

We restrict to spatially homogeneous configurations on the Lorentzian cylinder:
\[
  \phi(t,\Omega) = v(t).
\]
The reduced Lagrangian is
\begin{equation}
L_{\text{eff}}
= \frac{1}{2}\,\Omega_{d-1}\,R^{d-1}
\left( \dot v^2 - \mu^2\,v^2 \right),
\label{eq:ReducedLagrangian}
\end{equation}
where $\Omega_{d-1} = 2\pi^{d/2}/\Gamma(d/2)$ is the area of the unit
$(d-1)$-sphere.  This is the Lagrangian of a harmonic oscillator with frequency
$\mu$.

\subsubsection{Equation of motion}
\label{sec:freetheory:classicalsol:EOM}

Varying \eqref{eq:ReducedLagrangian} with respect to $v(t)$,
\begin{equation}
\ddot v + \mu^2\,v = 0.
\label{eq:CylEOM}
\end{equation}
The general solution is
\begin{equation}
v(t) = A\cos(\mu t + t_0),
\label{eq:HomogSol}
\end{equation}
where the amplitude $A > 0$ and the phase $t_0$ are integration constants;
$t_0$ reflects time-translation invariance and drops out of physical quantities.

\subsubsection{Energy of the classical orbit}
\label{sec:freetheory:classicalsol:energy}

The Hamiltonian corresponding to $L_{\text{eff}}$ is
\begin{equation}
H_{\text{eff}} = \frac{1}{2}\,\Omega_{d-1}\,R^{d-1}\bigl(\dot v^2 + \mu^2\,v^2\bigr).
\label{eq:CylHamiltonian}
\end{equation}
On the solution \eqref{eq:HomogSol}, $\dot v^2 + \mu^2 v^2 = A^2\mu^2$
(independent of $t$), so
\begin{equation}
E = \frac{1}{2}\,\Omega_{d-1}\,R^{d-1}\,A^2\mu^2.
\label{eq:CylEnergy}
\end{equation}

\subsection{Bohr--Sommerfeld quantization}
\label{sec:freetheory:BS}

\subsubsection{Action variable}
\label{sec:freetheory:BS:action}

The canonical momentum conjugate to $v$ is
\begin{equation}
\Pi = \frac{\partial L_{\text{eff}}}{\partial \dot v}
= \Omega_{d-1}\,R^{d-1}\,\dot v.
\label{eq:CanonicalMomentum}
\end{equation}
The action variable is the symplectic area enclosed in phase space,
\begin{equation}
I = \oint \Pi\,dv
= \Omega_{d-1}\,R^{d-1}\int_0^T \dot v^2\,dt,
\qquad T = \frac{2\pi}{\mu}.
\label{eq:ActionVariable}
\end{equation}
Substituting $\dot v = -A\mu\sin(\mu t + t_0)$,
\begin{equation}
I = \Omega_{d-1}\,R^{d-1}\,A^2\mu^2 \cdot \frac{\pi}{\mu}
= \pi\,\Omega_{d-1}\,R^{d-1}\,A^2\mu.
\label{eq:ActionVariableResult}
\end{equation}

\subsubsection{Quantization condition and fixed amplitude}
\label{sec:freetheory:BS:quantization}

The Bohr--Sommerfeld condition requires
\begin{equation}
I = 2\pi n, \qquad n \in \mathbb{Z}_{>0}.
\label{eq:BSCondition}
\end{equation}
Solving \eqref{eq:ActionVariableResult} for $A$,
\begin{equation}
A = 2\,\sqrt{\frac{n}{(d-2)\,\Omega_{d-1}\,R^{d-2}}}.
\label{eq:BSAmplitude}
\end{equation}
The fully explicit classical solution is
\begin{equation}
v(t) =
2\,\sqrt{\frac{n}{(d-2)\,\Omega_{d-1}\,R^{d-2}}}\,
\cos(\mu t + t_0).
\label{eq:BSClassicalSol}
\end{equation}
Note that the amplitude grows as $\sqrt{n}$, consistent with the field
rescaling $\phi = \sqrt{n}\,\varphi$ of Road~2.

\begin{figure}[h]
\centering
\fbox{%
\begin{tikzpicture}[
  cwa/.style={
    decoration={markings,
      mark=at position 0.25 with {\arrowreversed[scale=1.2]{Latex}}},
    postaction={decorate}},
  scale=1.2]
  \draw[->,thin] (-2.55,0) -- (2.20,0) node[right, font=\small]{$v$};
  \draw[->,thin] (0,-2.55) -- (0,2.65) node[above, font=\small]{$\Pi$};
  \fill[gray!20] (0,0) circle (1.732);
  \draw[red!60!black,    semithick, cwa] (0,0) circle (2.236); 
  \draw[orange!80!black, semithick, cwa] (0,0) circle (2.000); 
  \draw[green!55!black,  thick,     cwa] (0,0) circle (1.732); 
  \draw[teal!80!black,   semithick, cwa] (0,0) circle (1.414); 
  \draw[blue!70!black,   semithick, cwa] (0,0) circle (1.000); 
  \node at (0,0) {$\displaystyle I = \oint \Pi\,dv$};
  \draw[thin, rounded corners=2pt, fill=white]
    (2.55,-1.00) rectangle (3.52, 1.25);
  \node[font=\footnotesize\bfseries] at (3.035, 1.00) {$n$};
  \draw[red!60!black, semithick]
    (2.66, 0.65) -- (2.92, 0.65);
  \node[right, font=\footnotesize, red!60!black]
    at (2.96, 0.65) {$5$};
  \draw[orange!80!black, semithick]
    (2.66, 0.30) -- (2.92, 0.30);
  \node[right, font=\footnotesize, orange!80!black]
    at (2.96, 0.30) {$4$};
  \fill[gray!20] (2.66,-0.08) rectangle (2.92, 0.00);
  \draw[green!55!black, thick]
    (2.66,-0.04) -- (2.92,-0.04);
  \node[right, font=\footnotesize, green!55!black]
    at (2.96,-0.04) {$3$};
  \draw[teal!80!black, semithick]
    (2.66,-0.38) -- (2.92,-0.38);
  \node[right, font=\footnotesize, teal!80!black]
    at (2.96,-0.38) {$2$};
  \draw[blue!70!black, semithick]
    (2.66,-0.73) -- (2.92,-0.73);
  \node[right, font=\footnotesize, blue!70!black]
    at (2.96,-0.73) {$1$};
\end{tikzpicture}%
}
\caption{%
\textbf{Classical periodic orbits in phase space.}
Each closed orbit $(v,\Pi)$ satisfies $H(v,\Pi)=E_n$ and corresponds to
one period of the harmonic solution $v(t)=A\cos(\mu t+t_0)$; in rescaled
variables the orbits are concentric circles with radius $\propto\sqrt{n}$.
The Bohr--Sommerfeld condition $I=\oint\Pi\,dv=2\pi n$ selects the discrete
family (coloured by $n$, see legend); the shaded region ($n=3$, grey) has
area equal to the action variable~$I$ (labelled at centre).
Arrows indicate the clockwise direction of time evolution.}
\label{fig:phase_portrait}
\end{figure}
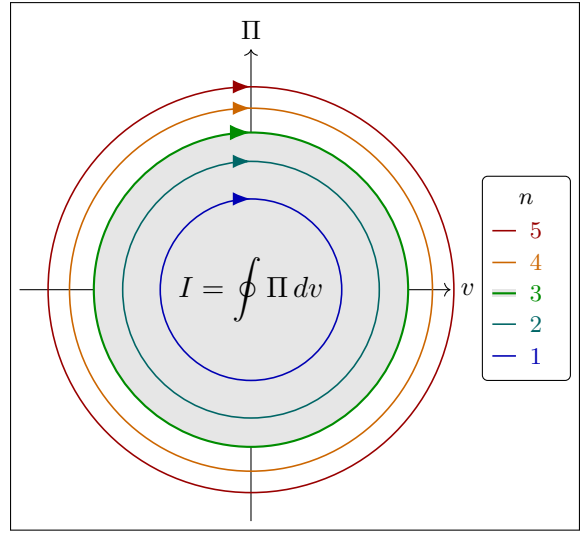

\subsection{Energy and scaling dimension}
\label{sec:freetheory:BSresult}

Substituting \eqref{eq:BSAmplitude} into \eqref{eq:CylEnergy},
\begin{align}
E &= \frac{1}{2}\,\Omega_{d-1}\,R^{d-1}\,A^2\mu^2 \nonumber\\
&= \frac{1}{2}\,\Omega_{d-1}\,R^{d-1}
   \cdot \frac{4n}{(d-2)\,\Omega_{d-1}\,R^{d-2}}
   \cdot \frac{(d-2)^2}{4R^2} \nonumber\\
&= \frac{n(d-2)}{2R}.
\label{eq:BSEnergy}
\end{align}
By the state--operator correspondence \eqref{eq:DeltaEnergy},
\begin{equation}
\Delta = R\,E = R\cdot\frac{n(d-2)}{2R},
\end{equation}
giving
\begin{equation}
\boxed{
\Delta_n = \frac{d-2}{2}\,n,
}
\label{eq:DeltaBS}
\end{equation}
in perfect agreement with Roads~1 and~2.

\paragraph{What Road~3 teaches.}
The harmonic oscillator structure with frequency $\mu = (d-2)/(2R)$ and energy levels
$E_n = n\mu$, is specific to the free theory.  As we shall see, in the interacting theory the
spatially homogeneous solution on the cylinder is a \emph{cnoidal wave}
${v(t) = x_0\,\cn(\omega t \,|\, m)}$~\eqref{eq:classical_orbit}, and the Floquet theory of the fluctuation operator
replaces simple harmonic-oscillator quantization.  Nevertheless the
Bohr--Sommerfeld framework --- compute $I(E)$, impose $I = 2\pi n$, invert to
get $E(n)$, apply $\Delta = RE$ --- is \emph{identical in structure}.

\section*{Summary: three roads, one result}

All three roads give
\[
  \Delta_n = \frac{d-2}{2}\,n.
\]
The table below collects the core logic of each derivation:

\vspace{0.4cm}
\begin{center}
\begin{tabular}{lll}
\hline
\textbf{Road} & \textbf{Central identity} & \textbf{Role of $n$} \\
\hline
1\ (Wick)    & $n!\,[G(x)]^n \sim |x|^{-n(d-2)}$ & combinatorial weight \\
2\ (Saddle)  & $[G(x_f-x_i)]^n$ from $e^{-n\Seff[v]}$ & semiclassical $1/\hbar$ \\
3\ (Cylinder)& $E_n = n\mu$,\ $\Delta = RE$          & Bohr--Sommerfeld level \\
\hline
\end{tabular}
\end{center}
\vspace{0.4cm}

\noindent
Roads~2 and~3 are the most important for what follows.  Road~2 teaches the
effective-action language; Road~3 teaches the classical-orbit language.  Both
generalize to the interacting theory with $\kappa = \lambda n$ fixed, and both
yield the same Bohr--Sommerfeld condition as the leading semiclassical
approximation.  The free theory is thus a perfect warm-up: it is simple enough
to admit three independent exact treatments, yet rich enough to expose the full
semiclassical structure in embryonic form.

\bigskip
\noindent\textbf{The QFT machinery.}\
With the free-theory intuition established, we now build the full semiclassical
apparatus.  The resolvent and its proper-time representation connect the energy
spectrum to a path integral over periodic configurations (the Gutzwiller trace
formula).  Hill's equation and Floquet theory handle the fluctuation spectrum
around the classical orbit; the Gel'fand--Yaglom theorem converts functional
determinants into stability angles.  Action variables and the Legendre transform
$\mathcal{S}_{\rm cl}(T)\leftrightarrow I(E)$ yield the final Bohr--Sommerfeld
quantization condition with quantum corrections~\cite{arXiv:1611.08915,arXiv:2111.11792}.

\section{The Action Variable in Quantum Mechanics}
\label{sec:action_variable_QM}

\begin{figure}[h]
\centering
\includegraphics[width=0.56\textwidth]{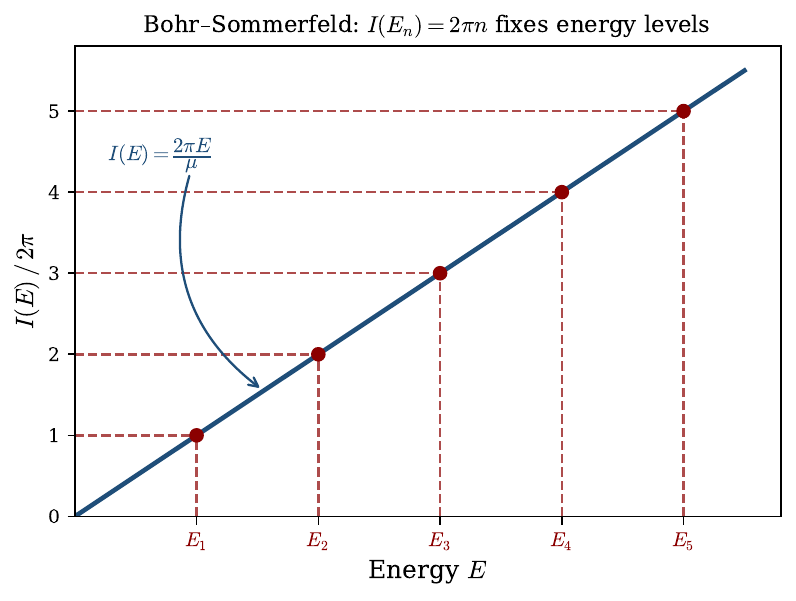}
\caption{%
\textbf{Action variable and Bohr--Sommerfeld quantization.}
For a harmonic oscillator with frequency $\mu$, $I(E)=2\pi E/\mu$ is linear in $E$.
The condition $I(E_n)=2\pi n$ (dashed lines) selects energy levels $E_n = n\mu$,
reproducing the exact quantum result. In the interacting theory the $I(E)$ curve is
nonlinear (set by the Jacobi-elliptic classical solution) but the same geometric
construction applies.}
\label{fig:bohr_sommerfeld}
\end{figure}

Consider a classical mechanical system with one degree of freedom: a particle
of mass~$m$ moving in a potential $V(q)$, where $q$ is the generalised
coordinate and $p = m\dot{q}$ the conjugate momentum.  The Hamiltonian is
\begin{equation*}
  H(q,p) \;=\; \frac{p^2}{2m} + V(q) \,,
\end{equation*}
and energy conservation $H = E$ constrains the motion to a curve in the
$(q,p)$ phase plane.  For a confining potential, this curve is a closed orbit
with classical turning points $q_\pm$ defined by $V(q_\pm) = E$ and with
period $\mathcal{T}(E) = \oint dq/\dot{q}$.

The \textbf{action variable} is the area enclosed by this orbit in phase space:
\begin{equation}
\boxed{
  I(E) \;=\; \oint p\,dq
}
\label{eq:action_variable_def_boxed}
\end{equation}
where the line integral runs counterclockwise around the closed orbit over
one complete period.  On the energy shell $H = E$, the momentum is
$p = \sqrt{2m[E - V(q)]}$ (taking the positive branch on the outward half),
so equivalently
\begin{equation}
  I(E) \;=\; \oint \sqrt{2m\bigl[E - V(q)\bigr]}\,dq \,.
\label{eq:action_variable_momentum}
\end{equation}
\textbf{Geometric interpretation.}
The action variable is literally the area of the region in the $(q,p)$ plane
bounded by the orbit.  Orbits at higher energy enclose more area; the
function $I(E)$ is therefore  increasing.%

\noindent
\textbf{Adiabatic invariance.}
Suppose the Hamiltonian depends on a slowly varying external parameter
$\lambda(t)$, with $|\dot{\lambda}/\lambda| \sim \tau_{\rm ext}^{-1}$ and
$\tau_{\rm ext} \gg \mathcal{T}$.  One can show that
\begin{equation}
  \frac{dI}{dt} \;=\; O\!\left(\frac{\mathcal{T}}{\tau_{\rm ext}}\right) \,, 
\end{equation}
so $I$ is approximately conserved under slow deformations of the
potential.\footnote{The proof follows from averaging the equations of motion
over one period; see, e.g., Landau \& Lifshitz \S49.}
In the QFT context this means that the integer $n = I/2\pi$ labelling the
composite operator is preserved along the RG flow: the same quantum number
identifies the operator at the free-theory fixed point and at the
Wilson--Fisher fixed point.

{ 
\subsection{Bohr--Sommerfeld Quantization}
\label{sec:bohr_sommerfeld_quantization}

Classical mechanics admits a continuous family of closed orbits, one for each
value of the energy $E$.  Quantum mechanics selects a discrete subset:
Bohr--Sommerfeld quantization is the semiclassical rule that determines which
classical orbits correspond to stationary quantum states.

\paragraph{The WKB wave function.}
In the classically allowed region $E > V(q)$, the time-independent
Schr\"{o}dinger equation $-(\hbar^2/2m)\psi'' + V(q)\psi = E\psi$ admits the
WKB (Wentzel--Kramers--Brillouin) approximate solution
\begin{equation}
  \psi(q) \;\approx\; \frac{C}{\sqrt{p(q)}}\,
  \exp\!\left(\frac{i}{\hbar}\int_{q_-}^{q} p(q')\,dq'\right)
  \;+\; \text{c.c.}\,,
\label{eq:WKB_solution}
\end{equation}
where $p(q) = \sqrt{2m[E-V(q)]}$.  The approximation is controlled by the
smallness parameter $\hbar|p'|/p^2 \ll 1$: the local de~Broglie wavelength
$\lambda_{\rm dB}(q) = 2\pi\hbar/p(q)$ must vary slowly on the scale set by
$p$ itself.  At each classical turning point $q_\pm$ (where $p\to 0$) this
condition fails.  Matching the oscillatory and evanescent solutions through
the turning points via \textbf{connection formulae} shows that the wave
function accumulates an additional phase of $-\pi/2$ at each turning point.

\paragraph{The Maslov-corrected quantization condition.}
For a smooth confining potential with exactly two turning points $q_-$ and
$q_+$, requiring $\psi$ to be single-valued after one complete oscillation and
accounting for the $-\pi/2$ phase shift at each turning point gives the
\textbf{Maslov-corrected} Bohr--Sommerfeld condition:
\begin{equation}
\boxed{
  I(E_n) \;=\; 2\pi\hbar\!\left(n + \tfrac{1}{2}\right), \qquad n = 0,1,2,\ldots
}
\label{eq:bohr_sommerfeld_maslov}
\end{equation}
The $+\tfrac{1}{2}$ is the \textbf{Maslov index} contribution: each of the
two turning points contributes $-\pi/2$ of phase, for a total of $-\pi$, which
in units of $2\pi$ amounts to $-\tfrac{1}{2}$.  Equivalently, the wave
function satisfies Dirichlet-like boundary conditions at the turning points to
leading semiclassical order, forcing the enclosed phase to be a half-integer
rather than an integer multiple of $2\pi\hbar$.

\paragraph{The working form.}
Setting $\hbar = 1$ and focusing on $n \gg 1$, the Maslov $+\tfrac{1}{2}$ is
a subleading correction at order $n^0$ in the scaling dimension $\Delta_n$.
In the saddle-point framework of Chapter~\ref{chap:composites} it arises automatically
from the sum of stability angles $\tfrac{1}{2}\sum_{\nu_\ell>0} n_\ell\nu_\ell$
(see \S\,\ref{sec:semiclassical_result}).  For the leading-order analysis we
therefore use the simpler form,
\begin{equation}
\boxed{
  I(E_n) \;=\; 2\pi n \,, \qquad n = 1, 2, 3, \ldots
}
\label{eq:bohr_sommerfeld_simple}
\end{equation}
which selects those orbits whose enclosed phase-space area is an integer
multiple of $2\pi$ (Fig.~\ref{fig:bohr_sommerfeld}).

\paragraph{Historical context.}
The condition $\oint p\,dq = nh$ (with $h = 2\pi\hbar$) originated in Bohr's
1913 model of the hydrogen atom (circular orbits) and was extended by Sommerfeld
in 1916 to elliptical orbits, correctly reproducing the fine-structure of
hydrogen without the full apparatus of quantum mechanics.  Einstein (1917)
formulated it in the covariant phase-space form $\oint p_i\,dq^i = nh$ valid
for any integrable system with $f$ degrees of freedom.  The rigorous derivation
via WKB connection formulae and the Maslov--Morse index was developed in the
1960s by Maslov, Keller, and others.

\paragraph{Worked example: harmonic oscillator.}
For the harmonic oscillator $V(q) = \tfrac{1}{2}m\omega^2 q^2$, the turning
points are $q_\pm = \pm\sqrt{2E/(m\omega^2)}$.  Substituting
$q = q_+\sin\vartheta$ into~\eqref{eq:action_variable_momentum}:
\begin{align}
  I(E) &\;=\; 2\int_{-q_+}^{q_+}
        \sqrt{2m\bigl(E - \tfrac{1}{2}m\omega^2 q^2\bigr)}\,dq
     \;=\; 2\sqrt{2mE}\,q_+ \int_{0}^{\pi}\cos^2\!\vartheta\,d\vartheta
     \nonumber \\
     &\;=\; 2\sqrt{2mE}\cdot\sqrt{\tfrac{2E}{m\omega^2}}\cdot\tfrac{\pi}{2}
     \;=\; \frac{2\pi E}{\omega}\,.
\label{eq:HO_action}
\end{align}
The mass cancels: $\sqrt{2mE}\cdot q_+ = 2E/\omega$.
The leading condition~\eqref{eq:bohr_sommerfeld_simple} gives
\begin{equation}
  I(E_n) = 2\pi n \;\implies\; E_n = n\omega \,.
\label{eq:HO_quantization}
\end{equation}
The Maslov-corrected condition~\eqref{eq:bohr_sommerfeld_maslov} shifts this to
$E_n = (n+\tfrac{1}{2})\omega$, which is the \emph{exact} quantum result for the
harmonic oscillator.  This coincidence (WKB = exact) is a special feature of the
quadratic potential; for generic anharmonic potentials the WKB spectrum acquires
higher-order $\hbar$ corrections beyond the Maslov term, which in our setting will 
contribute  to subleading orders in $1/n$ to $\Delta_n$.}

\subsection{The Key Identity $\dfrac{dI}{dE} = \mathcal{T}$}
\label{sec:proof_dI_dE}

The central identity relating the action variable to the period is:
\begin{equation}
\boxed{
  \frac{dI}{dE} \;=\; \mathcal{T}
}
\label{eq:dI_dE_equals_T_boxed}
\end{equation}

\paragraph{Proof.}
Write the action variable as the phase-space contour integral
\begin{equation}
  I(E) \;=\; \oint p(q,E)\,dq \,,
\label{eq:action_contour}
\end{equation}
where on the orbit at energy $E$ the momentum is
\begin{equation}
  p(q,E) \;=\; \sqrt{2m\bigl[E - V(q)\bigr]} \,.
\label{eq:momentum_on_orbit}
\end{equation}
Differentiating under the integral sign with respect to $E$ (the shape of
the orbit changes, but we differentiate $p$ at fixed $q$):
\begin{equation}
  \frac{dI}{dE} \;=\; \oint \frac{\partial p}{\partial E}\,dq \,.
\label{eq:dI_dE_first_form}
\end{equation}
From~\eqref{eq:momentum_on_orbit},
\begin{equation}
  \frac{\partial p}{\partial E}
  \;=\; \frac{m}{\sqrt{2m[E-V(q)]}}
  \;=\; \frac{m}{p} \,.
\label{eq:partial_p_E}
\end{equation}
Therefore
\begin{equation}
  \frac{dI}{dE} \;=\; m\oint \frac{dq}{p}
                \;=\; \oint \frac{dq}{\dot{q}}
                \;=\; \oint dt
\label{eq:dI_dE_intermediate}
\end{equation}
where we used $p = m\dot{q}$ and $dq/\dot{q} = dt$.  The integral of $dt$
around one orbit is precisely the period:
\begin{equation}
  \frac{dI}{dE} \;=\; \oint dt \;=\; \mathcal{T} \,. \quad\square
\label{eq:dI_dE_derivation}
\end{equation}

\paragraph{Check on the harmonic oscillator.}
From~\eqref{eq:HO_action},
\begin{equation}
  I(E) \;=\; \frac{2\pi E}{\omega}
\label{eq:HO_action_formula}
\end{equation}
gives immediately
\begin{equation}
  \frac{dI}{dE} \;=\; \frac{2\pi}{\omega} \,.
\label{eq:HO_dI_dE}
\end{equation}
The classical period of the harmonic oscillator is
\begin{equation}
  \mathcal{T} \;=\; \frac{2\pi}{\omega} \,,
\label{eq:HO_period}
\end{equation}
confirming $dI/dE = \mathcal{T}$.

\paragraph{Geometric interpretation.}
When the energy increases by $dE$, the orbit swells outward in phase space and
encloses an additional area $dI = \oint p\,dq$.  The proof shows that this
extra sliver equals $\mathcal{T}\,dE$: the orbit takes time $dt = dq/\dot{q}$
to traverse each infinitesimal arc $dq$, and summing over the full orbit gives
$\oint dt = \mathcal{T}$.  Equivalently: ``gaining one extra unit of energy and
letting the system run for one period produce the same increment of action.''

\paragraph{Action-angle coordinates.}
In action-angle variables $(I,\phi)$ the Hamiltonian depends only on $I$
(the integrability condition), so $\dot{\phi} = \partial H/\partial I = dE/dI$.
The angular velocity must equal the orbital frequency $\Omega = 2\pi/\mathcal{T}$,
which requires $dE/dI = 1/\mathcal{T}$, i.e.\ $dI/dE = \mathcal{T}$.
The identity therefore holds in every integrable system, regardless of the
shape of $V$.

\subsection{The Legendre Transform $\mathcal{S}_{\rm cl}(\mathcal{T})
            \leftrightarrow I(E)$}

In the saddle-point computation of Chapter~\ref{chap:composites}, the natural output
is the \textbf{classical action as a function of the period} $\mathcal{T}$:
\begin{equation}
  \mathcal{S}_{\rm cl}(\mathcal{T})
  \;=\; \int_0^{\mathcal{T}} L\,dt
  \;=\; \int_0^{\mathcal{T}} \bigl[p\dot{q} - H\bigr]\,dt
  \;=\; \int_0^{\mathcal{T}} \left[ \frac{\dot{q}^2}{2} - V(v)\right] \,dt \,,
\label{eq:classical_action_def}
\end{equation}
where $L = T_{\rm kin} - V$ is the Lagrangian and $v(t)$ denotes the classical
solution.  The Legendre transform switches to energy as the natural variable,
yielding $I(E)$:
\begin{equation}
\boxed{
  I(E) \;=\; \mathcal{T}(E)\,E + \mathcal{S}_{\rm cl}(\mathcal{T}(E))
}
\label{eq:legendre_relation_boxed}
\end{equation}

\paragraph{Derivation.}
Starting from the on-shell identity $L = p\dot{q} - E$ (since $H = E$ on the
orbit), integrate over one period:
\begin{equation}
  \mathcal{S}_{\rm cl}(\mathcal{T})
  \;=\; \int_0^{\mathcal{T}}(p\dot{q} - E)\,dt
  \;=\; \oint p\,dq - E\mathcal{T} \,.
\label{eq:classical_action_step1}
\end{equation}
Since $\oint p\,dq = I(E)$ by definition, this gives:
\begin{equation}
  \mathcal{S}_{\rm cl} \;=\; I - E\mathcal{T} \,.
\label{eq:classical_action_step2}
\end{equation}
Rearranging:
\begin{equation}
  I \;=\; \mathcal{S}_{\rm cl} + E\mathcal{T} \,.
\label{eq:action_relation}
\end{equation}
Differentiating~\eqref{eq:action_relation} with respect to $\mathcal{T}$,
with $I$ and $E$ both regarded as functions of $\mathcal{T}$:
\begin{equation}
  \frac{d\mathcal{S}_{\rm cl}}{d\mathcal{T}}
  \;=\; \frac{dI}{d\mathcal{T}} - E - \mathcal{T}\frac{dE}{d\mathcal{T}} \,.
\label{eq:differentiation_step}
\end{equation}
By the chain rule and the key identity $dI/dE = \mathcal{T}$
(\S\,\ref{sec:proof_dI_dE}):
\begin{equation}
  \frac{dI}{dE} \;=\; \mathcal{T}
\label{eq:key_relation}
\end{equation}
so $dI/d\mathcal{T} = (dI/dE)(dE/d\mathcal{T}) = \mathcal{T}\,dE/d\mathcal{T}$.
Substituting into~\eqref{eq:differentiation_step}:
\begin{equation}
  \frac{d\mathcal{S}_{\rm cl}}{d\mathcal{T}}
  \;=\; \mathcal{T}\frac{dE}{d\mathcal{T}} - E - \mathcal{T}\frac{dE}{d\mathcal{T}}
  \;=\; -E \,.
\label{eq:simplification}
\end{equation}
Therefore the energy is the conjugate variable to the period:
\begin{equation}
\boxed{
  E \;=\; -\frac{d\mathcal{S}_{\rm cl}}{d\mathcal{T}} \,.
}
\label{eq:energy_from_action}
\end{equation}
Substituting back into~\eqref{eq:action_relation} confirms~\eqref{eq:legendre_relation_boxed}:
\begin{equation}
  I \;=\; \mathcal{S}_{\rm cl} - \mathcal{T}\frac{d\mathcal{S}_{\rm cl}}{d\mathcal{T}}
       \;=\; \mathcal{T} E + \mathcal{S}_{\rm cl} \,. \quad\square
\label{eq:legendre_recovered}
\end{equation}

\paragraph{Thermodynamic analogy.}
The structure is identical to a Legendre transform in thermodynamics.
$\mathcal{S}_{\rm cl}(\mathcal{T})$ plays the role of the free energy $F(T)$
with $\mathcal{T}$ the ``temperature''; $E = -d\mathcal{S}_{\rm cl}/d\mathcal{T}$
mirrors $S = -\partial F/\partial T$ (entropy as conjugate to temperature);
and $I = \mathcal{T} E + \mathcal{S}_{\rm cl}$ is the analogue of the
thermodynamic internal energy.  In practice: the path-integral saddle gives
$\mathcal{S}_{\rm cl}(\mathcal{T})$ directly; the Legendre transform is the
exact dictionary to the energy spectrum via $I(E) = 2\pi n$.

\subsection{Relevance for QFT and the Scaling Dimension $\Delta_n$}

The machinery developed above is not merely a classical warm-up: it is the
computational engine for scaling dimensions.

\paragraph{The saddle-point → Bohr--Sommerfeld chain.}
On the cylinder $\mathbb{R}_\tau \times S^{d-1}_R$ the two-point correlator
of $\mathcal{O}_n$ is dominated by a periodic classical solution of the
Euclidean field equations, with period $\mathcal{T}$ and classical action
$\mathcal{S}_{\rm cl}(\mathcal{T})$.  Summing over all winding numbers
$k = 1, 2, 3, \ldots$ of this orbit (Chapter~\ref{chap:composites};
derived in \S\,\ref{sec:semiclassical_result}) produces a geometric series whose poles
occur at
\[
  I(E) \;=\; \mathcal{T}(E)\,E + \mathcal{S}_{\rm cl}(\mathcal{T}(E))
       \;=\; 2\pi n \,.
\]
This is precisely the Bohr--Sommerfeld condition~\eqref{eq:bohr_sommerfeld_simple}.
The scaling dimension then follows from the state--operator correspondence.
Two separate statements combine to give the result:
\begin{equation}
  \Delta_n \;=\; R\,E_n
  \qquad\text{(state--operator correspondence),}
\label{eq:delta_from_energy}
\end{equation}
where $E_n$ is the unique solution of the implicit equation
\begin{equation}
  I(E_n) \;=\; 2\pi n
  \qquad\text{(Bohr--Sommerfeld condition).}
\label{eq:BS_implicit}
\end{equation}
Equivalently, writing $I^{-1}$ for the inverse of $E\mapsto I(E)$:
\begin{equation}
  \boxed{\Delta_n \;=\; R\cdot I^{-1}(2\pi n).}
\label{eq:delta_from_I_inverse}
\end{equation}
This is non-trivial: $I(E)$ encodes the full classical dynamics of the
periodic orbit and depends on the coupling $\lambda$ through the shape of
the potential.  In the free theory $I(E)=2\pi E/\omega_0$ is linear, giving
$\Delta_n^{\rm free}=(d-2)n/2$; in the interacting theory $I(E)$ involves
Jacobi elliptic integrals and the inversion yields the non-trivial
$\kappa$-dependent coefficients $C_0(\kappa),C_1(\kappa),\ldots$ .

\paragraph{Level spacing and the classical period.}
Differentiating $I(E_n) = 2\pi n$ with respect to $n$ and using
$dI/dE = \mathcal{T}$:
\[
  \frac{dE_n}{dn} \;=\; \frac{2\pi}{\mathcal{T}(E_n)} \,.
\]
The spacing between successive energy levels is therefore fixed entirely by the
classical period — a strikingly sharp statement that \emph{no quantum
input is required for the leading-order spectrum}.

\chapter{The Interacting Theory Blueprint}\label{chap:doublescaling}\label{sec:semiclassical_program}

\noindent\textbf{Notation used in this chapter.}
Throughout, $n$ denotes simultaneously three things that the state--operator
correspondence identifies: (i) the \emph{field-degree} of the composite
operator $\phi^n$; (ii) the \emph{occupation number} (the number of quanta) of
the cylinder eigenstate; and (iii) the \emph{Bohr--Sommerfeld quantum number}
$I(E)=2\pi n$ of the classical orbit. The operator is neutral; $n$ counts
fields, not any global charge.
The period $\mathcal{T}$ always refers to the \emph{Euclidean} period of the
periodic saddle; it is related to the energy by $dI/dE=\mathcal{T}$.
The scaling dimension $\Delta$ and the cylinder energy $E$ are related by the
state--operator map $\Delta=RE$ with $R$ the sphere radius.

\medskip
We develop the semiclassical framework using a single real scalar field as
the prototype, since all the essential structure — periodic orbit, fluctuation
operator, Bohr--Sommerfeld quantization — is already present in this simplest
case.  The $O(N)$ generalization adds $N-1$ transverse fluctuation modes
but does not change the logic; it is discussed at the end of this section and
carried out in detail in Chapter~\ref{chap:interacting}.

\paragraph{What the free theory taught us.}
Chapter~\ref{chap:freetheory} computed the scaling dimension of the operator
$\phi^n$ (at $\lambda=0$) by three independent methods — Wick contractions,
saddle-point evaluation of the two-point function, and Bohr--Sommerfeld
quantization on the cylinder~$\mathbb{R}_\tau\times S^{d-1}$ of radius $R$ —
and obtained the same exact result in every case:
\begin{equation}
  \Delta_n^{\rm free}
  \;=\;
  n\,\frac{d-2}{2}\,.
\label{eq:free_delta_recall}
\end{equation}
Road~3 made the geometry transparent: on the cylinder, the free field admits
a spatially homogeneous solution oscillating at the conformal frequency
$\omega_0=\mu\equiv(d-2)/(2R)$.  The Bohr--Sommerfeld condition
$I(E)=2\pi n$ fixes the energy $E_{\rm cl}=n\omega_0$, and the
state--operator map $\Delta_n=R\,E_{\rm cl}$ immediately
gives~\eqref{eq:free_delta_recall}.  The goal of this section is to show
how each step of Road~3 generalizes once the quartic coupling $\lambda$ is
turned on.

\section{The Real Scalar \texorpdfstring{$\phi^4$}{phi4} Theory on the Cylinder}
\label{sec:interacting_cylinder}

After Weyl-mapping to $\mathbb{R}_\tau\times S^{d-1}_R$
(Section~\ref{app:soc}), the sphere curvature generates a conformal mass
$\mu^2=(d-2)^2/(4R^2)$ and the Euclidean action for a single real scalar
$\phi^4$ theory is
\begin{equation}
  S_E[\phi]
  \;=\;
  \int_0^{\mathcal{T}} \!d\tau
  \int_{S^{d-1}} \!d^{d-1}\!\Omega\;
  \Biggl[
    \frac{1}{2}\bigl(\partial_\tau\phi\bigr)^2
    + \frac{1}{2}\bigl|\nabla\phi\bigr|^2
    + \frac{\mu^2}{2}\,\phi^2
    + \frac{\lambda}{4}\,\phi^4
  \Biggr].
\label{eq:action_cyl_interacting}
\end{equation}
For the purposes of this formal setup we treat $\lambda>0$ as a free
parameter; it will be identified with the Wilson--Fisher fixed-point coupling
$\lambda_*$ in $d=4-\varepsilon$ when we specialize to the physical theory.

The Euler--Lagrange equation is
\begin{equation}
  -\partial_\tau^2\phi
  - \nabla^2_{S^{d-1}}\phi
  + \mu^2\,\phi
  + \lambda\,\phi^3
  \;=\; 0\,.
\label{eq:EOM_interacting}
\end{equation}

\paragraph{Spatially homogeneous ansatz.}
We seek classical solutions that are \emph{spatially homogeneous} on $S^{d-1}$,
i.e.\ $\phi(\tau,\Omega)=v(\tau)$ with $\nabla v=0$.  The gradient term drops
out and~\eqref{eq:EOM_interacting} reduces to the \textbf{quartic anharmonic
oscillator}:
\begin{equation}
  \ddot{v}(\tau)
  \;-\; \mu^2\,v(\tau)
  \;-\; \lambda\,v(\tau)^3
  \;=\; 0\,.
\label{eq:EOM_homogeneous}
\end{equation}
In Lorentzian time $t=-i\tau$ the signs flip and the potential becomes the
ordinary single-well $V_{\rm M}(v)=\tfrac{\mu^2}{2}v^2+\tfrac{\lambda}{4}v^4$,
which supports periodic oscillations around $v=0$ for all energies $E>0$.
 {We work directly on this oscillating orbit and continue to
denote its (real) time by $\tau$; on this contour the saddle obeys the single-well
equation}
\[
 {\ddot{v}(\tau)\;+\;\mu^2\,v(\tau)\;+\;\lambda\,v(\tau)^3\;=\;0\,,}
\]
 {whose} exact periodic solution is the Jacobi elliptic cosine:
\begin{equation}
  v_{\rm cl}(\tau)
  \;=\;
  x_0\;\cn\! \left(\omega\,\tau \big| m\right),
\label{eq:classical_orbit}
\end{equation}
where the amplitude $x_0$, frequency $\omega$, and elliptic modulus
$m\in[0,\tfrac{1}{2})$ are not independent but are fixed by the coupling
and conformal mass:
\begin{equation}
  \omega^2 \;=\; \frac{\mu^2}{1-2m}\,,
  \qquad
  x_0^2 \;=\; \frac{2m\,\mu^2}{\lambda(1-2m)}\,,
  \qquad
  \mathcal{T} \;=\; \frac{4\KK(m)}{\omega}\,.
\label{eq:orbit_params}
\end{equation}
Here $\KK(m)$ is the complete elliptic integral of the first kind.
The free-theory limit corresponds to $m\to 0$:  the frequency $\omega\to\mu$, the period $\mathcal{T}\to 2\pi/\mu$, and the
solution $v_{\rm cl}\to x_0\cos(\mu\tau)$ — recovering the harmonic orbit of
Road~3.  As $m\to\tfrac{1}{2}$ the orbit becomes increasingly anharmonic  and the semiclassical expansion must be handled with extra care. 

The modulus $m$ is the single parameter that encodes the deformation of the
orbit away from the free-theory circle.  In the double-scaling limit (discussed in detail below) 
  $n \to\infty,\quad \lambda\to 0$,  
$ \kappa \;\equiv\; \lambda\, n \;=\; \text{fixed}$ it becomes a fixed function $m=m(\kappa)$ of the scaling variable $\kappa$.

\begin{figure}[t]
  \centering
  \begin{minipage}[t]{0.47\textwidth}
    \centering
    \includegraphics[width=\textwidth]{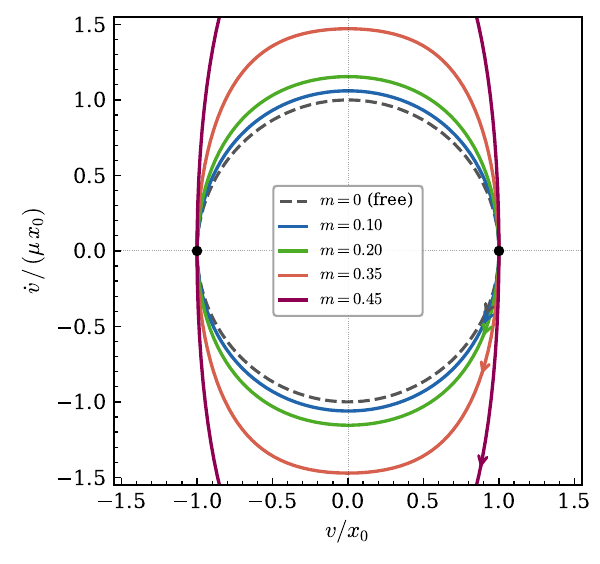}
    \caption*{\small(a) Phase-space portrait.}
  \end{minipage}
  \hfill
  \begin{minipage}[t]{0.47\textwidth}
    \centering
    \includegraphics[width=\textwidth]{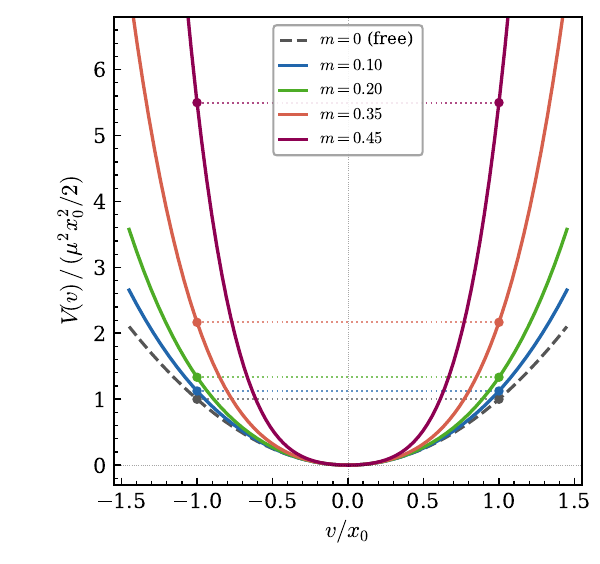}
    \caption*{\small(b) Lorentzian potential $V(v)=\tfrac{\mu^2}{2}v^2+\tfrac{\lambda}{4}v^4$.}
  \end{minipage}
  \caption{Classical orbit structure of the Jacobi-elliptic solution
    $v_{\rm cl}(\tau)=x_0\,\cn(\omega\tau|m)$~\eqref{eq:classical_orbit}
    for $m=0,\,0.10,\,0.20,\,0.35,\,0.45$
    (free theory to strongly coupled), at fixed amplitude $x_0$.
    \textbf{(a)} Orbits in the $(v/x_0,\,\dot{v}/\mu x_0)$ plane.
    The dashed circle is the free-theory harmonic orbit ($m=0$);
    all curves share the same turning points $v=\pm x_0$ (filled dots),
    and arrows show the direction of traversal in Euclidean time.
    As $m$ grows the orbit is deformed and the orbit
    develops the flat-topped shape of the Jacobi cosine.
    At fixed amplitude $x_0$ a larger $m$ means a larger coupling
    $\lambda=2m\mu^2/[x_0^2(1-2m)]$, hence more energy stored in the
    steeper potential at the turning points
    $E=({\mu^2 x_0^2}/{2})\cdot{(1-m)}/{(1-2m)}$.
    Since all of this energy is kinetic at $v=0$, the peak speed
    $|\dot{v}|_{\rm max}=\mu x_0\sqrt{(1-m)/(1-2m)}$ grows with $m$
    (the ratio $(1-m)/(1-2m)=1+m/(1-2m)$ is strictly increasing).
    \textbf{(b)} The corresponding potential (normalised by $\tfrac{1}{2}\mu^2 x_0^2$)
    becomes steeper with growing $\kappa=\lambda n$; dotted horizontal lines
    mark the energy $E_m=(1-m)/(1-2m)$ of each orbit at its turning points.}
  \label{fig:phase_portrait_cn}
\end{figure}

\section{Saddle-Point Expansion Around the Classical Orbit}
\label{sec:saddle_point_setup}

To compute the quantum spectrum we expand around the classical orbit,
\begin{equation}
  \phi(\tau,\Omega)
  \;=\;
  v_{\rm cl}(\tau)
  \;+\;
  \eta(\tau,\Omega)\,,
\label{eq:fluctuation_expansion}
\end{equation}
where $\eta$ is the quantum fluctuation, periodic in $\tau$ with period
$\mathcal{T}$.  Inserting into~\eqref{eq:action_cyl_interacting} and
expanding in $\eta$:
\begin{equation}
  S_E\bigl[v_{\rm cl}+\eta\bigr]
  \;=\;
  S_{\rm cl}(\mathcal{T})
  \;+\;
  \frac{1}{2}
    \int_0^{\mathcal{T}}\!d\tau
    \int_{S^{d-1}}\!d^{d-1}\!\Omega\;
    \eta\;\mathcal{M}\;\eta
  \;+\;
  O(\eta^3)\,,
\label{eq:fluctuation_action}
\end{equation}
where $S_{\rm cl}(\mathcal{T})=S_E[v_{\rm cl}]$ is the on-shell Euclidean action
and $\mathcal{M}$ is the \textbf{fluctuation operator} (second functional
derivative of $S_E$ at $v_{\rm cl}$):
\begin{equation}
  \mathcal{M}
  \;=\;
  -\partial_\tau^2
  \;-\; \nabla^2_{S^{d-1}}
  \;+\; \mu^2
  \;+\; 3\lambda\,v_{\rm cl}(\tau)^2\,.
\label{eq:fluctuation_operator}
\end{equation}
The $\tau$-dependent potential $3\lambda\,v_{\rm cl}^2 = 3\lambda x_0^2\,\cn^2(\omega\tau|m)$
is periodic with the same period $\mathcal{T}$ as the orbit, so $\mathcal{M}$ is a
\textbf{Hill's operator}.  Its spectral theory is governed by Floquet theory:
upon decomposing $\eta$ in spherical harmonics of angular momentum $\ell$ on
$S^{d-1}$ (with degeneracy $n_\ell$), each $\ell$-mode satisfies a Lamé
equation with Floquet stability angle $\nu_\ell$ (Section~\ref{energypath}).

\paragraph{Semiclassical path integral.}
Integrating out the Gaussian fluctuation $\eta$ yields the leading
semiclassical approximation to the $\mathcal{T}$-periodic path integral:
\begin{equation}
  Z(\mathcal{T})
  \;\approx\;
  e^{-S_{\rm cl}(\mathcal{T})}
  \cdot
  \bigl(\det\mathcal{M}\bigr)^{-1/2}
  \cdot
  \bigl(1 + O(1/n)\bigr)\,.
\label{eq:semiclassical_Z}
\end{equation}
The key point is the scaling: $S_{\rm cl}(\mathcal{T})\propto n$ (shown
below), so $e^{-S_{\rm cl}}\sim e^{-cn}$ is the leading classical weight and
each factor of $\bigl(\det\mathcal{M}\bigr)^{-1/2}$ contributes at relative
order $1/n$.  The fluctuation determinant is evaluated via the
Gel'fand--Yaglom theorem and Floquet theory in Section~\ref{energypath}~\cite{arXiv:1403.4545,arXiv:1210.4258,arXiv:1504.07997}.

\paragraph{Remark on the \texorpdfstring{$O(N)$}{O(N)} generalization.}
For the $O(N)$ theory with fields $\phi_a$, $a=1,\ldots,N$, one excites a
single component $\phi_1(\tau)=v_{\rm cl}(\tau)$ and sets $\phi_a=0$ for
$a\ge2$.  The EOM and its Jacobi-cn solution are unchanged.  The fluctuation
spectrum now splits into two sectors: the \emph{longitudinal} mode $\eta_1$
sees the same operator~\eqref{eq:fluctuation_operator}, while each of the
$N-1$ \emph{transverse} modes $\eta_a$ ($a\ge2$) sees a softer operator
$\mathcal{M}_T=-\partial_\tau^2-\nabla^2_{S^{d-1}}+\mu^2+\lambda\,v_{\rm cl}^2$
(one factor of $3\lambda v_{\rm cl}^2$ versus $\lambda v_{\rm cl}^2$, since the
transverse modes do not feel the longitudinal anharmonic restoring force).
The semiclassical path integral becomes
$Z(\mathcal{T})\approx e^{-S_{\rm cl}}(\det\mathcal{M}_L)^{-1/2}
(\det\mathcal{M}_T)^{-(N-1)/2}(1+O(1/n))$,
and the one-loop coefficient $C_1(\kappa)$ receives contributions from both
sectors.  This generalization is treated in Chapter~\ref{chap:interacting}.

\section{The Double-Scaling Limit and the \texorpdfstring{$1/n$}{1/n} Expansion}
\label{sec:double_scaling_limit}

At fixed $\lambda$ the perturbative expansion of $\Delta_n$ in powers of
$\lambda$ breaks down at order $k$ because the $k$-loop contribution
grows as $(\lambda n)^k$.  Both problems — the orbit deformation and the
breakdown of perturbation theory — are resolved by taking the
\textbf{double-scaling limit}
\begin{equation}
  n \to\infty,\quad \lambda\to 0,\quad
  \kappa \;\equiv\; \lambda\, n \;=\; \text{fixed}\,.
\label{eq:kappa_def}
\end{equation}
In this regime the classical action scales as $S_{\rm cl}\sim n$: indeed,
from~\eqref{eq:orbit_params} and the relation $I(E)=2\pi n$ we see that
$x_0^2\sim\lambda^{-1}\sim n/\kappa$, so the field amplitude grows as
$\phi\sim n^{1/2}$ and the quartic potential contributes $\lambda\phi^4\sim n$
per unit time — confirming $S_{\rm cl}\sim n$.  The modulus $m$ is then a
fixed function of $\kappa$ alone, $m=m(\kappa)$, determined by the
Bohr--Sommerfeld condition at leading order.  The scaling dimension admits the
uniform asymptotic expansion (cf.\ \eqref{eq:delta_expansion_program})
\begin{equation}
  \boxed{
  \Delta_n(\kappa)
  \;=\;
  n\,C_0(\kappa) + C_1(\kappa) + \frac{C_2(\kappa)}{n} + \cdots
  }
\label{eq:delta_expansion_program}
\end{equation}
where each $C_i(\kappa)$ resums all $(\lambda n)^k$ contributions at order
$n^{1-i}$.  In the free limit $\kappa\to 0$:
$C_0(0)=(d-2)/2$, $C_i(0)=0$ for $i\ge1$,
recovering~\eqref{eq:free_delta_recall}~\cite{arXiv:1309.5089,arXiv:1502.02033}.

\section{Blueprint: The Semiclassical Program in Five Steps}
\label{sec:semiclassical_summary}

The computation of the $C_i(\kappa)$ is a direct generalization of Road~3
from Chapter~\ref{chap:freetheory}:

\begin{enumerate}[leftmargin=2.2em,itemsep=4pt]
  \item \textbf{Find the classical periodic orbit.}
    The spatially homogeneous solution $v_{\rm cl}(\tau)=x_0\,\cn(\omega\tau|m)$
    of~\eqref{eq:EOM_homogeneous} is exact, with parameters~\eqref{eq:orbit_params}.
    The double-scaling limit fixes $m=m(\kappa)$; as $\kappa\to0$, $m\to0$
    and the orbit reduces to the free harmonic solution.

  \item \textbf{Compute the classical action and energy.}
    The on-shell action $S_{\rm cl}(\mathcal{T})$ and the energy $E_{\rm cl}$
    are related by the Legendre pair $I(E)=\mathcal{T}E-S_{\rm cl}(\mathcal{T})$,
    $dI/dE=\mathcal{T}$ (Section~\ref{sec:action_variable_QM}), which hold in the
    field theory without modification.  The action variable $I$ is computed from
    the orbit~\eqref{eq:classical_orbit} by $I=\int_0^\mathcal{T}\dot{v}_{\rm cl}^2\,d\tau$.

  \item \textbf{Apply the Bohr--Sommerfeld condition.}
    Setting $I(E_n)=2\pi n$ selects the quantum levels.  At leading order:
    \begin{equation*}
      C_0(\kappa)
      \;=\;
      \frac{R\,E_{\rm cl}}{n}
      \;=\;
      \frac{2\pi R}{\mathcal{T}(\kappa)}\,.
    \end{equation*}

  \item \textbf{Compute one-loop fluctuations.}
    Decompose $\eta(\tau,\Omega)$ in spherical harmonics $Y_{\ell m}$ on
    $S^{d-1}$.  For each $\ell$, the operator $\mathcal{M}$ restricted to the
    $\ell$-subspace becomes a Lam\'e equation with Floquet stability angle
    $\nu_\ell(\kappa)$ and degeneracy $n_\ell$; the mode can be occupied
    to any non-negative integer level $q_\ell\in\{0,1,2,\ldots\}$, the
    occupation number counting excitation quanta above the saddle.
    The Gel'fand--Yaglom
    theorem converts $\det\mathcal{M}$ into the stability angles; coupling
    renormalization in $d=4-\varepsilon$ produces a counterterm
    $\delta E_1(\kappa)$ (derived in \S\,\ref{sec:renorm};
    see~\eqref{eq:deltaE1_phi4}) that cancels the $1/\varepsilon$ poles in
    the stability-angle sums.  For the ground state (all $q_\ell=0$) the
    result is~\eqref{eq:C1_identification_boxed}; zero modes
    ($\nu_\ell=0$) are excluded and handled by collective coordinates.
  \item \textbf{Read off the scaling dimension.}
    By the state--operator correspondence (§\ref{app:soc}), $\Delta_n=R\,E_n$,
    so each order in the $1/n$ expansion of $E_n$ maps directly to a coefficient
    in~\eqref{eq:delta_expansion_program}.
\end{enumerate}

The path-integral implementation of steps 1--3 — the proper-time representation
of the resolvent, the periodic-orbit saddle, and the Gel'fand--Yaglom evaluation
of $\det\mathcal{M}$ — is developed in the next section~\cite{arXiv:1807.04434,arXiv:1601.03476,Kos:2013tga}.


\section{Cylinder Quantisation and the Action Variable}\label{chap:actionvariable}

\noindent
The preceding chapters established two pillars.
\textbf{Chapter~\ref{chap:freetheory}} (Road~3) showed that in the \emph{free}
theory the Bohr--Sommerfeld condition $I(E_n)=2\pi n$ on the cylinder
immediately gives $E_n = n\mu$, and hence $\Delta_n^{\rm free} = n(d{-}2)/2$,
without any path integral.  That derivation worked because the classical orbit
is a perfect circle in phase space, $I(E)=2\pi E/\mu$ is linear in $E$, and the
fluctuation operator is just a constant-frequency harmonic oscillator.
\textbf{Chapter~\ref{chap:doublescaling}} introduced the interacting theory on
the cylinder: the classical orbit is now the Jacobi-elliptic trajectory
$v_{\rm cl}(\tau)=x_0\,\mathrm{cn}(\omega\tau|m)$, whose shape and period
depend on the double-scaling coupling $\kappa=\lambda n$; the fluctuation
operator is a periodic Hill/Lam\'e operator rather than a simple harmonic
oscillator.

Three new ingredients appear in the interacting theory that were trivial or
absent in the free theory:
\begin{enumerate}
  \item \textbf{A nonlinear $I(E)$.}  The orbit is no longer a circle; its
  enclosed area $I(E)$ satisfies a nontrivial implicit equation involving
  elliptic integrals.  The Bohr--Sommerfeld condition $I(E_{\rm cl})=2\pi n$
  is still correct but now determines $E_{\rm cl}(n)$ only implicitly.
  \item \textbf{Stability angles $\nu_\ell$.}  The fluctuation operator
  $\mathcal{M}=-\partial_\tau^2-\nabla^2_{S^{d-1}}+\mu^2+3\lambda v_{\rm cl}^2$
  has a time-periodic coefficient.  By Floquet's theorem, its solutions pick up a
  phase $\nu_\ell$ over one period --- the \emph{stability angle} of mode $\ell$.
  In the free theory $\nu_\ell^{\rm free}=\omega_\ell\mathcal{T}$ is trivially
  linear in $\mathcal{T}$; in the interacting theory $\nu_\ell(m)$ is a
  nontrivial function of the Jacobi elliptic modulus $m\equiv m(\kappa)$.
  \item \textbf{Quantum zero-point and excitation energies from Floquet modes.}
  Each mode contributes a zero-point energy $\nu_\ell/(2\mathcal{T})$ and an
  excitation gap $q_\ell\,\nu_\ell/\mathcal{T}$ to the total cylinder energy ---
  the quantum analogues of $\tfrac{1}{2}\hbar\omega_\ell$ and
  $q_\ell\hbar\omega_\ell$ of the harmonic oscillator, with $\omega_\ell$
  replaced by $\nu_\ell/\mathcal{T}$.
\end{enumerate}

This section collects the central results of the semiclassical framework before
the full derivations in Chapter~\ref{chap:periodicsaddles}.
\begin{itemize}
  \item \S\,\ref{sec:free_vs_interacting} places free and interacting
  calculations side by side and presents the \emph{classical-to-quantum
  dictionary}.
  \item \S\,\ref{sec:blueprint_results} states the four key results
  --- quantization condition, energy formula, and $C_0$/$C_1$ identification
  --- without proof, with pointers to where each is derived.
\end{itemize}

\subsection{Free Theory vs.\ the Interacting Semiclassical Scheme}
\label{sec:free_vs_interacting}

Table~\ref{tab:free_vs_interacting} contrasts the free-theory calculation of
Chapter~\ref{chap:freetheory} (Road~3) with the general interacting
semiclassical scheme.  Reading across each row shows exactly \emph{which step}
is modified by the interaction and \emph{how}.

\begin{table}[ht]
\centering
\renewcommand{\arraystretch}{1.45}
\begin{tabular}{|p{3.8cm}|p{4.4cm}|p{4.8cm}|}
\hline
\textbf{Concept} &
  \textbf{Free theory} (Ch.~\ref{chap:freetheory}) &
  \textbf{Interacting theory} (\S\,\ref{chap:actionvariable}) \\
\hline
Classical orbit &
  Circle: $v_{\rm cl}=x_0\cos(\mu t)$;
  elliptic phase-space contour &
  Jacobi-cn: $v_{\rm cl}=x_0\,\mathrm{cn}(\omega t|m)$;
  deformed contour, shape $\propto\kappa=\lambda n$ \\
\hline
Action variable $I(E)$ &
  $\displaystyle\frac{2\pi E}{\mu}$ \;(linear; explicit $E_n=n\mu$) &
  Nonlinear elliptic integral; $I(E_{\rm cl})=2\pi n$ implicit \\
\hline
Fluctuation operator \& modes &
  $-\partial_t^2+\mu^2$ (constant);
  harmonic oscillator, $\omega_\ell=\sqrt{\mu^2+J_\ell^2/R^2}$ &
  $-\partial_t^2+\mu^2+3\lambda v_{\rm cl}^2$ (periodic);
  Lam\'e equation, Floquet band structure \\
\hline
Stability angle $\nu_\ell$ &
  $\omega_\ell\mathcal{T}$ \;(trivially linear) &
  Nontrivial $\nu_\ell(m)$ from monodromy \\
\hline
Zero-point energy &
  $\displaystyle\frac{1}{2}\sum_\ell n_\ell\,\omega_\ell$ &
  $\displaystyle\frac{1}{2\mathcal{T}}\sum_{\nu_\ell>0} n_\ell\,\nu_\ell$ \\
\hline
Leading $\Delta_n$ &
  $n(d{-}2)/2$ \;(closed form) &
  $nC_0(\kappa)+C_1(\kappa)+O(1/n)$ \\
\hline
\end{tabular}
\caption{Free theory vs.\ interacting semiclassical scheme.  The free-theory
limit is $m\to 0$ ($\kappa\to 0$): $\mathrm{cn}(z|0)=\cos z$, $\omega\to\mu$,
and the Lam\'e equation reduces to a constant-coefficient harmonic oscillator
with $\nu_\ell\to\omega_\ell\mathcal{T}$.}
\label{tab:free_vs_interacting}
\end{table}

\subsubsection{The Classical-to-Quantum Dictionary}
\label{sec:classical_quantum_dict}

Table~\ref{tab:classical_quantum_dict} gives the complete translation between
classical or semiclassical objects and their quantum meaning.  This dictionary
is the backbone of the semiclassical program; every formula in this section and
in Chapter~\ref{chap:interacting} is a consequence of it.

\begin{table}[ht]
\centering
\renewcommand{\arraystretch}{1.5}
\begin{tabular}{|p{4.4cm}|p{4.5cm}|p{4.6cm}|}
\hline
\textbf{Classical / semiclassical} &
  \textbf{Quantum meaning} &
  \textbf{Key formula} \\
\hline
Periodic orbit $v_{\rm cl}(t)$, period $\mathcal{T}$ &
  Saddle of the path integral; labels a tower of states $|n,\{q_\ell\}\rangle$ &
  EOM: $\ddot v+\mu^2 v+\lambda v^3=0$ \\
\hline
Action variable $I(E)=\oint\Pi\,dv$ &
  Phase accumulated per traversal; quantised in units of $2\pi$ &
  $I(E_{\rm cl})=2\pi n$ selects the orbit \\
\hline
Period $\mathcal{T}(E)$ &
  Slope of $I(E)$; conjugate of energy &
  $\mathcal{T}=dI/dE$ \\
\hline
Classical action $\mathcal{S}_{\rm cl}(\mathcal{T})$ &
  Legendre partner of $I$; controls saddle weight $e^{-\mathcal{S}_{\rm cl}}$ &
  $I=E\mathcal{T}+\mathcal{S}_{\rm cl}$ \\
\hline
Winding number $k$ &
  $k$-th term $e^{ik\mathcal{S}_{\rm cl}}$ in the propagator sum &
  $\sum_k e^{ik\mathcal{S}_{\rm cl}}$ resummed; poles at $I(E)=2\pi n$ (\S\,\ref{sec:semiclassical_result}) \\
\hline
Stability angle $\nu_\ell$ &
  Floquet phase: $\eta(t{+}\mathcal{T})=e^{i\nu_\ell}\eta(t)$ &
  Eigenvalue of monodromy matrix $M$ \\
\hline
$\nu_\ell/\mathcal{T}$ &
  Effective frequency of mode $\ell$ on the orbit &
  Replaces $\omega_\ell$ of simple harmonic oscillator \\
\hline
$\nu_\ell/(2\mathcal{T})$ &
  Zero-point energy of mode $\ell$ &
  Analogue of $\tfrac{1}{2}\hbar\omega_\ell$ \\
\hline
$q_\ell\,\nu_\ell/\mathcal{T}$ &
  Excitation energy from $q_\ell$ quanta in mode $\ell$ &
  Analogue of $q_\ell\hbar\omega_\ell$ \\
\hline
Occupation number $n=I/(2\pi)$ &
  Labels the composite operator $(\phi^{\otimes n})_{\rm primary}$ &
  Heavy / large-$n$ limit: $n\to\infty$ \\
\hline
Scaling dimension $\Delta_n$ &
  Cylinder energy times sphere radius &
  $\Delta_n = R\,E_{n,\{q_\ell\}}$ \\
\hline
\end{tabular}
\caption{Classical-to-quantum dictionary for semiclassical quantisation on the
cylinder.  The dictionary holds in any dimension $d$ and for any confining
periodic potential.  The stability angles $\nu_\ell$ are the central new object:
they encode the quantum mode structure of the fluctuations around the periodic
orbit.}
\label{tab:classical_quantum_dict}
\end{table} 
\subsubsection{Key Results}\label{sec:blueprint_results}

The ingredients above combine into the following central results, stated
here without proof.  Full derivations are supplied in
\S\,\ref{sec:resolvent}--\ref{sec:semiclassical_result} and
\S\,\ref{sec:hill_floquet}--\ref{sec:gelfand_yaglom},
all in Chapter~\ref{chap:periodicsaddles}.

\paragraph{Bohr--Sommerfeld quantization condition.}
The energy levels are selected by requiring that the total accumulated phase
equals $2\pi n$.  Including zero-point contributions from all stable Floquet
modes (see \S\,\ref{sec:semiclassical_result}):
\begin{equation}
\boxed{
I(E) + \sum_{\nu_\ell>0}\!\left(q_\ell+\tfrac{n_\ell}{2}\right)
\!\left(\mathcal{T}\frac{d\nu_\ell}{d\mathcal{T}}-\nu_\ell\right) = 2\pi n\,.
}
\label{eq:full_quantization_boxed}
\end{equation}

\paragraph{Energy eigenvalue formula.}
 \begin{equation}
\boxed{
E_{n,\{q_\ell\}} = E_{\rm cl}(n) + \delta E_1
+ \frac{1}{\mathcal{T}}\sum_{\nu_\ell>0}
\!\left(q_\ell+\tfrac{n_\ell}{2}\right)\nu_\ell
}
\label{eq:energy_final_boxed_new}
\end{equation}
Here $E_{\rm cl}(n)$ is determined implicitly by $I(E_{\rm cl})=2\pi n$;
$\delta E_1$ is the Gel'fand--Yaglom one-loop shift that cancels the UV
divergences in the mode sum (derived in \S\,\ref{sec:gelfand_yaglom};
see also \eqref{eq:deltaE1_phi4}).
The sum runs over angular-momentum multiplets
$\ell = 0, 1, 2, \ldots$ on $S^{d-1}$: for each $\ell$ there is a
\emph{single} Floquet stability angle $\nu_\ell(\kappa)$ — a function
of $\ell$ alone, not an independent variable — and a degeneracy
\[
  n_\ell \;=\; \frac{(2\ell+d-2)\,(\ell+d-3)!}{\ell!\,(d-2)!}
\]
counting the real spherical harmonics of angular momentum $\ell$ on
$S^{d-1}$ (e.g.\ $n_0=1$, $n_1=d$; for $d=4$: $n_\ell=(\ell+1)^2$;
full derivation in \S\,\ref{sec:hill_floquet}, \eqref{eq:degeneracy_formula}).
The integer $q_\ell\geq 0$ is the \emph{total occupation number} of
multiplet $\ell$: the sum of the individual quantum numbers across all
$n_\ell$ degenerate Floquet oscillators within that multiplet.  The
zero-point contribution $\tfrac{n_\ell}{2}\nu_\ell$ per multiplet is
the Casimir energy of the $n_\ell$ independent oscillators.
The condition $\nu_\ell>0$ is a \emph{filter}, not a second summation
index: it excludes zero modes, which are treated by collective
coordinates and contribute to $\delta E_1$.

\paragraph{Matching the $1/n$ coefficients.}
Since $E_{\rm cl}$ grows linearly in $n$ (from $I(E_{\rm cl})=2\pi n$ and
$dI/dE=\mathcal{T}$), the ground-state expansion
$\Delta_n = R\,E_{n,0} = nC_0(\kappa)+C_1(\kappa)+O(1/n)$ identifies:
\begin{equation}
\boxed{nC_0 = R\,E_{\rm cl}(n)\,,\qquad C_0 = \frac{2\pi R}{\mathcal{T}}\,.}
\label{eq:C0_identification_boxed}
\end{equation}
\begin{equation}
\boxed{
C_1 = R\,\delta E_1
+ \frac{R}{2\mathcal{T}}\sum_{\nu_\ell>0}
n_\ell\,\nu_\ell\,.
}
\label{eq:C1_identification_boxed}
\end{equation}
{For an excited state $\{q_\ell\}$ the same identification gives
\begin{equation*}
  C_1^{\{q_\ell\}} = R\,\delta E_1
  + \frac{R}{\mathcal{T}}\sum_{\nu_\ell>0}
  \Bigl(q_\ell+\tfrac{n_\ell}{2}\Bigr)\nu_\ell\,,
\end{equation*}
which reduces to the boxed (ground-state) expression at $q_\ell=0$; cf.\
\eqref{eq:energy_final_boxed_new}.}
The explicit computation of $E_{\rm cl}$, $\delta E_1$, and $\nu_\ell$ for
$O(N)$ $\phi^4$ at the Wilson--Fisher fixed point is carried out in
Chapter~\ref{chap:interacting}.

\part{The Semiclassical Derivation}

\chapter{Semiclassical Quantisation: Periodic Saddles, Fluctuations, and Floquet Theory}\label{chap:periodicsaddles}\label{energypath}\label{chap:floquet}

Section~\ref{chap:actionvariable} assembled the complete semiclassical
toolkit as a \emph{blueprint}: it introduced the action variable
$I(E)=\oint p\,dq$, the Bohr--Sommerfeld condition $I=2\pi n$, the
stability angles $\nu_\ell$, and the energy formula
\eqref{eq:energy_final_boxed_new}.  Additionally, section~\ref{sec:semiclassical_program} set up the semiclassical programme for a single real scalar
$\phi^4$ theory on $\mathbb{R}\times S^{d-1}$: the Euclidean action \eqref{eq:action_cyl_interacting},
the classical orbit $v_{\rm cl}(\tau)=x_0\cn(\omega\tau|m)$ \eqref{eq:classical_orbit}, the fluctuation
operator
\begin{equation*}
  \mathcal{M} = -\partial_\tau^2 - \nabla^2_{S^{d-1}} + \mu^2 + 3\lambda\,v_{\rm cl}(\tau)^2
\end{equation*}
(see \eqref{eq:fluctuation_operator}), and the semiclassical partition function
$Z(\mathcal{T})\approx e^{-S_{\rm cl}(\mathcal{T})}(\det\mathcal{M})^{-1/2}$ \eqref{eq:semiclassical_Z}.
The present section derives these objects from first principles via a path-integral argument, making
precise all steps 1--3 of the five-step programme summarised in Section~\ref{sec:semiclassical_program}.

 The present chapter proves the various steps, starting from
the Lorentzian path integral, and derive each ingredient in turn:

\begin{itemize}
  \item \S\,\ref{sec:resolvent} — the resolvent $G(E)=\Tr(H-E)^{-1}$
        and its proper-time (Schwinger) representation;
  \item \S\,\ref{sec:pi_pbc} — the thermal trace as a path integral with
        periodic boundary conditions;
  \item \S\,\ref{sec:semiclassical_expansion} — the saddle-point
        expansion around $v_{\rm cl}(\tau)$ and the fluctuation
        determinant $\det\mathcal{O}^{(2)}$;
  \item \S\,\ref{sec:semiclassical_result} — the Gutzwiller trace
        formula and the extraction of the quantization condition.
\end{itemize}

Sections~\ref{sec:hill_floquet}--\ref{sec:gelfand_yaglom} then derive the remaining input,
$\det\mathcal{O}^{(2)}$, via Floquet theory and the Gel'fand--Yaglom
theorem.

Section~\ref{sec:semiclassical_program} works in \emph{Euclidean} signature throughout: the Euclidean
time $\tau$ is related to Lorentzian time $t$ by $\tau = it$, and the weight in the path integral
is $e^{-S_E}$.  The present section develops the parallel derivation in \emph{Lorentzian} signature,
with real time $t$ and weight $e^{i\mathcal{S}}$; this is the natural language for the resolvent and
the trace formula.  The fluctuation operator studied here,
\begin{equation*}
  \mathcal{O}^{(2)} \equiv \frac{\delta^2\mathcal{S}}{\delta\phi\,\delta\phi}\bigg|_{\phi=v},
\end{equation*}
is the Lorentzian counterpart of $\mathcal{M}$: explicitly,
$\mathcal{O}^{(2)}_{\rm Lor} = \partial_t^2 - \nabla^2_{S^{d-1}} - \mu^2 - 3\lambda v_{\rm cl}^2$,
which maps to $-\mathcal{M}$ under $t\mapsto -i\tau$.  Similarly, the Lorentzian classical solution
$v(t,\Omega)$ reduces to the spatially homogeneous $v_{\rm cl}(\tau)$ of \eqref{eq:classical_orbit}
after analytic continuation.

\bigskip 
 The derivation proceeds in six steps:
\begin{enumerate}
    \item Express the energy spectrum through poles of the resolvent $G(E) = \Tr(H-E)^{-1}$ \eqref{eq:resolvent_def}.
    \item Rewrite the resolvent using a proper-time (Schwinger) representation \eqref{eq:proper_time_resolvent}.
    \item Recognize that the trace of the time-evolution operator equals a path integral with periodic
          boundary conditions \eqref{eq:trace_as_PI}.
    \item Evaluate this path integral semiclassically around periodic classical solutions—the
          $v_{\rm cl}$ of \eqref{eq:classical_orbit} \eqref{eq:semiclassical_trace}.
    \item Use the Gel'fand--Yaglom theorem and Floquet theory to compute $\det\mathcal{O}^{(2)}
          \equiv\det\mathcal{M}$ \eqref{eq:det_factorization}.
    \item Extract the quantization condition from the $\mathcal{T}$-integral saddle point (the classical
          Bohr--Sommerfeld condition $I=2\pi n$, recovered in Chapter~\ref{chap:interacting}).
\end{enumerate}

The resulting expression is a field-theoretic instance of the Gutzwiller trace formula \cite{Gutzwiller:1971fy},
which expresses the density of states in terms of periodic orbits.  This remarkable connection between
classical orbits and quantum spectra underlies much of modern semiclassical physics and has applications
ranging from quantum chaos to molecular spectroscopy.

\section{The Resolvent and Its Spectral Representation}\label{sec:resolvent}


The resolvent is a fundamental object in spectral theory because it encodes information about all energy eigenvalues simultaneously.
While the density of states $\rho(E)$ is hard to access directly from a path integral, the resolvent is naturally expressed as a
trace of the evolution operator, which has a beautiful path integral representation. This is the key bridge between the Hamiltonian
formalism and path integrals.

Consider a quantum field theory with Hamiltonian $H$. The resolvent is defined as
\begin{equation}
\boxed{
G(E) \equiv \Tr\frac{1}{H - E}
}
\label{eq:resolvent_def}
\end{equation}
where the trace is over the full Hilbert space. The resolvent is closely related to the Green's function: the resolvent {operator $(H-E)^{-1}$}
acting on a state gives the response of the system to a perturbation at energy $E$ {; its trace $G(E)$ is the corresponding spectral function, a meromorphic function of $E$ whose poles locate the eigenvalues}.

\subsection{Spectral Decomposition: Poles Locate Energy Eigenvalues}

Let $\{|n\rangle\}$ be a complete set of energy eigenstates with $H|n\rangle = E_n|n\rangle$. Then:
\begin{align}
G(E) &= \Tr\frac{1}{H - E} \nonumber\\[6pt]
&= \sum_n \langle n| \frac{1}{H - E} |n\rangle \nonumber\\[6pt]
&= \sum_n \frac{1}{E_n - E}
\label{eq:spectral_rep}
\end{align}
\noindent
 The poles of $G(E)$ are located precisely at the energy eigenvalues $E_n$. Near each pole:
\[
G(E) \approx \frac{g_n}{E_n - E} + \text{regular terms}
\]
where $g_n$ is the degeneracy of level $E_n$.

For a system with discrete spectrum (such as a QFT on a compact space like $\mathbb{R} \times S^{d-1}$),
$G(E)$ is meromorphic with simple poles at each $E_n$. The residue at each pole counts the degeneracy.

\subsection{The Inverse as a Proper-Time Integral}

We use the integral representation of the inverse:
\begin{equation}
\frac{1}{x} = i \int_0^\infty d\mathcal{T}\, e^{-ix\mathcal{T}}
\label{eq:inverse_rep}
\end{equation}
This formula is valid for $\text{Im}(x) < 0$ (or equivalently, with an implicit $-i\epsilon$ prescription), as verified by direct integration below.

Let's verify the formula explicitly:
\begin{align}
i \int_0^\infty d\mathcal{T}\, e^{-ix\mathcal{T}}
&= i \left[\frac{e^{-ix\mathcal{T}}}{-ix}\right]_0^\infty \nonumber\\[6pt]
&= \frac{i}{-ix} \left[\lim_{\mathcal{T} \to \infty} e^{-ix\mathcal{T}} - 1\right] \nonumber\\[6pt]
&= -\frac{1}{x} \left(0 - 1\right) \quad \text{(for Im}(x) < 0\text{)} \nonumber\\[6pt]
&= \frac{1}{x}
\end{align}

The convergence at large $\mathcal{T}$ is ensured by $\text{Im}(x) < 0$, which gives a decaying exponential.

\subsection{The Proper-Time Representation of the Resolvent}

Now apply this to the operator $(H - E)^{-1}$ with $x = H - E$. Since $H$ is Hermitian and we want $\text{Im}(H - E) < 0$,
we implicitly add $-i\epsilon$ to the energy: $E \to E + i\epsilon$. Then:
\begin{align}
\frac{1}{H - E - i\epsilon} &= i \int_0^\infty d\mathcal{T}\, e^{-i(H - E - i\epsilon)\mathcal{T}} \nonumber\\[6pt]
&= i \int_0^\infty d\mathcal{T}\, e^{-iH\mathcal{T}}\, e^{iE\mathcal{T}}\, e^{-\epsilon\mathcal{T}}
\end{align}

Taking the trace:
\begin{equation}
\boxed{
G(E) = i \int_0^\infty d\mathcal{T}\, e^{iE\mathcal{T}}\, \Tr\left(e^{-iH\mathcal{T}}\right)
}
\label{eq:proper_time_resolvent}
\end{equation}

The quantity $\Tr(e^{-iH\mathcal{T}})$ is the trace of the time-evolution operator over a time interval $\mathcal{T}$.
This is sometimes called the \textit{return amplitude} or \textit{propagator trace}.

\textbf{Physical interpretation of the $i\epsilon$ prescription:} The shift $E\to E+i\epsilon$ guarantees convergence of
the $\mathcal{T}$-integral at large $\mathcal{T}$: the factor $e^{-\epsilon\mathcal{T}}$ damps the integrand and selects
the retarded (causal) resolvent.  Physically, the prescription corresponds to adiabatic switching — the Hamiltonian is
turned on gradually via $e^{\epsilon\mathcal{T}}$, ensuring that the system settles into its stationary state rather than
accumulating oscillatory contributions from the infinite past.  This is the canonical way to promote the scalar identity
\eqref{eq:inverse_rep} to the operator resolvent $(H-E-i\epsilon)^{-1}$.

\textbf{Connection to thermodynamics:} With Euclidean time $\tau = it$, one writes $\Tr(e^{-H\tau})$, which is precisely
the partition function at inverse temperature $\beta = \tau$. In Lorentzian signature, $\mathcal{T}$ is real time. The
proper-time integral acts like a Laplace transform, converting information about the time-evolution operator into the
resolvent as a function of energy~\cite{arXiv:1611.02912,arXiv:1610.04495}.

\section{Path Integral with Periodic Boundary Conditions}\label{sec:pi_pbc}

\subsection{Deriving the Trace-to-Path-Integral Correspondence}

The trace of the evolution operator can be expressed as a path integral. For a scalar field $\phi$:

\begin{equation}
\Tr\left(e^{-iH\mathcal{T}}\right) = \int_{\phi(t+\mathcal{T},\Omega) = \phi(t,\Omega)} \mathcal{D}\phi\, e^{i\mathcal{S}[\phi]}
\label{eq:trace_as_PI}
\end{equation}

\textbf{Detailed derivation:} To establish this, we insert a complete set of field configurations at intermediate times.
Start with the definition of the trace:
\[
\Tr(e^{-iH\mathcal{T}}) = \int \mathcal{D}\phi_f\, \langle\phi_f| e^{-iH\mathcal{T}} |\phi_f\rangle
\]
where we use $|\phi_f\rangle$ for both the bra and ket because taking the trace requires the same state at the initial
and final times. The matrix element $\langle\phi_f| e^{-iH\mathcal{T}} |\phi_i\rangle$ is the propagation kernel, which
admits the path integral representation:
\[
\langle\phi_f| e^{-iH\mathcal{T}} |\phi_i\rangle = \int_{\phi(0)=\phi_i}^{\phi(\mathcal{T})=\phi_f} \mathcal{D}\phi\, e^{i\mathcal{S}[\phi]}
\]

Setting $\phi_i = \phi_f$ and integrating over all such configurations gives \eqref{eq:trace_as_PI}.

The crucial feature is that the path integral is over field configurations that are {periodic} in time:
\[
\phi(t + \mathcal{T}, \Omega) = \phi(t, \Omega) \qquad \forall\, \Omega \in S^{d-1}
\]
This is the Lorentzian analog of the thermal partition function's periodic Euclidean time. The periodicity arises naturally
because we are taking the trace, which imposes ``closing'' the path: the initial and final field configurations must be
identical~\cite{arXiv:1612.08985,arXiv:1707.00710}.

\textbf{Physical interpretation:} The periodicity condition is fundamental: it ensures that we are summing over all closed
field trajectories. In the semiclassical limit, the dominant contributions come from classical solutions that are periodic
with period $\mathcal{T}$, corresponding to closed orbits in the classical phase space.

\section{Semiclassical Expansion Around Periodic Solutions}\label{sec:semiclassical_expansion}

\subsection{Setup: Classical Periodic Solutions}

In the proper-time representation \eqref{eq:proper_time_resolvent}, $\mathcal{T}$ is a dummy integration variable — one integrates over all positive real values. In the semiclassical evaluation that follows, the path integral over field configurations is dominated by the classical periodic saddle, and the $\mathcal{T}$-integral itself is dominated by its saddle point, which picks out a specific value $\mathcal{T}=\mathcal{T}_{\rm cl}(E)$. At this saddle, the dummy variable coincides with the physical period of the classical orbit. The notation is the same by design: once the saddle point condition $dI/dE = \mathcal{T}$ is imposed, $\mathcal{T}$ ceases to be free and becomes the physical period.

We evaluate the path integral \eqref{eq:trace_as_PI} semiclassically. The dominant contribution comes from
classical solutions $v(t,\Omega)$ satisfying:
\begin{enumerate}
    \item The equations of motion: $\frac{\delta \mathcal{S}}{\delta \phi}\big|_{\phi=v} = 0$
    \item The periodicity condition: $v(t+\mathcal{T},\Omega) = v(t,\Omega)$
\end{enumerate}
For the real scalar $\phi^4$ theory studied in Section~\ref{sec:semiclassical_program}, these conditions
select the spatially homogeneous Jacobi-cn orbit $v(t,\Omega) = v_{\rm cl}(-it) = x_0\cn(-i\omega t|m)$
(analytic continuation of \eqref{eq:classical_orbit} to Lorentzian time), with period
$\mathcal{T}=4K(m)/\omega$ fixed by the quantisation condition.  The amplitude $x_0$ and frequency
$\omega$ are given in terms of the elliptic parameter $m$ and the coupling $\lambda$ by
\eqref{eq:orbit_params}.

 Additionally, a periodic classical solution is an orbit that closes after time $\mathcal{T}$.
In the path integral formulation, this corresponds to a trajectory that starts and ends at the same field configuration.
The existence of such solutions depends on the specific dynamics: for some systems and periods $\mathcal{T}$, no periodic
orbit exists (or exists only for special values of $\mathcal{T}$).

\subsection{Quadratic Expansion of the Action}

We expand the action around a periodic classical solution. Let $\phi = v + \eta$ where $\eta$ is a small fluctuation.
Using Taylor expansion:
\begin{align}
\mathcal{S}[v + \eta] &= \mathcal{S}[v] + \int d^d x\, \frac{\delta \mathcal{S}}{\delta \phi}\bigg|_v \eta
+ \frac{1}{2}\int d^d x\, d^d y\, \eta(x) \frac{\delta^2 \mathcal{S}}{\delta \phi(x)\delta\phi(y)}\bigg|_v \eta(y) + O(\eta^3) \nonumber\\[8pt]
\end{align}

The linear term vanishes by the equations of motion:
\begin{equation}
\int d^d x\, \frac{\delta \mathcal{S}}{\delta \phi}\bigg|_v \eta = 0
\end{equation}

Thus:
\begin{equation}
\mathcal{S}[v + \eta] = \mathcal{S}[v] + \frac{1}{2}\int d^d x\, d^d y\, \eta(x) \frac{\delta^2 \mathcal{S}}{\delta \phi(x)\delta\phi(y)}\bigg|_v \eta(y) + O(\eta^3)
\label{eq:action_expansion}
\end{equation}

Define the fluctuation operator (Hessian):
\begin{equation}
\mathcal{O}^{(2)} \equiv \frac{\delta^2 \mathcal{S}}{\delta \phi \delta \phi}\bigg|_{\phi=v}
\label{eq:hessian_def}
\end{equation}

This is the second functional derivative of the action evaluated at the classical solution. It is a differential operator
acting on the space of fluctuations.  For the real scalar $\phi^4$ theory, explicit evaluation gives
$\mathcal{O}^{(2)} = \partial_t^2 - \nabla^2_{S^{d-1}} - \mu^2 - 3\lambda v_{\rm cl}^2$, which is
related to the Euclidean operator $\mathcal{M}$ of \eqref{eq:fluctuation_operator} by the analytic
continuation $t\mapsto -i\tau$: $\mathcal{O}^{(2)}_{\rm Lor}=-\mathcal{M}_{\rm Euc}$. Their spectra
are therefore in bijection, and $\det\mathcal{O}^{(2)}\equiv\det\mathcal{M}$ (up to a sign convention
fixed by the $i\epsilon$ prescription).

\subsection{Gaussian Path Integral and the Fluctuation Determinant}

The path integral over fluctuations is Gaussian. Dropping cubic and higher-order terms:
\begin{align}
\int \mathcal{D}\eta\, \exp\left[\frac{i}{2}\int \eta \, \mathcal{O}^{(2)} \, \eta\right]
&= \left(\det \mathcal{O}^{(2)}\right)^{-1/2}
\label{eq:gaussian_integral}
\end{align}

  This follows from the infinite-dimensional generalization of the Gaussian integral formula:
\[
\int d^n x_1 \cdots d^n x_n\, \exp\left[-\frac{1}{2}x^T A x\right] = \frac{(2\pi)^{n/2}}{\sqrt{\det A}}
\]

For a functional integral with an operator $\mathcal{O}$:
\[
\int \mathcal{D}\eta\, \exp\left[\frac{i}{2}\int \eta \, \mathcal{O}^{(2)} \, \eta\right] = \frac{1}{\sqrt{\det \mathcal{O}^{(2)}}}
\]

The factor of $i$ in the exponent requires care: it means the determinant must be computed with the correct phase,
often specified via an $i\epsilon$ prescription. The absolute value of the determinant represents the suppression
(or enhancement) of zero-mode fluctuations.

\textbf{Physical interpretation:} The determinant measures how ``stiff'' the classical solution is to fluctuations.
If $\det \mathcal{O}^{(2)}$ is large (hard potential), fluctuations are suppressed and the semiclassical approximation
is good. If $\det \mathcal{O}^{(2)}$ is small (soft potential), fluctuations are important and one must be careful.

\section{Semiclassical Result}\label{sec:semiclassical_result}

Pulling together the path-integral representation of \S\,\ref{sec:pi_pbc} and the quadratic expansion of \S\,\ref{sec:semiclassical_expansion}, the trace of the evolution operator evaluates in the semiclassical limit to a sum over periodic classical solutions weighted by their classical actions and one-loop fluctuation determinants:
\begin{equation}
\boxed{
\Tr\left(e^{-iH\mathcal{T}}\right) \approx \sum_{\text{periodic orbits}} \mathcal{N}\, e^{i\mathcal{S}_{\rm cl}}\, \left|\detp \mathcal{O}^{(2)}\right|^{-1/2}
}
\label{eq:semiclassical_trace}
\end{equation}

The classical action $\mathcal{S}_{\rm cl} = \mathcal{S}[v]$ is evaluated on the periodic solution itself; it encodes the leading exponential weight $e^{i\mathcal{S}_{\rm cl}}$ that selects the dominant saddle in the large-$n$ limit.  The prefactor $\mathcal{N}$ collects normalization factors arising from the path-integral measure, the Jacobian of the collective-coordinate transformation (which promotes the continuous time-translation symmetry of the orbit to an explicit integration over the initial phase), and any additional symmetry factors.  The prime on the fluctuation determinant $\detp\mathcal{O}^{(2)}$ indicates that zero modes — directions in field space along which the action does not grow — have been removed from the product; each zero mode is treated separately as a collective coordinate and contributes a factor of $\mathcal{T}$ or a related geometric quantity rather than a Gaussian integral.  Finally, the sum ranges over all topologically distinct periodic classical solutions of the given period $\mathcal{T}$: in the $\phi^4$ theory on the cylinder there is a continuous family of such orbits, the Jacobi-$\mathrm{cn}$ solutions parameterized by the elliptic modulus $m\in[0,\tfrac{1}{2})$, and the $\mathcal{T}$-integral saddle selects the member of this family that satisfies the quantization condition $I(E)=2\pi n$~\cite{arXiv:1710.07336,arXiv:2102.12488}.


\section{Mode Decomposition on the Cylinder}

On the cylinder $\mathbb{R} \times S^{d-1}$, we can exploit the spatial translation and rotational symmetry.
Expand the fluctuation field in eigenmodes of the spatial Laplacian:
\begin{equation}
\eta(t, \Omega) = \sum_\ell \eta_\ell(t)\, Y_\ell(\Omega)
\label{eq:mode_expansion}
\end{equation}
where $Y_\ell(\Omega)$ are eigenfunctions of the Laplacian on $S^{d-1}$:
\begin{equation}
-\nabla^2_{S^{d-1}} Y_\ell = \Lambda(\ell)\, Y_\ell
\label{eq:sphere_eigenvalue}
\end{equation}

\textbf{Examples of eigenvalues:} For the sphere $S^{d-1}$ of unit radius, the eigenvalues are:
\begin{equation}
\Lambda(\ell) = \ell(\ell + d - 2), \quad \ell = 0, 1, 2, \ldots
\label{eq:sphere_eigenvalues}
\end{equation}

For concreteness:
\begin{itemize}
    \item In \textbf{3 spatial dimensions} ($S^2$): $\Lambda(\ell) = \ell(\ell+1)$, with $\ell=0,1,2,\ldots$ and degeneracy $n_\ell = 2\ell+1$.
    \item In \textbf{4 spatial dimensions} ($S^3$): $\Lambda(\ell) = \ell(\ell+2)$, with degeneracy {$n_\ell = (\ell+1)^2$}.
\end{itemize}

In generic dimensions, the degeneracy is:
\begin{equation}
n_\ell = \frac{(2\ell + d - 2)(\ell + d - 3)!}{\ell!(d-2)!}
\label{eq:degeneracy_formula}
\end{equation}

This counts the number of linearly independent spherical harmonics for a given $\ell$.

\subsection{Factorization of the Determinant}

For a \textbf{spatially homogeneous} saddle point $v = v(t)$ (constant on the sphere at each instant),
the fluctuation operator separates in mode space:
\begin{equation}
\mathcal{O}^{(2)} = -\partial_t^2 + V(t) + \nabla^2_{S^{d-1}}
\end{equation}

Acting on the mode expansion:
\begin{equation}
\mathcal{O}^{(2)}\eta = \sum_\ell \left[-\partial_t^2 + V(t) - \Lambda(\ell)\right]\eta_\ell(t)\, Y_\ell(\Omega)
\label{eq:mode_equation}
\end{equation}

Since the modes decouple, the determinant of the full operator factors as a product:
\begin{equation}
\boxed{
\detp \mathcal{O}^{(2)} = \prod_\ell \left(\detp \mathcal{O}^{(2)}_\ell\right)^{n_\ell}
}
\label{eq:det_factorization}
\end{equation}

where the one-dimensional operator for each mode is:
\begin{equation}
\mathcal{O}^{(2)}_\ell = -\partial_t^2 + V(t) - \Lambda(\ell)
\label{eq:1d_operator}
\end{equation}
  Each angular momentum mode $\ell$ contributes a factor $[\detp \mathcal{O}^{(2)}_\ell]^{n_\ell}$ to the
full determinant. The exponent $n_\ell$ is the degeneracy, accounting for the multiple angular momentum states with
the same eigenvalue. The prime indicates we must handle zero modes carefully (see below).

\section{Hill's Equation and Floquet Theory}
 \label{sec:hill_floquet}


The eigenvalue problem for fluctuation modes in one dimension leads to \textit{Hill's equation}:
\begin{equation}
\boxed{
\left[-\partial_t^2 + V(t)\right]\xi(t) = E\, \xi(t)
}
\label{eq:hills_equation}
\end{equation}
where $V(t)$ is periodic: $V(t + \mathcal{T}) = V(t)$.

\textbf{Physical example:} The simplest case is the \textit{Mathieu equation}:
\[
\xi''(t) + (a - 2q\cos(2t))\xi = 0
\]
This describes a parametrically driven harmonic oscillator, where the spring constant oscillates in time. Such systems
appear in particle accelerators and matter-wave optics. The parameter space $(a, q)$ has regions of stable (bounded)
solutions and unstable (exponentially growing) solutions—the famous Mathieu stability chart.

For each angular momentum mode $\ell$, we have the modified Hill equation:
\begin{equation}
\mathcal{O}^{(2)}_\ell \xi = 0 \quad \Leftrightarrow \quad
\left[-\partial_t^2 + V(t) - \Lambda(\ell)\right]\xi = 0
\label{eq:mode_hill}
\end{equation}

\subsection{Fundamental Matrix and Monodromy}

Consider the homogeneous equation \eqref{eq:mode_hill}. We define two linearly independent solutions
$\xi_{\ell,1}(t)$ and $\xi_{\ell,2}(t)$ by their initial conditions at $t = 0$:
\begin{align}
\xi_{\ell,1}(0) &= 1, \quad \xi_{\ell,1}'(0) = 0 \label{eq:xi1_initial}\\
\xi_{\ell,2}(0) &= 0, \quad \xi_{\ell,2}'(0) = 1 \label{eq:xi2_initial}
\end{align}

These are the two canonical solutions; any solution is a linear combination:
\begin{equation}
\xi_\ell(t) = c_1 \xi_{\ell,1}(t) + c_2 \xi_{\ell,2}(t)
\end{equation}

We can now define the fundamental matrix:
\begin{equation}
\Psi_\ell(t) = \begin{pmatrix} \xi_{\ell,1}(t) & \xi_{\ell,2}(t) \\ \xi_{\ell,1}'(t) & \xi_{\ell,2}'(t) \end{pmatrix}
\label{eq:fundamental_matrix}
\end{equation}

with the initial condition $\Psi_\ell(0) = \mathbb{I}$ (the identity matrix). The columns of $\Psi_\ell(t)$ are the
fundamental solutions; they form a basis of the solution space.

The \textit{monodromy matrix} describes how solutions transform under one complete period:
\begin{equation}
\mathcal{M}_\ell \equiv \Psi_\ell(\mathcal{T}) = \begin{pmatrix}
\xi_{\ell,1}(\mathcal{T}) & \xi_{\ell,2}(\mathcal{T}) \\
\xi_{\ell,1}'(\mathcal{T}) & \xi_{\ell,2}'(\mathcal{T})
\end{pmatrix}
\label{eq:monodromy_def}
\end{equation}

\textbf{Conservation of the Wronskian:} Since Hill's equation is a second-order ODE with no first-derivative term
(i.e., the equation is of the form $\xi'' + V(t)\xi = 0$), the Wronskian is conserved:
\begin{equation}
W[\xi_{\ell,1}, \xi_{\ell,2}] = \xi_{\ell,1} \xi_{\ell,2}' - \xi_{\ell,1}' \xi_{\ell,2} = \text{const}
\end{equation}

Evaluating at $t=0$:
\begin{equation}
W = (1)(1) - (0)(0) = 1
\end{equation}

This implies:
\begin{equation}
\boxed{\det \mathcal{M}_\ell = 1}
\label{eq:det_M_is_1}
\end{equation}

This is a crucial constraint: the monodromy matrix is always symplectic (it preserves the symplectic structure).

\subsection{Floquet Theory: Eigenvalues of the Monodromy Matrix}

\textbf{Floquet's Theorem:} For a linear ODE with periodic coefficients, there exist solutions of the form:
\begin{equation}
\xi_{\ell,\pm}(t) = e^{\pm i\nu_\ell t/\mathcal{T}}\, \chi_{\ell,\pm}(t)
\label{eq:floquet_solutions}
\end{equation}
where $\chi_{\ell,\pm}(t + \mathcal{T}) = \chi_{\ell,\pm}(t)$ are periodic functions with the same period. These are the
Floquet solutions. The exponents $\pm i\nu_\ell/\mathcal{T}$ are called the Floquet exponents.

\textbf{Derivation via eigenvalues:} Construct the characteristic polynomial of $\mathcal{M}_\ell$:
\begin{equation}
\det(\mathcal{M}_\ell - \lambda \mathbb{I}) = \lambda^2 - (\Tr \mathcal{M}_\ell)\lambda + \det \mathcal{M}_\ell = 0
\end{equation}

Using the constraint $\det \mathcal{M}_\ell = 1$:
\begin{equation}
\lambda^2 - (\Tr \mathcal{M}_\ell)\lambda + 1 = 0
\end{equation}

The solutions are:
\begin{equation}
\lambda_{\ell,\pm} = \frac{\Tr \mathcal{M}_\ell \pm \sqrt{(\Tr \mathcal{M}_\ell)^2 - 4}}{2}
\label{eq:eigenvalues}
\end{equation}

Note the important relation $\lambda_{\ell,+} \cdot \lambda_{\ell,-} = 1$.

\begin{figure}[htbp]
\centering
\includegraphics[width=0.88\textwidth]{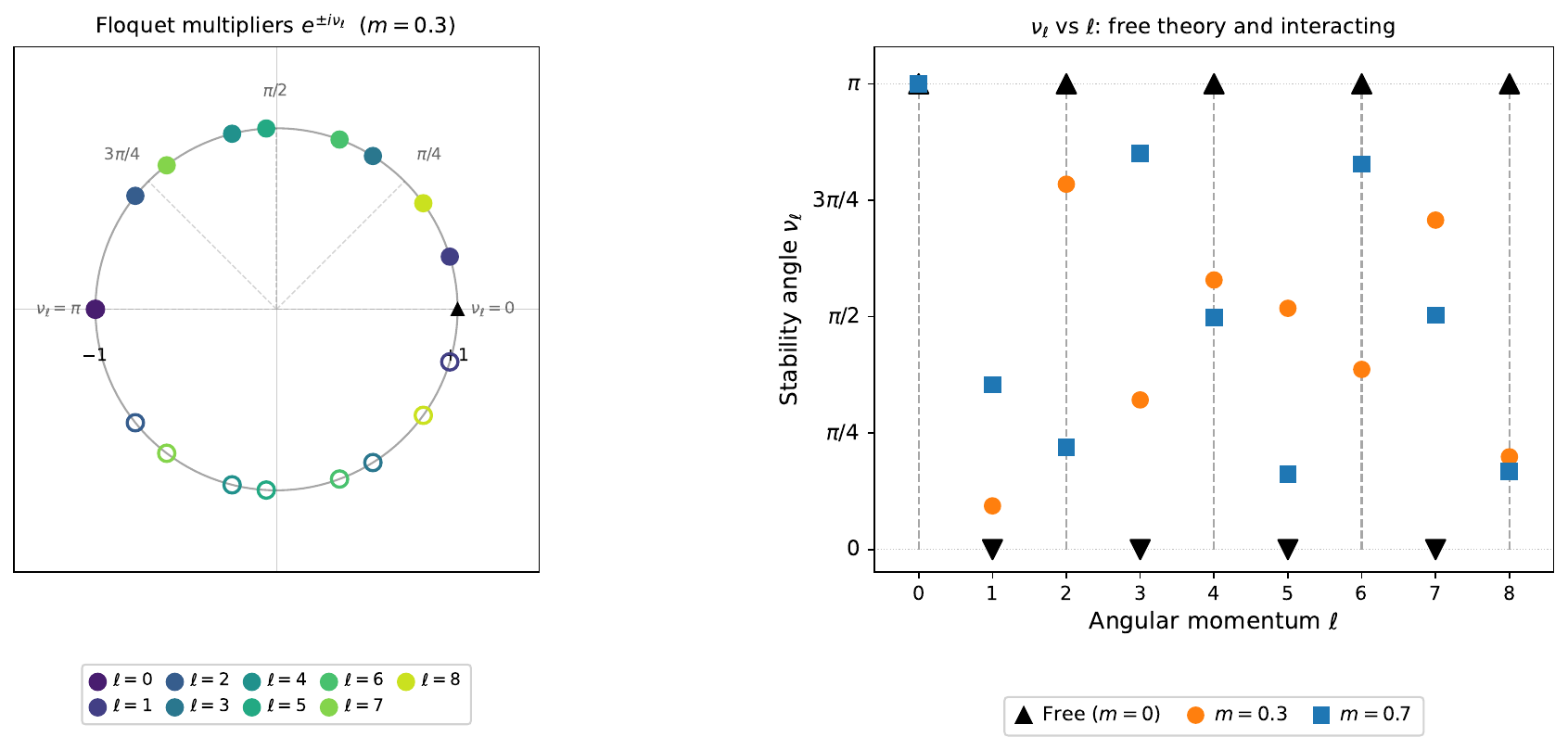}
\caption{%
\textbf{Floquet multipliers and stability angles.}
\textit{Left:} The monodromy eigenvalues $\lambda_{\ell,\pm}=e^{\pm i\nu_\ell}$ lie on the
unit circle for stable orbits.  Each angular-momentum mode $\ell$ acquires a distinct
phase $\nu_\ell\in(0,\pi)$ after one orbital period $\mathcal{T}$; the functional
determinant $\detp\mathcal{O}^{(2)}_\ell = 4\sin^2(\nu_\ell/2)$ measures the
``opening'' of that angle.  Filled dots are $e^{+i\nu_\ell}$; open dots are
$e^{-i\nu_\ell}$.
\textit{Right:} Stability angles $\nu_\ell\in[0,\pi]$ computed from the monodromy trace
for the transverse sector ($s=1$ Lam\'e equation, $d=4$).  Triangular markers on
the horizontal dashed lines at $\nu=0$ and $\nu=\pi$ show the free-theory values: by
eq.~\eqref{eq:nu_free}, $\cos\nu_\ell^{\text{free}}=(-1)^{\ell+1}$, so
$\nu_\ell^{\text{free}}=\pi$ for even $\ell$ and $0$ for odd $\ell$ — an
\emph{alternating comb} at the stability boundary ($|\Tr\mathcal{M}_\ell|=2$).
Solid curves show the interacting theory at $m=0.25,0.50,0.75,0.95$
($\kappa=\lambda n$ increases with $m$ via the Lam\'e modulus): interactions pull each
mode off the boundary into the interior of $(0,\pi)$.
Note that the $\ell=0$ mode ($A_0=1$, the lower edge of the upper Lam\'e band)
remains pinned at $\nu_0=\pi$ for all $m$, so it contributes nothing to
$C_1=\sum_\ell n_\ell(\nu_{\kappa,\ell}-\nu_{0,\ell})$; the non-trivial correction
comes entirely from $\ell\geq 1$.}
\label{fig:floquet}
\end{figure}

These eigenvalues are the \textbf{Floquet multipliers}:
\begin{equation}
\lambda_{\ell,\pm} = e^{\pm i\nu_\ell}
\label{eq:floquet_multipliers}
\end{equation}
where $\nu_\ell$ is the \textbf{stability angle} (or Floquet exponent) for mode $\ell$.  The geometric
picture — Floquet multipliers $e^{\pm i\nu_\ell}$ sitting on the unit circle, one pair per angular-momentum
mode — is illustrated in Figure~\ref{fig:floquet}.

\paragraph{Free-limit check: $\nu_\ell$ for a constant potential.}
As a consistency check and a reference point for the interacting theory, we compute $\nu_\ell$ exactly in the
\textbf{free limit} $m\to 0$ (equivalently $\kappa=\lambda n\to 0$).  In this limit the Lamé
potential $\kappa(\kappa+1)m\,\mathrm{sn}^2(z|m)\to 0$, the rescaled Hill equation (cf.\
\eqref{eq:Lame_operator}) reduces to
\begin{equation}
  \xi''(z) + A_\ell\,\xi(z) = 0,
  \qquad
  A_\ell \;=\; \left(1 + \frac{2\ell}{d-2}\right)^{\!2},
\label{eq:hills_free}
\end{equation}
and the half-period shrinks to $2K(0)=\pi$.  Since $A_\ell>0$ the solutions are purely oscillatory.
The two canonical solutions with initial data $(\xi_1(0),\xi_1'(0))=(1,0)$ and
$(\xi_2(0),\xi_2'(0))=(0,1)$ are
\begin{equation}
  \xi_1(z) = \cos\!\bigl(\sqrt{A_\ell}\,z\bigr),
  \qquad
  \xi_2(z) = \frac{\sin\!\bigl(\sqrt{A_\ell}\,z\bigr)}{\sqrt{A_\ell}}.
\label{eq:free_solutions}
\end{equation}
Evaluating the fundamental matrix at $z = \pi$ gives the monodromy:
\begin{equation}
  \mathcal{M}_\ell^{\text{free}}
  = \begin{pmatrix}
      \cos\!\bigl(\pi\sqrt{A_\ell}\bigr) & \dfrac{\sin\!\bigl(\pi\sqrt{A_\ell}\bigr)}{\sqrt{A_\ell}} \\[8pt]
      -\sqrt{A_\ell}\sin\!\bigl(\pi\sqrt{A_\ell}\bigr) & \cos\!\bigl(\pi\sqrt{A_\ell}\bigr)
    \end{pmatrix},
\label{eq:monodromy_free}
\end{equation}
whose determinant equals $\cos^2(\pi\sqrt{A_\ell})+\sin^2(\pi\sqrt{A_\ell})=1$, confirming the
symplectic constraint \eqref{eq:det_M_is_1}.  The trace yields
\begin{equation}
  \frac{\Tr\mathcal{M}_\ell^{\text{free}}}{2} = \cos\!\bigl(\pi\sqrt{A_\ell}\bigr) = \frac{\lambda_{\ell,+} + \lambda_{\ell,-}}{2} = \cos \nu_\ell ,
\end{equation}
and therefore the free-limit stability angle is
\begin{equation}
\boxed{
  \nu_\ell^{\text{free}} = \pi\sqrt{A_\ell}
  = \pi\!\left(1+\frac{2\ell}{d-2}\right).
}
\label{eq:nu_free}
\end{equation}
In $d=4$ this simplifies to $\nu_\ell^{\text{free}}=(\ell+1)\pi$: each successive angular-momentum
mode accumulates one additional half-winding per period.  {This boxed value is the
\emph{accumulated} phase --- the quantity that enters $C_1$ through the free-theory subtraction
$\nu_{0,\ell}$; the trace inversion $\cos\nu_\ell=\tfrac12\Tr\mathcal{M}_\ell$ and
Figure~\ref{fig:floquet} instead display its reduction to the principal branch $[0,\pi]$ (mod $2\pi$):
$\pi$ for even $\ell$ and $0$ for odd $\ell$.}  The result is consistent with the general
formula $\nu_{0,\ell}=\mu\sqrt{A_\ell}\,\mathcal{T}$ (quoted in Section~\ref{sec:C1} for the
free-theory subtraction in $C_1$), since in the free limit $\mathcal{T}^{\text{free}}=\pi/\mu$.
Interactions ($m>0$) deform these angles away from the integer multiples of $\pi$, and it is
precisely this deformation that generates the non-trivial one-loop correction $C_1$ to $\Delta_n$.

\subsection{The Stability Angle and Physical Interpretation}

\textbf{Derivation from the Floquet multipliers \eqref{eq:floquet_multipliers}.}
Since $\mathcal{M}_\ell$ is a $2\times 2$ matrix, its trace equals the sum of its eigenvalues.
Using \eqref{eq:floquet_multipliers}, these are $\lambda_{\ell,+}=e^{+i\nu_\ell}$ and
$\lambda_{\ell,-}=e^{-i\nu_\ell}$, so
\begin{equation}
\Tr \mathcal{M}_\ell \;=\; \lambda_{\ell,+} + \lambda_{\ell,-}
\;=\; e^{+i\nu_\ell} + e^{-i\nu_\ell} \;=\; 2\cos\nu_\ell\,.
\label{eq:trace_from_multipliers}
\end{equation}
For a stable orbit the right-hand side is real and $|\Tr\mathcal{M}_\ell|<2$, so the equation
$\Tr\mathcal{M}_\ell = 2\cos\nu_\ell$ can be inverted for $\nu_\ell\in(0,\pi)$.
Substituting the explicit entries of the monodromy matrix \eqref{eq:monodromy_def} then gives:
\begin{equation}
\boxed{
\cos\nu_\ell = \frac{\Tr \mathcal{M}_\ell}{2} = \frac{\xi_{\ell,1}(\mathcal{T}) + \xi_{\ell,2}'(\mathcal{T})}{2}
}
\label{eq:stability_angle_def}
\end{equation}

The Floquet multipliers are $e^{\pm i\nu_\ell}$, and the Floquet solutions satisfy:
\begin{equation}
\xi_{\ell,\pm}(t + \mathcal{T}) = e^{\pm i\nu_\ell}\, \xi_{\ell,\pm}(t)
\end{equation}

\textbf{Physical interpretation:} The stability angle $\nu_\ell$ measures the phase rotation acquired by
fluctuations in mode $\ell$ after one orbital period $\mathcal{T}$. If $\nu_\ell = 0$, the perturbations return
to their original phase each period (neutral stability). If $0 < \nu_\ell < \pi$, the perturbations precess;
if $\nu_\ell = \pi$, the perturbations are ``out of phase'' but still oscillatory. If $\nu_\ell$ is complex,
the solutions are exponentially growing (instability).

This angle is the analog of the classical action variable for the transverse oscillations. In a sense, it
quantifies how ``closed'' the classical periodic orbit is in the full phase space.

\section{Gel'fand--Yaglom Theorem}\label{sec:gelfand_yaglom}

Computing functional determinants from first principles (finding all eigenvalues explicitly) is usually impossible.
The Gel'fand--Yaglom theorem provides an elegant shortcut that expresses the determinant in terms of solutions
to the associated differential equation. This is one of the most powerful tools in semiclassical physics.

The Gel'fand--Yaglom theorem \cite{Dunne:2007rt} provides a powerful method to compute functional determinants of
one-dimensional operators without explicitly finding all eigenvalues. It works beautifully for periodic boundary conditions,
which are precisely what we encounter in the path integral with periodic classical solutions.

\subsection{Dirichlet Boundary Conditions: Simple Case}

Consider the Sturm--Liouville operator $\mathcal{O} = -\partial_t^2 + V(t)$ on the
interval $[0,\mathcal{T}]$ with \textbf{Dirichlet boundary conditions}
$\xi(0)=\xi(\mathcal{T})=0$.  Introduce the $\lambda$-deformed initial-value problem
\begin{equation}
  \bigl[-\partial_t^2 + V(t) - \lambda\bigr]\,\xi_2(t;\lambda) = 0,
  \qquad \xi_2(0;\lambda)=0,\quad \partial_t\xi_2(0;\lambda)=1,
  \label{eq:xi2_def}
\end{equation}
and the reference (free) operator $\mathcal{O}_0=-\partial_t^2$, whose solution
$\xi_2^{(0)}(t;\lambda)$ satisfies the same initial conditions with $V=0$.
The Gel'fand--Yaglom theorem \cite{Dunne:2007rt} states:
\begin{equation}
  \frac{\det\mathcal{O}}{\det\mathcal{O}_0}
  \;=\; \frac{\xi_2(\mathcal{T};\,0)}{\xi_2^{(0)}(\mathcal{T};\,0)}\,.
  \label{eq:gf_dirichlet_result}
\end{equation}
Since $\xi_2^{(0)}(t;0)=t$, the reference value is simply
$\xi_2^{(0)}(\mathcal{T};0)=\mathcal{T}$, and one often writes the result as
$\det\mathcal{O}\propto\xi_2(\mathcal{T};0)$, absorbing the (universal, divergent)
free determinant into the path-integral normalisation.

\vskip  1em
To derive the formula above,  define $F(\lambda)\equiv\xi_2(\mathcal{T};\lambda)$.
Any Dirichlet eigenfunction satisfying $\psi(0)=0$ must be proportional to
$\xi_2(t;\lambda)$ (since that is the unique solution with $\xi_2(0;\lambda)=0$),
so the right-endpoint condition $\psi(\mathcal{T})=0$ becomes simply
\[
  F(\lambda_n) = \xi_2(\mathcal{T};\lambda_n) = 0.
\]
The Dirichlet eigenvalues $\{\lambda_n\}$ are exactly the zeros of $F(\lambda)$.
By the Picard iteration, ODE solutions depend analytically on parameters, so
$F(\lambda)$ is an \emph{entire function} of $\lambda$.

We now argue that  the Dirichlet eigenvalues grow as $\lambda_n\sim n^2$.  For the operator $\mathcal{O}=-\partial_t^2+V(t)$ on $[0,\mathcal{T}]$ with Dirichlet boundary
conditions, the large-$n$ asymptotics of the eigenvalues are governed by the kinetic
term alone.  Intuitively, the $n$-th eigenfunction oscillates $n$ times on $[0,\mathcal{T}]$,
so its second derivative contributes $\sim (n\pi/\mathcal{T})^2$.  For large $n$ the
potential $V(t)$ is negligible compared with this kinetic energy, and one obtains:
\[
  \lambda_n \;=\; \left(\frac{n\pi}{\mathcal{T}}\right)^2 + O(1), \qquad n\to\infty.
\]
This is confirmed explicitly by the constant-potential example of \S\ref{sec:gelfand_yaglom}:
$\lambda_n = \Omega^2+(n\pi/\mathcal{T})^2$, where $\Omega^2$ is exactly the subleading
$O(1)$ shift.  The $n^2$ growth implies
$\sum_{n=1}^\infty 1/|\lambda_n|\sim\sum_{n=1}^\infty 1/n^2 < \infty$. 

The convergence condition allow us to employ the \textbf{ Hadamard factorization of $F$ and $F_0$ separately.}  
This means $F(\lambda)$ is an entire function of \emph{order less than one}, for
which the Hadamard factorization theorem takes its simplest form where no exponential
convergence factors are needed:
\[
  F(\lambda) = F(0)\prod_{n=1}^\infty\!\left(1-\frac{\lambda}{\lambda_n}\right),
  \qquad
  F_0(\lambda) = F_0(0)\prod_{n=1}^\infty\!\left(1-\frac{\lambda}{\lambda_n^{(0)}}\right).
\]
Rewriting each factor as $(1-\lambda/\lambda_n)=(\lambda_n-\lambda)/\lambda_n$ and
collecting the denominators:
\[
  F(\lambda)
  = \frac{F(0)}{\displaystyle\prod_n\lambda_n}
    \prod_{n=1}^\infty(\lambda_n-\lambda)
  = \frac{F(0)}{\det\mathcal{O}}
    \prod_{n=1}^\infty(\lambda_n-\lambda),
\]
and likewise $F_0(\lambda)=F_0(0)\,\prod_n(\lambda_n^{(0)}-\lambda)/\det\mathcal{O}_0$.
Dividing gives
\begin{equation}
  \frac{F(\lambda)}{F_0(\lambda)}
  = \frac{F(0)}{F_0(0)}\cdot\frac{\det\mathcal{O}_0}{\det\mathcal{O}}
    \cdot\prod_{n=1}^\infty\frac{\lambda_n-\lambda}{\lambda_n^{(0)}-\lambda}\,.
  \label{eq:gfy_ratio}
\end{equation}

 For $|\lambda|\to\infty$ the potential $V(t)$ is negligible and the ODE
$(-\partial_t^2-\lambda)\xi\approx 0$ governs both cases, giving the same
leading behaviour
\[
  \xi_2(\mathcal{T};\lambda)\;\sim\;\frac{\sin(\sqrt\lambda\,\mathcal{T})}{\sqrt\lambda}
  \;\sim\;\xi_2^{(0)}(\mathcal{T};\lambda),
  \qquad |\lambda|\to\infty.
\]
Hence $F(\lambda)/F_0(\lambda)\to 1$ in this limit.
The infinite product $\prod_n(\lambda_n-\lambda)/(\lambda_n^{(0)}-\lambda)\to 1$
as well (each factor tends to 1).  Substituting into \eqref{eq:gfy_ratio}:
\[
  1 = \frac{F(0)}{F_0(0)}\cdot\frac{\det\mathcal{O}_0}{\det\mathcal{O}}\cdot 1.
\]
Rearranging immediately yields \eqref{eq:gf_dirichlet_result}:
\[
  \frac{\det\mathcal{O}}{\det\mathcal{O}_0}
  = \frac{F(0)}{F_0(0)}
  = \frac{\xi_2(\mathcal{T};\,0)}{\xi_2^{(0)}(\mathcal{T};\,0)}\,.\qquad\square
\]

\subsection{An Explicit Example: Constant Potential}

To verify the theorem concretely, take the constant potential $V(t)=\Omega^2$.
The operator is $\mathcal{O}=-\partial_t^2+\Omega^2$ and the solution of
\eqref{eq:xi2_def} at $\lambda=0$ is
\[
  \xi_2(t;0) = \frac{\sinh(\Omega t)}{\Omega},
  \qquad \xi_2^{(0)}(t;0) = t,
\]
so the Gel'fand--Yaglom formula \eqref{eq:gf_dirichlet_result} predicts
\begin{equation}
  \frac{\det(-\partial_t^2+\Omega^2)}{\det(-\partial_t^2)}
  = \frac{\sinh(\Omega\mathcal{T})}{\Omega\mathcal{T}}\,.
  \label{eq:gfy_const_check}
\end{equation}

To independently check the results, recall that for the constant-potential operator the Dirichlet eigenvalues are explicit:
\[
  \lambda_n = \Omega^2 + \left(\frac{n\pi}{\mathcal{T}}\right)^2, \qquad n=1,2,\ldots,
  \qquad
  \lambda_n^{(0)} = \left(\frac{n\pi}{\mathcal{T}}\right)^2.
\]
Hence
\[
  \frac{\det\mathcal{O}}{\det\mathcal{O}_0}
  = \prod_{n=1}^\infty\frac{\lambda_n}{\lambda_n^{(0)}}
  = \prod_{n=1}^\infty\left(1+\frac{\Omega^2\mathcal{T}^2}{n^2\pi^2}\right)
  = \frac{\sinh(\Omega\mathcal{T})}{\Omega\mathcal{T}}\,,
\]
where the last step uses Euler's infinite-product formula for the hyperbolic sine,
$\sinh(x)/x=\prod_{n=1}^\infty(1+x^2/n^2\pi^2)$.
This agrees with \eqref{eq:gfy_const_check} exactly.  The free-particle limit $\Omega\to 0$ gives ratio $\to 1$ (as it must, since
$\mathcal{O}\to\mathcal{O}_0$), and $\xi_2(\mathcal{T};0)\to\mathcal{T}=\xi_2^{(0)}(\mathcal{T};0)$
confirms the normalisation.

\subsection{Periodic Boundary Conditions}

For \textbf{periodic boundary conditions}, we need solutions satisfying:
\begin{equation}
\xi_\ell(t + \mathcal{T}) = \xi_\ell(t), \quad \xi_\ell'(t + \mathcal{T}) = \xi_\ell'(t)
\label{eq:periodic_bc}
\end{equation}

The condition for a nontrivial periodic solution is that the monodromy matrix must have eigenvalue 1:
\begin{equation}
\mathcal{M}_\ell \begin{pmatrix} \xi_\ell(0) \\ \xi_\ell'(0) \end{pmatrix} = \begin{pmatrix} \xi_\ell(0) \\ \xi_\ell'(0) \end{pmatrix}
\label{eq:monodromy_eigenvalue_1}
\end{equation}

This requires:
\begin{equation}
\det(\mathbb{I} - \mathcal{M}_\ell) = 0
\label{eq:periodic_condition}
\end{equation}

The generalization of  \textbf{Gel'fand--Yaglom for periodic boundary conditions}  gives:
\begin{equation}
\boxed{
\detp \mathcal{O}^{(2)}_\ell \propto \det(\mathbb{I} - \mathcal{M}_\ell)
}
\label{eq:det_periodic}
\end{equation}

The prime indicates that we remove the zero mode (eigenvector with zero eigenvalue) before taking the determinant.
For a periodic potential with a closed classical orbit, there is generically one true zero mode (from time translation),
which we exclude.
 
Equations~\eqref{eq:periodic_condition} and~\eqref{eq:det_periodic} can look contradictory at first: the first says
$\det(\mathbb{I}-\mathcal{M}_\ell)=0$, while the second says $\detp\mathcal{O}^{(2)}_\ell$ is \emph{proportional}
to that same vanishing quantity, yet the calculation below yields $4\sin^2(\nu_\ell/2)\neq 0$.  The resolution is
that the two equations apply to \emph{different mode sectors}:

\begin{itemize}
  \item \textbf{The zero-mode sector (longitudinal, $\ell=0$).}  Time-translation symmetry of the periodic orbit
    guarantees that $\dot\phi_{\rm cl}(\tau)$ is a periodic solution of the linearised equation
    $\mathcal{O}^{(2)}_0\,\xi=0$.  Hence $\lambda=0$ is an eigenvalue of $\mathcal{O}^{(2)}_0$, the monodromy
    matrix $\mathcal{M}_0$ has eigenvalue $1$, and $\det(\mathbb{I}-\mathcal{M}_0)=0$.  This is precisely what
    equation~\eqref{eq:periodic_condition} states: a zero mode \emph{exists}.  For this sector the full
    (unprimed) determinant vanishes; the primed determinant $\detp\mathcal{O}^{(2)}_0$ removes that zero eigenvalue
    and is treated separately via the Faddeev--Popov collective-coordinate method, trading the zero-mode Gaussian
    integral for an integral over the orbit's initial phase and producing a factor of $\mathcal{T}$.

  \item \textbf{The transverse sectors ($\ell\geq1$, and $O(N)$ transverse modes).}  For these sectors the
    orbit has no zero mode.  The stability angle $\nu_\ell\neq 0$ generically, so
    $\det(\mathbb{I}-\mathcal{M}_\ell)=4\sin^2(\nu_\ell/2)\neq 0$.
    Equation~\eqref{eq:det_periodic} applies directly: the prime is trivial (no zero eigenvalue to
    remove), and the functional determinant is finite and non-zero.
\end{itemize}

In summary: \eqref{eq:periodic_condition} \emph{diagnoses} the zero mode in the longitudinal sector;
\eqref{eq:det_periodic} evaluated to $4\sin^2(\nu_\ell/2)$ gives the physical result for all
non-zero-mode sectors.  There is no contradiction.

\subsection{Characteristic Function and Primed Determinant}

The periodic Gel'fand--Yaglom theorem is most cleanly stated in terms of the
$\lambda$-dependent \emph{characteristic function}
\begin{equation}
  F_\ell(\lambda) \;\equiv\; \det\bigl(\mathbb{I}-\mathcal{M}_\ell(\lambda)\bigr),
  \label{eq:hill_characteristic_function}
\end{equation}
where $\mathcal{M}_\ell(\lambda)$ is the monodromy matrix of the $\lambda$-shifted
operator $\mathcal{O}^{(2)}_\ell - \lambda$.  Up to a normalization independent of
the background, the \emph{unprimed} functional determinant is proportional to its
value at $\lambda=0$:
\begin{equation}
  \det\mathcal{O}^{(2)}_\ell \;\propto\; F_\ell(0)
  \;=\; \det\bigl(\mathbb{I}-\mathcal{M}_\ell(0)\bigr).
  \label{eq:det_periodic_unprimed}
\end{equation}
In a sector with no zero mode this formula can be used directly.  In the
longitudinal ($\ell=0$) sector, however, $\lambda=0$ is a periodic eigenvalue, so
$F_0(0)=0$ and the unprimed determinant vanishes.  The \emph{primed} determinant
is obtained by extracting the first nonvanishing coefficient in the small-$\lambda$
expansion of $F_\ell(\lambda)$:
\begin{equation}
  F_\ell(\lambda) = \lambda\,F_\ell'(0) + O(\lambda^2),
  \qquad
  \detp\mathcal{O}^{(2)}_\ell \;\propto\; F_\ell'(0).
  \label{eq:primed_det_from_slope}
\end{equation}
This makes precise what ``removing the zero eigenvalue'' means analytically:
the primed determinant is the slope of the characteristic function at $\lambda=0$,
\emph{not} the value $F_\ell(0)=0$.  The Faddeev--Popov collective-coordinate
treatment of the zero-mode Gaussian integral in the path integral produces a
factor of $\mathcal{T}$ that absorbs this slope into the physical amplitude.

\paragraph{Identification of the zero mode.}
The zero mode in the longitudinal sector arises from the time-translation
symmetry of the periodic orbit $v(t)$.  Shifting the orbit by an infinitesimal
phase $t_0$,
\[
  v(t+t_0) = v(t) + t_0\,\dot{v}(t) + O(t_0^2),
\]
shows that the infinitesimal fluctuation generated by time translation is
\begin{equation}
  \eta_{\ell=0}(t) \;\propto\; \dot{v}(t).
  \label{eq:translation_zero_mode_vdot}
\end{equation}
Since the orbit is periodic, $\dot{v}(t+\mathcal{T})=\dot{v}(t)$, so
$\eta_{\ell=0}$ satisfies periodic boundary conditions and is a bona fide
periodic zero mode of $\mathcal{O}^{(2)}_0$.  It is this mode that forces
$\det(\mathbb{I}-\mathcal{M}_0)=0$ and that must be treated by the
collective-coordinate method.

\section{Explicit Evaluation: From Monodromy to Determinant}

We now compute $\det(\mathbb{I} - \mathcal{M}_\ell)$ explicitly. Write the monodromy matrix as:
\begin{equation}
\mathcal{M}_\ell = \begin{pmatrix} A & B \\ C & D \end{pmatrix}, \quad \text{where} \quad \det \mathcal{M}_\ell = 1, \quad A+D = \Tr \mathcal{M}_\ell
\end{equation}

Then:
\begin{equation}
\mathbb{I} - \mathcal{M}_\ell = \begin{pmatrix} 1-A & -B \\ -C & 1-D \end{pmatrix}
\end{equation}

The determinant is:
\begin{align}
\det(\mathbb{I} - \mathcal{M}_\ell)  
&= 1 - (A+D) + (AD - BC)  = 1 - \Tr \mathcal{M}_\ell + \det \mathcal{M}_\ell
\end{align}

Using $\det \mathcal{M}_\ell = 1$:
\begin{equation}
\det(\mathbb{I} - \mathcal{M}_\ell) = 2 - \Tr \mathcal{M}_\ell
\label{eq:det_I_M_intermediate}
\end{equation}
Now use the definition of the stability angle, $\cos \nu_\ell = \frac{\Tr \mathcal{M}_\ell}{2}$:
\begin{align}
\det(\mathbb{I} - \mathcal{M}_\ell) &= 2 - 2\cos\nu_\ell = 4\sin^2(\nu_\ell/2)
\label{eq:det_explicit_full}
\end{align}
Thus:
\begin{equation}
\boxed{
\detp \mathcal{O}^{(2)}_\ell = 4\sin^2\left(\frac{\nu_\ell}{2}\right)
}
\label{eq:det_sin_squared}
\end{equation}
where $\nu_\ell$ is the stability angle for mode $\ell$.

  This  formula connects the functional determinant, which measures quantum fluctuations around
the periodic orbit, to the monodromy matrix eigenvalues. The latter describes the classical stability of the orbit. It is
a nice illustration of the deep connection between classical and quantum mechanics in the semiclassical limit.

\section{ Fluctuation Determinant}

\subsection{Product Over All Modes}

The full determinant is the product over all angular momentum multiplets and their spherical-harmonic degeneracies:
\begin{equation}
\detp \mathcal{O}^{(2)} = \prod_{\ell=0}^{\infty} \left[4\sin^2\left(\frac{\nu_\ell}{2}\right)\right]^{n_\ell}
\label{eq:full_det_product}
\end{equation}
where $n_\ell$ is the number of real spherical harmonics of angular momentum $\ell$ on $S^{d-1}$
(see \eqref{eq:degeneracy_formula}), and the factor$  4\sin^2\!\left(\tfrac{\nu_\ell}{2}\right) $ s the just computed Gel'fand--Yaglom determinant of the one-dimensional radial operator $\mathcal{O}^{(2)}_\ell$. It already encodes {both} monodromy eigenvalues $e^{\pm i\nu_\ell}$.

The restriction to $\nu_\ell>0$ simply removes zero modes (broken-symmetry directions for which
$\nu_\ell=0$, treated by collective coordinates); it does not introduce any additional multiplicity.
Hence the primed determinant is
\begin{equation}
\detp \mathcal{O}^{(2)} = \prod_{\nu_\ell>0} \left[4\sin^2\left(\frac{\nu_\ell}{2}\right)\right]^{n_\ell},
\label{eq:det_positive_nu}
\end{equation}
where the restriction $\nu_\ell>0$ excludes zero modes ($\nu_\ell=0$, treated by collective coordinates),
and the exponent $n_\ell$ is the same spherical-harmonic degeneracy as in \eqref{eq:full_det_product}.

For the semiclassical path integral, we need $(\detp \mathcal{O}^{(2)})^{-1/2}$. 

 For $0< \nu  < 2 \pi$:
\begin{equation}
\left[4\sin^2\left(\frac{\nu}{2}\right)\right]^{-1/2} =  i\, e^{-i\nu/2}(1 - e^{-i\nu})^{-1}
\label{eq:det_inverse_sqrt}
\end{equation}

Thus our final form for the determinant factor is:
\begin{equation}
\boxed{
\prod_{\nu_\ell>0} \left[4\sin^2\left(\frac{\nu_\ell}{2}\right)\right]^{-n_\ell/2} \sim \prod_{\nu_\ell>0} e^{-i\nu_\ell n_\ell/2} \prod_{\nu_\ell>0}(1 - e^{-i\nu_\ell})^{-n_\ell}
}
\label{eq:full_det_inverse_sqrt}
\end{equation}

\section{The Gutzwiller Trace Formula}

\subsection{Combining All Pieces}

Substituting the inverse determinant \eqref{eq:full_det_inverse_sqrt} into the semiclassical trace \eqref{eq:semiclassical_trace},
we obtain the Gutzwiller trace formula \cite{Gutzwiller:1971fy}, one of the central results connecting classical periodic
orbits to the quantum spectrum:
\begin{equation}
\Tr(e^{-iH\mathcal{T}}) = \sum_{\text{periodic orbits}} \exp\left[i\left(\mathcal{S}_{\rm cl} - \frac{1}{2}\sum_{\nu_\ell>0} n_\ell \nu_\ell\right)\right] \Delta_1 \Delta_2
\label{eq:gutzwiller_formula}
\end{equation}

where:

\noindent
\textbf{Zero-mode contribution} $\Delta_1$: This factor arises from collective coordinate integration. For a closed periodic
orbit, there is always a zero mode corresponding to time translation (shifting all time labels by a constant amount).
When we integrate over the period $\mathcal{T}$, this zero mode produces the factor $\Delta_1$. In practice, $\Delta_1$
is determined by carefully handling the measure of the path integral and the collective coordinates (time, center of mass,
etc.), see \cite{Dashen:1974ci}. It is often absorbed into a normalization.

\noindent
\textbf{Excitation contribution} $\Delta_2$: This factor arises from the non-zero modes of the fluctuation operator.
It represents quantum excitations built on the classical periodic orbit.

\subsection{Structure of $\Delta_2$: Excitation Expansion}

The excitation contribution is defined as:
\begin{equation}
\Delta_2 = \prod_{\nu_\ell>0} (1 - e^{-i\nu_\ell})^{-n_\ell}
\label{eq:delta2_def}
\end{equation}
where the product is over angular momentum modes with positive stability angles.

 For one multiplet, define
$x_\ell=e^{-i\nu_\ell}$.  Then
\[
(1-e^{-i\nu_\ell})^{-n_\ell}
=
(1-x_\ell)^{-n_\ell}
=
\prod_{a=1}^{n_\ell}
\left(\sum_{m_a=0}^{\infty}x_\ell^{m_a}\right).
\]
Expanding the product gives
\[
(1-x_\ell)^{-n_\ell}
=
\sum_{m_1,\ldots,m_{n_\ell}\geq0}
x_\ell^{m_1+\cdots+m_{n_\ell}} .
\]
Grouping terms with fixed total occupation
\[
q_\ell=m_1+\cdots+m_{n_\ell},
\]
the coefficient of $x_\ell^{q_\ell}$ is the number of
non-negative integer solutions of
\[
m_1+\cdots+m_{n_\ell}=q_\ell .
\]
This number is
\[
\binom{q_\ell+n_\ell-1}{n_\ell-1}.
\]
Hence
\begin{equation}
(1-e^{-i\nu_\ell})^{-n_\ell}
=
\sum_{q_\ell=0}^{\infty}
\binom{q_\ell+n_\ell-1}{n_\ell-1}
e^{-iq_\ell\nu_\ell}.
\label{eq:delta2_expansion_single}
\end{equation}
The binomial coefficient is therefore purely combinatorial: it depends only on
$q_\ell$ and $n_\ell$, while $\nu_\ell$ enters only through the phase.

Taking the product over all positive stability angles,
\[
\Delta_2
=
\prod_{\nu_\ell>0}
(1-e^{-i\nu_\ell})^{-n_\ell}
=
\prod_{\nu_\ell>0}
\left[
\sum_{q_\ell=0}^{\infty}
\binom{q_\ell+n_\ell-1}{n_\ell-1}
e^{-iq_\ell\nu_\ell}
\right].
\]
Expanding the product of sums amounts to choosing one
$q_\ell\geq0$ for each multiplet:
\[
\Delta_2
=
\sum_{\{q_\ell\geq0\}}
\prod_{\nu_\ell>0}
\left[
\binom{q_\ell+n_\ell-1}{n_\ell-1}
e^{-iq_\ell\nu_\ell}
\right].
\]
Separating the phases,
\[
\prod_{\nu_\ell>0}e^{-iq_\ell\nu_\ell}
=
\exp\!\left[
-i\sum_{\nu_\ell>0}q_\ell\nu_\ell
\right],
\]
we obtain
\begin{equation}
\Delta_2
=
\sum_{\{q_\ell\geq0\}}
\left[
\prod_{\nu_\ell>0}
\binom{q_\ell+n_\ell-1}{n_\ell-1}
\right]
\exp\!\left[
-i\sum_{\nu_\ell>0}q_\ell\nu_\ell
\right] \ .
\label{eq:delta2_expansion}
\end{equation}
Here $q_\ell \geq 0$ is the \emph{total occupation number} of multiplet $\ell$, summed over all $n_\ell$ degenerate modes.  The degeneracy factor $\binom{q_\ell+n_\ell-1}{n_\ell-1}$ counts distinct quantum states of the $n_\ell$ oscillators with the same total occupation $q_\ell$; it equals $1$ when $n_\ell=1$ and grows as $q_\ell^{n_\ell-1}/(n_\ell-1)!$ for large $q_\ell$. 

\subsection{Physical Interpretation: Excitations on Periodic Orbits}

The factor $\Delta_2 = \prod_{\nu_\ell>0}(1-e^{-i\nu_\ell})^{-n_\ell}$ is the generating function for the entire
tower of quantum states built on the classical periodic orbit. Each term in the expansion
\eqref{eq:delta2_expansion} is labelled by a set of non-negative integers $\{q_\ell\}$, one for each
angular-momentum channel with $\nu_\ell>0$. Via the state--operator correspondence, every such term
corresponds to a distinct composite operator in the CFT.

\paragraph{The $n_\ell$ independent oscillators and the meaning of $q_\ell$.}
For each angular momentum $\ell$, there are $n_\ell = (2\ell+d-2)\Gamma(\ell+d-2)/[\Gamma(d-1)\Gamma(\ell+1)]$
(see \eqref{eq:degeneracy_formula})
independent real spherical harmonics on $S^{d-1}$, each giving rise to an independent quantum harmonic
oscillator with Floquet quasi-frequency $\omega_\ell = \nu_\ell/\mathcal{T}$.  The factor
$(1-e^{-i\nu_\ell})^{-n_\ell}$ in $\Delta_2$ is the product of $n_\ell$ independent geometric series,
one per harmonic-oscillator mode.  The integer $q_\ell \geq 0$ labelling each term of the expansion
\eqref{eq:delta2_expansion} is the \emph{total occupation number} of multiplet $\ell$: the sum of the
individual occupation numbers across all $n_\ell$ degenerate modes.  The factor $\binom{q_\ell+n_\ell-1}{n_\ell-1}$
counts the number of distinct quantum states of the $n_\ell$ oscillators that share the same total
occupation $q_\ell$.

\paragraph{Role of the quantum numbers $n$ and $\{q_\ell\}$.}
Two sets of integers label a state in the semiclassical spectrum, and they play entirely different roles:

\begin{itemize}
\item $n\in\mathbb{Z}_{>0}$ is the \emph{Bohr--Sommerfeld integer} from~\eqref{eq:full_quantization_boxed}.  It selects the classical periodic orbit and sets the leading scale $nC_0$ of the energy.  In the $O(N)$ $\phi^4$ application (Chapter~\ref{chap:interacting}), $n$ is identified with the \emph{degree} of the composite operator — the total number of field insertions (e.g.\ $\mathcal{O}_n\sim(\phi_a\phi_a)^{n/2}$); in the present generic semiclassical framework it is simply the orbit label.  All states with the same $n$ share the same leading dimension $nC_0$: the spectrum is \emph{completely degenerate at leading order}.

\item $\{q_\ell\geq 0\}$ are the \emph{Floquet excitation numbers}: non-negative integers, one per multiplet $\ell$ with $\nu_\ell>0$, labelling the quantum state built on top of that orbit.  They first appear at subleading order and lift the degeneracy.  The full scaling dimension~\eqref{eq:energy_final_boxed_new} (via $\Delta=RE$) is
\[
   \Delta_{n,\{q_\ell\}} = nC_0 + R\,\delta E_1
   + \frac{R}{\mathcal{T}}\sum_{\nu_\ell>0}\!\Bigl(q_\ell+\tfrac{n_\ell}{2}\Bigr)\nu_\ell\,.
\]
The ground state has $q_\ell=0$ for all $\ell$; its dimension is $\Delta_{n,0}=nC_0+C_1$ with $C_1=R\,\delta E_1+\frac{R}{2\mathcal{T}}\sum_{\nu_\ell>0}n_\ell\nu_\ell$.  Excited states have $\delta\Delta = \frac{R}{\mathcal{T}}\sum_{\nu_\ell>0}q_\ell\nu_\ell$ above the ground state.
\end{itemize}

Table~\ref{tab:q_ell_cases} summarises the key special cases, using $\ell_0$ for any fixed angular-momentum channel chosen to be excited.

\begin{table}[h]
\centering
\renewcommand{\arraystretch}{1.6}
\begin{tabular}{lll}
\hline
\textbf{State} & \textbf{$\{q_\ell\}$} & \textbf{$\delta\Delta$ above ground state} \\
\hline
Ground state          & $q_\ell=0$ for all $\ell$                         & $0$ \\
Single-channel $\ell_0$, $k$ quanta
                      & $q_{\ell_0}=k\geq 1$,\ $q_\ell=0$ ($\ell\neq\ell_0$)
                                                                           & $kR\nu_{\ell_0}/\mathcal{T}$ \\
Descendant ($\ell_0=1$, $k$ quanta)
                      & $q_1=k\geq 1$,\ $q_\ell=0$ ($\ell\geq 2$)        & $k$ \emph{exactly} \\
General state         & $\{q_\ell\}$ arbitrary                            & $\dfrac{R}{\mathcal{T}}\displaystyle\sum_{\nu_\ell>0}q_\ell\nu_\ell$ \\
\hline
\end{tabular}
\caption{Semiclassical states for fixed Bohr--Sommerfeld integer $n$.  The ground-state dimension is $\Delta_{n,0}=nC_0+C_1$.  The descendant shift is exact because $\nu_1=\mathcal{T}/R$.}
\label{tab:q_ell_cases}
\end{table}

\noindent
The degeneracy of the single-channel state at level $k$ is $\binom{k+n_{\ell_0}-1}{n_{\ell_0}-1}$; for descendants ($\ell_0=1$, $n_1=d$) this gives $\binom{k+d-1}{d-1}$, matching the $d$ components of the momentum operator $P_\mu$ at level $k$.  For a general state $\{q_\ell\}$ the total degeneracy is $\prod_{\nu_\ell>0}\binom{q_\ell+n_\ell-1}{n_\ell-1}$.  The Lorentz spin $s$ and number of d'Alembertian insertions $p$ of the corresponding CFT operator satisfy $2p+s=\sum_{\ell}q_\ell\cdot\ell$.

\paragraph{The Floquet phase as a quasi-momentum accumulation.}
The factor $\exp(-iq_\ell \nu_\ell)$ in each term of $\Delta_2$ \eqref{eq:delta2_expansion} is the
Bohr--de Broglie phase accumulated over one period by the $q_\ell$ quanta in multiplet $\ell$:
the quasi-momentum of the $\ell$-th Floquet mode is $p_\ell = \nu_\ell/\mathcal{T}$,
so in time $\mathcal{T}$ a state carrying $q_\ell$ quanta acquires phase $q_\ell\nu_\ell = q_\ell p_\ell\mathcal{T}$.
The requirement that the full path-integral amplitude be consistent after one traversal of the orbit is
precisely the Bohr--Sommerfeld condition \eqref{eq:full_quantization_boxed}: it quantises $n$
as the integer that counts how many times the classical orbit is traversed.

\paragraph{Primaries versus descendants.}
The stability angle of the $\ell=1$ channel satisfies $\nu_1 = \mathcal{T}/R$ (the mode corresponding
to a single spatial derivative; see \S\,\ref{subsubsec:descendant}).  A necessary condition for a
state $\{q_\ell\}$ to correspond to a conformal \emph{primary} operator is therefore $q_1=0$: with $q_1=0$
the spin-1 channel contributes only its zero-point energy $\tfrac{n_1}{2}\nu_1$ to $\Delta$, unchanged
from the ground state.  A state with $q_1=k>0$ acquires additional scaling dimension
$kR\nu_1/\mathcal{T}=k$, so it lies in the conformal multiplet of a primary with one fewer unit of
angular momentum: it is a \emph{descendant} generated by $k$ successive applications of the momentum
operator $P_\mu$.

\section{Saddle-Point Analysis of the $\mathcal{T}$-Integral}
\label{sec:saddle_point}

\subsection{Full Expression for the Resolvent}

We now return to the proper-time integral for the resolvent \eqref{eq:proper_time_resolvent}. Substituting the
Gutzwiller formula \eqref{eq:gutzwiller_formula}:
\begin{equation}
G(E) = i \int_0^\infty d\mathcal{T}\, \Delta_1\, \exp\left[i\left(\mathcal{S}_{\rm cl} - \sum_{\nu_\ell>0} \left(q_\ell + \frac{n_\ell}{2}\right) \nu_\ell + E\mathcal{T}\right)\right]
\label{eq:G_full_integral}
\end{equation}

This is a sum over all periodic classical solutions (via the implicit sum in the trace formula), and for each
solution, a sum over all excitation quantum numbers $\{q_\ell\}$ (from the $\Delta_2$ expansion). The integral over
$\mathcal{T}$ is a proper-time integral that will be evaluated by saddle points.

\subsection{The Saddle-Point Condition}

The $\mathcal{T}$-integral is dominated by saddle points of the exponent:
\begin{equation}
\mathcal{F}(\mathcal{T}) = \mathcal{S}_{\rm cl}(\mathcal{T}) - \sum_{\nu_\ell>0} \left(q_\ell + \frac{n_\ell}{2}\right) \nu_\ell(\mathcal{T}) + E\mathcal{T}
\label{eq:exponent_def}
\end{equation}

The saddle condition is:
\begin{equation}
\frac{d\mathcal{F}}{d\mathcal{T}} = 0 \quad \Rightarrow \quad
\frac{d\mathcal{S}_{\rm cl}}{d\mathcal{T}} - \sum_{\nu_\ell>0} \left(q_\ell + \frac{n_\ell}{2}\right) \frac{d\nu_\ell}{d\mathcal{T}} + E = 0
\label{eq:saddle_detailed}
\end{equation}

Rearranging:
\begin{equation}
\boxed{
E = -\frac{d\mathcal{S}_{\rm cl}}{d\mathcal{T}} + \sum_{\nu_\ell>0} \left(q_\ell + \frac{n_\ell}{2}\right) \frac{d\nu_\ell}{d\mathcal{T}}
}
\label{eq:quantization_condition}
\end{equation}

\textbf{Physical interpretation:} This is the {quantum mechanical energy quantization condition} for states
built on the periodic classical orbit. Let's interpret each term:

\begin{itemize}
    \item The first term $-\frac{d\mathcal{S}_{\rm cl}}{d\mathcal{T}}$ is the classical energy of the orbit. Recall that
    the action $\mathcal{S}_{\rm cl}$ depends on the period $\mathcal{T}$ because a longer-period orbit has a different
    energy. The derivative $\frac{d\mathcal{S}_{\rm cl}}{d\mathcal{T}}$ is related to the classical frequency or action
    via the Legendre transform.

    \item The second term $\sum_{\nu_\ell>0} \left(q_\ell + \frac{n_\ell}{2}\right) \frac{d\nu_\ell}{d\mathcal{T}}$ represents
    quantum corrections from excited modes. Here $q_\ell \geq 0$ is the total occupation number of multiplet $\ell$
    (the sum of quanta across all $n_\ell$ degenerate modes), while the $\frac{n_\ell}{2}$ term is the zero-point
    contribution from all $n_\ell$ independent Floquet oscillators. The stability angle $\nu_\ell$ encodes the
    effective quasi-frequency of oscillations in mode $\ell$ around the orbit.
\end{itemize}

\subsection{Interpretation as Bohr-Sommerfeld Quantization}

The saddle-point condition \eqref{eq:quantization_condition} is a field-theoretic generalization of the Bohr-Sommerfeld
quantization rule from quantum mechanics. In 1D quantum mechanics, the WKB quantization rule is:
\[
\oint p \, dq = 2\pi n + \text{phase corrections}
\]

Here, in the field-theoretic setting, we have a generalization where the action $\mathcal{S}_{\rm cl}$ plays the role of
$\oint p \, dq$, and the stability angles $\nu_\ell$ encode the phase corrections from quantum fluctuations.

The term $\frac{1}{2}\sum_{\nu_\ell>0} n_\ell \nu_\ell$ is often called the \textit{Maslov phase} correction, it accounts for the
focusing and defocusing of the classical trajectory due to quantum potential barriers.

\subsection{Energy as a Function of Period}

We can invert the saddle-point condition to find $\mathcal{T}$ as a function of $E$ (or vice versa):
\begin{equation}
\mathcal{T}(\{q_\ell\}, E) : \quad E = -\frac{d\mathcal{S}_{\rm cl}}{d\mathcal{T}} + \sum_{\nu_\ell>0} \left(q_\ell + \frac{n_\ell}{2}\right) \frac{d\nu_\ell}{d\mathcal{T}}
\label{eq:T_of_E}
\end{equation}

For each choice of excitation quantum numbers $\{q_\ell\}$, there is (generically) a unique period $\mathcal{T}^*$
at which the saddle condition is satisfied. This $\mathcal{T}^*$ corresponds to the periodic orbit that, when excited
with the given quantum numbers, has energy $E$.

Thus, the Gutzwiller formula \eqref{eq:gutzwiller_formula} combined with the saddle-point analysis provides a complete
description of the quantum energy spectrum in terms of classical periodic orbits and their stability properties.

 {Anticipating the renormalisation of Chapter~\ref{chap:interacting}:
the divergent bare action and the divergent zero-point sum combine into a finite
regularised action $\mathcal{S}_{\rm reg}$ with
$-\partial\mathcal{S}_{\rm reg}/\partial\mathcal{T}=E_{\rm cl}+\delta E_1$, so the
spectrum takes the final form
\begin{equation*}
  E_{n,\{q_\ell\}} = E_{\rm cl} + \delta E_1
  + \frac{1}{\mathcal{T}}\sum_{\nu_\ell>0}\Bigl(q_\ell+\tfrac{n_\ell}{2}\Bigr)\nu_\ell\,,
\end{equation*}
which is precisely the boxed result \eqref{eq:energy_final_boxed_new} assembled in
the blueprint.}

\part{A Physical Application}

\chapter{ The four-dimensional $\phi^4$ theory}
\label{chap:interacting}

\section{The $O(N)$ $\phi^4$ theory  }\label{sec:phi4}

Having developed the full semiclassical machinery in the previous sections, we now apply it to a concrete
and physically important example: the critical $O(N)$ $\lambda(\phi_a\phi_a)^2$ theory in $d=4-\epsilon$ dimensions.
This theory describes a rich variety of universality classes: the Ising model ($N=1$),
the XY model ($N=2$), the Heisenberg magnet ($N=3$), and the Higgs sector of the Standard Model ($N=4$); see~\cite{Pelissetto:2000ek} for a comprehensive review of the physical applications of the $O(N)$ universality classes.  Complementary methodologies to determine the spectrum include the state-of-the-art $\varepsilon$-expansion~\cite{Henriksson:2025hwi,Henriksson:2025vyi,Schnetz:2022nsc,Kompaniets:2017yct} and the numerical conformal bootstrap in $d=3$~\cite{Simmons-Duffin:2016wlq,Kos:2013tga,Rattazzi:2008pe}.  A comprehensive review of the known conformal data of the $O(N)$ CFT is given in~\cite{Henriksson:2022rnm}.

The Lagrangian is deceptively simple:
\begin{equation}
\mathcal{L} = \frac{1}{2}(\partial\phi_a)^2 - \frac{\lambda}{4}(\phi_a\phi_a)^2\,, \qquad a = 1,\dots, N\,.
\label{eq:phi4Lag}
\end{equation}
\noindent
The renormalization-group $\beta$-function $\beta(\lambda)=\mu\,d\lambda/d\mu$ controls the scale
dependence of the coupling.  In $d=4-\epsilon$ the coupling carries mass dimension $\epsilon$, so at
tree level $\beta(\lambda)=-\epsilon\lambda$.  Including quantum corrections to two loops in the
$\overline{\rm MS}$ scheme one finds
\begin{equation}
  \beta(\lambda)
  = -\epsilon\,\lambda
  + \frac{N+8}{8\pi^2}\,\lambda^2
  - \frac{3(3N+14)}{64\pi^4}\,\lambda^3
  + \mathcal{O}(\lambda^4)\,.
  \label{eq:beta_phi4}
\end{equation}
The one-loop coefficient $\beta_0=(N+8)/8\pi^2>0$ makes the coupling marginally irrelevant in four
dimensions ($\epsilon=0$), so the only fixed point is the Gaussian one $\lambda=0$.  For $\epsilon>0$
the linear term tilts the $\beta$-function and opens a second zero at $\lambda_*>0$: this is the
Wilson--Fisher (WF) fixed point \cite{Wilson:1971dc}.   Because $-\epsilon\lambda$ is the dominant term near the origin while
the $\lambda^2$ term is quadratically suppressed there, the location of the WF fixed point is directly
controlled by $\epsilon$:

 \noindent
\noindent
To make the signs explicit, write \eqref{eq:beta_phi4} as
\begin{equation}
  \beta(\lambda)
  =
  -\epsilon\lambda
  +
  \beta_0\lambda^2
  -
  \beta_1\lambda^3
  +
  \mathcal{O}(\lambda^4),
  \qquad
  \beta_0=\frac{N+8}{8\pi^2},
  \qquad
  \beta_1=\frac{3(3N+14)}{64\pi^4}>0 .
\end{equation}
 Setting $\beta(\lambda_*)=0$ and expanding
\begin{equation}
  \lambda_* = A_1\epsilon + A_2\epsilon^2
  + \mathcal{O}(\epsilon^3),
\end{equation}
one finds
\begin{equation}
  \beta(\lambda_*)
  =
  \left(-A_1+\beta_0 A_1^2\right)\epsilon^2
  +
  \left(-A_2+2\beta_0 A_1A_2-\beta_1 A_1^3\right)\epsilon^3
  +
  \mathcal{O}(\epsilon^4).
\end{equation}
The vanishing of the coefficients of $\epsilon^2$ and $\epsilon^3$ gives
\begin{equation}
  -A_1+\beta_0 A_1^2=0,
  \qquad
  -A_2+2\beta_0 A_1A_2-\beta_1 A_1^3=0.
\end{equation}
For the interacting fixed point, the first equation gives
\begin{equation}
  A_1=\frac{1}{\beta_0}.
\end{equation}
Substituting this into the second equation gives
\begin{equation}
  A_2=\frac{\beta_1}{\beta_0^3}.
\end{equation}
Therefore
\begin{equation}
  \lambda_*
  =
  \frac{\epsilon}{\beta_0}
  +
  \frac{\beta_1}{\beta_0^3}\epsilon^2
  +
  \mathcal{O}(\epsilon^3),
  \qquad
  A_1 = \frac{8\pi^2}{N+8},
  \qquad
  A_2 =
  \frac{24\pi^2(3N+14)}{(N+8)^3}.
  \label{eq:lstar_expanded}
\end{equation}
Equivalently,
\begin{equation}
  \lambda_*
  =
  \frac{8\pi^2\epsilon}{N+8}
  +
  \frac{24\pi^2(3N+14)\epsilon^2}{(N+8)^3}
  +
  \mathcal{O}(\epsilon^3).
  \label{eq:WFfixedpoint}
\end{equation}

\noindent
  In the leading-order limit (omitting the higher-order correction):
\begin{equation}
  \lambda_*\simeq\frac{\epsilon}{\beta_0}=\frac{8\pi^2\,\epsilon}{N+8}+\mathcal{O}(\epsilon^2)\,.
\label{eq:lstar_LO}
\end{equation}
In other words, \emph{$\epsilon$ is the distance (in the space of couplings) between the Gaussian and
Wilson--Fisher fixed points}, measured in units of $\beta_0^{-1}$.

\begin{figure}[ht]
  \centering
  \includegraphics[width=\linewidth]{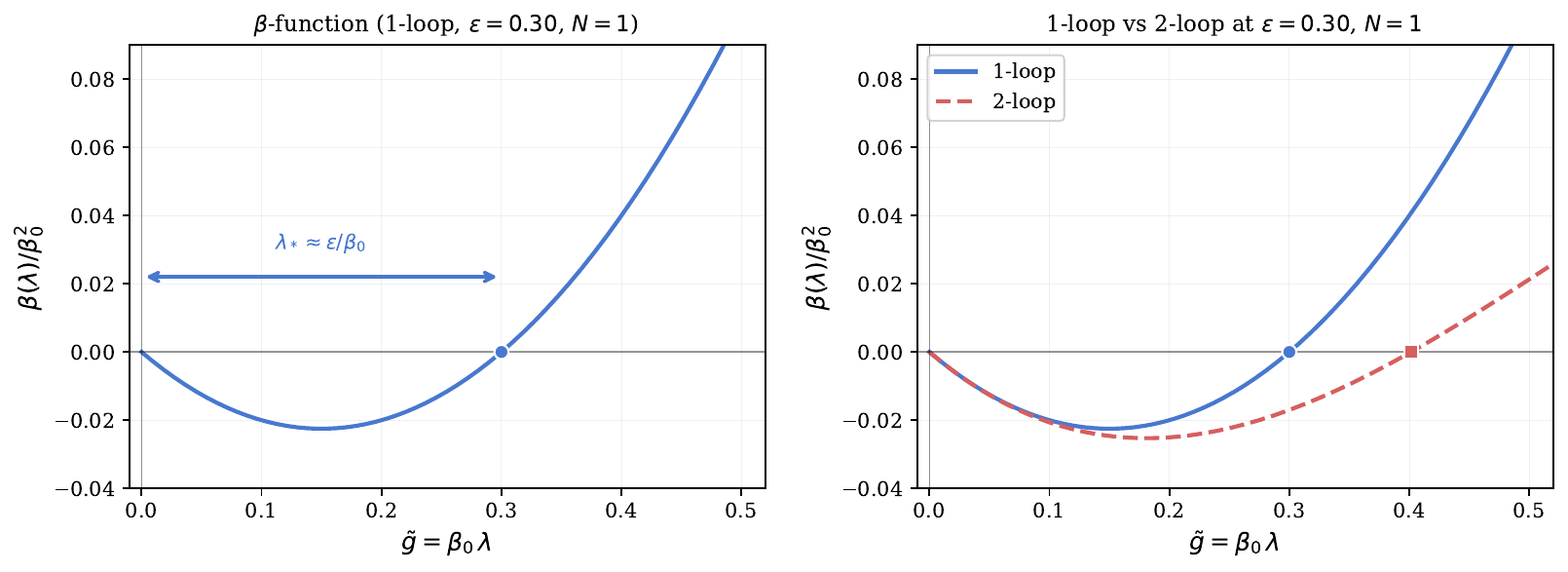}
  \caption{%
    $\beta$-function for the $O(N)$ $\phi^4$ theory ($N=1$, Ising) in units of
    $\tilde g=\beta_0\lambda$, $\beta_0=(N+8)/8\pi^2$.
    \textbf{Left:} one-loop $\beta$-function (blue) at $\epsilon=0.30$.  The filled circle
    marks the Wilson--Fisher fixed point at $\tilde g^*\approx\epsilon$; the double-headed
    arrow shows that $\epsilon$ is the coupling-space distance from the Gaussian fixed point
    at the origin.
    \textbf{Right:} one-loop (blue, solid) vs.\ two-loop (red, dashed) at $\epsilon=0.30$.
    The two-loop correction shifts the fixed point to
    $\tilde g^*\approx\epsilon+c_1\epsilon^2$.%
  }
  \label{fig:beta_phi4}
\end{figure}

Figure~\ref{fig:beta_phi4} illustrates this for $N=1$ (Ising universality class) using the
dimensionless coupling $\tilde g=\beta_0\lambda$.  The left panel shows the one-loop $\beta$-function
at $\epsilon=0.30$: the curve (blue) has a zero at $\tilde g=0$ (Gaussian FP) and a second zero at
$\tilde g^*\approx\epsilon$ (WF FP), with the double-headed arrow making explicit that $\epsilon$
is the coupling-space distance between the two fixed points.
The right panel compares the one-loop (blue, solid) and two-loop (red, dashed) results at the same
$\epsilon$; the two-loop correction shifts the WF coupling by
$\delta\tilde g^*=c_1\epsilon^2+\mathcal{O}(\epsilon^3)$ with $c_1=\beta_1/\beta_0^2=17/27$ for $N=1$.

\noindent
 As  noted in~\cite{Antipin:2020rdw}, the Wilson--Fisher fixed point can be achieved for either real or complex values of the coupling, and the semiclassical approach applies equally in both cases (the resulting CFT may be non-unitary, but the scaling dimensions are still well defined).

\bigskip
The central result of this chapter is the semiclassical formula for the scaling dimension of the lowest operator of degree $n$ (i.e.\ $\mathcal{O}_n\sim(\phi_a\phi_a)^{n/2}$) in the $O(N)$ $\phi^4$ theory at the Wilson--Fisher fixed point:
\begin{equation}
  \Delta_n(\kappa) \;=\; n\,C_0(\kappa) \;+\; C_1(\kappa) \;+\; \mathcal{O}(1/n)\,,
  \label{eq:Delta_target_box}
\end{equation}
where $\kappa = \lambda_* n$ is the double-scaling parameter held fixed as $n\to\infty$, $\lambda_*\to 0$.
The two coefficient functions are:
\begin{itemize}
  \item $C_0(\kappa)$: determined by the \emph{classical periodic saddle} $v_{\rm cl}(\tau) = x_0\,\cn(\omega\tau|m)$
    of the Euclidean action, via the Bohr--Sommerfeld quantisation $I(E_{\rm cl}) = 2\pi n$ and the
    state--operator map $\Delta = R\,E$.  It encodes the leading, extensive contribution.
  \item $C_1(\kappa)$: the \emph{one-loop fluctuation determinant} around the saddle, computed via
    Floquet theory (stability angles $\nu_\ell$) and the Gel'fand--Yaglom theorem.
    It is the $n^0$ correction.
\end{itemize}
The result interpolates between the perturbative regime ($\kappa\to 0$, recovers the $\varepsilon$-expansion) and the semiclassical regime (non-vanishing values of $\kappa$).  The explicit Jacobi elliptic / Lam\'e technology in the sections below is the concrete machinery that evaluates $C_0$ and $C_1$ as closed-form functions of $\kappa$.

\subsection{Classical solution and leading order: $C_0$}

On the cylinder $\mathbb{R}\times S^{d-1}$ with sphere radius $R$, the Lagrangian takes the form
\begin{equation}
\mathcal{L}^{(\mathrm{cyl})} = \frac{1}{2}(\partial\phi_a)^2
- \frac{\mu^2}{2}\phi_a\phi_a
- \frac{\lambda}{4}(\phi_a\phi_a)^2\,,
\label{eq:Lcyl}
\end{equation}
where $\mu = (d-2)/(2R)$ is the conformal mass induced by the coupling to the Ricci curvature of the sphere (see Section~\ref{sec:conformal_coupling}).

We look for \emph{spatially homogeneous} classical solutions by setting
\[
\phi_1(t,\Omega)=v(t)\,,\qquad 
\phi_a(t,\Omega)=0\,,\qquad a=2,\ldots,N .
\]
For nonzero \(v(t)\), this choice selects a direction in field space and leaves
an \(O(N-1)\) subgroup manifest. The classical equation of motion reduces to
\begin{equation}
\ddot v+\mu^2 v+\lambda v^3=0\,.
\label{eq:anharmonic_oscillator}
\end{equation}
This is the equation of the quartic anharmonic oscillator.

The corresponding conserved energy is
\begin{equation}
E=\frac12 \dot v^2+\frac{\mu^2}{2}v^2+\frac{\lambda}{4}v^4 .
\end{equation}
For \(\mu^2>0\) and \(\lambda>0\), the potential is a single-well confining
potential. Hence bounded periodic motion exists for positive energy, with
turning points \(v=\pm x_0\) determined by
\[
E=\frac{\mu^2}{2}x_0^2+\frac{\lambda}{4}x_0^4 .
\]

The periodic solution can be written in terms of the Jacobi elliptic cosine as
\begin{equation}
v(t)=x_0\,\cn(\omega t\,|\,m)\,,
\label{eq:cn_solution}
\end{equation}
where the amplitude \(x_0\), frequency \(\omega\), and elliptic modulus \(m\)
are related by
\begin{equation}
x_0=\mu\sqrt{\frac{2m}{\lambda(1-2m)}}\,,\qquad
\omega=\frac{\mu}{\sqrt{1-2m}}\,,\qquad
0\leq m<\frac12 .
\label{eq:params_cn}
\end{equation}

To verify this, use the identity
\[
\frac{d^2}{dz^2}\cn(z\,|\,m)
=
(2m-1)\cn(z\,|\,m)-2m\,\cn^3(z\,|\,m).
\]
Substitution of \eqref{eq:cn_solution} gives
\begin{align}
\ddot v
&=
x_0\omega^2
\left[
(2m-1)\cn(\omega t\,|\,m)
-
2m\,\cn^3(\omega t\,|\,m)
\right]  \nonumber\\
&=
-\mu^2 x_0\cn(\omega t\,|\,m)
-
\frac{2m\mu^2 x_0}{1-2m}\cn^3(\omega t\,|\,m),
\end{align}
where we used \(\omega^2=\mu^2/(1-2m)\). Therefore
\[
\ddot v+\mu^2 v+\lambda v^3=0
\]
provided
\[
\lambda x_0^2=\frac{2m\mu^2}{1-2m},
\]
which is precisely the first relation in \eqref{eq:params_cn}. Thus
\eqref{eq:cn_solution} solves the equation of motion exactly.

The period of the solution is
\begin{equation}
\mathcal{T}
=
\frac{4\KK(m)}{\omega}
=
\frac{4\KK(m)\sqrt{1-2m}}{\mu}\,,
\label{eq:period_phi4}
\end{equation}
where \(\KK(m)\) is the complete elliptic integral of the first kind.

\begin{figure}[h]
\centering
\includegraphics[width=0.90\textwidth]{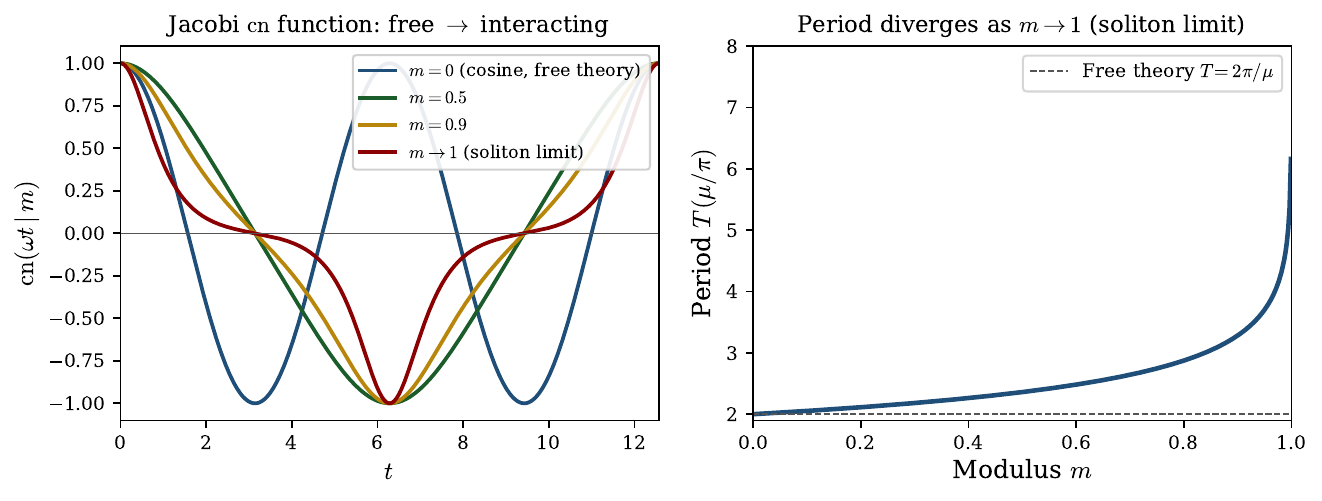}
\caption{%
\textbf{Jacobi elliptic cosine solution of the quartic oscillator.}
The homogeneous classical solution can be written as
\(v(t)=x_0\cn(\omega t\,|\,m)\), with \(0\le m<1/2\).
At \(m=0\), the solution reduces to the harmonic oscillator limit.
As \(m\) increases, the quartic interaction becomes increasingly important.
The period is \(\mathcal{T}=4K(m)/\omega\), with
\(\omega=\mu/\sqrt{1-2m}\). In this parametrization the endpoint
\(m=1/2\) is singular and is not included.}
\label{fig:jacobi_cn}
\end{figure}

In what follows we use dimensionless cylinder time $t \rightarrow t/R$. In these units the
conformal mass in \(d\) dimensions is
\[
\mu= \frac{d-2}{2} \ .
\]
The physical energy \(E\) enters the scaling dimension through the
dimensionless combination $\Delta=R\,E$. Equivalently, all formulas below are written in units of the cylinder radius.
The factors of \(R\) can be restored by dimensional analysis.

\subsection{Bohr--Sommerfeld condition and the parameter \(m\)}

According to the general framework of Section~\ref{energypath}, the parameter
\(m\) is fixed by the classical Bohr--Sommerfeld condition
\[
I=2\pi n .
\]
The action variable is
\[
I=\oint p\,dq .
\]
For the homogeneous mode, the canonical momentum is proportional to \(\dot v\),
with the proportionality factor given by the volume of the spatial sphere.
Therefore, on the unit three-sphere,
\[
I
=
\Omega_3\int_0^{\mathcal T}\dot v^{\,2}\,dt,
\qquad
\Omega_3=2\pi^2 .
\]
Equivalently, the factor \(\Omega_3\) is the remnant of the spatial integral
over the homogeneous classical configuration.

Using the periodic solution
\[
v(t)=x_0\,\cn(\omega t\,|\,m),
\]
with
\[
x_0=\mu\sqrt{\frac{2m}{\lambda(1-2m)}}\,,
\qquad
\omega=\frac{\mu}{\sqrt{1-2m}}\,,
\qquad
0\leq m<\frac12 ,
\]
we have
\[
\dot v(t)
=
-x_0\omega\,\sn(\omega t\,|\,m)\dn(\omega t\,|\,m).
\]
Hence
\[
\dot v^{\,2}
=
x_0^2\omega^2
\sn^2(\omega t\,|\,m)\dn^2(\omega t\,|\,m).
\]
Since the period is
\[
\mathcal T=\frac{4\KK(m)}{\omega},
\]
we obtain
\[
I
=
\Omega_3 x_0^2\omega
\int_0^{4\KK(m)}
\sn^2(u\,|\,m)\dn^2(u\,|\,m)\,du .
\]
Using the standard elliptic integral identity
\[
\int_0^{4\KK(m)}
\sn^2(u\,|\,m)\dn^2(u\,|\,m)\,du
=
\frac{4}{3m}
\left[
(1-m)\KK(m)-(1-2m)\ELE(m)
\right],
\]
we find
\[
I
=
\frac{8\Omega_3\mu^3}{3\lambda(1-2m)^{3/2}}
\left[
(1-m)\KK(m)-(1-2m)\ELE(m)
\right].
\]
Equivalently,
\[
I
=
\frac{8\Omega_3\mu^3}{3\lambda(1-2m)^{3/2}}
\left[
(2m-1)\ELE(m)+(1-m)\KK(m)
\right].
\]
Since \(\Omega_3=2\pi^2\), this becomes
\begin{equation}
I
=
\frac{16\pi^2\mu^3}{3\lambda(1-2m)^{3/2}}
\left[
(2m-1)\ELE(m)+(1-m)\KK(m)
\right].
\label{eq:I_phi4}
\end{equation}

The Bohr--Sommerfeld condition \(I=2\pi n\) therefore gives
\begin{equation}
\lambda n
=
\frac{8\pi\mu^3}{3}
\frac{
(2m-1)\ELE(m)+(1-m)\KK(m)
}
{(1-2m)^{3/2}} .
\label{eq:BS_m_relation}
\end{equation}
This equation implicitly defines \(m=m(\lambda n/\mu^3)\).

In the small-\(\lambda n\) regime, the condition may be inverted
perturbatively. Expanding \eqref{eq:BS_m_relation} at small \(m\) gives
\[
\lambda n
=
2\pi^2\mu^3 m
+\frac{21\pi^2\mu^3}{4}m^2
+\frac{403\pi^2\mu^3}{32}m^3
+\frac{14765\pi^2\mu^3}{512}m^4
+\cdots .
\]
Therefore
\begin{equation}
m
=
\frac{\lambda n}{2\pi^2\mu^3}
-\frac{21(\lambda n)^2}{32\pi^4\mu^6}
+\frac{479(\lambda n)^3}{512\pi^6\mu^9}
-\frac{22745(\lambda n)^4}{16384\pi^8\mu^{12}}
+\cdots .
\label{eq:m_small}
\end{equation}
For \(d=4\), where \(\mu=1\), this reduces to the expansion quoted in the
main text. This small-\(m\) expansion reproduces ordinary perturbation theory
order by order.

\subsection{Classical energy: \(C_0\)}

The classical contribution to the scaling dimension is
\[
\Delta_{\rm cl}=RE_{\rm cl}\equiv nC_0 .
\]
Since we work in dimensionless cylinder variables, this quantity is obtained by
integrating the dimensionless Hamiltonian density over the unit spatial sphere.

The classical Hamiltonian density is
\[
T_{00}
=
\frac12\dot v^{\,2}
+\frac{\mu^2}{2}v^2
+\frac{\lambda}{4}v^4 .
\]
Because the energy is conserved, it may be evaluated at a turning point of the
motion. At the turning point \(v=x_0\) and \(\dot v=0\), so
\[
T_{00}
=
\frac{\mu^2}{2}x_0^2
+\frac{\lambda}{4}x_0^4 .
\]
Using
\[
x_0^2
=
\frac{2m\mu^2}{\lambda(1-2m)},
\]
we obtain
\[
T_{00}
=
\frac{\mu^4}{\lambda}
\frac{m(1-m)}{(1-2m)^2}.
\]
Multiplying by the volume of the unit three-sphere,
\[
\Omega_3=2\pi^2,
\]
gives
\begin{equation}
nC_0
=
\Delta_{\rm cl}
=
RE_{\rm cl}
=
\Omega_3 T_{00}
=
\frac{2\pi^2\mu^4}{\lambda}
\frac{m(1-m)}{(1-2m)^2}.
\label{eq:C0_phi4}
\end{equation}
For \(d=4\), where \(\mu=1\), this reduces to
\[
nC_0
=
\frac{2\pi^2m(1-m)}
{\lambda(1-2m)^2}.
\]

Substituting the small-\(m\) expansion \eqref{eq:m_small} and then tuning the
coupling to the Wilson--Fisher fixed point \eqref{eq:WFfixedpoint}, one obtains,
in \(d=4-\epsilon\) and for \(N=1\),
\begin{equation}
nC_0
=
n\left(
1
+\frac{1}{6}(\epsilon n)
-\frac{17}{324}(\epsilon n)^2
+\frac{125}{3888}(\epsilon n)^3
-\frac{3563}{139968}(\epsilon n)^4
+\cdots
\right).
\label{eq:C0_perturbative}
\end{equation}
Viewed as a function of the double-scaling variable \(\kappa=\lambda n\), the
coefficient \(C_0\) is independent of \(N\) --- as expected at the classical
level, since the background trajectory involves only a single component of the
\(O(N)\) vector. Indeed, in terms of \(\kappa\) the same expansion reads
\[
C_0
=
1
+\frac{3}{16\pi^2}\kappa
-\frac{17}{256\pi^4}\kappa^2
+\frac{375}{8192\pi^6}\kappa^3
-\cdots ,
\]
with no reference to \(N\). The \(N\)-dependence visible in
\eqref{eq:C0_perturbative} is therefore not intrinsic to \(C_0\): it enters
only through the leading-order Wilson--Fisher relation
\(\lambda_*\simeq 8\pi^2\epsilon/(N+8)\), which trades \(\kappa\) for
\(\epsilon n\). Restoring it explicitly,
\begin{equation}
nC_0
=
n\left(
1
+\frac{3}{2(N+8)}(\epsilon n)
-\frac{17}{4(N+8)^2}(\epsilon n)^2
+\frac{375}{16(N+8)^3}(\epsilon n)^3
-\frac{10689}{64(N+8)^4}(\epsilon n)^4
+\cdots
\right),
\label{eq:C0_perturbative_genN}
\end{equation}
which reduces to \eqref{eq:C0_perturbative} at \(N=1\) (the Ising universality
class quoted above).

In the large-\(\lambda n\) regime, the Bohr--Sommerfeld condition drives
\[
m\to\frac12^- .
\]
Writing
\[
m=\frac12-\delta,
\qquad
\delta\to0^+,
\]
one finds from \eqref{eq:BS_m_relation}
\begin{equation}
m
=
\frac12
-
\pi
\left(
\frac{\mu^3\Gamma(1/4)}
{6\Gamma(3/4)}
\right)^{2/3}
(\lambda n)^{-2/3}
+
\mathcal O\!\left((\lambda n)^{-4/3}\right).
\label{eq:m_large}
\end{equation}
Here \(\KK(m)\) and \(\ELE(m)\) remain finite as \(m\to1/2\); in particular,
\[
\KK(1/2)=\frac{\Gamma(1/4)^2}{4\sqrt{\pi}}.
\]
Thus this limit corresponds to the large-amplitude
limit of the single-well quartic oscillator.

\begin{figure}[ht]
\centering
\includegraphics[width=\linewidth]{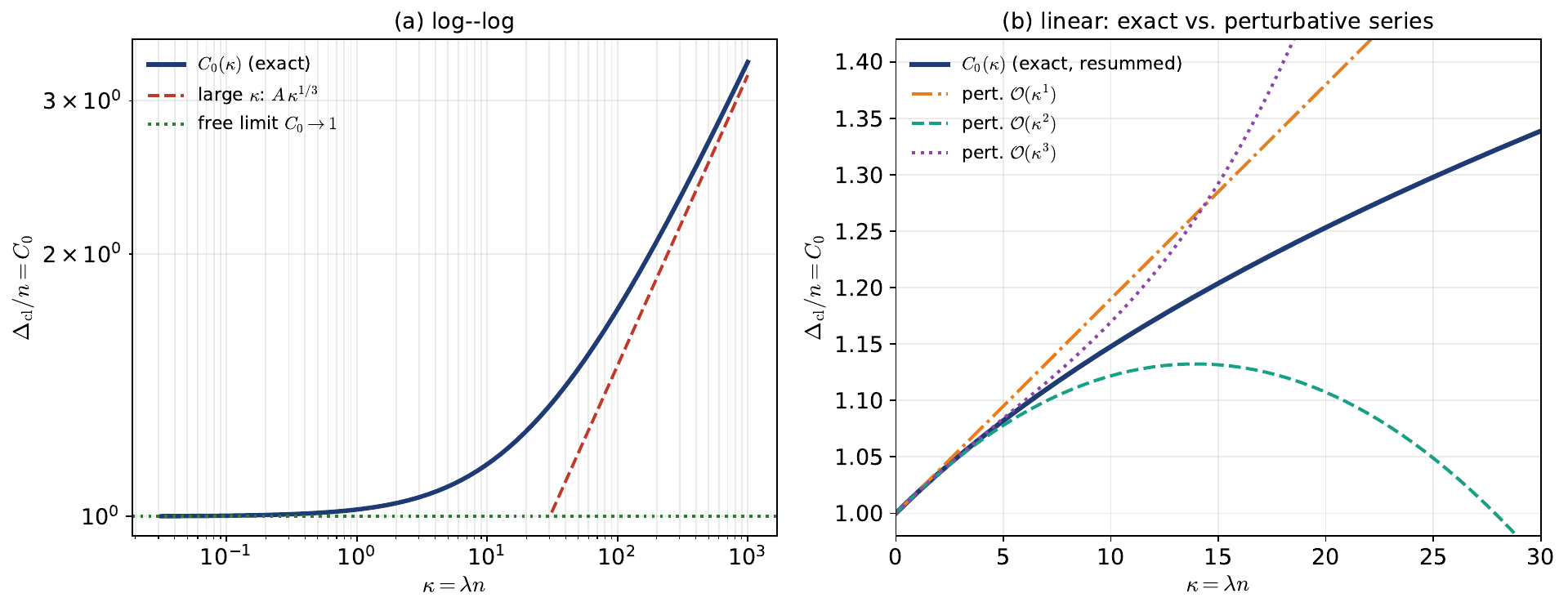}
\caption{%
\textbf{Leading classical coefficient $C_0(\kappa)=\Delta_{\rm cl}/n$ in $d=4$.}
The exact result (solid blue) is obtained by solving the Bohr--Sommerfeld
condition \eqref{eq:BS_m_relation} for the elliptic modulus $m(\kappa)$ and
substituting into \eqref{eq:C0_phi4}, with $\kappa=\lambda n$.
\textbf{(a)} Log--log view over four decades in $\kappa$: $C_0\to1$ in the free
limit $\kappa\to0$ (dotted green) and crosses over to the large-charge
asymptote $C_0\sim A\,\kappa^{1/3}$ (dashed red), with
$A=\bigl(3\Gamma(3/4)/2^{5/4}\Gamma(1/4)\bigr)^{4/3}$ from
\eqref{eq:C0_large}, reproducing $\Delta_n\sim n^{d/(d-1)}=n^{4/3}$.
\textbf{(b)} Linear view comparing the exact (resummed) curve with the
small-$\kappa$ perturbative truncations
$C_0=1+\tfrac{3}{16\pi^2}\kappa-\tfrac{17}{256\pi^4}\kappa^2
+\tfrac{375}{8192\pi^6}\kappa^3-\cdots$ from \eqref{eq:C0_perturbative}:
the series tracks the exact answer only for $\kappa\lesssim3$ and then
diverges term by term, signalling its asymptotic character and the need for
the semiclassical resummation at finite $\kappa$.}
\label{fig:C0_classical}
\end{figure}

Substituting the large-\(\lambda n\) expansion of \(m\) into
\eqref{eq:C0_phi4} gives
\begin{equation}
nC_0
=
\left(
\frac{3\Gamma(3/4)}
{2^{5/4}\Gamma(1/4)}
\right)^{4/3}
\lambda^{1/3}n^{4/3}
+
4\pi^3
\left(
\frac{6\Gamma(3/4)}{\Gamma(1/4)^7}
\right)^{2/3}
\mu^2
\lambda^{-1/3}n^{2/3}
+
\mathcal O(n^0).
\label{eq:C0_large}
\end{equation}
 The leading power \(n^{4/3}\) agrees with the general large-charge limit
\[
\Delta_n\sim n^{d/(d-1)}
\]
in \(d=4\). Figure~\ref{fig:C0_classical} displays $C_0(\kappa)$ across the
full range, showing the free limit, the large-charge $\kappa^{1/3}$ growth,
and the breakdown of the perturbative expansion at finite $\kappa$.

\subsection{Renormalization of the action}\label{sec:renorm}

To compute the next-to-leading correction $C_1$, we must renormalize the classical action.
Denote the bare coupling by $\lambda_0$. At one loop:
\begin{equation}
\lambda_0 = \lambda M^\epsilon\,e^{\beta_0\lambda/\epsilon}\,,
\qquad
\beta_0 = \frac{N+8}{8\pi^2}\,,
\label{eq:bare_coupling}
\end{equation}
where $\beta_0$ is the one-loop coefficient of the beta function and $M$ is the RG scale.

The one-loop $\beta_0$ coefficient arises from 1PI bubble diagrams: there are $N$ tadpole insertions of $\phi_a^2$ corresponding to the $N$ field components, plus $5$ more from the box diagram structure with $\phi_a\phi_a$, giving the numerator $N+8$ relative to the standard normalization. The exponential form $e^{\beta_0\lambda/\epsilon}$ resums the leading-log UV divergences. Expanding in $\lambda$, each power $\lambda^k$ comes with a factor $1/\epsilon^k$ from $k$-loop diagrams with $k-1$ loop insertions.

The bare classical action in $d = 4-\epsilon$ is:
\begin{equation}
\mathcal{S}_{\mathrm{cl}}(\lambda_0)
= -\frac{\pi^{2-\epsilon/2}(\epsilon-2)^3 R^{-\epsilon}s(m)}{\lambda_0\,\Gamma(2-\epsilon/2)}\,,
\label{eq:Scl_bare}
\end{equation}
where
\begin{equation}
s(m) = \frac{(m-1)(3m-2)\KK(m) + (4m-2)\ELE(m)}{3(1-2m)^{3/2}}\,.
\end{equation}
The form of $\mathcal{S}_{\mathrm{cl}}(\lambda_0)$ follows from the fact that on the cylinder, the classical action $\int_0^\mathcal{T}[\frac{1}{2}\dot{v}^2 - V_{\mathrm{cyl}}(v)]dt$ evaluates to a function of $m$ and $\lambda$ alone after inserting the elliptic-function solution and integrating; the function $s(m)$ originates from the quartic average $\int_0^{4\KK}\cn^4$. The full derivation is given in Appendix~\ref{app:Scl_derivation}. In $d=4-\epsilon$ the coupling carries dimension $\mu^\epsilon$, producing the overall factor $R^{-\epsilon}$ and the combination $\Gamma(2-\epsilon/2)$. (Recall that the sphere radius was set to $R=1$ in Section~\ref{sec:phi4}, so $R^{-\epsilon}=1$; we keep it explicit to track the cylinder dimensions.)

After expressing $\lambda_0$ in terms of $\lambda$ using \eqref{eq:bare_coupling},
the renormalized action acquires a correction at $\mathcal{O}(\lambda^0)$:
\begin{equation}
\mathcal{S}_{\mathrm{cl}}(\lambda)
= -\frac{\pi^{2-\epsilon/2}(\epsilon-2)^3 M^{-\epsilon}R^{-\epsilon}s(m)}{\Gamma(2-\epsilon/2)}
\left(\frac{1}{\lambda} - \frac{\beta_0}{\epsilon} + \mathcal{O}(\lambda)\right)\,.
\end{equation}

The term proportional to $\beta_0/\epsilon$ is a quantum correction containing a \textbf{$1/\epsilon$ pole}:
\begin{equation}
\frac{\pi^{2-\epsilon/2}(\epsilon-2)^3 M^{-\epsilon}R^{-\epsilon}s(m)}{\Gamma(2-\epsilon/2)}\frac{\beta_0}{\epsilon}
= -8\pi^2\beta_0 s(m)\left(\frac{1}{\epsilon}
- \frac{1}{2}\left(2+\gamma_E + 2\log(\sqrt{\pi}MR)\right) + \mathcal{O}(\epsilon)\right)\,,
\label{eq:pole_from_renorm}
\end{equation}
where $\gamma_E$ is Euler's constant. This pole \emph{cancels} the UV divergences arising from the sum over stability angles. The pole structure is as follows: the renormalized action contributes $+8\pi^2\beta_0 s(m)/\epsilon$, while the fluctuation sum produces $-8\pi^2\beta_0 s(m)/\epsilon$ from both the $\kappa=1$ sum (multiplied by $N-1$ transverse modes) and the $\kappa=2$ sum. The function $s(m)$ appearing in \eqref{eq:sum_nu2} and \eqref{eq:sum_nu1} matches exactly, ensuring finiteness of the combination $\mathcal{S}_{\mathrm{cl}} - \frac{1}{2}\sum_{\nu_\ell>0}\nu_\ell$.

Taking the derivative with respect to the period and evaluating at the fixed point, one identifies:
\begin{equation}
R\,\delta E_1 = -\frac{\lambda_* n}{2}\beta_0\,C_0(\lambda_* n)\,.
\label{eq:deltaE1_phi4}
\end{equation}
This is the one-loop renormalization contribution $\delta E_1$ appearing in the general energy formula.

\section{Fluctuation operators and the Lam\'e equation}\label{sec:fluctuations}

To determine the stability angles, we expand the fields around the classical trajectory:
\[
\phi_1 = v(t) + \eta(\vec{x},t)\,, \qquad
\phi_a = \tilde\phi_a(\vec{x},t) \quad \text{for } a=2,\dots,N\,,
\]
\noindent
 Substituting into the Lagrangian \eqref{eq:Lcyl} and collecting quadratic terms, note that $(\phi_a\phi_a)^2 = (v+\eta)^4 + 2(v+\eta)^2\sum_{a=2}^N\tilde\phi_a^2 + \ldots$. The second variation of the quartic term with respect to $\tilde\phi_a^2$ at $\phi_1=v$ gives a mass term $-\lambda v^2$, while the second variation with respect to $\eta$ gives $-3\lambda v^2$ (since the second derivative of $v^4$ is $12v^2$). This yields:
\begin{equation}
\mathcal{L}_2 = \sum_{a=2}^N \frac{1}{2}\tilde\phi_a\,\cO_1\,\tilde\phi_a
+ \frac{1}{2}\eta\,\cO_2\,\eta\,,
\label{eq:L2_fluct}
\end{equation}
where the two fluctuation operators are:
\begin{equation}
\cO_\kappa = -\partial_t^2 + \Delta_S - \mu^2
- \frac{\kappa(\kappa+1)}{2}\lambda\,v^2(t)\,,
\qquad \kappa = 1,2\,.
\label{eq:Okappa}
\end{equation}
Here $\Delta_S$ is the Laplacian on $S^{d-1}$, and the exponent $\kappa(\kappa+1)/2 = 1$ for $\kappa=1$ (transverse) and $= 3$ for $\kappa=2$ (longitudinal), matching the mass terms above. The operator $\cO_1$ governs the $N-1$ Goldstone-like transverse fluctuations $\tilde\phi_a$, while $\cO_2$ governs the longitudinal (radial) fluctuation $\eta$.

\subsection{Reduction to the Lam\'e equation}

 To reduce $\mathcal{O}_\kappa$ to standard Lamé form, decompose the fluctuation into spherical harmonics $Y_{\ell m}$ on $S^{d-1}$, so that $\Delta_S Y_{\ell m} = -J_\ell^2 Y_{\ell m}$ with $J_\ell^2 = \ell(\ell+d-2)/R^2$. In $d=4$: $J_\ell^2 = \ell(\ell+2)/R^2 = \ell(\ell+2)$ (using $R=1$). The mode equation becomes
\[
[-\partial_t^2 - J_\ell^2 - \mu^2 - \frac{\kappa(\kappa+1)}{2}\lambda x_0^2\,\cn^2(\omega t|m)]\psi = 0\,.
\]

Now substitute $z = \omega t$ and use $v^2(t) = x_0^2\,\cn^2(z|m) = x_0^2[1 - \sn^2(z|m)]$:
\[
\frac{\kappa(\kappa+1)}{2}\lambda x_0^2\,\cn^2(z|m) = \frac{\kappa(\kappa+1)}{2}\cdot\frac{2m\mu^2}{1-2m}\cdot[1-\sn^2(z|m)] = \kappa(\kappa+1)m\frac{\mu^2}{1-2m} - \kappa(\kappa+1)m\frac{\mu^2}{1-2m}\sn^2(z|m)\,.
\]

The constant piece combines with $\mu^2/(1-2m)$ (from rescaling $\partial_t^2 = \omega^2\partial_z^2$) and with $J_\ell^2$ to produce the eigenvalue $\Lambda_\kappa(\ell)$. The $\sn^2$ term becomes the Lamé potential. Thus, by introducing the rescaled variable $z = \mu t/\sqrt{1-2m}$, both operators can be written as:
\begin{equation}
\cO_\kappa = \frac{\mu^2}{1-2m}\,L_\kappa\,,
\label{eq:Okappa_rescaled}
\end{equation}
where $L_\kappa$ is the \textbf{$\kappa$-gap Lam\'e operator}:
\begin{equation}
L_\kappa = -\partial_z^2 + \kappa(\kappa+1)\,m\,\sn^2(z\,|\,m) - \Lambda_\kappa(\ell)\,,
\label{eq:Lame_operator}
\end{equation}
with
\begin{equation}
\Lambda_\kappa(\ell) = \kappa(\kappa+1)m + (1-2m)A_\ell\,,
\qquad
A_\ell = \left(1 + \frac{2\ell}{d-2}\right)^2\,.
\label{eq:Lambda_kappa}
\end{equation}

The key feature of the Lam\'e equation is that it is exactly solvable: the $\kappa$-gap Lam\'e potential has $\kappa$ spectral gaps and the Bloch solutions (Floquet solutions) can be expressed in closed form using Jacobi theta and zeta functions.

\subsection{Band structure of the Lam\'e operator}

For $\kappa=1$ (transverse modes), the spectrum has a single gap and \textbf{two allowed bands}:
\[
\{m,\, 1\} \quad \text{and} \quad \{1+m,\, \infty\}\,.
\]

For $\kappa=2$ (longitudinal mode), the spectrum has two gaps and \textbf{three allowed bands}:
\begin{align}
&\left\{2\left(m - \sqrt{1-m+m^2}+1\right),\; 1+m\right\}\,,\nonumber\\
&\left\{1+4m,\; 4+m\right\}\,,\nonumber\\
&\left\{2\left(m + \sqrt{1-m+m^2}+1\right),\; \infty\right\}\,.
\end{align}
When $\Lambda_\kappa(\ell)$ falls within a gap, the corresponding stability angle becomes complex, signaling an instability of the periodic orbit.

Physically, the allowed bands correspond to propagating (oscillatory) Bloch waves, while the gaps correspond to evanescent solutions that grow exponentially over one period. When $\Lambda_\kappa(\ell)$ falls in a gap, the Floquet exponent $\nu_{\kappa,\ell}$ becomes purely imaginary, meaning that small fluctuations around the periodic orbit grow exponentially — i.e., the orbit is unstable.

\begin{figure}[htbp]
\centering
\begin{minipage}[t]{0.48\textwidth}
  \centering
  \includegraphics[width=\textwidth]{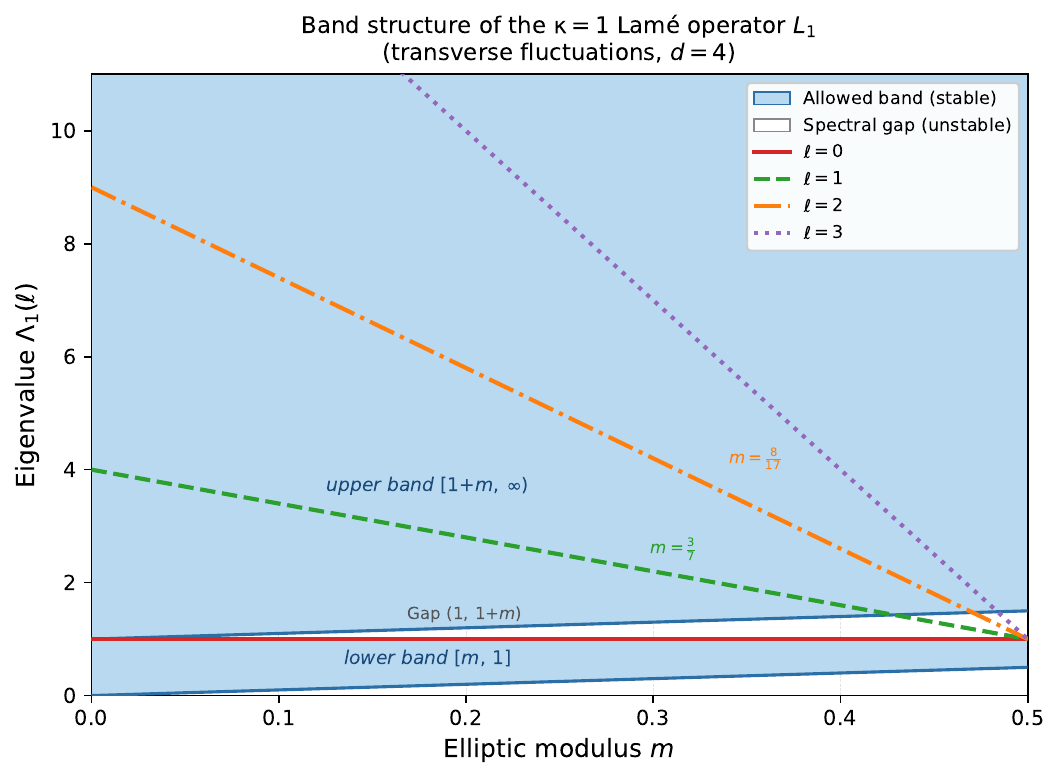}
  \caption{Band structure of the $\kappa=1$ (single-gap) Lam\'e operator $L_1$
  governing the $N{-}1$ transverse fluctuations in $d=4$.
  Blue shading: allowed bands $[m,1]$ and $[1{+}m,\infty)$.
  White strip: spectral gap $(1,1{+}m)$.
  Coloured lines: eigenvalues $\Lambda_1(\ell)=2m+(1{-}2m)(1{+}\ell)^2$
  for $\ell=0,1,2,3$.
  The $\ell=0$ zero mode (red) sits on the upper band edge for all $m$.
  The $\ell=1$ and $\ell=2$ modes enter the gap at $m=\tfrac{3}{7}$
  and $m=\tfrac{8}{17}$, respectively, signalling instability.}
  \label{fig:bandslame1}
\end{minipage}
\hfill
\begin{minipage}[t]{0.48\textwidth}
  \centering
  \includegraphics[width=\textwidth]{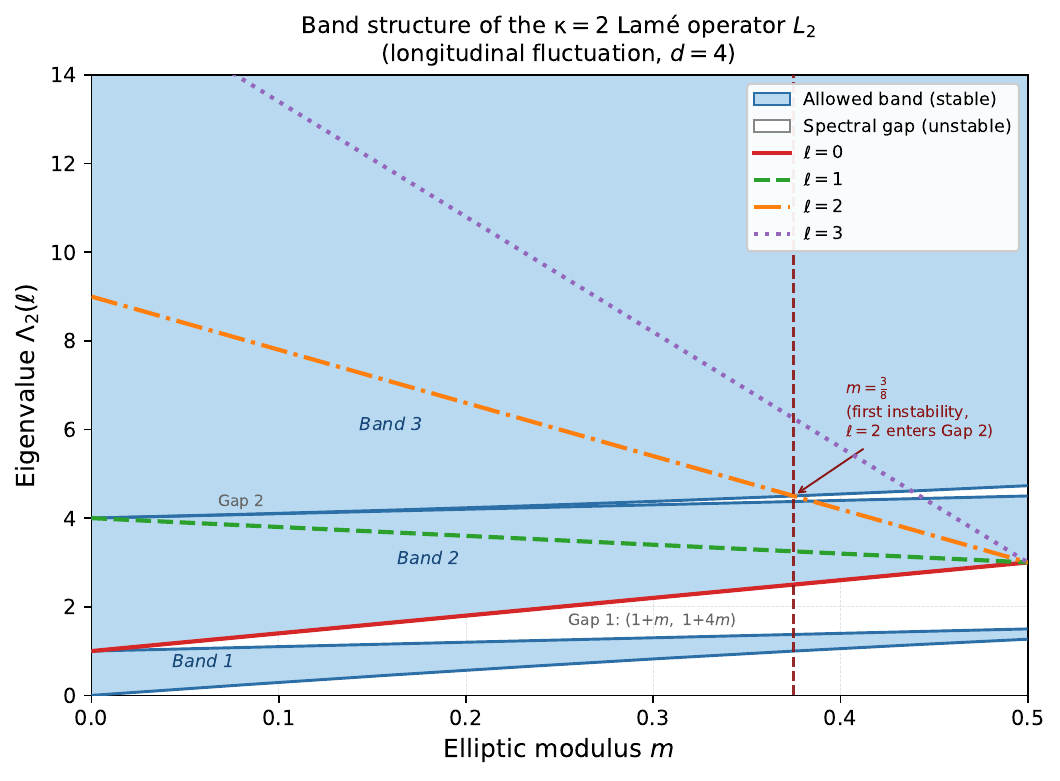}
  \caption{Band structure of the $\kappa=2$ (two-gap) Lam\'e operator $L_2$
  governing the longitudinal fluctuation in $d=4$.
  Three allowed bands (blue) separated by two spectral gaps (white).
  Coloured lines: eigenvalues $\Lambda_2(\ell)=6m+(1{-}2m)(1{+}\ell)^2$
  for $\ell=0,1,2,3$.
  The $\ell=0$ zero mode (red) runs along the lower edge of Band~2
  ($\Lambda_2(0)=1{+}4m$).
  The $\ell=2$ mode (orange) crosses into Gap~2 at $m=\tfrac{3}{8}$
  ($\lambda n\approx 50$), marking the onset of the first classical instability.}
  \label{fig:bandslame2}
\end{minipage}
\end{figure}

\subsection{Stability angles from Bloch solutions}

The stability angle $\nu_{\kappa,\ell}$ is defined via the monodromy of the Bloch solution $\xi(z+2K|m) = e^{i\nu_{\kappa,\ell}}\xi(z)$, which encodes the behavior over one period of the potential.

For $\kappa=1$, the two independent Bloch solutions are:
\begin{equation}
\xi_{\ell,\pm}(z) = \frac{H(z\pm\alpha\,|\,m)}{\Theta(z\,|\,m)}\,
e^{\mp z\,Z(\alpha\,|\,m)}\,,
\label{eq:Bloch_kappa1}
\end{equation}
where $H$, $\Theta$, $Z$ are the Jacobi Eta, Theta, and Zeta functions, and $\alpha$ solves:
\begin{equation}
\sn(\alpha\,|\,m) = \sqrt{\frac{1+m-\Lambda_1(\ell)}{m}}\,.
\label{eq:alpha_eq}
\end{equation}
From the periodicity properties of the Jacobi functions, the stability angles are:
\begin{equation}
\nu_{1,\ell} = -4i\,\KK(m)\,Z(\alpha\,|\,m)\,.
\label{eq:nu1}
\end{equation}

For $\kappa=2$, the Bloch solutions involve two parameters $\alpha_\pm$~\cite{Pawellek:2008st}:
\begin{equation}
\xi_{\ell,\pm}(z) = \frac{H(z\pm\alpha_+\,|\,m)\,H(z\pm\alpha_-\,|\,m)}{\Theta(z\,|\,m)^2}\,
e^{\mp z\left[Z(\alpha_+\,|\,m)+Z(\alpha_-\,|\,m)\right]}\,,
\label{eq:Bloch_kappa2}
\end{equation}
where $\alpha_\pm$ solve:
\begin{equation}
\sn^2(\alpha_\pm\,|\,m) = \frac{4(1+m)-\Lambda_2(\ell)}{6m}
\pm \frac{1}{2m}\sqrt{\frac{4}{3}(1-m+m^2)
- \frac{1}{3}\left(\Lambda_2(\ell) - 2(1+m)\right)^2}\,.
\label{eq:alpha_pm}
\end{equation}
The corresponding stability angles are:
\begin{equation}
\nu_{2,\ell} = 2\pi - 4i\,\KK(m)\left[Z(\alpha_+\,|\,m) + Z(\alpha_-\,|\,m)\right]\,.
\label{eq:nu2}
\end{equation}

A useful identity relating the Jacobi Zeta function to the complete elliptic integral of the third kind $\Pi$ is:
\begin{equation}
Z\!\left(\sn^{-1}\!\left(\frac{a}{\sqrt{m}}\,\Big|\,m\right)\,\Big|\,m\right)
= i\sqrt{1-a^2}\sqrt{1-\frac{m}{a^2}}
\left(1-\frac{\Pi(a^2\,|\,m)}{\KK(m)}\right)\,,
\label{eq:Z_Pi_identity}
\end{equation}
which allows practical evaluation of the stability angles without solving the transcendental equations \eqref{eq:alpha_eq} and \eqref{eq:alpha_pm} directly.

\subsection{Zero modes}\label{sec:zeromodes}

When a symmetry of the theory is broken by the classical configuration, the corresponding stability angle vanishes.

\paragraph{Time translation zero mode ($\nu_{2,0}=0$).}
For $\ell=0$ and $d=4$, we have $\Lambda_2(0)=1+4m$, which lies at a band edge of the $L_2$ Lam\'e operator. The equations for $\alpha_\pm$ reduce to:
\[
\sn^2(\alpha_+\,|\,m) = \frac{1}{m}\,, \qquad \sn^2(\alpha_-\,|\,m) = 0\,.
\]
Using the identity \eqref{eq:Z_Pi_identity}:
\[
-4i\KK(m)\,Z\!\left(\sn^{-1}(1/\sqrt{m}\,|\,m)\,|\,m\right) = 2\pi\,,
\qquad
Z(0\,|\,m) = 0\,,
\]
and therefore $\nu_{2,0} = 2\pi - 2\pi - 0 = 0$.

To verify this directly: since $v(t)$ satisfies the equation of motion $\ddot{v}+\mu^2 v+\lambda v^3=0$, differentiating with respect to $t$ gives $\dddot{v}+\mu^2\dot{v}+3\lambda v^2\dot{v}=0$, i.e., $\mathcal{O}_2\dot{v}=0$ (where the term $3\lambda v^2$ is the potential of $\mathcal{O}_2$). Thus $\dot{v}$ is indeed the zero eigenfunction of the radial fluctuation operator. This is the zero mode associated with time-translation symmetry.

\paragraph{$O(N)$ zero modes ($\nu_{1,0}=0$).}
For $\ell=0$ and $d=4$, $\Lambda_1(0)=1$, which lies at the band edge of the $\kappa=1$ Lam\'e operator. Since $\sn^{-1}(1\,|\,m)=\KK(m)$ and $Z(\KK(m)\,|\,m)=0$, we get $\nu_{1,0}=0$.

Physically these zero modes correspond to the $N-1$ rotations in $O(N)$ that rotate the classical configuration $(\phi_1,\ldots,\phi_N)=(v,0,\ldots,0)$ into a different direction in field space. These are the Goldstone bosons of the spontaneously broken $O(N)\to O(N-1)$ symmetry. The corresponding eigenfunctions are the $N-1$ constant fields $\tilde\phi_a = \mathrm{const}$, i.e., the $\ell=0$ transverse modes.

\subsection{Leading quantum correction: $C_1$}\label{sec:C1}

According to the general formulas \eqref{eq:C0_identification_boxed} and \eqref{eq:C1_identification_boxed}, the leading quantum correction is:
\begin{equation}
C_1 = R\,\delta E_1 + \frac{R}{2\mathcal{T}}
\sum_{\ell=0}^\infty \left(
n_\ell\left[(N{-}1)\nu_{1,\ell}+\nu_{2,\ell} - N\nu_{0,\ell}\right]
+ 2q_{1,\ell}\nu_{1,\ell} + 2q_{2,\ell}\nu_{2,\ell}
\right)\,,
\label{eq:C1_general}
\end{equation}
Here the occupation numbers $q_{\kappa,\ell}\in\mathbb{Z}_{\ge 0}$ label how many quanta of the mode $(\kappa,\ell)$ are excited above the vacuum. The $\frac{1}{2}\nu_{\kappa,\ell}$ terms are zero-point contributions from each mode. The free-theory subtraction (where $n_\ell = (2\ell+d-2)\Gamma(\ell+d-2)/[\Gamma(d-1)\Gamma(\ell+1)]$ is the eigenvalue degeneracy on the sphere) removes the contribution of $N$ free scalars, which is the reference theory. The renormalization contribution $R\,\delta E_1$ is computed in Section~\ref{sec:renorm}~\cite{arXiv:1606.09210,arXiv:1606.08598}.

Since $\cO_0$ has a static potential, its Bloch solutions are plane waves with $\nu_{0,\ell}=\mu\sqrt{A_\ell}\,\mathcal{T}$, and the free-theory contribution vanishes in dimensional regularization:
\[
\sum_{\ell=0}^\infty n_\ell\sqrt{A_\ell} = 0\,.
\]
Therefore:
\begin{equation}
C_1 = R\,\delta E_1 + \frac{R}{2\mathcal{T}}
\sum_{\ell=0}^\infty \left(
n_\ell\left[(N{-}1)\nu_{1,\ell}+\nu_{2,\ell}\right]
+ 2q_{1,\ell}\nu_{1,\ell} + 2q_{2,\ell}\nu_{2,\ell}
\right)\,.
\label{eq:C1_simplified}
\end{equation}

\subsection{Regularization of the sum over stability angles}

Each sum over $\ell$ is UV divergent and must be regularized. The strategy is:
\begin{enumerate}
\item Expand $n_\ell\nu_{\kappa,\ell}$ at large $\ell$ as $\sum_{k=1}^\infty c_k\,\ell^{d-k}$.
\item Subtract the first five divergent terms (which diverge in $d=4$), compute the finite sum directly in $d=4$.
\item Add back the subtracted terms in $\zeta$-regularized form, using $\sum_\ell \ell^{s} \to \zeta(-s)$.
\end{enumerate}

Concretely, the regularized sum for the $\kappa=2$ stability angles (starting from $\ell=1$ since $\nu_{2,0}=0$ is a zero mode) gives:
\begin{align}
\frac{1}{2}\sum_{\ell=1}^\infty n_\ell\nu_{2,\ell}
&= -\frac{3\left[(m{-}1)(3m{-}2)\KK(m)+(4m{-}2)\ELE(m)\right]}{(1-2m)^{3/2}\epsilon}
\nonumber\\
&\quad + \frac{1}{2}\sum_{\ell=2}^\infty \sigma_2(\ell)
\nonumber\\
&\quad - \frac{((29{-}5m)m{-}14)\KK(m) - 5\pi(1{-}2m)^{3/2} + (24{-}48m)\ELE(m)}{(1-2m)^{3/2}}
+ \mathcal{O}(\epsilon)\,,
\label{eq:sum_nu2}
\end{align}
where $\sigma_2(\ell) = n_\ell\nu_{2,\ell}\big|_{d=4}$ minus its large-$\ell$ asymptotic expansion (ensuring convergence). Similarly, the sum for $\kappa=1$ yields:
\begin{align}
\frac{1}{2}\sum_{\ell=0}^\infty n_\ell\nu_{1,\ell}
&= -\frac{(m{-}1)(3m{-}2)\KK(m)+(4m{-}2)\ELE(m)}{3(1-2m)^{3/2}\epsilon}
\nonumber\\
&\quad + \pi + \frac{2(m\KK(m)-\ELE(m))}{\sqrt{1-2m}}
+ \frac{1}{2}\sum_{\ell=1}^\infty \sigma_1(\ell)
+ \mathcal{O}(\epsilon)\,.
\label{eq:sum_nu1}
\end{align}

The $1/\epsilon$ poles in eqs.~\eqref{eq:sum_nu2} and \eqref{eq:sum_nu1} are proportional to $s(m)/(1-2m)^{3/2}$, where $s(m)$ is defined in Section~\ref{sec:renorm}. Their residues are
\[
[\text{pole in }\eqref{eq:sum_nu1}]=-\frac{s(m)}{\epsilon}\,,
\qquad
[\text{pole in }\eqref{eq:sum_nu2}]=-\frac{9\,s(m)}{\epsilon}\,.
\]
Because the half from the zero-point energy is already contained in the
left-hand sides of \eqref{eq:sum_nu1}--\eqref{eq:sum_nu2}, the total fluctuation
pole is obtained by weighting these with the mode multiplicities ($N-1$
transverse, one longitudinal) \emph{without any further factor of $\tfrac12$}:
\[
(N-1)\times[\text{pole in }\eqref{eq:sum_nu1}] + [\text{pole in }\eqref{eq:sum_nu2}]
= -(N+8)\,\frac{s(m)}{\epsilon}
= -8\pi^2\beta_0\, s(m)\cdot\frac{1}{\epsilon}\,,
\]
where $\beta_0=(N+8)/8\pi^2$. This precisely cancels the pole in eq.~\eqref{eq:pole_from_renorm}. This cancellation is a non-trivial consistency check: it confirms that the combination $\mathcal{S}_{\rm cl}^{\rm ren} - \frac{1}{2}\sum_{\kappa,\ell} n_\ell\nu_{\kappa,\ell}$ is UV-finite, as required for the scaling dimension to be well-defined~\cite{arXiv:1309.5621}.

\subsubsection{The descendant mode $\nu_{2,1}$}
\label{subsubsec:descendant}

An important observation is that the $\ell=1$ mode of the $\kappa=2$ operator has stability angle:
\begin{equation}
\nu_{2,1} = \frac{\mathcal{T}}{R}\,,
\end{equation}
which is \emph{proportional to the period} (and equals $\mathcal{T}$ since $R=1$). According to the corrected energy formula \eqref{eq:energy_final_boxed_new}, in which the excitation term is $(q_\ell+n_\ell/2)\nu_\ell/\mathcal{T}$ (with $q_\ell$ the \emph{total} occupation of multiplet $\ell$), exciting the $\ell=1$ channel of the $\kappa=2$ operator by $q_{2,1}=1$ quantum shifts the energy by $\delta E = \nu_{2,1}/\mathcal{T} = 1/R$, hence shifts the scaling dimension by
\[
  \delta\Delta = R\,\delta E = \frac{R}{R} = 1\,.
\]
This is precisely the unit shift expected from acting with $\partial_\mu$ on the primary, confirming that the $\ell=1$ mode with $\nu_{2,1}=\mathcal{T}/R$ corresponds to the \textbf{descendant channel}. Consequently:
\begin{itemize}
\item A necessary condition for a state to be primary is $q_{2,1}=0$ (no quanta in the spin-1 channel).
\item Descendants are generated by increasing $q_{2,1}$: the state with $q_{2,1}=k$ is a level-$k$ descendant with $\delta\Delta=k$ relative to the primary.
\end{itemize}

\subsubsection{Final result for $C_1$}

Collecting all contributions and taking the $\epsilon\to 0$ limit:
\begin{align}
C_1 &= \frac{1}{8(1-2m)^2\KK(m)}\bigg[
\KK(m)\Big(m(-7mN+26m+3N-70)+28\Big)
\nonumber\\
&\quad + 2\pi(1-2m)^{3/2}(N+4)
+ 4(2m-1)(N+11)\ELE(m)\bigg]
\nonumber\\
&\quad + \frac{1}{8\sqrt{1-2m}\KK(m)}\left(
\sum_{\ell=2}^\infty \sigma_2(\ell)
+ (N{-}1)\sum_{\ell=1}^\infty \sigma_1(\ell)
+ 2\sum_{\ell=1}^\infty \Big(q_{1,\ell}\nu_{1,\ell}+q_{2,\ell}\nu_{2,\ell}\Big)\bigg|_{d=4}
\right)\,.
\label{eq:C1_final}
\end{align}
This is the main technical result  of \cite{Antipin:2025ilv,Antipin:2025rsr} for the $O(N)$ CFT. The sums over $\ell$ can be evaluated numerically for any $m(\lambda n)$ or analytically in limiting regimes.

\subsection{Perturbative semiclassics}\label{sec:perturbative}

We now expand $C_1$ at small $\lambda n$ (equivalently small $m$) to recover the perturbative $\epsilon$-expansion and generate new predictions.

The small-$m$ expansion is organized as a power series in $\lambda n/(16\pi^2)$, which is the natural loop-counting parameter. At the Wilson-Fisher fixed point $\lambda_* = 8\pi^2\epsilon/(N+8) + \ldots$, this becomes $\lambda_* n/(16\pi^2) \approx \epsilon n/(2(N+8))$. The double-scaling limit of Chapter~\ref{chap:doublescaling} holds when both $n\to\infty$ and $\epsilon\to 0$ with $\epsilon n$ fixed.

\subsubsection{Small-$m$ expansion of the stability angle sums}

Expanding the finite sums:
\begin{align}
\frac{1}{2}\sum_{\ell=1}^\infty \sigma_1(\ell)
&= \frac{11\pi m^2}{64} + \mathcal{O}(m^3)
= 11\pi\left(\frac{\lambda n}{16\pi^2}\right)^2 + \mathcal{O}\!\left(\left(\frac{\lambda n}{16\pi^2}\right)^3\right)\,,
\\[4pt]
\frac{1}{2}\sum_{\ell=2}^\infty \sigma_2(\ell)
&= \frac{51\pi m^2}{64} + \mathcal{O}(m^3)
= 51\pi\left(\frac{\lambda n}{16\pi^2}\right)^2 + \mathcal{O}\!\left(\left(\frac{\lambda n}{16\pi^2}\right)^3\right)\,.
\end{align}

The individual stability angle contributions, for $\ell\ge1$, expand as
(the $\ell=0$ channels are the exact zero modes $\nu_{1,0}=\nu_{2,0}=0$ and are
excluded):
\begin{align}
\frac{R\,\nu_{1,\ell}}{\mathcal{T}}\bigg|_{d=4}
&= \ell - \frac{2(3\ell+1)}{\ell+1}\left(\frac{\lambda n}{16\pi^2}\right)
\nonumber\\
&\quad - \left(\frac{20}{\ell+1} + \frac{8}{(\ell+1)^3}
+ \frac{2}{\ell+2} + \frac{2}{\ell} - 51\right)
\left(\frac{\lambda n}{16\pi^2}\right)^2
+ \mathcal{O}\!\left(\left(\frac{\lambda n}{16\pi^2}\right)^3\right)\,,
\label{eq:nu1_small}
\end{align}
\begin{align}
\frac{R\,\nu_{2,\ell}}{\mathcal{T}}\bigg|_{d=4}
&= \ell - \frac{6(\ell-1)}{\ell+1}\left(\frac{\lambda n}{16\pi^2}\right)
\nonumber\\
&\quad - \left(\frac{36}{\ell+1} + \frac{72}{(\ell+1)^3}
+ \frac{18}{\ell+2} + \frac{18}{\ell} - 51\right)
\left(\frac{\lambda n}{16\pi^2}\right)^2
+ \mathcal{O}\!\left(\left(\frac{\lambda n}{16\pi^2}\right)^3\right)\,.
\label{eq:nu2_small}
\end{align}

As a consistency check, at $\ell=1$ one finds from \eqref{eq:nu2_small} that $R\nu_{2,1}/\mathcal{T} = 1 + 0\cdot(\lambda n/16\pi^2) + \ldots = 1$, confirming that the $\ell=1$ longitudinal mode produces descendants with $\Delta\to\Delta+1$.

\subsection{Full scaling dimension at NLO}

Collecting all contributions and evaluating at the fixed point \eqref{eq:WFfixedpoint}, the scaling dimension in the perturbative limit reads:
\begin{align}
\Delta_{n,q_\ell}
&= nC_0(\lambda_* n) + C_1(\lambda_* n) + \mathcal{O}(1/n)
\nonumber\\[6pt]
&= n\left(1-\frac{\epsilon}{2}\right) + \sum_{\ell=1}^\infty(q_{1,\ell}+q_{2,\ell})\ell
\nonumber\\[4pt]
&\quad + \left[\frac{3n^2}{2(N+8)}
- \left(\frac{4-N}{2(N+8)}
+ \sum_{\ell=1}^\infty \frac{q_{1,\ell}(1+3\ell)+3q_{2,\ell}(\ell-1)}{(\ell+1)(N+8)}\right)n
+ \mathcal{O}(n^0)\right]\epsilon
\nonumber\\[4pt]
&\quad + \bigg[-\frac{17n^3}{4(N+8)^2}
+ \bigg(\frac{-11N^2+10N+604}{4(N+8)^3}
\nonumber\\
&\qquad + \sum_{\ell=1}^\infty \frac{q_{1,\ell}\left(3\ell(\ell+2)(\ell(\ell(17\ell+43)+35)+5)-4\right)}
{4\ell(\ell+1)^3(\ell+2)(N+8)^2}
\nonumber\\
&\qquad + \sum_{\ell=1}^\infty \frac{3(\ell-1)\left(\ell(\ell(\ell(17\ell+78)+135)+98)+12\right)q_{2,\ell}}
{4\ell(\ell+1)^3(\ell+2)(N+8)^2}
\bigg)n^2 + \mathcal{O}(n)\bigg]\epsilon^2
+ \mathcal{O}(\epsilon^3)\,.
\label{eq:Delta_full}
\end{align}
This is the central result of this section. It provides a \emph{unified formula} for the scaling dimensions of all traceless symmetric Lorentz operators in the $O(N)$ singlet sector, parametrized by two sets of non-negative integers $\{q_{1,\ell}\}$ and $\{q_{2,\ell}\}$. The NLO result for the Ising model ($N=1$) was first presented in~\cite{Antipin:2025ilv}. The perturbative strategy of combining the semiclassical NLO result with available multi-loop data to fix subleading coefficients has been systematically employed in~\cite{Antipin:2025ilv,Bednyakov:2023iuj}.

\subsection{Examples: identifying operators from quantum numbers}

\subsubsection{Ising CFT ($N=1$)}

For $N=1$, only the $q_{2,\ell}$ quantum numbers appear (since there are no transverse modes). Setting $q_{2,\ell}=0$ gives the ground state $\phi^n$:
\begin{equation}
\Delta_n = n\left(1-\frac{\epsilon}{2}\right)
+ \frac{n(n-1)}{6}\epsilon
- \frac{n^2(17n-67)}{324}\epsilon^2
+ \mathcal{O}(\epsilon^2 n,\, \epsilon^3)\,.
\end{equation}

The first excited state $q_{2,\ell}=\delta_{\ell,2}$ corresponds to $\partial^2\phi^n$ (spin-2):
\begin{equation}
\Delta_{n,\delta_{\ell,2}} = 2 + n\left(1-\frac{\epsilon}{2}\right)
+ \frac{n(3n-5)}{18}\epsilon
- \frac{n^2(102n-539)}{1944}\epsilon^2
+ \mathcal{O}(\epsilon n^0,\,\epsilon^2 n,\,\epsilon^3)\,.
\end{equation}

Exciting the same mode twice, $q_{2,\ell}=2\delta_{\ell,2}$, gives spin-4, spin-2, and spin-0 operators {together with a mixed-symmetry $[2,2]$ multiplet (in $d=4$: $45=25+9+1+10$)}, with degenerate scaling dimension (the NLO degeneracy is lifted at NNLO):
\begin{equation}
\Delta_{n,2\delta_{\ell,2}} = 4 + n\left(1-\frac{\epsilon}{2}\right)
+ \frac{n(3n-7)}{18}\epsilon
- \frac{n^2(51n-338)}{972}\epsilon^2
+ \mathcal{O}(\epsilon n^0,\,\epsilon^2 n,\,\epsilon^3)\,.
\end{equation}

These expressions agree with the known 2-loop $\epsilon$-expansion results for the Ising model when expanded at fixed $n$. For instance, at $n=2$: $\Delta_2 = 2(1-\epsilon/2) + (2/6)\epsilon - (4\cdot 17-67\cdot 2)/162\cdot\epsilon^2 + \ldots$, matching the known anomalous dimension of $\phi^2$.

\subsubsection{General $O(N)$ model}

The ground state is the singlet operator $(\phi_a\phi_a)^{n/2}$ with $q_{1,\ell}=q_{2,\ell}=0$:
\begin{equation}
\Delta_n = n\left(1-\frac{\epsilon}{2}\right)
+ \frac{(3n+N-4)n}{2(N+8)}\epsilon
+ \frac{n^2(-17n(N+8)+N(10-11N)+604)}{4(N+8)^3}\epsilon^2
+ \mathcal{O}(\epsilon^2 n,\, \epsilon^3)\,.
\end{equation}

Note the $N$-dependence enters first at order $\epsilon$ through the term $N-4$ in the numerator: for $N<4$ this is a negative correction (the WF fixed point shifts operators down), while for $N>4$ it is positive. At $N=4$ (Higgs sector) the $\mathcal{O}(\epsilon)$ coefficient of $n$ vanishes, reflecting an enhanced symmetry at that point.

For $N>1$, the $\ell=1$ transverse mode has $\nu_{1,1}\neq \mathcal{T}/R$, so exciting it yields a \emph{primary} (not a descendant).
 {A single such excitation, $q_{1,\ell}=\delta_{\ell,1}$, furnishes a spin-1
candidate. Were a spin-1 singlet of dimension $\Delta=d-1$ to appear it would be a
conserved \emph{virial current}, whose presence would signal a theory that is
scale- but not conformally invariant. No such operator arises here: the leading
candidate of this type, $\partial\Box\phi^6$, sits at $\Delta=9+O(\epsilon)\gg d-1$,
so full conformal invariance is maintained~\cite{Antipin:2025rsr}.}
Exciting it twice, $q_{1,\ell}=2\delta_{\ell,1}$, gives spin-2 and spin-0 operators ($\partial^2\phi^n$ and $\Box\phi^n$):
\begin{align}
\Delta_{n,2\delta_{\ell,1}} &= 2 + n\left(1-\frac{\epsilon}{2}\right)
+ \frac{3n^2+(N-12)n}{2(N+8)}\epsilon
\nonumber\\
&\quad - \frac{\epsilon^2 n^2(51n(N+8)+33N^2-254N-3604)}{12(N+8)^3}
+ \mathcal{O}(\epsilon n^0,\,\epsilon^2 n,\,\epsilon^3)\,.
\end{align}

\subsubsection{The spin tower and the NLO spectrum table}
\label{subsubsec:spectrum_table}

Exciting a single $\ell=s$ longitudinal mode ($q_{2,\ell}=\delta_{\ell,s}$)
produces the operator $\partial^s\phi^n$ in the single, non-degenerate
$(s/2,s/2)$ Lorentz representation. Its scaling dimension follows in closed form
from \eqref{eq:Delta_full}. Writing
$\Delta_{n,\delta_{\ell,s}} = n\!\left(1-\tfrac{\epsilon}{2}\right) + s
+ \gamma_{n,\delta_{\ell,s}}$, the anomalous part is, for general $N$,
\begin{equation}
\gamma_{n,\delta_{\ell,s}}
= \frac{n\left[\,3n+N-\dfrac{10s-2}{s+1}\,\right]}{2(N+8)}\,\epsilon
+ \frac{n^2\left[\,{-}17n(N+8)+N(10-11N)+604+R(s)\,(N+8)\,\right]}{4(N+8)^3}\,\epsilon^2,
\label{eq:spintower_ON}
\end{equation}
with
\[
R(s)=\frac{3(s-1)\left(17s^4+78s^3+135s^2+98s+12\right)}{s(s+1)^3(s+2)}\,.
\]
Equation~\eqref{eq:spintower_ON} is the $O(N)$ generalisation of the spin-tower
formula of~\cite{Antipin:2025ilv} (recovered at $N=1$); for $s=2$ its $N=1$
limit, combined with multi-loop data, fixes the full two-loop dimension of the
$\partial^2\phi^n$ tower~\cite{Henriksson:2025vyi},  {in which the
anomalous part carries an overall $(n-2)$ factor}. At $n=2,\,s=2$ \ {this
anomalous part therefore vanishes and} one recovers the conserved stress tensor with
$\Delta_{T_{\mu\nu}}=d=4-\epsilon$ {; the displayed large-$n$ formula
\eqref{eq:spintower_ON} reproduces this only up to the $O(n^0)$ constant supplied by
the perturbative matching}.

Table~\ref{tab:ON_spectrum} collects the NLO anomalous dimensions
$\gamma_{n,q_\ell}$ for the lowest operators, organised by the occupation
pattern $\{q_{2,\ell}\}$. Single-mode rows give the spin tower
$\partial^s\phi^n$ (here extended to $s=10$); multi-mode rows give the degenerate
multiplets obtained from the $SO(3,1)$ tensor-product decomposition, with the
schematic operator content and (in parentheses) the multiplicity of operators
sharing the listed form. Each entry is reproduced from the semiclassical master
formula \eqref{eq:Delta_full}; only terms surviving at NLO are shown
($\mathcal{O}(\epsilon n^0)$ and $\mathcal{O}(\epsilon^2 n)$ are dropped).

\begin{table}[ht]
\centering
\renewcommand{\arraystretch}{1.6}
\resizebox{\textwidth}{!}{%
\begin{tabular}{|c|l|l|}
\hline
$q_{2,\ell}$ & Operators & Anomalous dimension $\gamma_{n,q_\ell}$ \\ \hline
$0$ & $\phi^n$ &
$\dfrac{n\epsilon(3n+N-4)}{2(N+8)}+\dfrac{n^2\epsilon^2\big({-}17n(N+8)+N(10-11N)+604\big)}{4(N+8)^3}$ \\ \hline
$\delta_{\ell,2}$ & $\partial^2\phi^n$ &
$\dfrac{n\epsilon(3n+N-6)}{2(N+8)}+\dfrac{n^2\epsilon^2\big({-}102n(N+8)+N(197-66N)+4720\big)}{24(N+8)^3}$ \\
$\delta_{\ell,3}$ & $\partial^3\phi^n$ &
$\dfrac{n\epsilon(3n+N-7)}{2(N+8)}+\dfrac{n^2\epsilon^2\big({-}680n(N+8)+N(1651-440N)+34168\big)}{160(N+8)^3}$ \\
$\delta_{\ell,4}$ & $\partial^4\phi^n$ &
$\dfrac{n\epsilon(15n+5N-38)}{10(N+8)}+\dfrac{n^2\epsilon^2\big({-}4250n(N+8)+N(11431-2750N)+222448\big)}{1000(N+8)^3}$ \\
$\delta_{\ell,5}$ & $\partial^5\phi^n$ &
$\dfrac{n\epsilon(3n+N-8)}{2(N+8)}+\dfrac{n^2\epsilon^2\big({-}1785n(N+8)+N(5092-1155N)+95756\big)}{420(N+8)^3}$ \\
$\delta_{\ell,6}$ & \textcolor{red}{$\partial^6\phi^n$} &
$\dfrac{n\epsilon(21n+7N-58)}{14(N+8)}+\dfrac{n^2\epsilon^2\big({-}23324n(N+8)+N(69145-15092N)+1272088\big)}{5488(N+8)^3}$ \\
$\delta_{\ell,7}$ & \textcolor{red}{$\partial^7\phi^n$} &
$\dfrac{n\epsilon(6n+2N-17)}{4(N+8)}+\dfrac{n^2\epsilon^2\big({-}7616n(N+8)+N(23201-4928N)+420360\big)}{1792(N+8)^3}$ \\
$\delta_{\ell,8}$ & \textcolor{red}{$\partial^8\phi^n$} &
$\dfrac{n\epsilon(9n+3N-26)}{6(N+8)}+\dfrac{n^2\epsilon^2\big({-}27540n(N+8)+N(85619-17820N)+1533832\big)}{6480(N+8)^3}$ \\
$\delta_{\ell,9}$ & $\partial^9\phi^n$ &
$\dfrac{n\epsilon(15n+5N-44)}{10(N+8)}+\dfrac{n^2\epsilon^2\big({-}23375n(N+8)+N(73826-15125N)+1311108\big)}{5500(N+8)^3}$ \\
$\delta_{\ell,10}$ & $\partial^{10}\phi^n$ &
$\dfrac{n\epsilon(33n+11N-98)}{22(N+8)}+\dfrac{n^2\epsilon^2\big({-}226270n(N+8)+N(723707-146410N)+12764096\big)}{53240(N+8)^3}$ \\ \hline
$2\delta_{\ell,2}$ & $\partial^4\phi^n,\ \partial^2\Box\phi^n,\ \Box^2\phi^n$ &
$\dfrac{n\epsilon(3n+N-8)}{2(N+8)}+\dfrac{n^2\epsilon^2\big({-}51n(N+8)+N(167-33N)+2908\big)}{12(N+8)^3}$ \\
$\delta_{\ell,2}+\delta_{\ell,3}$ & $\partial^5\phi^n,\ \partial^3\Box\phi^n,\ \partial\Box^2\phi^n$ &
$\dfrac{n\epsilon(3n+N-9)}{2(N+8)}+\dfrac{n^2\epsilon^2\big({-}2040n(N+8)+N(7693-1320N)+124424\big)}{480(N+8)^3}$ \\
$\delta_{\ell,2}+\delta_{\ell,4}$ & \textcolor{red}{$\partial^6\phi^n$}$,\ \partial^4\Box\phi^n,\ \partial^2\Box^2\phi^n$ &
$\dfrac{n\epsilon(15n+5N-48)}{10(N+8)}+\dfrac{n^2\epsilon^2\big({-}6375n(N+8)+N(25709-4125N)+402172\big)}{1500(N+8)^3}$ \\
$2\delta_{\ell,3}$ & \textcolor{red}{$\partial^6\phi^n$}$,\ \partial^4\Box\phi^n,\ \partial^2\Box^2\phi^n,\ \Box^3\phi^n$ &
$\dfrac{n\epsilon(3n+N-10)}{2(N+8)}+\dfrac{n^2\epsilon^2\big({-}340n(N+8)+N(1451-220N)+22088\big)}{80(N+8)^3}$ \\
$3\delta_{\ell,2}$ & \textcolor{red}{$\partial^6\phi^n$}$,\ \partial^4\Box\phi^n\,(2),\ \partial^2\Box^2\phi^n\,(3),\ \Box^3\phi^n$ &
$\dfrac{n\epsilon(3n+N-10)}{2(N+8)}+\dfrac{n^2\epsilon^2\big({-}34n(N+8)+N(157-22N)+2304\big)}{8(N+8)^3}$ \\ \hline
$\delta_{\ell,2}+\delta_{\ell,5}$ & \textcolor{red}{$\partial^7\phi^n$}$,\ \partial^5\Box\phi^n,\ \partial^3\Box^2\phi^n$ &
$\dfrac{n\epsilon(3n+N-10)}{2(N+8)}+\dfrac{n^2\epsilon^2\big({-}1190n(N+8)+N(4993-770N)+76624\big)}{280(N+8)^3}$ \\
$\delta_{\ell,3}+\delta_{\ell,4}$ & \textcolor{red}{$\partial^7\phi^n$}$,\ \partial^5\Box\phi^n,\ \partial^3\Box^2\phi^n,\ \partial\Box^3\phi^n$ &
$\dfrac{n\epsilon(15n+5N-53)}{10(N+8)}+\dfrac{n^2\epsilon^2\big({-}17000n(N+8)+N(76999-11000N)+1139992\big)}{4000(N+8)^3}$ \\
$\delta_{\ell,2}+\delta_{\ell,6}$ & \textcolor{red}{$\partial^8\phi^n$}$,\ \partial^6\Box\phi^n,\ \partial^4\Box^2\phi^n$ &
$\dfrac{n\epsilon(21n+7N-72)}{14(N+8)}+\dfrac{n^2\epsilon^2\big({-}69972n(N+8)+N(301417-45276N)+4568120\big)}{16464(N+8)^3}$ \\
$\delta_{\ell,3}+\delta_{\ell,5}$ & \textcolor{red}{$\partial^8\phi^n$}$,\ \partial^6\Box\phi^n,\ \partial^4\Box^2\phi^n,\ \partial^2\Box^3\phi^n$ &
$\dfrac{n\epsilon(3n+N-11)}{2(N+8)}+\dfrac{n^2\epsilon^2\big({-}14280n(N+8)+N(67007-9240N)+976216\big)}{3360(N+8)^3}$ \\ \hline
\end{tabular}}
\caption{NLO ($\mathcal{O}(\epsilon^2)$) anomalous dimensions $\gamma_{n,q_\ell}$
for traceless-symmetric Lorentz operators of the $O(N)$ CFT, labelled by the
longitudinal occupation pattern $\{q_{2,\ell}\}$ and obtained from the
semiclassical master formula \eqref{eq:Delta_full}; the full dimension is
$\Delta_{n,q_\ell}=n(1-\tfrac{\epsilon}{2})+\sum_\ell q_{2,\ell}\,\ell
+\gamma_{n,q_\ell}$. The single-mode block is the spin tower
$\partial^s\phi^n$ of \eqref{eq:spintower_ON}, shown here through $s=10$. The
multi-mode block lists, for each total derivative level $L=\sum_\ell q_{2,\ell}\,\ell$,
the degenerate fully-symmetric multiplet $\partial^{s}\Box^{(L-s)/2}\phi^n$
obtained from the $SO(3,1)\!\cong\!SU(2)\!\times\!SU(2)$ decomposition of
$\bigotimes_\ell\mathrm{Sym}^{q_{2,\ell}}\!\big[(\tfrac{\ell}{2},\tfrac{\ell}{2})\big]$;
parenthesised numbers give the multiplicity of operators of identical schematic
form. \textcolor{red}{Red} marks the pure operator $\partial^{L}\phi^n$, which
recurs across every pattern of a given level $L$ with a \emph{different}
dimension --- a visual reminder that the schematic form alone does not fix the
operator, the occupation pattern does. The rows $s=2$--$6$ and the
$L=4,5,6$ multi-mode patterns reproduce Ref.~\cite{Antipin:2025ilv}; the
$s=7$--$10$ tower rows, the $L=7,8$ distinct-mode rows
($\delta_{\ell,2}+\delta_{\ell,5}$, $\delta_{\ell,3}+\delta_{\ell,4}$,
$\delta_{\ell,2}+\delta_{\ell,6}$, $\delta_{\ell,3}+\delta_{\ell,5}$), and the
closed form \eqref{eq:spintower_ON} extend it. Repeated-mode patterns at
$L\ge7$ (e.g.\ $2\delta_{\ell,2}+\delta_{\ell,3}$, $2\delta_{\ell,4}$) also carry
mixed-symmetry multiplets and are omitted. NNLO/higher terms
($\mathcal{O}(\epsilon n^0)$ at one loop, $\mathcal{O}(\epsilon^2 n)$ at two
loops) are not displayed.}
\label{tab:ON_spectrum}
\end{table}

\subsubsection{Operator construction rules}

The connection between occupation numbers and operators is:
\begin{itemize}
\item Exciting the $\ell^*$ mode adds $\ell^*$ derivatives in the $(\ell^*/2,\,\ell^*/2)$ Lorentz representation.
\item The total number of derivatives satisfies $\sum_\ell(q_{1,\ell}+q_{2,\ell})\ell = 2p + s$.
\item Multiple excitations of the same mode require taking tensor products of the corresponding representations, leading to multiplets of operators with degenerate NLO scaling dimensions.
\item For example, the $\ell=2$ longitudinal excitation $q_{2,2}=1$ adds a symmetric traceless spin-2 tensor: the corresponding operator is $\phi_a\phi_a\,T_{\mu\nu}\phi_b\phi_b$ (schematically), which is a spin-2 primary with scaling dimension $\Delta_{n,\delta_{\ell,2}}$ given above.
\item The $q_{2,1}$ mode produces descendants; hence $q_{2,1}=0$ is necessary for primaries.
\item Remaining degeneracies are generically lifted at NNLO and higher orders.
\end{itemize}

\subsubsection{{Explicit Lorentz decompositions and the hyperfine analogy}}
{
The schematic content listed in Table~\ref{tab:ON_spectrum} follows from a single rule
of $\mathfrak{so}(3,1)\cong\mathfrak{su}(2)\oplus\mathfrak{su}(2)$ representation theory:
a quantum in the mode $\ell$ carries the traceless-symmetric rank-$\ell$ representation
$(\tfrac{\ell}{2},\tfrac{\ell}{2})$, and a multi-mode excitation lives in the
(symmetrised) tensor product of the corresponding factors; decomposing that product
into irreducibles assigns each piece a definite spin and derivative structure. Three
cases make the dictionary explicit.}

 {\emph{Two $\ell=2$ quanta} ($q_{2,\ell}=2\delta_{\ell,2}$): the content is
$(1,1)\otimes(1,1)$, and with $1\otimes1=2\oplus1\oplus0$ in each $\mathfrak{su}(2)$
factor the diagonal entries $(2,2)\oplus(1,1)\oplus(0,0)$ are the traceless-symmetric
operators $\partial^4\phi^n$ (spin $4$), $\partial^2\Box\phi^n$ (spin $2$) and
$\Box^2\phi^n$ (spin $0$), while the off-diagonal entries $(2,1)\oplus(1,2)$,
$(2,0)\oplus(0,2)$ and $(1,0)\oplus(0,1)$ form mixed-symmetry multiplets --- all at the
single NLO dimension $\Delta_{n,2\delta_{\ell,2}}$.}

 {\emph{One $\ell=2$ and one $\ell=3$ quantum} ($q_{2,\ell}=\delta_{\ell,2}+\delta_{\ell,3}$):
the product $(1,1)\otimes(\tfrac32,\tfrac32)$ with $1\otimes\tfrac32=\tfrac52\oplus\tfrac32\oplus\tfrac12$
gives the diagonal tower $(\tfrac52,\tfrac52)\oplus(\tfrac32,\tfrac32)\oplus(\tfrac12,\tfrac12)$,
i.e.\ $\partial^5\phi^n$, $\partial^3\Box\phi^n$ and $\partial\Box^2\phi^n$.}

 {\emph{Three $\ell=2$ quanta} ($q_{2,\ell}=3\delta_{\ell,2}$): now
$1\otimes1\otimes1=3\oplus(2)^{2}\oplus(1)^{3}\oplus0$, so the spin-$6,4,2,0$ operators
$\partial^6\phi^n,\ \partial^4\Box\phi^n,\ \partial^2\Box^2\phi^n,\ \Box^3\phi^n$ appear
with multiplicities $1,2,3,1$, exactly as in the last $L=6$ row of
Table~\ref{tab:ON_spectrum}.}

{\emph{Degeneracy and its lifting --- a hyperfine analogy.} At leading order every
operator built on a given saddle shares the classical dimension $nC_0$, an exact
degeneracy that had been observed diagrammatically~\cite{Antipin:2025ilv} and is here
immediate, since all these operators are excitations of the \emph{same} classical
orbit. The NLO fluctuation sum resolves them into the Lorentz multiplets labelled by
$\{q_\ell\}$, lifting the degeneracy much as relativistic and spin--orbit corrections
split a hydrogenic level into fine and hyperfine multiplets; higher semiclassical
orders split the residual multiplets further. Because each $|n,\{q_\ell\}\rangle$ is
built as an energy eigenstate on the cylinder, the construction returns dilatation
eigenoperators directly: no operator-mixing problem need be solved, in contrast to the
diagrammatic approach where nearly degenerate operators must first be disentangled.}

\subsection{Instabilities at large $\lambda n$: classical scars}\label{sec:instabilities}

The stability angles $\nu_{2,\ell}$ become complex when $\Lambda_2(\ell)$ falls in a gap of the $L_2$ Lam\'e operator, i.e., when:
\begin{equation}
\sqrt{\frac{4-5m}{1-2m}} \le \ell+1 \le \sqrt{2}\sqrt{1+\frac{\sqrt{1-m+m^2}}{1-2m}}\,.
\end{equation}
For small $\lambda n$ (small $m$), no integer $\ell$ satisfies this inequality, and the orbit is stable.
As $\lambda n$ increases, the first instability appears for the $\ell=2$ mode at $m=3/8$ ($\lambda n\approx 50$).

The instability survives in the large-$N$ limit because the $\nu_{1,\ell}$ modes become complex when
\[
\ell \le \frac{1-2m-\sqrt{2m^2-3m+1}}{2m-1}\,,
\]
and these modes have multiplicity $N{-}1\sim N$. The first complex $\nu_{1,\ell}$ appears at $m=3/7$, corresponding to $\lambda n \approx 128$.

These unstable periodic orbits that nonetheless leave a strong imprint on the energy eigenstate wavefunctions are known as \textbf{classical scars}. In the large-$\lambda n$ limit ($m\to 1/2$), the instability produces an imaginary part in the energy, related to the decay rate of the state.

Physically, these instabilities arise because as the amplitude $x_0$ grows (large $\lambda n$), the anharmonic potential becomes so curved that some fluctuation modes experience parametric resonance over one oscillation period. The classical scar terminology comes from quantum chaos: the periodic orbit leaves an imprint on the wavefunction even though it is classically unstable.

Remarkably, the NLO correction modifies the large-$n$ behavior of the scaling dimension from $\Delta_n \sim n^{4/3}$ to
\begin{equation}
\Delta_n \sim n^{d/(d-1)}\,,
\end{equation}
mimicking the generic nonperturbative expectation for the scaling dimension of the lowest operator of degree $n$ in CFTs with a heavy-operator sector~\cite{arXiv:1702.04148}.

\section{Benchmarking the spectrum: comparison with other methodologies}
\label{sec:comparison}

The decisive test of the semiclassical framework is the physically relevant
three-dimensional Ising CFT, reached by setting $N=1$ and extrapolating to
$\epsilon=1$. There, several complementary methodologies are available for the
scaling dimensions $\Delta_n$ of the lowest scalar operators $\phi^n$: the
numerical conformal bootstrap~\cite{Simmons-Duffin:2016wlq,Henriksson:2022gpa,Kos:2013tga},
the most accurate tool at small $n$ but so far carried out only for
$n=1,2,4,5,6$ (there is no $n=3$ primary); the resummed $\epsilon$-expansion,
built on multi-loop diagrammatic data~\cite{Kompaniets:2017yct,arXiv:1606.08598,Henriksson:2025hwi}
together with Borel/hypergeometric resummation~\cite{Shalaby:2020xvv}; and the
semiclassical expansion at LO~\cite{Antipin:2024ekk} and NLO (this chapter).

A direct dividend of the semiclassical result is that it \emph{generates} new
perturbative data. At $k$-loop order the perturbative dimension is a polynomial
$P_k(n)$ of degree $k+1$ in $n$; the NLO calculation fixes its two leading
coefficients — the $C_0$ and $C_1$ resummations of \eqref{eq:C0_perturbative}
and \eqref{eq:Delta_full} — and matching the remaining $k$ coefficients onto
existing $k$-loop data at $k$ distinct values of $n$ then determines $P_k(n)$
completely. Carried out with the current five-loop input, this yields the full
$\mathcal{O}(\epsilon^5)$ dimension of $\phi^n$ in the Ising CFT~\cite{Antipin:2025ilv},
{\small
\begin{align}
\Delta_n &= \Big(1-\tfrac{\epsilon}{2}\Big)n + \tfrac{n}{6}(n-1)\,\epsilon
- \tfrac{n}{324}\big(17n^2-67n+47\big)\,\epsilon^2 \nonumber\\
&\ + \big(0.03215\,n^4 - 0.12338\,n^3 + 0.07348\,n^2 + 0.02710\,n\big)\epsilon^3 \nonumber\\
&\ + \big({-}0.02546\,n^5 + 0.11381\,n^4 - 0.20197\,n^3 \nonumber\\
&\qquad + 0.34865\,n^2 - 0.23920\,n\big)\epsilon^4 \nonumber\\
&\ + \big(0.02317\,n^6 - 0.12108\,n^5 + 0.32742\,n^4 - 0.53835\,n^3 \nonumber\\
&\qquad - 0.06733\,n^2 + 0.38900\,n\big)\epsilon^5\,,
\label{eq:fiveloop_phin}
\end{align}}%
which for $n\ge8$ improves on previously available lower-order results. The
$n^{k+1}$ and $n^{k}$ coefficients of each $\epsilon^k$ term are exactly the
$C_0$ and $C_1$ semiclassical predictions; the lower powers come from the
diagrammatic matching. For $n>5$, where direct multi-loop bootstrap data are
unavailable, the three-dimensional estimates follow from the $[2/3]$ Pad\'e
approximant of \eqref{eq:fiveloop_phin}.

\begin{figure}[ht]
\centering
\includegraphics[width=0.72\textwidth]{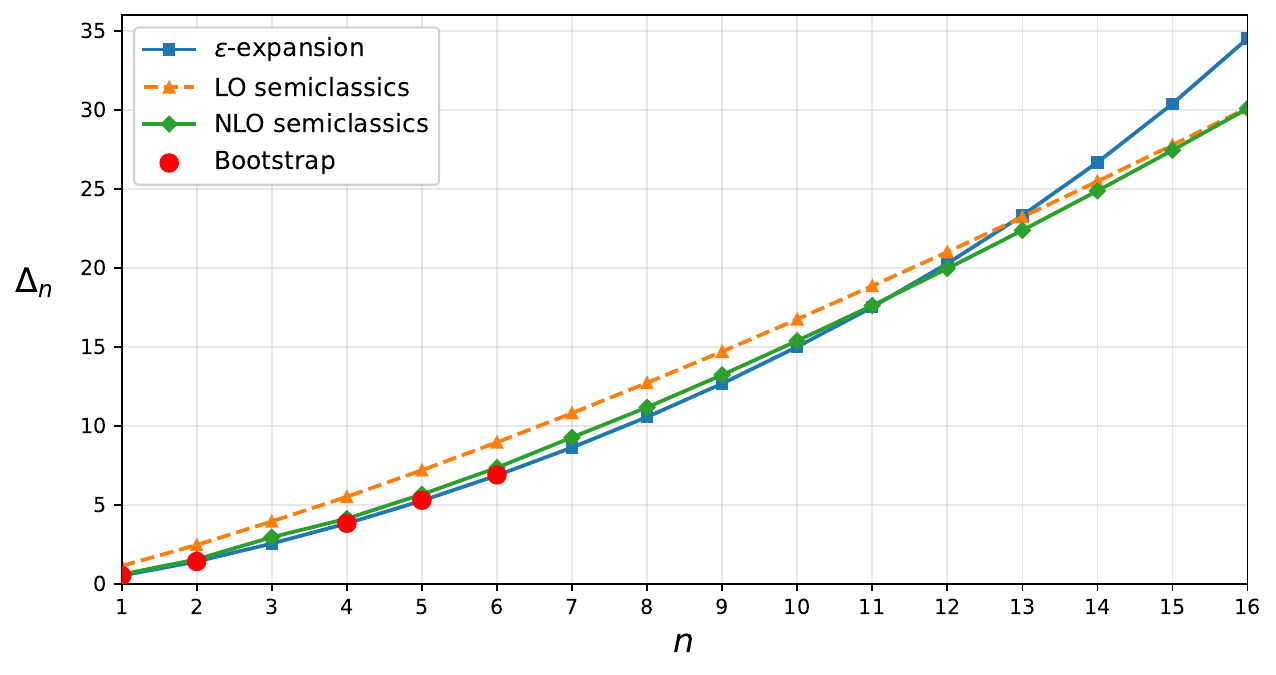}
\caption{\textbf{$\Delta_n$ in the three-dimensional Ising
CFT for $n\le16$}, from the numerical bootstrap, the resummed $\epsilon$-expansion,
and semiclassics at LO and NLO. The NLO curve tracks the bootstrap wherever the
latter exists, with agreement \emph{improving} as $n$ grows; the $n=6$ NLO point
is absent because the $\ell=2$ stability angle has gone complex
(§\ref{sec:instabilities}). For $n\gtrsim12$ the most accurate prediction is
expected to be the semiclassical NLO. Adapted from Ref.~\cite{Antipin:2025ilv}.}
\label{fig:moneyplot}
\end{figure}

Figure~\ref{fig:moneyplot}  already shown in \cite{Antipin:2025ilv},  collects these predictions for $n\le16$.
Three features stand out. First, the NLO curve follows the bootstrap data
closely, and the agreement \emph{improves} with $n$: the relative discrepancy
falls from about $14\%$ at $n=1$ to about $6\%$ at $n=5$. This is  exactly what the double-scaling logic
predicts, since $n$ is the semiclassical large parameter. Second, the $n=6$ NLO
prediction is missing: this is precisely the first classical instability of
§\ref{sec:instabilities}, where the $\ell=2$ stability angle becomes complex at
$m=3/8$ ($\lambda n\approx50$), so the homogeneous saddle no longer controls the
$n=6$ state. Third, for $n\gtrsim12$ the LO curve lies closer to the
$\epsilon$-expansion than the NLO one. This reflects the degradation of the
resummed $\epsilon$-expansion at large $n$ (its optimal truncation drops to one
loop already for $n>2$), so proximity to it is not a measure of accuracy there.

%

Taken together, and given that the semiclassical expansion is by construction
controlled at large $n$, these observations indicate that the NLO semiclassical
prediction is expected to become the most accurate available determination — the
\emph{de facto} state of the art — for $n\gtrsim\mathcal{O}(10)$, a regime that
lies beyond the reach of both the bootstrap and the resummed $\epsilon$-expansion.
Independent high-$n$ numerical input (for instance Monte Carlo) for $n>6$ would
be valuable to test this expectation directly.

\subsection{{Fuzzy sphere and lattice radial quantization}}
\label{sec:benchmarking_lattice}

{ 
The fuzzy-sphere method~\cite{arXiv:2210.13482} provides the most direct numerical
realization of the cylinder setup underlying this entire chapter.  At the Ising critical
point on $S^2$, exact diagonalization of the Hamiltonian yields the spectrum $\{E_k\}$;
via $\Delta_k = RE_k$ one reads off scaling dimensions simultaneously for all primary
operators with a given $SO(3)$ quantum number.  For $n = 1, 2, 4, 5$ these results agree
with the bootstrap data in Figure~\ref{fig:moneyplot} to within a few percent. The $n=6$
point, marked absent in the figure due to the $\ell=2$ instability in the semiclassical
calculation (\S\,\ref{sec:instabilities}), can in principle be accessed by the fuzzy
sphere, and a comparison there would directly test whether the semiclassical instability
is a genuine breakdown or an artefact of the homogeneous-saddle approximation.  For
the $O(N)$ universality classes the fuzzy-sphere framework has recently been applied
for $N=2,3,4$~\cite{arXiv:2512.02234}, yielding operator spectra consistent with the
conformal bootstrap and providing the first independent numerical check of the
$O(N)$ semiclassical predictions of Section~\ref{sec:phi4}.

Lattice radial quantization~\cite{arXiv:1212.1757,arXiv:2006.15636,arXiv:2311.01100}
works on a simplicial discretization of $\mathbb{R}\times S^2$ and in principle accesses
any $n$-particle sector, but current implementations are also limited to small $n$ by
computational resources.  Both the fuzzy sphere and lattice radial quantization share
with the semiclassical framework the essential feature that the state--operator dictionary
$\Delta = RE$ is exact: there is no $a/R$ or $1/N_{\rm mat}$ correction to the map
itself, only to the eigenvalues.  They therefore provide scheme-independent benchmarks
against which the semiclassical $1/n$ series can be tested without any perturbative
ambiguity.

The emerging picture is a three-way
complementarity:
\begin{itemize}
  \item \emph{Fuzzy sphere and lattice radial quantization}: reliable for $n \lesssim 6$
  in $d=3$; exact (at given cutoff) but computationally costly at large $n$.
  \item \emph{Resummed $\varepsilon$-expansion}: reliable for small $n$ at moderate
  $\varepsilon$; breaks down at large $n$ where the optimal truncation drops to one loop.
  \item \emph{Semiclassical NLO}: reliable for $n \gtrsim 6$--$8$ in $d=3$; becomes
  increasingly accurate as $n$ grows.
\end{itemize}
The transition region $7 \le n \le 12$ is where all three methods overlap, and where
new fuzzy-sphere or lattice data would provide the most stringent test.   }

\bigskip
\noindent
This chapter turned the general machinery of the preceding ones into a working
calculation. For the $O(N)$ $\phi^4$ theory at the Wilson--Fisher point, the
classical saddle and the Bohr--Sommerfeld condition fixed the leading
coefficient $C_0(\kappa)$, while the Lam\'e fluctuation spectrum and its Floquet
stability angles delivered the one-loop coefficient $C_1$, together resumming
infinite classes of Feynman diagrams into closed elliptic functions of
$\kappa=\lambda n$. The NLO term lifts the leading-order degeneracy and resolves
the full operator content $\partial^s\Box^p\phi^n$ through the occupation
numbers $\{q_{\ell}\}$, reproducing the known multi-loop data, predicting 
higher-loop results, and --- extrapolated to $d=3$ --- competing with and ultimately
surpassing other methods at large $n$. The same calculation also marks the
boundary of the approach: beyond $\lambda n\approx50$ the homogeneous saddle
becomes unstable, and the strongly coupled regime calls for the more elaborate
saddle analysis left to future work. The construction is by no means tied to the
$O(N)$ scalar at Wilson--Fisher: related models, and the same logic in different
space-time dimensions, have already been explored in
Refs.~\cite{Antipin:2025ilv,Antipin:2025rsr}, and the framework
 {extends naturally to theories with fermions and gauge-Yukawa
interactions, where composite towers such as $(\bar\psi\psi)^n$ and mixed
scalar--fermion operators arise; the semiclassical treatment of these fermionic
sectors is developed in work in preparation.} The Conclusions that follow place these results in the
wider arc of the lecture.

\chapter{Conclusions and Outlook}\label{chap:conclusions}

\begin{figure}[htbp]
\centering
\begin{tikzpicture}[
  classical/.style={draw=blue!70!black, fill=blue!10, thick, rounded corners, text width=5.2cm, minimum height=1.1cm, align=center, font=\small},
  quantum/.style={draw=red!70!black, fill=red!10, thick, rounded corners, text width=5.2cm, minimum height=1.1cm, align=center, font=\small},
  result/.style={draw=green!50!black, fill=green!10, very thick, rounded corners, text width=5.2cm, minimum height=1.1cm, align=center, font=\small},
  decision/.style={draw=orange!70!black, fill=orange!10, thick, rounded corners, text width=5.2cm, minimum height=1.1cm, align=center, font=\small},
  arrow/.style={->, very thick, >=latex},
  plabel/.style={font=\footnotesize\itshape, text=gray!70!black},
]
\node[classical] (step1) at (0, 0) {\textbf{1.} Place the CFT on $\mathbb{R}_\tau\times S^{d-1}_R$\\(state--operator: $\Delta = R\,E$)};
\node[classical] (step2) at (7, 0) {\textbf{2.} Homogeneous periodic\\saddle $v(t)$ of period $\mathcal{T}$};
\node[decision] (step3) at (7, -2.2) {\textbf{3.} Bohr--Sommerfeld:\\$I(E_{\rm cl})=\oint p\,dq=2\pi n$};
\node[classical] (step4) at (0, -2.2) {\textbf{4.} Classical energy\\$nC_0 = R\,E_{\rm cl}$,\ \ $C_0 = 2\pi R/\mathcal{T}$};
\node[quantum] (step5) at (0, -4.4) {\textbf{5.} Expand $\phi = v + \eta$\\(fluctuations)};
\node[quantum] (step6) at (7, -4.4) {\textbf{6.} Mode decomposition on $S^{d-1}$\\$\to$ Hill/Lam\'e equations};
\node[quantum] (step7) at (7, -6.6) {\textbf{7.} Monodromy $\mathcal{M}_\ell$,\ $\Tr\mathcal{M}_\ell=2\cos\nu_\ell$\\$\to$ stability angles $\nu_\ell$};
\node[quantum] (step8) at (0, -6.6) {\textbf{8.} Gel'fand--Yaglom:\\$\det(\mathbb{I}-\mathcal{M}_\ell) = 4\sin^2(\nu_\ell/2)$};
\node[quantum] (step9) at (0, -8.8) {\textbf{9.} Gutzwiller trace formula\\$\to$ one-loop shift $\delta E_1$};
\node[quantum] (step10) at (7, -8.8) {\textbf{10.} Renormalisation:\\$\delta E_1$ cancels $1/\epsilon$ poles};
\node[result] (step11) at (3.5, -11) {\textbf{11.} $E = E_{\rm cl} + \delta E_1 + \dfrac{1}{\mathcal{T}}\displaystyle\sum_{\nu_\ell>0}\Big(q_\ell+\tfrac{n_\ell}{2}\Big)\nu_\ell$};
\node[result] (step12) at (3.5, -13) {\textbf{12.} State--operator: $\Delta_{n,\{q_\ell\}} = R\,E = nC_0 + C_1 + O(1/n)$};
\draw[arrow] (step1) -- (step2);
\draw[arrow] (step2) -- (step3);
\draw[arrow] (step3) -- (step4);
\draw[arrow] (step4) -- (step5);
\draw[arrow] (step5) -- (step6);
\draw[arrow] (step6) -- (step7);
\draw[arrow] (step7) -- (step8);
\draw[arrow] (step8) -- (step9);
\draw[arrow] (step9) -- (step10);
\draw[arrow] (step10) |- (step11);
\draw[arrow] (step11) -- (step12);
\node[plabel, rotate=90, anchor=south] at (-3.2, -1.1) {\textsc{Classical}};
\node[plabel, rotate=90, anchor=south] at (-3.2, -6.6) {\textsc{Quantum Fluctuations}};
\node[plabel, rotate=90, anchor=south] at (-3.2, -12.0) {\textsc{Result}};
\draw[decorate, decoration={brace, amplitude=6pt, mirror}, thick, blue!50!black]
  (-2.8, 0.7) -- (-2.8, -3.0);
\draw[decorate, decoration={brace, amplitude=6pt, mirror}, thick, red!50!black]
  (-2.8, -3.6) -- (-2.8, -9.6);
\draw[decorate, decoration={brace, amplitude=6pt, mirror}, thick, green!50!black]
  (-2.8, -10.2) -- (-2.8, -13.8);
\end{tikzpicture}
\caption{The semiclassical pipeline for heavy composite operators, organised into
its classical (blue), quantum-fluctuation (red), and result (green) phases. The
classical input fixes the leading coefficient through $nC_0=R\,E_{\rm cl}(n)$,
$C_0=2\pi R/\mathcal{T}$; the Floquet/Gel'fand--Yaglom analysis supplies the
subleading coefficient $C_1 = R\,\delta E_1 + \frac{R}{2\mathcal{T}}\sum_{\nu_\ell>0} n_\ell\,\nu_\ell$,
with $n_\ell$ the degeneracy of angular momentum $\ell$ on $S^{d-1}$ and $q_\ell\ge0$
the Floquet excitation numbers that lift the leading degeneracy.}
\label{fig:concl_pipeline}
\end{figure}
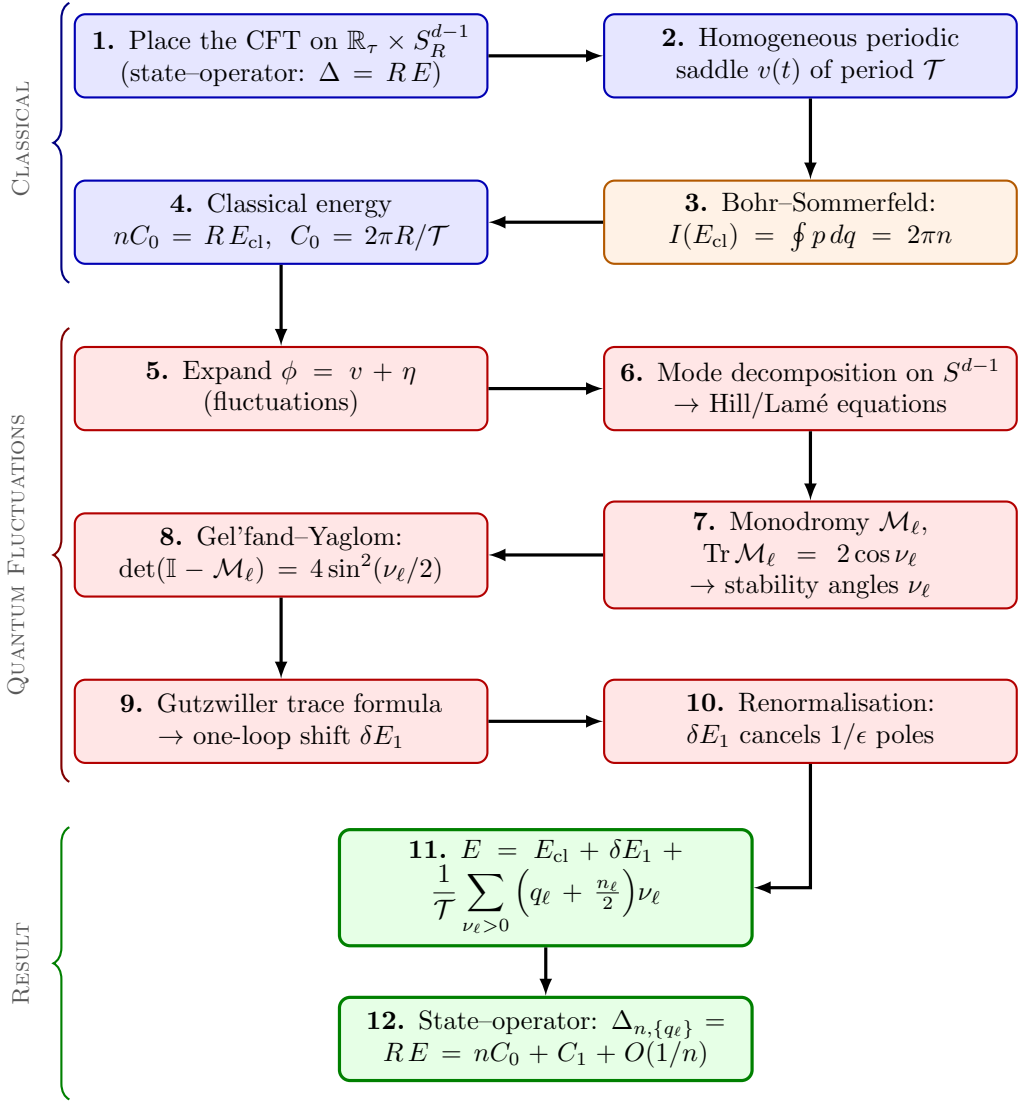

 A critical system is described by a conformal field theory; its local operators
carry the universal data, and their scaling dimensions are the response of the
system to small disturbances. These lecture notes set out to show that the scaling dimensions of  neutral heavy
composite operators, an important piece of conformal-field-theory data, can be
computed as the energies of concrete classical-plus-quantum motion.   The state--operator correspondence turns each such
operator into a state on the cylinder $\mathbb{R}_\tau\times S^{d-1}_R$, with its scaling dimension related to the energy on the cylinder of radios $R$ given by 
$   \Delta = R\,E$,  so that an operator dimension becomes a cylinder energy. When the operator is
made parametrically heavy the corresponding state is highly occupied and behaves
classically, and its energy becomes accessible by semiclassical means even where
ordinary perturbation theory is not the natural language.
 The naively infinite series of Feynman diagrams is
then resummed into a single classical saddle plus a one-loop determinant.
Figure~\ref{fig:concl_pipeline} collects the resulting pipeline in one place; the
rest of this chapter retraces the logical thread part by part and closes with an
outlook on open directions.

\section{The conformal toolbox (Part~I)}

The starting point---Part~I---was a self-contained treatment of conformal
symmetry, built from the infinitesimal conformal Killing equation
\begin{equation}
  \partial_\mu\epsilon_\nu + \partial_\nu\epsilon_\mu
  = \tfrac{2}{d}(\partial\cdot\epsilon)\,\delta_{\mu\nu}\,,
\end{equation}
whose solutions are at most quadratic in $x$ for $d>2$.  The resulting
$\tfrac{(d+1)(d+2)}{2}$ independent generators, translations $P_\mu$,
rotations $M_{\mu\nu}$, dilatations $D$, and special conformal
transformations $K_\mu$, span the Lie algebra $\mathfrak{so}(d+1,1)$,
with the explicit embedding
$\{P_\mu,K_\mu,D,M_{\mu\nu}\}\hookrightarrow J_{AB}$ given in
\eqref{eq:so_emb}.  The finite transformation law of a scalar primary
$\mathcal{O}_\Delta$ under a conformal map $x\mapsto x'$ with local
rescaling factor $\Omega(x)$,
\begin{equation}
  \mathcal{O}'(x') = \Omega(x)^{-\Delta}\,\mathcal{O}(x)\,,
\end{equation}
was then used to fix the form of two- and three-point functions entirely
in terms of the operator dimensions and OPE coefficients, while
four-point functions are constrained up to a function of the two
conformally invariant cross-ratios.

A crucial bridge, established in Chapter~\ref{chap:soc}, is the
\emph{state--operator correspondence}.  The Weyl map from flat space
$\mathbb{R}^d$ to the cylinder $\mathbb{R}_\tau\times S^{d-1}_R$ converts
the radial coordinate into Euclidean time, and the dilatation generator
$D$ into the cylinder Hamiltonian.  Every local operator $\mathcal{O}_\Delta$
inserted at the origin creates a state of definite cylinder energy $E$, with
\begin{equation}
  \Delta = R\,E\,,
\end{equation}
the relation that converts an energy eigenvalue into a scaling dimension.
This single equation is the engine of the entire semiclassical programme:
computing $\Delta_n$ reduces to computing the ground-state energy of the
quantum field theory in a finite volume at fixed operator degree $n$.

Chapter~\ref{chap:composites} addressed the mixing problem that arises in
the presence of interactions: at each operator dimension, finitely many
primary operators with the same quantum numbers mix under renormalization,
and their physical scaling dimensions are the eigenvalues of the anomalous
dimension matrix.  For large $n$ this matrix is exponentially large, yet the
semiclassical approach sidesteps the diagonalization entirely by working
directly with the energy eigenstates of the cylinder Hamiltonian.

\section{The semiclassical blueprint (Parts~II and~III)}

Parts~II and~III developed the technical machinery that makes the
large-$n$ limit tractable.

\paragraph{Three roads in free theory.}
Chapter~\ref{chap:freetheory} verified the free-theory result
\begin{equation}
  \Delta_n^{\rm free} = \frac{d-2}{2}\,n
\end{equation}
via three independent routes: (i)~explicit Wick-contraction counting of
the two-point function $\langle\phi^n(x)\phi^n(0)\rangle$,
(ii)~a flat-space saddle-point calculation in which the operator insertion
induces a non-trivial field configuration, and (iii)~a Bohr--Sommerfeld
quantization of the homogeneous classical solution on the cylinder.  The
agreement of all three routes provides a stringent consistency check of
the formalism before interactions are switched on.

\paragraph{The interacting setup and double-scaling limit.}
Once the coupling $\lambda$ is turned on, the cylinder Hamiltonian for
the O$(N)$ $\phi^4$ theory acquires an anharmonic potential, and the
saddle-point equation becomes non-trivial.  The key observation,
elaborated in Chapter~\ref{chap:doublescaling}, is that in the double-scaling limit 
$n \to\infty,\quad \lambda\to 0$,  
$ \kappa \;\equiv\; \lambda\, n \;=\; \text{fixed}$ the saddle-point approximation becomes
controllable: the action scales as $n$, corrections are suppressed in
$1/n$, and the effective coupling $\kappa = \lambda n$ is kept finite so
that genuinely non-perturbative effects in $\lambda$ at fixed $n$ are
resummed.

\paragraph{Action variable and Bohr--Sommerfeld quantization.}
The quantization of the periodic classical orbit is organized by the
action variable
\begin{equation}
  I(E) = \oint p\,dq\,,
\end{equation}
whose Bohr--Sommerfeld quantization condition $I(E)=2\pi n$ selects the
discrete energy levels.  The Legendre pair
\begin{equation}
  I(E) = \mathcal{T}\,E - S_{\rm cl}(\mathcal{T})
\end{equation}
connects $I(E)$ to the on-shell action $S_{\rm cl}(\mathcal{T})$ evaluated
at period $\mathcal{T}$, and the identity
\begin{equation}
  \frac{dI}{dE} = \mathcal{T}(E)
\end{equation}
(derived in Chapter~\ref{chap:freetheory} via the action-variable formalism)
ensures that the saddle-point integration over $\mathcal{T}$ reproduces the
Bohr--Sommerfeld condition exactly.

\paragraph{Resolvent, path integral, and the Gutzwiller formula.}
Chapter~\ref{chap:periodicsaddles} derived the semiclassical expression for
the resolvent---the generating function of energy eigenvalues---by combining:
\begin{enumerate}
\item a proper-time representation of the operator $(E-H)^{-1}$ as a path
      integral with periodic boundary conditions in imaginary time;
\item a quadratic expansion of the action around each periodic classical
      solution to isolate the one-loop fluctuation determinant;
\item a saddle-point integration over the period $\mathcal{T}$, whose
      poles reproduce the quantum energy levels.
\end{enumerate}
The result is the \emph{Gutzwiller trace formula}: the semiclassical density
of states is a sum over periodic orbits, weighted by the classical action
and the fluctuation determinant.

\paragraph{Floquet theory and Gel'fand--Yaglom.}
The fluctuation operator around a periodic orbit $v(\tau)$ is a Hill
operator of the form $-\partial_\tau^2 + \omega(\tau)$, where $\omega(\tau)$
inherits the period $\mathcal{T}$ of the orbit.  Sections~\ref{sec:hill_floquet}--\ref{sec:gelfand_yaglom}
of Chapter~\ref{chap:periodicsaddles} analyse this operator in full generality: Floquet's theorem guarantees
that the monodromy matrix $\mathcal{M}$ encodes all spectral information
through its eigenvalues $e^{\pm i\nu_\ell}$, where $\nu_\ell$ is the
\emph{stability angle} of mode $\ell$.  The Gel'fand--Yaglom theorem then
converts the functional determinant of the fluctuation operator (with periodic
boundary conditions) into a closed-form expression in the monodromy matrix,
\begin{equation}
  \detp\!\left(-\partial_\tau^2+\omega\right)\Big|_{\rm PBC}
  = 4\sin^2\!\bigl(\nu_\ell/2\bigr)\,,
\end{equation}
sidestepping the notoriously difficult problem of computing infinite products
of eigenvalues directly.  The full fluctuation determinant over all spatial
modes $\ell$ (each with degeneracy $n_\ell$ on $S^{d-1}$) then gives the
one-loop quantum correction to the energy:
\begin{equation}
  \delta E_1 = -\frac{1}{2\mathcal{T}}\sum_\ell n_\ell
  \ln\!\left[4\sin^2\!\bigl(\nu_\ell/2\bigr)\right]
  + \text{counterterms}\,.
\end{equation}
Including the contribution of oscillator excitations labelled by occupation
numbers $\{q_\ell\}$, the full semiclassical energy reads
\begin{equation}
  E = E_{\rm cl}(n) + \delta E_1
  + \frac{1}{\mathcal{T}}\sum_{\nu_\ell>0}
    \!\left(q_\ell+\tfrac{n_\ell}{2}\right)\nu_\ell(n)\,,
  \label{eq:concl_energy}
\end{equation}
and the corresponding scaling dimension follows from $\Delta = R\,E$.

\section{The interacting result: O$(N)$ $\phi^4$ at Wilson--Fisher (Part~IV)}

Part~IV applied the full machinery to the critical O$(N)$ $\phi^4$ theory
in $d=4-\varepsilon$, evaluated at the Wilson--Fisher fixed-point coupling
$\lambda_*$.

\paragraph{Classical solution.}
The Euler--Lagrange equation on the cylinder admits a homogeneous periodic
solution expressible in terms of a Jacobi elliptic function,
\begin{equation}
  v(\tau) = x_0\,\cn(\omega\tau\,|\,m)\,,
\end{equation}
parametrized by the elliptic modulus $m\in[0,1/2)$.  The degree $n$ and
the energy $E$ are both determined by $m$, which thereby plays the role of
the sole classical free parameter.  In the small-$\kappa$ limit
($m\to 0$) the solution reduces to a cosine; in the large-$\kappa$ limit
($m\to 1/2$) it approaches a half-period rectangle wave, signalling a
bifurcation in the orbit structure.  The leading coefficient in the
$1/n$ expansion of the scaling dimension,
\begin{equation}
  C_0(\kappa) = \frac{\Delta_n}{n}\bigg|_{\rm leading}\,,
\end{equation}
is entirely determined by $E_{\rm cl}$ evaluated on the elliptic solution.

\paragraph{Fluctuation spectrum: the Lam\'e equation.}
The linearized fluctuation equation around the elliptic orbit takes the form
of the \emph{Lam\'e equation}---a Hill equation with an elliptic-function
coefficient---whose stability angles $\nu_\ell$ are known in closed form.
The renormalization of the fluctuation determinant requires the cancellation
of ultraviolet $1/\varepsilon$ poles from the path integral against the
Wilson--Fisher counterterms; after renormalization, the finite remainder
gives the next-to-leading coefficient
\begin{equation}
  C_1(\kappa) = R\,\delta E_1(\kappa)
  + \frac{R}{2\mathcal{T}}\sum_{\nu_\ell>0} n_\ell\,\nu_\ell
  \;\Big|_{\rm renormalized}\,.
\end{equation}
The scaling dimension then takes the compact form
\begin{equation}
  \Delta_n = n\,C_0(\kappa) + C_1(\kappa) + O(1/n)\,,
  \label{eq:concl_Deltan}
\end{equation}
which at small $\kappa$ matches the perturbative result for $\langle\phi^n\rangle$
and at large $\kappa$ encodes genuinely non-perturbative resummations of
the $\lambda$-expansion.

\paragraph{Operator identification and excitation spectrum.}
The occupation numbers $\{q_\ell\}$ in \eqref{eq:concl_energy} classify
the entire tower of operators above the ground state in a given charge
sector: each non-zero $q_\ell$ corresponds to exciting a particular
angular mode on $S^{d-1}$ and shifts $\Delta_n$ by a quantized amount
proportional to $\nu_\ell$.  Spin-$s$ primaries, descendants, and
mixed-symmetry operators all arise as specific choices of occupation
numbers, giving a complete semiclassical dictionary for the operator
spectrum in the heavy-operator regime.

\paragraph{Instabilities and classical scars.}
At sufficiently large $\kappa$, the stability angle $\nu_{2,\ell}$ of
the $L_2$ Lam\'e sector becomes imaginary for certain modes, signalling
parametric resonance over one classical period.  The resulting unstable
periodic orbits are \emph{classical scars}: they leave a strong imprint
on the energy eigenfunctions even though they are not stable saddle
points of the action.  The onset of the first instability at
$\kappa\approx 50$ (for $\ell=2$) marks the boundary of the region of
validity of the perturbative $1/n$ expansion.

\section{Outlook}

The framework developed in these notes opens several directions for
further research.

\paragraph{Higher orders in $1/n$.}
The two-loop correction $C_2(\kappa)$ to the scaling dimension
\eqref{eq:concl_Deltan} requires evaluating two-loop diagrams around the
elliptic background, which brings in two-loop renormalization,
non-Gaussian fluctuations, and higher-order Floquet-spectrum corrections.
Progress in this direction would extend the semiclassical expansion to
three terms and enable a sharper comparison with conformal bootstrap data.

\paragraph{Other theories and symmetry groups.}
The semiclassical canovaccio is not specific to O$(N)$ $\phi^4$.  Any
CFT containing heavy composite operators in a double-scaling limit admits a
tractable semiclassical description, and the same five-step programme---cylinder embedding,
classical periodic orbit, Bohr--Sommerfeld quantization, Gel'fand--Yaglom
fluctuation determinant, state--operator correspondence---can be applied.
Examples include $\mathbb{CP}^{N-1}$ models, non-linear sigma models, and
theories with fermionic matter, each of which leads to a different classical
orbit and a different Floquet spectrum.

\paragraph{Finite-$N$ corrections and large-$N$ cross-checks.}
In the O$(N)$ theory the $N$-dependence enters at every loop order through
 {the fluctuation-channel multiplicities ($N-1$ transverse and one
longitudinal mode for each $\ell$)} and through the renormalization-group
coefficients{; the spherical-harmonic degeneracies $n_\ell$ are
 geometric}.  At large $N$, the semiclassical result should be reproducible
by a $1/N$ expansion, providing a non-trivial cross-check of both methods
in their overlapping regime of validity.

 \paragraph{{Semiclassics, conformal windows, and safe QFTs.}}
{The semiclassical methods developed in these lectures also connect naturally
with the long-standing problem of charting conformal windows in QCD-like gauge
theories.  {The  notion of an infrared conformal window
originates in the perturbative Banks--Zaks fixed point, the interacting infrared
fixed point that appears just below the loss of asymptotic freedom once the
two-loop term of the gauge $\beta$-function is retained
\cite{Caswell:1974gg,Banks:1981nn}.  As the number of flavours is lowered the
fixed-point coupling grows until chiral symmetry breaks and conformality is lost;
locating this lower edge and characterising the associated quantum phase transition
has been a central question ever since.  Ladder and gap-equation analyses point to
a conformal phase transition, in which the infrared fixed point
merges with an ultraviolet one and the order parameter switches on with an
essential singularity \cite{Miransky:1996pd,Appelquist:1996dq}, while an
alternative scenario has conformality lost discontinuously, through a ``jump''
\cite{Sannino:2012wy}.  A complementary, weakly coupled handle on the same regime
is provided by conjectured magnetic dual gauge theories, which organise the
low-lying spectrum near the lower edge and yield, for instance, a prediction for
the electroweak $S$ parameter \cite{Sannino:2010fh}.} 
The basic physical idea of conformal window is generalized in the
orientifold and higher-representation constructions, where fermions in two-index
symmetric or antisymmetric representations can drive the theory close to
conformality with a small number of flavours \cite{hep-ph/0405209}.  This
perspective was subsequently widened into a systematic phase-diagram analysis of
nonsupersymmetric $SU(N)$ gauge theories with fermions in higher-dimensional
representations, showing how the onset of infrared conformality depends on colour,
flavour, and representation \cite{hep-ph/0611341}, a picture now
under sustained scrutiny on the lattice \cite{Cacciapaglia:2020kgq}.  {Defects are relevant tools to investigate the dynamics of the conformal window:  the presence of a heavy flavoured probe introduces a conformal defect whose spectrum encodes how heavy mesons dress and
propagate across the QCD conformal window \cite{DiRisi:2024nus}.} 
Fixed-charge semiclassics provides a complementary probe of this dynamics. Near the lower edge
of the conformal window, the large-charge sector can be described by an effective
theory of Goldstone modes and a light dilaton, giving analytic access to the ground
state, the excitation spectrum, and the would-be scaling dimensions of the lowest
charged operators \cite{2003.08396}. The same framework can be extended to nonzero
$\theta$ angle and axion dynamics, where the fixed-charge expansion tracks how
topological terms and the dilaton potential affect near-conformal physics
\cite{2208.09227}. The recent extension of semiclassical control
to heavy neutral operators and further discussed here is especially relevant in this context
\cite{Antipin:2024ekk}. Since the dilaton is itself a neutral scalar excitation
associated with the approximate breaking of scale invariance, neutral sectors may
provide a direct diagnostic of dilatonic dynamics, including the structure of the
effective potential, the mass gap, and the interpolation between walking behaviour
and genuine infrared conformality.   On the ultraviolet side, asymptotically safe gauge--Yukawa
theories offer an equally natural arena: their interacting fixed points are
perturbatively controlled in the Veneziano limit \cite{1406.2337}, while the
large-charge sector exposes the symmetry-breaking pattern, Goldstone spectrum, and
scaling dimensions directly at the safe fixed point \cite{1905.00026}. In this
sense, semiclassics provides a common organizing language for walking dynamics,
infrared conformal windows, dilaton effective theories, heavy neutral sectors, and
ultraviolet-complete asymptotically safe field theories, including the broader
large-$N_f$ safe-QCD scenario \cite{1709.02354}.}

\paragraph{Connection to integrability and the bootstrap.}
For low-dimensional CFTs, the conformal bootstrap --- and, in the special case of
integrable theories, exact methods such as the Thermodynamic Bethe
Ansatz~\cite{Zamolodchikov:1989cf,Gromov:2013pga} --- can
yield rigorous information on the operator spectrum that is inaccessible
perturbatively.  Comparing the semiclassical $1/n$ expansion against bootstrap
bounds~\cite{arXiv:1403.4545,arXiv:1504.07997} (and against exact results wherever a
theory happens to be integrable) is an
ongoing programme that may reveal universal features of the double-scaling limit
independent of the specific model~\cite{arXiv:2010.00407}.

\paragraph{Quantum information and entanglement.}
Entanglement measures are among the sharpest probes of CFT
data~\cite{Calabrese:2009qy}, and the cylinder/state--operator technology that
underlies these lectures connects to them directly. By the Casini--Huerta--Myers
map, the entanglement entropy of a spherical region equals the thermal entropy of
the CFT on a hyperbolic spatial slice~\cite{arXiv:1102.0440}, computed from exactly
the kind of fluctuation spectrum around a homogeneous saddle that delivered $C_1$
here. This link has very recently been made concrete in the cognate large-charge
expansion, where the semiclassical effective theory yields the charged
(symmetry-resolved) R\'enyi entanglement entropy of strongly coupled fixed points
such as Wilson--Fisher --- one of the first holography-free entanglement
computations in an interacting CFT~\cite{Watanabe:2025mnc}. The heavy
neutral-operator framework developed here is well suited to extend such results
beyond the charged sector: to how entanglement scales with the operator degree
$n$, and to the symmetry-resolved entanglement of the $O(N)$ charge
sectors~\cite{Goldstein:2017bua}. Heavy states moreover probe thermalization ---
the subsystem eigenstate thermalization hypothesis predicts that the reduced
density matrix of a heavy primary is approximately thermal~\cite{He:2017vyf} ---
so the very instabilities and ``classical scars'' encountered at large $\lambda n$
may mark where a single saddle ceases to capture the entanglement of typical heavy
eigenstates, tying the semiclassical breakdown to quantum chaos in the CFT
spectrum.
 
\paragraph{{A semiclassical lens on learning.}}{
The structures developed in these lectures also appear, in a different language,
in the theory of wide neural networks.  Our double-scaling limit isolates a
dominant classical saddle and organises quantum fluctuations into a controlled
$1/n$ expansion; the infinite-width limit plays the same role for a network,
suppressing fluctuations so that its prior becomes a Gaussian process
\cite{Lee:2017qzq} and, in the neural-tangent-kernel regime, its training
collapses to a linear kernel evolution \cite{Jacot:2018ivh}.  The dynamical
content then lives in the corrections: just as the subleading $1/n$ terms here
resum into the interacting effective action, finite-width corrections deform the
limiting Gaussian or kernel description by genuine interactions
\cite{Roberts:2021fes,Yaida:2019sjo}.  The parallel is more than qualitative.
The Gel'fand--Yaglom one-loop determinant that fixes our fluctuation prefactor is
the field-theoretic counterpart of the Laplace--Occam factor in Bayesian model
evidence \cite{MacKay:1992bayes}, and the harmonic decomposition on $S^{d-1}$
that diagonalises the fluctuation operator mirrors the spectral analysis of
dot-product kernels and the resulting spectral bias of wide networks
\cite{Bordelon:2020spectral}.  These are structural correspondences, including the often-invoked analogy between the
renormalisation group and representation learning, which remains heuristic
\cite{Mehta:2014xqf}.  Still, the parallel is useful: it suggests that the
semiclassical toolkit assembled here, saddles, action variables, Floquet
spectra, and functional determinants, provides a transferable language for
organising fluctuations and effective descriptions well beyond its original
field-theoretic setting.  A more systematic development is currently under investigation.
}

\bigskip
In summary, the semiclassical approach to heavy operators provides a
controlled, systematically improvable method for computing CFT data
in a regime---large $n$, fixed $\kappa = \lambda n$---that is out of reach
of conventional perturbation theory.  The central formula \eqref{eq:concl_Deltan}
distils a rich interplay of classical mechanics (periodic orbits, action
variables), spectral theory (Hill's equation, Floquet theory, Gel'fand--Yaglom),
and renormalization (ultraviolet finiteness of the one-loop correction)
into a compact and physically transparent result.  It is our hope that these
lectures serve as a useful starting point for students and researchers wishing
to explore this active and rapidly developing corner of conformal field theory. 

\vskip 1cm 

\section*{Acknowledgments}
It is a pleasure to thank Oleg Antipin and Jahmall Bersini for many years of fruitful collaboration, and for teaching me many aspects of semiclassical methods in conformal field theory. I am grateful to Clelia Gambardella, Jacob Holzendorff Hafjall and Giulia Muco for their comments on the lectures and for many stimulating discussions on conformal field theories, both with and without defects. I thank Vigilante Di Risi and Davide Iacobacci for our collaboration on heavy-quark conformal effective field theory and for carefully reading the manuscript. I am also grateful to Manuel Del Piano, Jonas Neuser, and Mattia Damia Paciarini for their comments on the manuscript. I am indebted to Michele Della Morte, Gerald Dunne, Bjarke Gudnason, David Kyed, Viljami Juhani Leino, Hugh Osborn, Claudio Pica, Antonio Rago, Thomas Ryttov, Apoorv Tiwari, Jessica Turner, and Matthias Oliver Wilhelm for valuable comments and suggestions. I also warmly thank Domenico Orlando and Susanne Reffert for numerous illuminating discussions on large-charge expansions, semiclassical methods, and related developments in conformal field theory. Finally, I thank all the collaborators whose questions, insights, and enthusiasm have shaped the lectures on which these notes are based.

\paragraph{AI use.} Large language model tools were used for language editing, LaTeX/bibliography support, and reference cross-checking. All scientific content was produced and verified by the author, who bears sole responsibility for it.

\part{Appendices}

\appendix

\chapter{Identity components of topological groups}
\label{app:identity_component}

In Section~\ref{app:conf:so_iso} we wrote
$\mathrm{SO}(p,q) = \{\ldots\}_{0}$ and
$\mathrm{Conf}(\mathbb{R}^d)_{0} \cong \mathrm{SO}(d+1,1)$, with a subscript
$0$ denoting ``the identity component''. This appendix collects the
definition and the most relevant facts.

\paragraph{Definition.}
Given a topological group $G$ with identity element $e$, the
\emph{identity component} $G_{0}$ is by definition the connected component
of $G$ that contains $e$; equivalently, $G_{0}$ is the set of all group
elements that can be reached from $e$ by a continuous path lying entirely
inside $G$. The identity component is automatically a connected, closed,
normal Lie subgroup of $G$, and it has the same dimension as $G$
itself---the other components are disjoint translates of $G_{0}$ that share
its dimension but are unreachable from $e$ by continuous deformation.

\paragraph{Why the subscript matters.}
Whenever a group has more than one connected component, the subscript is
necessary to specify which one we mean. A familiar example: the orthogonal
group $\mathrm{O}(n)$ has \emph{two} connected components, distinguished by
the sign of $\det\Lambda$; the identity component $\mathrm{O}(n)_{0}$
consists of the rotations (determinant $+1$) and is the usual
$\mathrm{SO}(n)$, while the other component consists of the reflections
(determinant $-1$).

For positive-definite signature ($q=0$) the determinant-one condition in
the definition of $\mathrm{SO}(p,q)$ already selects a single connected
piece, so the subscript $0$ is redundant; one simply recovers the familiar
rotation group $\mathrm{SO}(p)$. For indefinite signature ($p,q\ge 1$),
however, the determinant-one subset of $\mathrm{O}(p,q)$ is itself
disconnected---it splits into two components, one containing the identity
$e$ and one in which the orientation of the negative-eigenvalue subspace
is reversed---so the subscript $0$ does real work: it picks out the
component containing $e$. Some references write this stricter object as
$\mathrm{SO}^{+}(p,q)$; throughout Section~\ref{app:conf:so_iso} we mean
it whenever we write $\mathrm{SO}(p,q)$. The next paragraph proves
disconnectedness.

\paragraph{Proof of disconnectedness for indefinite signature.}
For $p,q\ge 1$ the disconnectedness of
$\{\Lambda\in\mathrm{O}(p,q):\det\Lambda=+1\}$ can be established by
explicit construction. Adopt a basis in which
$\eta=\mathrm{diag}(+\mathbf{1}_p,\,-\mathbf{1}_q)$, and write any
$\Lambda\in\mathrm{O}(p,q)$ in $(p|q)$ block form
\begin{equation}
  \Lambda = \begin{pmatrix} A & B \\ C & D \end{pmatrix}\,,\qquad
  A\in\mathbb{R}^{p\times p},\quad D\in\mathbb{R}^{q\times q}\,.
\end{equation}
The defining relation $\Lambda^T\eta\,\Lambda = \eta$ expanded block by
block yields
\begin{equation}
  A^T A - C^T C = \mathbf{1}_p\,,\qquad
  D^T D - B^T B = \mathbf{1}_q\,.
\end{equation}
Since $C^T C$ and $B^T B$ are positive semi-definite, both equalities
give $A^T A \succeq \mathbf{1}_p$ and $D^T D \succeq \mathbf{1}_q$, where
$X \succeq Y$ denotes the L\"owner partial order on symmetric matrices
($X \succeq Y$ iff $X-Y$ is positive semi-definite, equivalently
$v^T(X-Y)v \ge 0$ for every vector $v$, equivalently every eigenvalue of
$X-Y$ is non-negative). Concretely, $A^T A \succeq \mathbf{1}_p$ means
every eigenvalue of $A^T A$ is at least $1$, and likewise for $D^T D$. Hence
\begin{equation}
  (\det A)^2 \,=\, \det(A^T A) \,\ge\, 1\,,\qquad
  (\det D)^2 \,=\, \det(D^T D) \,\ge\, 1\,,
\end{equation}
for every $\Lambda\in\mathrm{O}(p,q)$. In particular $\det A$ and $\det D$
are continuous functions that never vanish on $\mathrm{O}(p,q)$, so
$\mathrm{sign}(\det A)\in\{+1,-1\}$ and $\mathrm{sign}(\det D)\in\{+1,-1\}$
are locally constant: they must be constant on each connected component of
$\mathrm{O}(p,q)$.

For the identity $\Lambda=\mathbf{1}_{p+q}$ both signs are $+1$. Consider
now the element that simultaneously flips one $p$-direction and one
$q$-direction,
\begin{equation}
  \Lambda_{\star}
  \;=\;
  \mathrm{diag}\bigl(\underbrace{-1,\,+1,\ldots,+1}_{p\ \text{entries}}\,;\;
                     \underbrace{-1,\,+1,\ldots,+1}_{q\ \text{entries}}\bigr)\,.
\end{equation}
Off-diagonal blocks vanish, so
$\Lambda_{\star}^T\eta\,\Lambda_{\star}=\eta$ is immediate; and
$\det\Lambda_{\star}=(\det A_\star)(\det D_\star)=(-1)(-1)=+1$, so
$\Lambda_{\star}$ does lie in the determinant-one subset of
$\mathrm{O}(p,q)$. However its block determinants are
$\det A_\star=-1$ and $\det D_\star=-1$. By local constancy of
$\mathrm{sign}(\det A)$, $\Lambda_{\star}$ and $\mathbf{1}_{p+q}$ cannot
be joined by any continuous path in $\mathrm{O}(p,q)$, let alone one
staying inside the determinant-one subset. The determinant-one subset is
therefore disconnected: it contains at least the identity component (where
both block-determinant signs are $+1$) and the component containing
$\Lambda_{\star}$ (where both signs are $-1$). The full $\mathrm{O}(p,q)$
then has four connected components, labelled by the two independent
signs $(\mathrm{sign}\det A,\,\mathrm{sign}\det D)\in\{\pm 1\}^2$; the one
containing the identity, with both signs $+1$, is what we call
$\mathrm{SO}(p,q)$ in this lecture.

\paragraph{Application: the conformal inversion has $\det=-1$.}
For the Euclidean conformal group we have $(p,q)=(d+1,1)$. We now show
that the inversion $x^\mu\mapsto x^\mu/|x|^2$ corresponds to an element of
$\mathrm{O}(d+1,1)$ with determinant $-1$, so it sits not even in the
determinant-one subset, let alone in the identity component
$\mathrm{SO}(d+1,1)\subset\{\det=+1\}$.

The conformal group acts linearly on the ambient $\mathbb{R}^{d+1,1}$ via
the standard projective light-cone construction. With the metric
$\eta=\mathrm{diag}(-1,+1,\ldots,+1)$ used in Section~\ref{app:conf:so_iso}
(timelike direction labelled $A=-1$), embed $x\in\mathbb{R}^d$ as
\begin{equation}
  X^{-1} \;=\; \tfrac{1}{2}(1+|x|^2)\,,\qquad
  X^{0}  \;=\; \tfrac{1}{2}(1-|x|^2)\,,\qquad
  X^{\mu}\;=\; x^\mu\,,
\end{equation}
which lies on the light cone
$\eta_{AB}X^A X^B = -(X^{-1})^2+(X^0)^2+|x|^2 = 0$. Introducing the
null combinations $X^\pm = X^{-1}\pm X^0$, the embedding takes the simple
form $X^{+}=1,\ X^{-}=|x|^2,\ X^{\mu}=x^\mu$, and the inner product reads
$\eta_{AB}X^A X^B = -X^+ X^- + |X^\mu|^2$.

Under the inversion $x^\mu\mapsto x'^\mu=x^\mu/|x|^2$, the new ambient
representative is
$(X'^+, X'^-, X'^\mu) = (1,\,1/|x|^2,\,x^\mu/|x|^2)$,
which after rescaling projectively by $|x|^2$ becomes the equivalent
representative
$(|x|^2,\, 1,\, x^\mu) = (X^-,\, X^+,\, X^\mu)$. Inversion therefore acts
on the ambient coordinates by simply exchanging the two null directions
$X^+\leftrightarrow X^-$ while leaving each $X^\mu$ untouched. In the
$(X^{-1},X^0)$ basis this exchange fixes $X^{-1}$ and flips
$X^0\mapsto -X^0$, so the corresponding element of $\mathrm{O}(d+1,1)$ is
\begin{equation}
  \Lambda_{\mathrm{inv}}
  \;=\;
  \mathrm{diag}\bigl(\,+1\,,\, -1\,,\, +1,\,\ldots,\,+1\,\bigr)\,,
\end{equation}
with entries ordered as $(X^{-1},X^0,X^1,\ldots,X^d)$. One checks at
once that $\Lambda_{\mathrm{inv}}^T\eta\,\Lambda_{\mathrm{inv}}=\eta$, and
\begin{equation}
  \det\Lambda_{\mathrm{inv}} \;=\; -1\,.
\end{equation}
Hence $\Lambda_{\mathrm{inv}}$ lies in the determinant-$(-1)$ subset of
$\mathrm{O}(d+1,1)$, which is disjoint from the determinant-$(+1)$ subset
containing the identity, and a fortiori disjoint from the identity
component $\mathrm{SO}(d+1,1)\subset\{\det=+1\}$. Since $\det\Lambda$ is
continuous on $\mathrm{O}(d+1,1)$, the inversion cannot be reached from
$\mathbf{1}_{d+2}$ by any continuous path of conformal transformations
(see also Figure~\ref{fig:cft_sct}). The isomorphism of
Section~\ref{app:conf:so_iso} is therefore correctly stated between the
two identity components,
\begin{equation}
  \mathrm{Conf}(\mathbb{R}^d)_{0}\;\cong\;\mathrm{SO}(d+1,1)\,.
\end{equation}

\chapter{The conformal scalar and its effective mass on the cylinder}
\label{app:conformal_coupling}

This appendix establishes the geometric origin of the cylinder ``conformal mass''
quoted in Section~\ref{app:soc:conformal_scalar}: a scalar that is conformally
invariant in flat space must be supplemented by a curvature coupling on a curved
background, and on $\mathbb{R}\times S^{d-1}_R$ that coupling becomes a mass term~\cite{arXiv:1012.3210,arXiv:1102.0440,arXiv:1105.4598}.

\section{Why the scalar needs the conformal curvature coupling}
\label{app:soc:why_conformal_coupling}

A free massless scalar in flat space has action
\begin{equation}
  S_{\rm flat}
  =
  \frac12\int d^d x\,\partial_\mu\phi\,\partial^\mu\phi .
\end{equation}
This action is scale invariant if the scalar has engineering dimension
\begin{equation}
  \Delta_\phi=\frac{d-2}{2}.
\end{equation}
Indeed, under $x\to \lambda x$,
\begin{equation}
  d^d x\to \lambda^d d^d x,
  \qquad
  \partial_\mu\to \lambda^{-1}\partial_\mu,
  \qquad
  \phi\to \lambda^{-\Delta_\phi}\phi,
\end{equation}
so
\begin{equation}
  \int d^d x\,(\partial\phi)^2
  \to
  \lambda^{d-2-2\Delta_\phi}
  \int d^d x\,(\partial\phi)^2.
\end{equation}
Thus scale invariance requires
\begin{equation}
  d-2-2\Delta_\phi=0,
  \qquad
  \Delta_\phi=\frac{d-2}{2}.
\end{equation}

However, when the scalar is placed on a curved background, the minimally coupled
action
\begin{equation}
  S_{\rm min}
  =
  \frac12\int d^d x\,\sqrt{g}\,
  g^{\mu\nu}\partial_\mu\phi\,\partial_\nu\phi
  \label{eq:min_scalar_action}
\end{equation}
is not Weyl invariant in dimensions $d>2$. To see this explicitly, consider the
local Weyl transformation
\begin{equation}
  g_{\mu\nu}\to g'_{\mu\nu}=e^{2\sigma(x)}g_{\mu\nu},
  \qquad
  \phi\to \phi'=e^{-\frac{d-2}{2}\sigma(x)}\phi .
  \label{eq:weyl_scalar_transform}
\end{equation}
Then
\begin{equation}
  \sqrt{g}\to \sqrt{g'}=e^{d\sigma}\sqrt{g},
  \qquad
  g^{\mu\nu}\to g'^{\mu\nu}=e^{-2\sigma}g^{\mu\nu}.
\end{equation}
Define
\begin{equation}
  a\equiv \frac{d-2}{2}.
\end{equation}
The derivative of the transformed scalar is
\begin{equation}
  \partial_\mu\phi'
  =
  e^{-a\sigma}
  \left(
    \partial_\mu\phi
    -
    a\,\phi\,\partial_\mu\sigma
  \right).
\end{equation}
Therefore
\begin{align}
  \sqrt{g'}\,g'^{\mu\nu}
  \partial_\mu\phi'\partial_\nu\phi'
  &=
  \sqrt{g}\,
  e^{d\sigma}e^{-2\sigma}e^{-2a\sigma}
  g^{\mu\nu}
  \left(
    \partial_\mu\phi-a\phi\partial_\mu\sigma
  \right)
  \left(
    \partial_\nu\phi-a\phi\partial_\nu\sigma
  \right)
  \nonumber\\
  &=
  \sqrt{g}\,
  g^{\mu\nu}
  \left[
    \partial_\mu\phi\,\partial_\nu\phi
    -
    2a\,\phi\,\partial_\mu\phi\,\partial_\nu\sigma
    +
    a^2\phi^2\,\partial_\mu\sigma\,\partial_\nu\sigma
  \right],
  \label{eq:minimal_weyl_transform}
\end{align}
because
\begin{equation}
  d-2-2a=d-2-(d-2)=0.
\end{equation}
The second and third terms in \eqref{eq:minimal_weyl_transform} do not vanish for
local $\sigma(x)$. Hence the minimally coupled scalar action is not Weyl invariant.

To restore Weyl invariance, add a curvature coupling:
\begin{equation}
  S_E
  =
  \frac12
  \int d^d x\,\sqrt{g}\,
  \left(
    g^{\mu\nu}\partial_\mu\phi\,\partial_\nu\phi
    +
    \xi R\phi^2
  \right).
  \label{eq:curved_scalar_action_general_xi}
\end{equation}
We now determine $\xi$.

Under a Weyl transformation,
\begin{equation}
  R'
  =
  e^{-2\sigma}
  \left[
    R
    -
    2(d-1)\nabla^2\sigma
    -
    (d-1)(d-2)(\nabla\sigma)^2
  \right],
  \label{eq:Ricci_scalar_weyl}
\end{equation}
where
\begin{equation}
  (\nabla\sigma)^2
  =
  g^{\mu\nu}\partial_\mu\sigma\partial_\nu\sigma,
  \qquad
  \nabla^2\sigma
  =
  g^{\mu\nu}\nabla_\mu\nabla_\nu\sigma.
\end{equation}
Also,
\begin{equation}
  \phi'^2=e^{-2a\sigma}\phi^2.
\end{equation}
Thus
\begin{align}
  \sqrt{g'}\,R'\phi'^2
  &=
  \sqrt{g}\,
  e^{d\sigma}
  e^{-2\sigma}
  e^{-2a\sigma}
  \left[
    R
    -
    2(d-1)\nabla^2\sigma
    -
    (d-1)(d-2)(\nabla\sigma)^2
  \right]\phi^2
  \nonumber\\
  &=
  \sqrt{g}\,
  \left[
    R\phi^2
    -
    2(d-1)\phi^2\nabla^2\sigma
    -
    (d-1)(d-2)\phi^2(\nabla\sigma)^2
  \right],
  \label{eq:curvature_term_weyl_transform}
\end{align}
again because $d-2-2a=0$.

Using integration by parts,
\begin{equation}
  \int d^d x\sqrt{g}\,\phi^2\nabla^2\sigma
  =
  -\int d^d x\sqrt{g}\,\nabla_\mu(\phi^2)\nabla^\mu\sigma,
\end{equation}
and
\begin{equation}
  \nabla_\mu(\phi^2)=2\phi\nabla_\mu\phi,
\end{equation}
we find
\begin{equation}
  -2(d-1)\int d^d x\sqrt{g}\,\phi^2\nabla^2\sigma
  =
  4(d-1)\int d^d x\sqrt{g}\,\phi\,\nabla_\mu\phi\,\nabla^\mu\sigma .
  \label{eq:curvature_ibp}
\end{equation}

The full transformed action differs from the original one by terms proportional to
\begin{equation}
  \phi\,\nabla_\mu\phi\,\nabla^\mu\sigma
  \qquad
  \text{and}
  \qquad
  \phi^2(\nabla\sigma)^2.
\end{equation}
From the kinetic term \eqref{eq:minimal_weyl_transform}, the extra contribution is
\begin{equation}
  \frac12\int d^d x\sqrt{g}\,
  \left[
    -2a\,\phi\,\nabla_\mu\phi\,\nabla^\mu\sigma
    +
    a^2\phi^2(\nabla\sigma)^2
  \right].
\end{equation}
From the curvature term, using \eqref{eq:curvature_term_weyl_transform} and
\eqref{eq:curvature_ibp}, the extra contribution is
\begin{equation}
  \frac12\xi
  \int d^d x\sqrt{g}\,
  \left[
    4(d-1)\phi\,\nabla_\mu\phi\,\nabla^\mu\sigma
    -
    (d-1)(d-2)\phi^2(\nabla\sigma)^2
  \right].
\end{equation}
Therefore Weyl invariance requires the coefficient of
$\phi\,\nabla_\mu\phi\,\nabla^\mu\sigma$ to vanish:
\begin{equation}
  -2a+4\xi(d-1)=0.
\end{equation}
Using $a=(d-2)/2$, this gives
\begin{equation}
  \xi
  =
  \frac{a}{2(d-1)}
  =
  \frac{d-2}{4(d-1)}.
\end{equation}
Hence
\begin{equation}
  \boxed{
  \xi_c=\frac{d-2}{4(d-1)}.
  }
  \label{eq:xi_conformal}
\end{equation}
The coefficient of $\phi^2(\nabla\sigma)^2$ then also vanishes, because
\begin{equation}
  a^2-\xi_c(d-1)(d-2)
  =
  \frac{(d-2)^2}{4}
  -
  \frac{d-2}{4(d-1)}(d-1)(d-2)
  =
  0.
\end{equation}
Thus the action
\begin{equation}
  \boxed{
  S_E
  =
  \frac12
  \int d^d x\,\sqrt{g}\,
  \left(
    g^{\mu\nu}\partial_\mu\phi\,\partial_\nu\phi
    +
    \frac{d-2}{4(d-1)}R\phi^2
  \right)
  }
  \label{eq:conformal_scalar_action}
\end{equation}
is locally Weyl invariant~\cite{arXiv:1308.2337}.

\section{The conformal mass on the cylinder}
\label{app:soc:conformal_mass}

On the cylinder
\begin{equation}
  \mathbb{R}\times S^{d-1}_{R},
\end{equation}
the scalar curvature is entirely due to the sphere:
\begin{equation}
  R_{\rm cyl}
  =
  \frac{(d-1)(d-2)}{R^2}.
\end{equation}
The curvature term in \eqref{eq:conformal_scalar_action} therefore gives
\begin{equation}
  \xi_c R_{\rm cyl}
  =
  \frac{d-2}{4(d-1)}
  \frac{(d-1)(d-2)}{R^2}
  =
  \frac{(d-2)^2}{4R^2}.
\end{equation}
Hence the scalar on the cylinder has an effective conformal mass
\begin{equation}
  \boxed{
  \mu^2
  =
  \frac{(d-2)^2}{4R^2},
  \qquad
  \mu
  =
  \frac{d-2}{2R}.
  }
  \label{eq:conformal_mass_soc}
\end{equation}
This mass is not an explicit breaking of conformal invariance. It is the geometric
effect required by Weyl invariance when the flat-space conformal scalar is mapped
to the curved cylinder~\cite{arXiv:1409.1937,arXiv:1111.6290}.

\chapter{The flat metric in radial and generalized spherical coordinates}
\label{app:metric_spherical}

This appendix derives the result used in Section~\ref{app:soc:weyl_map}: the flat
Euclidean metric of $\mathbb{R}^d$ written first in radial coordinates and then in
fully explicit generalized spherical coordinates~\cite{arXiv:1108.6194,arXiv:1305.1321}.

To derive the metric in radial coordinates, start from the Cartesian flat metric
\begin{equation}
  ds^2_{\mathbb{R}^d}
  =
  \delta_{\mu\nu}\,dx^\mu dx^\nu .
\end{equation}
Introduce polar coordinates by writing
\begin{equation}
  x^\mu = r\,\hat n^\mu,
  \qquad
  r=|x|,
  \qquad
  \hat n^\mu \hat n_\mu =1 .
\end{equation}
Differentiating $x^\mu=r\hat n^\mu$ gives
\begin{equation}
  dx^\mu
  =
  \hat n^\mu\,dr
  +
  r\,d\hat n^\mu .
\end{equation}
Substituting into the Cartesian line element,
\begin{align}
  ds^2_{\mathbb{R}^d}
  &=
  \delta_{\mu\nu}
  \left(
    \hat n^\mu dr + r\,d\hat n^\mu
  \right)
  \left(
    \hat n^\nu dr + r\,d\hat n^\nu
  \right)
  \nonumber\\
  &=
  \delta_{\mu\nu}\hat n^\mu\hat n^\nu\,dr^2
  +
  2r\,\delta_{\mu\nu}\hat n^\mu d\hat n^\nu\,dr
  +
  r^2\delta_{\mu\nu}d\hat n^\mu d\hat n^\nu .
\end{align}
Now use
\begin{equation}
  \delta_{\mu\nu}\hat n^\mu\hat n^\nu
  =
  \hat n^\mu\hat n_\mu
  =
  1.
\end{equation}
Moreover, differentiating the constraint $\hat n^\mu\hat n_\mu=1$ gives
\begin{equation}
  d(\hat n^\mu\hat n_\mu)=0,
\end{equation}
hence
\begin{equation}
  2\hat n_\mu d\hat n^\mu=0,
  \qquad
  \hat n_\mu d\hat n^\mu=0.
\end{equation}
Therefore the cross term vanishes:
\begin{equation}
  2r\,\delta_{\mu\nu}\hat n^\mu d\hat n^\nu\,dr=0.
\end{equation}
Thus
\begin{equation}
  ds^2_{\mathbb{R}^d}
  =
  dr^2
  +
  r^2\,\delta_{\mu\nu}d\hat n^\mu d\hat n^\nu .
\end{equation}
The final term is precisely the standard metric on the unit sphere $S^{d-1}$:
\begin{equation}
  d\Omega_{d-1}^2
  \equiv
  \delta_{\mu\nu}d\hat n^\mu d\hat n^\nu,
  \qquad
  \hat n^\mu\hat n_\mu=1.
\end{equation}
Therefore the flat Euclidean metric in radial coordinates is
\begin{equation}
  \boxed{
  ds^2_{\mathbb{R}^d}
  =
  dr^2+r^2d\Omega_{d-1}^2 .
  }
\end{equation} 
To make the angular metric $d\Omega_{d-1}^2$ explicit, introduce generalized
spherical coordinates on $\mathbb{R}^d$.  Write
\begin{align}
  x^1 &= r\cos\theta_1, \nonumber\\
  x^2 &= r\sin\theta_1\cos\theta_2, \nonumber\\
  x^3 &= r\sin\theta_1\sin\theta_2\cos\theta_3, \nonumber\\
  &\hspace{0.5cm}\vdots \nonumber\\
  x^{d-1} &= r\sin\theta_1\sin\theta_2\cdots \sin\theta_{d-2}\cos\theta_{d-1}, \nonumber\\
  x^d &= r\sin\theta_1\sin\theta_2\cdots \sin\theta_{d-2}\sin\theta_{d-1}.
  \label{eq:general_spherical_coordinates}
\end{align}
The angular ranges are
\begin{equation}
  0\leq \theta_i\leq \pi,
  \qquad
  i=1,\ldots,d-2,
  \qquad
  0\leq \theta_{d-1}<2\pi.
\end{equation}

Equivalently,
\begin{equation}
  x^\mu = r\,\hat n^\mu(\theta_1,\ldots,\theta_{d-1}),
  \qquad
  \hat n^\mu \hat n_\mu =1.
\end{equation}
The flat metric is
\begin{equation}
  ds_{\mathbb{R}^d}^2
  =
  dr^2+r^2 d\Omega_{d-1}^2,
\end{equation}
where
\begin{equation}
  d\Omega_{d-1}^2
  =
  \delta_{\mu\nu}\,d\hat n^\mu d\hat n^\nu .
\end{equation}

We now derive the explicit angular form recursively.  Separate the first
coordinate from the remaining $d-1$ coordinates:
\begin{equation}
  x^1=r\cos\theta_1,
  \qquad
  (x^2,\ldots,x^d)=r\sin\theta_1\,\hat m,
\end{equation}
where $\hat m\in S^{d-2}$ satisfies
\begin{equation}
  \hat m^a\hat m_a=1,
  \qquad
  a=1,\ldots,d-1.
\end{equation}
Thus
\begin{equation}
  \hat n
  =
  \left(
    \cos\theta_1,
    \sin\theta_1\,\hat m
  \right).
\end{equation}
Differentiating,
\begin{equation}
  d\hat n
  =
  \left(
    -\sin\theta_1\,d\theta_1,
    \cos\theta_1\,d\theta_1\,\hat m
    +
    \sin\theta_1\,d\hat m
  \right).
\end{equation}
Therefore
\begin{align}
  d\Omega_{d-1}^2
  &=
  d\hat n\cdot d\hat n
  \nonumber\\
  &=
  \sin^2\theta_1\,d\theta_1^2
  +
  \left(
    \cos\theta_1\,d\theta_1\,\hat m
    +
    \sin\theta_1\,d\hat m
  \right)
  \cdot
  \left(
    \cos\theta_1\,d\theta_1\,\hat m
    +
    \sin\theta_1\,d\hat m
  \right)
  \nonumber\\
  &=
  \sin^2\theta_1\,d\theta_1^2
  +
  \cos^2\theta_1\,d\theta_1^2\,(\hat m\cdot\hat m)
  +
  2\sin\theta_1\cos\theta_1\,d\theta_1\,(\hat m\cdot d\hat m)
  +
  \sin^2\theta_1\,d\hat m\cdot d\hat m .
\end{align}
Using
\begin{equation}
  \hat m\cdot\hat m=1,
\end{equation}
and differentiating the constraint $\hat m\cdot\hat m=1$,
\begin{equation}
  d(\hat m\cdot\hat m)=2\hat m\cdot d\hat m=0,
\end{equation}
we obtain
\begin{equation}
  \hat m\cdot d\hat m=0.
\end{equation}
Hence the cross term vanishes and
\begin{align}
  d\Omega_{d-1}^2
  &=
  \left(\sin^2\theta_1+\cos^2\theta_1\right)d\theta_1^2
  +
  \sin^2\theta_1\,d\hat m\cdot d\hat m
  \nonumber\\
  &=
  d\theta_1^2
  +
  \sin^2\theta_1\,d\Omega_{d-2}^2 .
  \label{eq:sphere_metric_recursion}
\end{align}
Thus the unit-sphere metric satisfies the recursive relation
\begin{equation}
  \boxed{
  d\Omega_{d-1}^2
  =
  d\theta_1^2
  +
  \sin^2\theta_1\,d\Omega_{d-2}^2 .
  }
\end{equation}

Iterating this recursion gives the explicit generalized solid-angle metric:
\begin{equation}
  \boxed{
  d\Omega_{d-1}^2
  =
  d\theta_1^2
  +
  \sin^2\theta_1\,d\theta_2^2
  +
  \sin^2\theta_1\sin^2\theta_2\,d\theta_3^2
  +
  \cdots
  +
  \left(
    \prod_{j=1}^{d-2}\sin^2\theta_j
  \right)
  d\theta_{d-1}^2 .
  }
  \label{eq:general_solid_angle_metric}
\end{equation}
Equivalently,
\begin{equation}
  \boxed{
  d\Omega_{d-1}^2
  =
  \sum_{k=1}^{d-1}
  \left(
    \prod_{j=1}^{k-1}\sin^2\theta_j
  \right)
  d\theta_k^2,
  }
  \label{eq:general_solid_angle_metric_compact}
\end{equation}
where the empty product for $k=1$ is defined to be $1$~\cite{arXiv:1011.1485,arXiv:1809.05111}.

The corresponding volume element on the unit sphere is obtained from the square
root of the determinant of the angular metric:
\begin{equation}
  d\Omega_{d-1}
  =
  \sqrt{\det g_{S^{d-1}}}\,
  d\theta_1\cdots d\theta_{d-1}.
\end{equation}
Since the metric is diagonal, one finds
\begin{equation}
  \boxed{
  d\Omega_{d-1}
  =
  \sin^{d-2}\theta_1\,
  \sin^{d-3}\theta_2\,
  \cdots\,
  \sin\theta_{d-2}\,
  d\theta_1\,d\theta_2\cdots d\theta_{d-1}.
  }
  \label{eq:general_solid_angle_volume_element}
\end{equation}
Thus the flat Euclidean metric in generalized spherical coordinates is
\begin{equation}
  \boxed{
  ds_{\mathbb{R}^d}^2
  =
  dr^2
  +
  r^2
  \left[
  \sum_{k=1}^{d-1}
  \left(
    \prod_{j=1}^{k-1}\sin^2\theta_j
  \right)
  d\theta_k^2
  \right].
  }
  \label{eq:flat_metric_general_spherical}
\end{equation}

\chapter{Classical action of the interacting saddle}
\label{app:Scl_derivation}

This appendix derives the bare classical action \eqref{eq:Scl_bare} used in
Section~\ref{sec:renorm}. We work in the units of Section~\ref{sec:phi4}
($R=1$); the dimensionful $R^{-\epsilon}$ is restored at the end.

\paragraph{Action of the homogeneous saddle.}
The saddle $v(t)=x_0\,\cn(\omega t\,|\,m)$ is spatially homogeneous, so the
integral over $S^{d-1}$ contributes the sphere volume $\Omega_{d-1}R^{d-1}$ with
$\Omega_{d-1}=2\pi^{d/2}/\Gamma(d/2)$, and
\begin{equation}
\mathcal{S}_{\rm cl}
=\Omega_{d-1}R^{d-1}\int_0^{\mathcal T}\!dt\left[\tfrac12\dot v^2-\tfrac{\mu^2}{2}v^2-\tfrac{\lambda}{4}v^4\right],
\label{eq:app_Scl_setup}
\end{equation}
with $\mu=(d-2)/2$ and, from \eqref{eq:params_cn},
$\omega^2=\mu^2/(1-2m)$, $x_0^2=2m\mu^2/[\lambda(1-2m)]$, and period
$\mathcal T=4\KK(m)/\omega$.

\paragraph{Period integrals.}
Setting $u=\omega t$ and using $\dot v=-x_0\omega\,\sn\,\dn$, one needs the
averages over $u\in[0,4\KK]$:
\begin{equation}
\int_0^{4\KK}\!\cn^2\,du = \frac{4}{m}\big[\ELE-(1-m)\KK\big],
\qquad
\int_0^{4\KK}\!\sn^2\dn^2\,du = \frac{4}{3m}\big[(1-m)\KK-(1-2m)\ELE\big],
\label{eq:app_period_integrals}
\end{equation}
together with the quartic average, obtained by reducing
$\cn^4=1-2\sn^2+\sn^4$ with the standard $\int\sn^2$, $\int\sn^4$:
\begin{equation}
\int_0^{4\KK}\!\cn^4\,du
=\frac{4}{3m^2}\big[(3m-2)(m-1)\KK+(4m-2)\ELE\big]
=\frac{4(1-2m)^{3/2}}{m^2}\,s(m).
\label{eq:app_cn4_s}
\end{equation}
The bracket in \eqref{eq:app_cn4_s} is exactly the numerator of $s(m)$: the
quartic average is the sole origin of $s(m)$.

\paragraph{Collapse of the bracket.}
Write $\mathcal{S}_{\rm cl}=(\Omega_{d-1}R^{d-1}/\omega)\,B$ with
\[
B=\tfrac12 x_0^2\omega^2\!\int_0^{4\KK}\!\!\sn^2\dn^2\,du
-\tfrac{\mu^2}{2}x_0^2\!\int_0^{4\KK}\!\!\cn^2\,du
-\tfrac{\lambda}{4}x_0^4\!\int_0^{4\KK}\!\!\cn^4\,du .
\]
Inserting $\omega^2=\mu^2/(1-2m)$ and $x_0^2=2m\mu^2/[\lambda(1-2m)]$, the
$\KK$- and $\ELE$-dependent pieces from the three integrals combine and reduce
to a single term,
\begin{equation}
\frac{B}{\omega}=\frac{4\mu^3}{\lambda}\,s(m),
\qquad\Longrightarrow\qquad
\mathcal{S}_{\rm cl}=\Omega_{d-1}R^{d-1}\,\frac{4\mu^3}{\lambda}\,s(m).
\label{eq:app_Bcollapse}
\end{equation}

\paragraph{Prefactor and result.}
For $d=4-\epsilon$ one has $d/2=2-\epsilon/2$ and $\mu=(d-2)/2=(2-\epsilon)/2$,
hence $4\mu^3=(2-\epsilon)^3/2=-(\epsilon-2)^3/2$ and
\[
\Omega_{d-1}\cdot 4\mu^3
=\frac{2\pi^{2-\epsilon/2}}{\Gamma(2-\epsilon/2)}\left(-\frac{(\epsilon-2)^3}{2}\right)
=-\frac{\pi^{2-\epsilon/2}(\epsilon-2)^3}{\Gamma(2-\epsilon/2)}.
\]
Restoring the engineering dimension of the bare coupling, $[\lambda_0]=\epsilon$
(which supplies the net factor $R^{-\epsilon}$ when $R$ is reinstated), gives
\eqref{eq:Scl_bare}:
\begin{equation}
\mathcal{S}_{\rm cl}(\lambda_0)
=-\frac{\pi^{2-\epsilon/2}(\epsilon-2)^3\,R^{-\epsilon}\,s(m)}{\lambda_0\,\Gamma(2-\epsilon/2)}.
\label{eq:app_Scl_result}
\end{equation}
As a check, at $d=4$ ($\epsilon\to0$) this reduces to
$\mathcal{S}_{\rm cl}=8\pi^2 s(m)/\lambda$, which agrees with the direct numerical
evaluation of $\Omega_3\int_0^{\mathcal T}(\tfrac12\dot v^2-V_{\rm cyl})\,dt$.

\bibliographystyle{JHEP}
\bibliography{arxiv-final}
\end{document}